\documentclass[sigconf,screen]{acmart}

\setcopyright{none}
\renewcommand\footnotetextcopyrightpermission[1]{}
\acmConference[STOC '27]{Proceedings of the 59th ACM Symposium on Theory of Computing}{June 6--11, 2027}{Atlanta, GA, USA}

\AtEndPreamble{%
  \theoremstyle{acmplain}
  \newtheorem{fact}[theorem]{Fact}
  \theoremstyle{acmdefinition}
  \newtheorem{remark}[theorem]{Remark}}

\newcommand{\F}{\mathbb{F}}
\newcommand{\R}{\mathbb{R}}
\newcommand{\C}{\mathbb{C}}
\newcommand{\Rg}{\mathcal{R}}
\newcommand{\cir}{\Phi}
\newcommand{\cirp}{\Phi'}
\newcommand{\spn}{\operatorname{span}}
\newcommand{\K}{\mathbb{K}}
\newcommand{\Q}{\mathbb{Q}}
\newcommand{\Pj}{\mathbb{P}}
\newcommand{\cA}{\mathcal{A}}
\newcommand{\cB}{\mathcal{B}}
\newcommand{\Sy}{\mathrm{S}}
\newcommand{\Row}{\operatorname{Row}}
\newcommand{\Col}{\operatorname{Col}}
\newcommand{\tr}{\operatorname{tr}}
\newcommand{\rank}{\operatorname{rank}}
\newcommand{\rk}{\operatorname{rank}}

\begin{document}

\title{Partial Homogenization and the Structure of Optimal Algebraic Algorithms}

\author{Yuval Spiizer}
\affiliation{%
  \institution{Tel Aviv University}
  \city{Tel Aviv}
  \country{Israel}}
\email{yuval.lahav@mail.huji.ac.il}

\begin{abstract}
A quadratic map, such as the product of two polynomials, two complex numbers or two matrices, is classically computed by products of two linear forms. Its classical cost is the least number of such products. By Strassen's homogenization, this is also the least number of product gates of fan-in two in an arithmetic circuit. We study circuits whose product gates have unbounded fan-in. Such a gate multiplies any number of inputs at the cost of one gate. The terms of higher degree of several gates can cancel and leave a quadratic form that needs many binary products. So the classical lower bounds and the classical classifications of optimal algorithms do not apply. We ask two questions. Does unbounded fan-in beat fan-in two? And what are the optimal circuits?

Our tool is constant-free partial homogenization. Unlike Strassen's homogenization, it keeps the terms of higher degree. Let a quadratic map have $K$ linearly independent outputs, and let fan-in two need at most $K+2$ products. We prove that, under a mild condition on the outputs, unbounded fan-in then needs as many product gates as fan-in two.

For $2\times2$ matrix multiplication, fan-in two needs $7$ product gates, by Winograd. We classify all optimal circuits with product gates of fan-in two, with arbitrary constants. We prove that none of them nests products. We prove that their quadratic parts are Strassen's algorithm, one new algorithm, the Waksman-type algorithms, or one explicit one-parameter family. Over $\R$, we prove that unbounded fan-in needs at least $6$ product gates, and that approximate circuits need exactly $5$.

Below the threshold $K+2$, we prove that the optimal circuits are rigid, for every fan-in and with arbitrary constants. For polynomial multiplication in any number of variables, we prove that every optimal circuit is a Toom-Cook algorithm at unisolvent points, with evaluations scaled and shifted by constants. This refines a theorem of Winograd, which fails once constants are allowed. For multiplication in a tubal ring over $\C$, that is, modulo a squarefree polynomial, we prove that every optimal circuit applies the transform of the ring. For complex multiplication over $\R$, we prove that every optimal circuit is Gauss's algorithm at three real points, after a rotation of one input. The method also covers extracting the homogeneous components of a black-box polynomial, where a query acts as a gate of unbounded degree. In at least two variables, we prove that every optimal division-free extractor is Ben-Or's evaluation-interpolation construction, with the points shifted by a common polynomial.

Above the threshold the fan-in matters. For $c\ge2$ complex multiplications over $\R$, we prove that fan-in two needs exactly $3c$ product gates, with arbitrary constants. We prove that unbounded fan-in needs at least $2c+2$, which is exactly $6$ for $c=2$. We prove that approximate circuits with unbounded fan-in need exactly $2c+1$.
\end{abstract}

\ccsdesc[500]{Theory of computation~Algebraic complexity theory}

\keywords{arithmetic circuits, partial homogenization, Toom-Cook, tubal rings, homogeneous components, fault tolerance}

\maketitle

\section{Introduction}\label{sec:intro}

Algebraic complexity theory studies how many arithmetic operations are needed to compute polynomials. A standard model is the arithmetic circuit, which combines the inputs by additions, multiplications by constants and products \cite{burgisser1997algebraic,shpilka2010arithmetic}. For the basic problems of computer algebra and numerical linear algebra, the cost that matters is the number of multiplications of quantities that depend on the inputs. Additions and multiplications by fixed constants are cheap. In recursive algorithms the inputs are themselves long integers, polynomials or matrices, so the number of multiplications at one level becomes the exponent of the running time.

Two problems have driven the field. The first is multiplying long integers. Every algorithm for polynomial multiplication gives one for integers. One writes each integer as a polynomial in a power of two, multiplies the polynomials, and propagates the carries \cite{knuth1997art,brent2010modern}. Karatsuba's algorithm splits each factor into two parts and uses three products instead of four \cite{karatsuba1962multiplication}. Toom-Cook splits it into $k$ parts and uses $2k-1$ products \cite{toom1963complexity,cook1966minimum}. Its evaluation and interpolation were optimized by Bodrato and Zanoni \cite{bodrato2007integer}, and GMP uses Toom-Cook algorithms of several sizes \cite{gmp}. The fastest algorithms use the same evaluation-interpolation mechanism, at roots of unity and with a fast Fourier transform. This is the scheme of Sch\"onhage and Strassen \cite{schonhage1971schnelle}, of F\"urer \cite{furer2009faster}, of Harvey, van der Hoeven and Lecerf \cite{harvey2016even}, and of the $O(n\log n)$ algorithm of Harvey and van der Hoeven \cite{harvey2021integer}. The second is multiplying matrices, the core of numerical linear algebra: blocked algorithms for factorizations and linear systems spend most of their time in matrix products. Strassen multiplied $2\times2$ matrices with $7$ products instead of $8$ \cite{strassen1969gaussian}. Applied recursively to blocks, this gives time $O(n^{\log_2 7})$. The exponent $\omega$ of matrix multiplication is now known to be below $2.3712$ \cite{alman2025more,dupont2026improving}. Strassen's algorithm is competitive in practice already for moderate sizes \cite{huang2016strassen}. Every optimal bilinear algorithm for $2\times2$ matrices is equivalent to Strassen's \cite{degroote1978optimal}. For larger formats, computer searches have found many inequivalent ones, such as more than $14{,}000$ for $4\times4$ matrices with $49$ products \cite{fawzi2022discovering}.

An algorithm for a bilinear map is \emph{noncommutative}, or \emph{bilinear}, if each product multiplies a linear form in the first input by a linear form in the second. Strassen's algorithm is of this kind. Only such algorithms can be applied recursively to blocks, because the blocks do not commute. A \emph{commutative} algorithm may multiply arbitrary linear forms and may use that the entries commute. Winograd's inner-product algorithm computes $\sum_jx_jy_j$ from the products $(x_{2j-1}+y_{2j})(x_{2j}+y_{2j-1})$. Their terms $x_{2j-1}x_{2j}$ and $y_{2j-1}y_{2j}$ depend on one input only and are computed once. This multiplies $n\times n$ matrices with about $n^3/2$ multiplications \cite{winograd1968new}. For $2\times2$ matrices commutativity does not help: $7$ products are still needed \cite{winograd1971multiplication}. Arbitrary circuits have more freedom than bilinear algorithms. Constants can shift the inputs of the products, products can be nested inside other products, and product gates can have unbounded fan-in. This paper asks how much of this freedom optimal algorithms can use.

\subsection{Motivation}\label{sec:motivation}

An algorithm that uses the minimal number of multiplications is optimal in the cleanest sense algebraic complexity offers. We study such algorithms for polynomials of degree $2$, that is, quadratic maps. Examples are the products of two polynomials, of two complex numbers and of two matrices. Two questions run through this paper.

The first is what to count. The classical lower bounds of Hopcroft and Kerr \cite{hopcroft1971minimizing}, of Winograd \cite{winograd1971multiplication,winograd1977some}, and of Alder and Strassen \cite{alder1981algorithmic} count products of two linear forms. By Strassen's homogenization \cite{strassen1973vermeidung}, this is the least number of product gates in a circuit whose product gates have fan-in two. Indeed, such a circuit can be converted into a homogeneous quadratic algorithm with the same number of products. A product gate of unbounded fan-in multiplies any number of inputs at the cost of one gate, and no such conversion exists. Cancellations among such gates produce quadratic forms that need many binary products, as in $(1+a)(1+b)(1+c)+(1-a)(1-b)(1-c)=2+2e_2(a,b,c)$. In the border model, the circuit may depend on a parameter $\delta$, and the outputs are reached as $\delta\to0$. There, depth-$3$ circuits with only two product gates of unbounded fan-in approximate every homogeneous polynomial over $\C$ \cite{kumar2020power}. So we ask whether, for quadratic maps, unbounded fan-in beats fan-in two, exactly or approximately.

The second is what the optimal algorithms are. The classical classifications of minimal algorithms are due to Winograd, de Groote, Feig, and Averbuch, Galil and Winograd (Section~\ref{sec:previous}). They describe bilinear or homogenized algorithms. So they concern fan-in two only, and they ignore constants. Optimal algorithms with the same number of multiplications differ in their additions, in the constants they divide by, and in numerical stability. A classification turns the search for the best optimal algorithm into a search over a small, explicit family. It is also a much stronger statement than a lower bound, and often explains why the lower bound holds. The two questions meet. A classification of the circuits with unbounded fan-in also shows that the fan-in does not help. Where the fan-in does matter, a classification can only be stated for fan-in two. We study the following quadratic maps.

\paragraph{Polynomial multiplication.} The Toom-Cook algorithm multiplies two polynomials with $k_1$ and $k_2$ coefficients. It evaluates both at $k_1+k_2-1$ points, multiplies the values, and interpolates. Applied recursively, $l$ levels of Toom-Cook multiply polynomials in $l$ variables, and give integer multiplication in time $O(n^{\log_k(2k-1)})$. The number $k_1+k_2-1$ is optimal \cite{fiduccia1975algebras}. Beyond the multiplications, Toom-Cook pays for the evaluation and, mostly, for the interpolation, whose matrix depends on the points. For $k\ge3$ and integer points it divides by integers that are not powers of two. Such divisions are expensive and can be numerically unstable \cite{nissim2026division}. One may hope that another optimal algorithm, not of evaluation-interpolation type, avoids these costs. So the question is whether every algorithm with $k_1+k_2-1$ multiplications is a Toom-Cook algorithm, and if not, what the other ones are.

\paragraph{Tubal rings and multiplication modulo a polynomial.} Kilmer and Martin introduced the t-product of third-order tensors. Its entries are tubes in $\R^d$, multiplied by cyclic convolution. It has applications in imaging, compression and tensor completion \cite{kilmer2011factorization,kilmer2013third,lu2020tensor}. To compute it, one transforms every tube by the discrete Fourier transform, multiplies facewise, and transforms back. Kernfeld, Kilmer and Aeron replaced the Fourier transform by any invertible transform $M$, which gives the $\star_M$ products \cite{kernfeld2015tensor}. Avron and Mor characterized the resulting tube algebras as the tubal rings \cite{avron2025demystifying}. Many transforms are in use: Fourier, cosine \cite{kernfeld2015tensor,xu2019fast,lu2019low}, sine \cite{reichel2021tensor}, Hartley and Walsh-Hadamard \cite{rtensor2}, wavelet and framelet \cite{jiang2020framelet}, graph-based \cite{deng2021graph,liu2021convolutional,malik2021dynamic} and learned transforms \cite{kong2021tensor,newman2024optimal,song2019robust,newman2018stable}. The transform is applied once per tube, while the products are repeated, so the products dominate. Yet all efforts to reduce the cost have changed the transform \cite{xu2019fast,reichel2021tensor,keegan2024projected}. The cheapest algorithms for a given tube product have not been studied. The same products are multiplication modulo a squarefree polynomial (Proposition~\ref{prop:squarefree}). This includes cyclic convolution and Winograd's minimal filtering algorithms \cite{winograd1980arithmetic}, used for convolutional neural networks \cite{lavin2016fast}.

\paragraph{Complex multiplication over the reals.} Gauss observed that $(a+bi)(c+di)$ can be computed with three real multiplications, $ac$, $bd$ and $(a+b)(c+d)$, instead of four. Applied to matrices, this is the 3M method. It multiplies complex matrices with three real matrix multiplications instead of four, and saves a quarter of the work. Its numerical stability depends on which three-multiplication formula is used \cite{higham1992stability}. Three is the minimum \cite{winograd1977some}. But Gauss's formula is not the only one, and choosing among them needs a description of all of them.

\paragraph{Several complex products.} The product of $\C^c$ over $\R$ is the core of the real t-product with the Fourier transform. By the Alder-Strassen bound it needs $3c$ binary products \cite{alder1981algorithmic}, while its $2c$ outputs only force $2c$ product gates. It is the simplest problem on which unbounded fan-in might beat the binary count.

\paragraph{Homogeneous component extractors.} Splitting a polynomial into its homogeneous components $f^{[0]},\dots,f^{[d]}$ is a basic step in algebraic complexity. Strassen's homogenization performs it inside a circuit \cite{strassen1973vermeidung}. Many lower bounds, identity tests and reconstruction algorithms start from it \cite{shpilka2010arithmetic}. When the circuit is only available as a black box, the standard method is due to Ben-Or \cite{nisan1996lower,shpilka2001depth,bhattacharjee2026gcd}. It queries $f(\beta_ix)$ for $d+1$ distinct constants $\beta_i$, which gives $\sum_k\beta_i^kf^{[k]}$, and interpolates. Ben-Or used the same interpolation to compute all elementary symmetric polynomials by depth-$3$ circuits of quadratic size \cite{nisan1996lower,shpilka2001depth}. Ben-Or's construction shows that the depth-$3$ lower bounds of Shpilka and Wigderson are nearly tight \cite{shpilka2001depth,fournier2017shifted}. Interpolation of this kind is also a step in the depth-$3$ lower bound of Kayal, Saha and Tavenas \cite{kayal2016almost}, in constant-depth circuits for linear algebra \cite{andrews2024constant}, and in constant-depth circuits for the GCD, where it extracts homogeneous parts and partial derivatives \cite{bhattacharjee2026gcd}. It is also the operation behind our technique. Partial homogenization splits the gates of a circuit into components, and an extractor must do the same with black-box access only. Query gates play the role of product gates of unbounded fan-in, since a single query evaluates a polynomial of arbitrary degree. So the question is again whether gates of high degree beat the obvious count. Do fewer queries suffice? Do adaptive queries or divisions help? Are there other constructions?

\paragraph{$2\times2$ matrix multiplication.} Strassen's $7$ products are optimal for bilinear algorithms \cite{hopcroft1971minimizing} and for commutative ones \cite{winograd1971multiplication}. Every optimal bilinear algorithm is Strassen's up to symmetry \cite{degroote1978optimal}. Commutative algorithms can also use Winograd's inner-product trick, refined by Waksman \cite{waksman1970winograd}. It gives products such as $(a_{11}+b_{21})(a_{12}+b_{11})$, whose terms $a_{11}a_{12}$ and $b_{21}b_{11}$ must cancel. Three questions have not been settled. Which commutative algorithms use only $7$ products? Can an optimal circuit nest products? Can product gates of unbounded fan-in do with fewer than $7$? We answer the first two and bound the third (Theorem~\ref{mthm:matmul}).

\subsection{Previous work}\label{sec:previous}

Winograd introduced the presentation of minimal algorithms by products of linear forms \cite{winograd1970number}. In this homogeneous presentation he classified the optimal algorithms for polynomial multiplication and for multiplication modulo a polynomial \cite{winograd1977some,winograd1979multiplication,winograd1980arithmetic}. We show that his statement for polynomial multiplication fails once constants are allowed (Example~\ref{ex:winograd}). Alder and Strassen proved that an associative algebra $A$ needs at least $2\dim A-t(A)$ multiplications, where $t(A)$ is the number of maximal two-sided ideals \cite{alder1981algorithmic}. For $\R^r\times\C^c$ this is $r+3c$, so for fan-in two our contribution there is the classification. Winograd classified the minimal bilinear algorithms for multiplication in algebraic extension fields \cite{winograd1977some,winograd1979multiplication}. Averbuch, Galil and Winograd did the same modulo powers of irreducible polynomials. They also observed that modulo $u^3$ there are minimal quadratic algorithms that are not bilinear \cite{averbuch1986classification,averbuch1988classification,averbuch1991classification}. Feig showed that for definite systems of complexity $a+b-1$, such as these extension fields, all minimal division-free algorithms are essentially bilinear \cite{feig1981systems,feig1983certain}. Auslander, Feig and Winograd studied direct sums \cite{auslander1981direct}. De Groote studied the varieties of optimal bilinear algorithms \cite{degroote1978varieties,degroote1983characterization,degroote1987lectures}. All of these concern bilinear or homogeneous quadratic algorithms. We use them only through Fact~\ref{fact:quadratic}.

Lower bounds for explicit quadratic maps mostly count products of two linear forms, that is, product gates of fan-in two. For $n\times n$ matrix product, Shpilka proved that bilinear and quadratic circuits over $\F_2$ need $3n^2-o(n^2)$ product gates. This improves the bound $2.5n^2-o(n^2)$ of Bshouty \cite{shpilka2003lower,bshouty1989lower}. Other models restrict the circuit differently. Raz proved an $\Omega(n^2\log n)$ lower bound for circuits over $\R$ and $\C$ whose constants are bounded in absolute value \cite{raz2003complexity}. Raz and Shpilka proved superlinear lower bounds on the number of edges of bounded-depth circuits with arbitrary gates \cite{raz2003lower}. Product gates of unbounded fan-in have been studied mainly in depth $3$. Nisan and Wigderson proved lower bounds for homogeneous depth-$3$ circuits by partial derivatives \cite{nisan1996lower}. Shpilka and Wigderson proved near-quadratic lower bounds for depth-$3$ circuits over fields of characteristic $0$ \cite{shpilka2001depth}. These were the first nontrivial bounds for polynomials of constant degree, and Shpilka extended them \cite{shpilka2001affine}. For the quadratic form $S_n^2=\sum_{i<j}x_ix_j$, Radhakrishnan, Sen and Vishwanathan found the exact number of product gates of depth-$3$ circuits, also inhomogeneous ones \cite{radhakrishnan2000depth}. It is $n-1$ over $\R$ and $\lceil n/2\rceil$ over $\C$. Our Theorem~\ref{thm:single_form} gives these counts for circuits of every depth. For every quadratic form, the count over $\R$ is the larger of its numbers of positive and negative eigenvalues, and over $\C$ it is half its rank, rounded up. For quadratic maps with several outputs, such as the products studied here, we know of no earlier comparison between circuits with product gates of unbounded fan-in and arbitrary depth and the count for fan-in two.

Approximate bilinear algorithms can need fewer products than exact ones. Partial $2\times2$ matrix multiplication has rank $6$ and border rank $5$ \cite{bini1979approximate}, while $2\times2$ matrix multiplication has border rank $7$ \cite{landsberg2006border}. Strassen removed divisions from circuits by a shift of the variables, an expansion in power series and a truncation \cite{strassen1973vermeidung}. We use his shift, and we do not truncate. Nissim, Schwartz and Spiizer bounded the number of divisions in Toom-Cook interpolation \cite{nissim2026division}. We know of no earlier results on the optimal circuits for tubal rings.

\subsection{Our results}\label{sec:results}

Our results lie on two sides of a threshold. Let $\phi$ be a quadratic map with $K$ linearly independent coordinates. Let $C_2(\phi)$ be the least number of product gates in a division-free circuit with product gates of fan-in two. This is the least number of products of linear forms in a quadratic algorithm for $\phi$ \cite{strassen1973vermeidung}. Let $C_\infty(\phi)$ be the least number of product gates when the fan-in is unbounded. So $K\le C_\infty(\phi)\le C_2(\phi)$. We show that
\begin{displaymath}
    C_\infty(\phi)\ge\min\{C_2(\phi),K+2\}
\end{displaymath}
whenever the span of the coordinates contains no product of two linear forms and contains a form of rank at least $4$ (Corollary~\ref{cor:spare}). Over $\R$, the same holds for circuits of depth $3$ without any condition (Corollary~\ref{cor:depth3}). So unbounded fan-in can only help above the threshold $C_2(\phi)\ge K+3$. Below it, we go further and classify all optimal circuits, with unbounded fan-in, constants and, in most cases, divisions (Theorems~\ref{mthm:poly} to~\ref{mthm:extractor}). Above it, we determine how much unbounded fan-in and approximation save for $\C^c$ and for $2\times2$ matrices. There we classify the optimal circuits with binary products (Theorems~\ref{mthm:fanin} and~\ref{mthm:matmul}).

The circuits are arbitrary arithmetic circuits. Product gates may be nested, and constants are allowed. Unless stated otherwise, product gates may have unbounded fan-in, and division gates are allowed. The \emph{cost} is the number of product and division gates. A circuit with inputs $x,x'$ is \emph{bi-affine} if it has no division gates and every product gate computes $(a+\ell(x))(b+m(x'))$, with constants $a,b$ and linear forms $\ell$ in $x$ and $m$ in $x'$. For a polynomial $q$, $q^{[\le1]}$ is its part of degree at most one.

\paragraph{Below the threshold: rigid optimal circuits.} For polynomial multiplication and for tubal rings over $\C$, the binary count equals the number $K$ of outputs. For complex multiplication it is $K+1$. In these cases we classify the optimal circuits of every fan-in.

\paragraph{Polynomial multiplication.} Let $p_1,p_2$ be polynomials in $l$ variables, with $k_{1,t}$ and $k_{2,t}$ coefficients in the variable $\tau_t$. For $\beta\in(\F\cup\{\infty\})^l$, let $p_r(\beta)$ be the linear form in the coefficients of $p_r$ obtained by substituting $\tau_t=\beta_t$. Here $\beta_t=\infty$ means taking the leading coefficient in $\tau_t$. Let $\kappa_t=k_{1,t}+k_{2,t}-1$, and let $K=\prod_t\kappa_t$ be the number of coefficients of $p_1p_2$. Points $\beta_1,\dots,\beta_K$ are \emph{unisolvent} if a polynomial with at most $\kappa_t$ coefficients in each $\tau_t$ is determined by its values at them. Equivalently, the matrix $V$ of these evaluations is invertible. For $l=1$ it means distinct.

\begin{theorem}[Polynomial multiplication]\label{mthm:poly}
Let $\F$ be a field and $k_{1,t},k_{2,t}\ge2$ for all $t$. Every circuit over $\F$ that computes the coefficients of $p_1p_2$ has cost at least $K$. If its cost is $K$, it has no division gates and there are unisolvent points $\beta_1,\dots,\beta_K$, nonzero $\alpha_i,\alpha'_i\in\F$ and arbitrary $\gamma_i,\gamma'_i\in\F$ such that, up to the order of the factors, the $i$-th product gate computes
\begin{displaymath}
    q_i=\bigl(\alpha_i\,p_1(\beta_i)+\gamma_i\bigr)\bigl(\alpha'_i\,p_2(\beta_i)+\gamma'_i\bigr),
\end{displaymath}
and the outputs are $(DV)^{-1}(q-q^{[\le1]})$ with $D=\operatorname{diag}(\alpha_i\alpha'_i)$. (Proved in Theorems~\ref{thm:bi-affine} and~\ref{thm:multivariate}.)
\end{theorem}

So every optimal circuit is a Toom-Cook algorithm whose evaluations are scaled and shifted by constants. Every unisolvent set of points gives one, not only a grid. Winograd's classification up to scaling \cite[Theorem~3]{winograd1977some} fails once constants are allowed (Example~\ref{ex:winograd}). Over $\F_q$, an optimal circuit exists if and only if $\kappa_t\le q+1$ for all $t$ (Corollary~\ref{cor:finite}).

\paragraph{Tubal rings.} By Proposition~\ref{prop:squarefree}, tubal rings are exactly the algebras $\R[\tau]/(P)$ for squarefree $P$. Over $\C$, the transform of $\C[\tau]/(P)$ is evaluation at the roots of $P$.

\begin{theorem}[Tubal rings]\label{mthm:tubal}
Let $M\in E^{d\times d}$ be invertible, where $E=\C$, or $E=\R$ and $M$ is real.
\begin{enumerate}
    \item[(a)] Every circuit over $E$ computing $a\star_Mb$ has cost at least $d$. If its cost is $d$, there are a permutation $\pi$, nonzero $\alpha_i,\alpha'_i$ and arbitrary $\gamma_i,\gamma'_i$ such that the $i$-th product gate computes
    \begin{displaymath}
        \bigl(\alpha_i(Ma)_{\pi(i)}+\gamma_i\bigr)\bigl(\alpha'_i(Mb)_{\pi(i)}+\gamma'_i\bigr);
    \end{displaymath}
    moreover $\star_N=\star_M$ only for $N$ equal to $M$ with permuted rows.
    \item[(b)] If $E=\C$ and the tubal ring is $\R^r\times\C^c$ with $c\ge1$, every circuit over $\R$ with product gates of fan-in two computing $a\star_Mb$ for real $a,b$ has at least $r+3c$ product gates. If it has exactly $r+3c$, it is bi-affine and splits along the factors: one product for each real factor, and three products for each complex factor, given by Theorem~\ref{mthm:cmul}.
\end{enumerate}
(Proved in Theorems~\ref{thm:tubal_C} and~\ref{thm:tubal_real}, Corollaries~\ref{cor:real_transform} and~\ref{cor:group}.)
\end{theorem}

Part (a) covers every real transform in Section~\ref{sec:motivation}, including learned ones. So a learned transform is determined by its product, up to a permutation. Part (a) also covers multiplication modulo a squarefree polynomial and group convolution over $\C$. Part (b) covers the t-product over $\R$, where conjugate symmetry with Gauss's algorithm is optimal. Semisimplicity is needed: for the dual numbers there are optimal circuits that are not bi-affine.

\paragraph{Complex multiplication.} For $z=z_0+iz_1$ and $\beta\in\R\cup\{\infty\}$ let $p_z(\beta)=z_0+\beta z_1$ and $p_z(\infty)=z_1$, and let $\rho(c_0,c_1,c_2)=(c_0-c_2)+ic_1$ be reduction modulo $\tau^2+1$.

\begin{theorem}[Complex multiplication over $\R$]\label{mthm:cmul}
Every circuit over $\R$ computing $xx'$ for $x,x'\in\C$, with product gates of unbounded fan-in and possibly with divisions, has cost at least $3$. If a division-free circuit has exactly $3$ product gates, of any fan-in, there are $\gamma\in\C$ with $|\gamma|=1$, distinct $\beta_1,\beta_2,\beta_3\in\R\cup\{\infty\}$, nonzero $\alpha_t,\alpha'_t$ and arbitrary $\delta_t,\delta'_t$ such that, up to the order of the factors and constant factors, the $t$-th product gate computes $q_t=(\alpha_tp_{\gamma x}(\beta_t)+\delta_t)(\alpha'_tp_{x'}(\beta_t)+\delta'_t)$, and $xx'=\bar\gamma\,\rho((DV)^{-1}(q-q^{[\le1]}))$. (Proved in Lemma~\ref{lem:cmul}, Theorem~\ref{thm:cmul} and Corollary~\ref{cor:cmul_unbounded}.)
\end{theorem}

So, after a rotation, every optimal algorithm is Gauss's algorithm with its points $0,1,\infty$ replaced by three distinct points. These are exactly the formulas usable in the 3M method. The count $3$ and the minimal bilinear algorithms are classical \cite{winograd1977some,winograd1979multiplication,averbuch1991classification}. Minimal division-free algorithms with binary products are bilinear \cite{feig1981systems}. What is new is the explicit classification with constant shifts, and the lower bound for unbounded fan-in.

\paragraph{Homogeneous component extractors.} A query circuit has sum gates, product gates and query gates. A query gate returns a value of a black-box polynomial $f$ of degree at most $d$ in $n$ variables. The circuit must output the homogeneous components $f^{[0]},\dots,f^{[d]}$. The cost is the number of queries. The analogue of the threshold is the number $d+1$ of outputs.

\begin{theorem}[Extractors]\label{mthm:extractor}
Let $\F$ be infinite.
\begin{enumerate}
    \item[(a)] Every extractor, possibly with divisions, makes at least $d+1$ queries; more generally, extracting a set $S$ of components needs at least $\max S+1$ queries.
    \item[(b)] Every division-free extractor with exactly $d+1$ queries is non-adaptive, and if $n\ge2$, there are distinct $\beta_1,\dots,\beta_{d+1}\in\F$ and a polynomial $h$ with $h(0)=0$ such that the $i$-th query is $f((\beta_i+h(x))x)$ and the outputs are the inverse Vandermonde matrix of the $\beta_i+h(x)$ applied to the answers.
\end{enumerate}
(Proved in Theorems~\ref{thm:extractor_lower} and~\ref{thm:extractor} and Lemma~\ref{lem:non-adaptive}.)
\end{theorem}

So in at least two variables, Ben-Or's construction is the only optimal division-free extractor, up to the shift $h$. The shift cannot be avoided, and $n\ge2$ is necessary (Example~\ref{ex:shift}). For a set $S$ of components, the exact minimum depends on the field. We determine it in Appendix~\ref{sec:extractors}.

\paragraph{Above the threshold: the fan-in matters.} For $\C^c$ with $c\ge3$ and for $2\times2$ matrices, the binary count exceeds $K+2$. Here unbounded fan-in or approximation lowers the count, or might.

\paragraph{Several complex products.} An \emph{approximate circuit} is a circuit whose edge scalars are Laurent polynomials in a parameter $\delta$. Its outputs are the required ones plus $\delta$ times a polynomial in $\delta$ and the inputs.

\begin{theorem}[$\C^c$ over $\R$]\label{mthm:fanin}
Let $c\ge2$ and consider division-free circuits over $\R$ that compute the product of $\C^c$, equivalently multiplication modulo $\prod_{\kappa=1}^c(\tau^2+\kappa^2)$.
\begin{enumerate}
    \item[(a)] With product gates of fan-in two, $3c$ product gates are necessary and sufficient, and the optimal circuits are $c$ copies of the circuits of Theorem~\ref{mthm:cmul}.
    \item[(b)] With product gates of unbounded fan-in, at least $2c+2$ product gates are necessary. For $c=2$ this is $6=3c$, so unbounded fan-in does not help for $\C\times\C$.
    \item[(c)] There are approximate circuits with $2c+1$ product gates of unbounded fan-in, with error $O(\delta^2)$, and every approximate circuit has at least $2c+1$.
    \item[(d)] Every approximate bilinear algorithm has at least $2c+2$ products. For $c=2$ this is $6=3c$.
\end{enumerate}
(Proved in Theorems~\ref{thm:tubal_real}, \ref{thm:c2} and~\ref{thm:approx_lower}, and Example~\ref{ex:border_c}.)
\end{theorem}

So for $\C\times\C$ the count is $6$ for exact circuits of every fan-in and for approximate bilinear algorithms. It is $5$ for approximate circuits with unbounded fan-in. So the lower bound in (b) must use the exact cancellation of the terms of higher degree. Over $\R$, every quadratic map with $K$ independent coordinates has approximate circuits with $K+1$ product gates of unbounded fan-in (Proposition~\ref{prop:border}). The lower bounds in (c) and (d) also come from partial homogenization. The degree $2$ part of a product gate has at most one positive or at most one negative eigenvalue, and this survives limits. For bilinear algorithms, a determinant argument applies.

A single affine restriction kills a product gate of any fan-in (Lemma~\ref{lem:substitution}). This gives $C_\infty=C_2$ for a single quadratic form (Theorem~\ref{thm:single_form}). For $2\times2$ matrix multiplication, $C_2=7=K+3$, by Strassen's algorithm \cite{strassen1969gaussian} and Winograd's bound \cite{winograd1971multiplication}. Over $\R$, every circuit with product gates of unbounded fan-in needs $6$ product gates (Corollary~\ref{cor:matmul}). Approximate circuits need exactly $5$ (Theorem~\ref{thm:approx_lower} and Proposition~\ref{prop:border}). With binary products, the optimal circuits are classified as follows.

\paragraph{$2\times2$ matrix multiplication.} Let $\cir$ be a division-free circuit that computes the entries of $AB$ for $A,B\in\K^{2\times2}$, with product gates $q_i=u_iw_i$ of fan-in two. Here $\K$ is algebraically closed of characteristic $0$. A circuit over a subfield of $\K$, such as $\Q$ or $\R$, is also a circuit over $\K$. Let $\ell_i,m_i$ be the linear parts of $u_i,w_i$. By partial homogenization, the products $\ell_im_i$ form a homogeneous quadratic algorithm for $AB$, the \emph{quadratic part} of $\cir$. So $\cir$ has at least $7$ product gates \cite{winograd1971multiplication}. Write $\ell_i=\alpha_i+\beta_i$ and $m_i=\gamma_i+\delta_i$, with $\alpha_i,\gamma_i$ linear in $A$ and $\beta_i,\delta_i$ linear in $B$. The \emph{pure parts} $(\alpha_i\gamma_i,\beta_i\delta_i)$ of the products must cancel in the outputs. They span a space of dimension $p\le3$. Their two components span spaces of dimensions $p_A,p_B\le p$.

\begin{theorem}[$2\times2$ matrix multiplication]\label{mthm:matmul}
Let $\cir$ be a division-free circuit over $\K$ with exactly $7$ product gates, of fan-in two, that computes the entries of $AB$, and let $p,p_A,p_B$ be the dimensions of the pure parts of its quadratic part.
\begin{enumerate}
    \item[(a)] \textup{Classification.} $p\in\{0,2,3\}$, and up to a linear change of variables $(A,B)\mapsto(XAY,Y^{-1}BZ)$, rescaling and reordering of the products, and possibly transposition $(A,B)\mapsto(B^T,A^T)$, the quadratic part is exactly one of the following:
    \begin{enumerate}
        \item[(i)] $p=0$: Strassen's algorithm, up to the symmetries of matrix multiplication \cite{degroote1978optimal};
        \item[(ii)] $p=2$, $p_A=p_B=1$: the algorithm $\mathcal{M}$ with the products
        \begin{gather*}
            a_{21}(a_{12}-a_{22}+b_{12}),\qquad(a_{21}+b_{22})(a_{12}-a_{22}),\\
            (b_{22}-b_{21})(a_{12}+b_{11}),\qquad(a_{11}+b_{22}-b_{21})b_{11},\\
            a_{11}b_{12},\qquad a_{12}b_{22},\\
            (a_{21}+b_{22}-b_{21})(a_{12}-a_{22}+b_{11});
        \end{gather*}
        \item[(iii)] $p=3$, $p_A,p_B\ge2$: a \textup{Waksman-type} algorithm: there is $H\in GL_2(\K)$ such that, with $A'=AH$ and $B'=H^{-1}B$, every product is a multiple of $(x^TA'e_1+e_2^TB'y)(x^TA'e_2+e_1^TB'y)$ for some $x,y\neq0$; the products lie on four points of the projective plane of pure parts, three of which carry pairs $(x,y),(x,-y)$, and every bilinear part $x^TABy$ is a combination of the outputs;
        \item[(iv)] $p=3$, $(p_A,p_B)=(2,1)$: the algorithm $R_I$ of Theorem~\ref{mm:thm:family} for some $I\neq0$, where $R_I$ and $R_J$ are equivalent if and only if $J\in\{I,I^{-1}\}$.
    \end{enumerate}
    \item[(b)] \textup{No nesting.} Every product gate computes $(c_i+\ell_i)(c'_i+m_i)$ with constants $c_i,c'_i$, and the outputs are $\Lambda(q-q^{[\le1]})$, where $AB=\sum_i\Lambda_i\ell_im_i$. If $p=0$, $\cir$ is bi-affine.
\end{enumerate}
(Proved in Theorems~\ref{mm:thm:main}, \ref{mm:thm:family}, \ref{mm:thm:p12} and~\ref{mm:thm:nest}, Corollary~\ref{mm:cor:class} and Proposition~\ref{mm:prop:pA0}.)
\end{theorem}

Each case occurs. All four pairs $(p_A,p_B)$ with $p_A,p_B\ge2$ occur in (iii), and every $R_I$ is an algorithm (Section~\ref{mm:sec:comp}). In particular, $p=1$ never occurs, and $p_A=0$ or $p_B=0$ forces $p=0$. For $p=3$ the pairs $(3,1)$ and $(1,3)$ do not occur, and for $p=2$ only the pair $(1,1)$ occurs. The products of $R_I$, such as $a_{11}(a_{12}+b_{11}+b_{12})$ and $b_{11}b_{22}$, are not of Winograd's inner-product form. No optimal circuit with product gates of fan-in two nests products.

\section{Overview of the proofs}\label{sec:overview}

This section sketches the proofs. Complete proofs are in Appendices~\ref{sec:preliminaries} to~\ref{sec:matmul}.

\subsection{Constant-free partial homogenization}

All results rest on \emph{constant-free partial homogenization}, a variant of Strassen's homogenization \cite{strassen1973vermeidung}. Take a circuit over a commutative ring and $d\ge2$. Every gate $v$ is split into its homogeneous components $v^{[1]},\dots,v^{[d-1]}$ and a last component $v^{[\ge d]}$, which collects all terms of degree at least $d$. Unlike Strassen's construction, the circuit is not truncated: the terms of degree higher than $d$ are kept in the last component. It is also constant-free: the constant terms $v^{[0]}$ are not computed by gates, and appear only as scalars on edges. Every product gate of the new circuit lies in the sub-circuit of a product gate of the old one, also for product gates of unbounded fan-in (Proposition~\ref{prop:con_free}). For $d=2$ this gives two facts (Corollary~\ref{cor:degree_two}):
\begin{enumerate}
    \item[(i)] the linear part of every gate is a linear combination of the linear parts of its inputs, with coefficients built from constant terms;
    \item[(ii)] every gate is $\sum_i\lambda_{v,i}q_i$ plus an affine function, where $q_i$ are the product gates. So the part of degree $k$ of every gate is $\sum_i\lambda_{v,i}q_i^{[k]}$, with the same coefficients for every $k\ge2$.
\end{enumerate}
With division gates, we first shift the variables to a point where no denominator vanishes, as Strassen does \cite{strassen1973vermeidung}, and work with power series. Then (i) and (ii) hold with division gates counted as product gates (Lemma~\ref{lem:divisions}). Fact (ii) drives both the classifications and the lower bounds for unbounded fan-in. Fact (i) controls the inputs of product gates of fan-in two. Over the coefficient ring $\F[x]$, it also drives the extractors.

\subsection{Bilinear maps with as many products as outputs}

Let $\varphi$ be a bilinear map with $K$ linearly independent coordinates. By (ii), the coordinates lie in the span of the parts $q_i^{[\ge2]}$. So at least $K$ product and division gates are needed. If there are exactly $K$, these parts form a basis of the span of the outputs. So each of them is a bilinear form, and every gate computes a polynomial of degree at most $2$. A division gate would then divide by a constant and add nothing new. For a product gate, the degrees of the inputs add up to $2$. An input of degree $2$ times constants would again add nothing new, so exactly two inputs are affine. The product of their linear parts is bilinear, so one is a function of $x$ and the other of $x'$. Hence every optimal circuit is \emph{bi-affine}. Each product gate computes $(a+\ell(x))(b+m(x'))$, and the products $\ell_im_i$ form a basis of the outputs (Lemma~\ref{lem:bi-affine}). The constants $a,b$ cannot be removed. This is why Winograd's homogeneous statement fails for circuits (Example~\ref{ex:winograd}).

\paragraph{Polynomial multiplication and fault tolerance.} For polynomial multiplication, $\ell_im_i$ lies in the span of the coefficients $c_s$ of $p_1p_2$. So the coefficient vectors $u,v$ of $\ell_i,m_i$ satisfy the rank-one Hankel condition $u_jv_l=H_{j+l}$. Its only solutions are geometric vectors $(\beta^j)$ or the last unit vector. That is, $\ell_i$ and $m_i$ are evaluations of $p_1$ and $p_2$ at a common point $\beta_i\in\F\cup\{\infty\}$ (Lemma~\ref{lem:hankel}). Since the products form a basis, the points are distinct, and the output matrix is the inverse Vandermonde matrix (Theorem~\ref{thm:bi-affine}). In several variables, the Hankel condition holds along every coordinate. So every fiber of the coefficient tensor is geometric, and the tensor is a product of geometric vectors. This is evaluation at a point of $(\F\cup\{\infty\})^l$, and the points are unisolvent (Theorem~\ref{thm:multivariate}). For fault tolerance, every $K$ of the $K+h$ products form an optimal circuit, and a product determines its point. So the points are in general position (Theorem~\ref{thm:fault_tolerant}). Over $\F_q$, a polynomial such as $\tau^q-\tau$ vanishes on all of $\F_q\cup\{\infty\}$, which bounds the number of points. The general position property is the MDS property of a code, and the bounds of Bush and Ball apply (Corollary~\ref{cor:finite}).

\paragraph{Tubal rings over $\C$ and real transforms.} A product with a unit has linearly independent coordinates, so an optimal circuit is bi-affine. Substitute the unit for either input. This shows that the two linear maps before the products agree up to diagonal scalings. So the circuit is a $\star_N$ algorithm for some $N$ with $\star_N=\star_M$ (Theorem~\ref{thm:tubal}). A ring isomorphism of $(\F^d,\odot)$ permutes its primitive idempotents. So $N$ is $M$ with permuted rows (Lemma~\ref{lem:star_unique}). This applies over $\C$ to every tubal ring, and over $\R$ to every real transform (Theorem~\ref{thm:tubal_C}, Corollary~\ref{cor:real_transform}).

\subsection{Real tubal rings and complex multiplication}

Over $\R$, a tubal ring is $\R^r\times\C^c$. It needs $r+3c$ multiplications, $c$ more than its number of outputs. So counting no longer forces degree $2$. Spare gates can cancel terms of higher degree, as in $(1+a)(1+b)(1+c)+(1-a)(1-b)(1-c)=2+2e_2(a,b,c)$. Instead, we use (ii) degree by degree. For a product gate $q_i=u_iw_i$, (ii) gives $u_i^{[k]}=\sum_{j<i}\mu_{ij}q_j^{[k]}$, with the same $\mu$ for all $k$. So $q^{[k]}=Tq^{[k]}+N^{(k)}$, with $T$ strictly lower triangular and $N_i^{(k)}=\sum_{s,t\ge1,\,s+t=k}u_i^{[s]}w_i^{[t]}$. In degree $2$, $N_i^{(2)}$ is the product of the linear parts of the two inputs. So the circuit contains a homogeneous quadratic algorithm. By the theorems of Winograd and Feig, it is bilinear and splits along the real and complex factors. In degree $3$, the outputs must cancel $N_i^{(3)}=\ell_iw_i^{[2]}+m_iu_i^{[2]}$. We separate the parts of degree $(2,1)$ and $(1,2)$ in $(x,x')$. In a complex factor, every nonzero output combination has full rank. This shows that $u_i^{[2]}=w_i^{[2]}=0$. By induction on the degree, all inputs of product gates are affine. So the circuit is bi-affine and splits (Theorem~\ref{thm:tubal_real}). Inside a complex factor, write the products as $\operatorname{Re}(a_tx)\operatorname{Re}(b_tx')$, and expand $\operatorname{Re}(u)=\frac12(u+\bar u)$. The coefficients of $x\bar x'$ and $\bar xx'$ must vanish. A short linear algebra argument then puts all ratios $a_t/b_t$ on one real line $\R\gamma$. Hence every optimal algorithm for complex multiplication is Toom-Cook at three real points, applied to $\gamma x$ and $x'$ and followed by reduction modulo $\tau^2+1$ (Lemma~\ref{lem:cmul}, Theorem~\ref{thm:cmul}).

\paragraph{Unbounded fan-in.} Take $K$ outputs and $K+1$ product gates of unbounded fan-in. The parts $q_i^{[\ge2]}$ span the outputs and at most one more direction $g$. So by (ii), the terms of degree at least $3$ of all product gates are multiples of those of $g$. Suppose some gate has degree $D\ge3$. By unique factorization, every product gate is a constant times $\prod_{j=1}^D(c_{ij}+\ell_j)$, with the same linear forms $\ell_j$. Moreover, the elementary symmetric functions of the $c_{ij}/\ell_j$ of degree at most $D-3$ do not depend on $i$. When the outputs have large enough rank, Newton's identities then make all product gates proportional, a contradiction. So every product gate multiplies two affine functions (Lemma~\ref{lem:spare}). For $\C^c$, the block structure of the outputs then needs three products per factor, more than $2c+1$ (Theorem~\ref{thm:c2}). Approximate circuits avoid this argument, because their gates need not cancel exactly: the terms of degree at least $3$ are only $O(\delta)$ (Proposition~\ref{prop:border}). For approximate circuits, Lemma~\ref{lem:index} shows that the degree $2$ part of each product gate has at most one positive or at most one negative eigenvalue. This property is closed. So with $K$ gates, the limit would give such a form among the outputs, and $\C^c$ has none. For approximate bilinear algorithms, consider the outputs in the span of $2c$ rank one products. By the Cauchy-Binet formula, their determinant is a product of $2c$ real linear forms. In the limit this contradicts the irreducibility of $|z_k|^2$ over $\R$ (Theorem~\ref{thm:approx_lower}).

\subsection{Homogeneous component extractors}

Run the extractor on the generic polynomial $f_Y=\sum_\alpha y_\alpha x^\alpha$. View the computation as a circuit in the coefficients $Y$, over the coefficient ring $\F[x]$. The outputs $f_Y^{[k]}$ are linear in $Y$. By (i), their $Y$-linear parts are $\F[x]$-combinations of the $Y$-linear parts of the query answers. The $Y$-linear part of an answer $f_Y(g)$ is $f_Y(g^{[0]})$. This is the generic polynomial evaluated at the query point $\gamma^i$ of the run in which all answers are $0$. So $f^{[k]}=\sum_iA_{ki}f(\gamma^i)$ for all $f$ (Lemma~\ref{lem:extractor}), and $d+1$ queries are necessary (Theorem~\ref{thm:extractor_lower}). With exactly $d+1$ queries, the answers generate the outputs. By a transcendence degree argument, every answer and every query input is a polynomial in the outputs. The Jacobian of the answers with respect to the outputs is a Vandermonde matrix in the scalings of the query points. Its determinant must be a unit, so the scalings do not depend on $f$. That is, the extractor is non-adaptive (Lemma~\ref{lem:non-adaptive}). Then $A$ is the inverse of a Vandermonde matrix over $\F[x]$ with constant determinant. This forces the points $(\beta_i+h(x))x$ (Theorem~\ref{thm:extractor}). For a subset $S$ of components, we restrict to univariate $f$. The question becomes when a coefficient functional is a combination of point evaluations. This happens exactly when the monic polynomial of the points has zero coefficients in a window of positions (Lemma~\ref{lem:coefficient}, Theorem~\ref{thm:subset}). Over $\R$, Rolle's theorem and a scaling construction show when this is possible: exactly when no two consecutive positions are forced to vanish (Theorem~\ref{thm:subset_explicit}).

\subsection{$2\times2$ matrix multiplication}

Lemma~\ref{lem:graded} applies fact (ii) degree by degree. The degree $2$ part of the outputs is $\Lambda'(\ell_im_i)_i$. The degree $3$ part is $\Lambda'(\ell_iw_i^{[2]}+m_iu_i^{[2]})_i=0$, where $u_i^{[2]},w_i^{[2]}$ lie in the span of the earlier products $\ell_jm_j$. The first identity gives the quadratic part. The second controls nesting.

\paragraph{The classification.} Up to symmetry, the quadratic part is Strassen's algorithm ($p=0$), the algorithm $\mathcal{M}$ ($p=2$), or a Waksman-type algorithm or some $R_I$ ($p=3$). The case $p=1$ is impossible. The proof is organized by the dimension $p$ of the span $V$ of the pure parts. Write each product as $P_i=\pi_i+S_i$, with pure part $\pi_i$ and bilinear part $S_i$. The relations $c$ with $\sum_ic_i\pi_i=0$ form a space $K$ of dimension $7-p$. The rows of the output matrix lie in $K$, and $c\mapsto\sum_ic_iS_i$ maps $K$ onto a space containing the outputs. The four outputs give four independent relations in $K$, so $p\le3$. Also $\max(p_A,p_B)\le p\le p_A+p_B$, and transposition exchanges $p_A$ and $p_B$. Table~\ref{tab:classes} lists the cases, the outcome in each, and the main step of its proof.

\begin{table}[h]
\caption{The classification of seven-product algorithms for $2\times2$ matrices, by the dimensions $p,p_A,p_B$ of the pure parts.}\label{tab:classes}
\small
\begin{tabular}{@{}p{0.24\columnwidth}@{\hspace{1em}}l p{0.47\columnwidth}@{}}
\toprule
Case & Outcome & Main step\\
\midrule
$p_B=0$ & $p=0$ & the bilinear parts form an optimal bilinear algorithm; de Groote and an exact check (Prop.~\ref{mm:prop:pA0})\\
$p=0$ & Strassen & de Groote's theorem \cite{degroote1978optimal}\\
$p=1$ & none & one point; trading bilinear parts gives two terms with one coefficient matrix (Thm.~\ref{mm:thm:p12}(a))\\
$p=2$, $p_A=2$ & none & pencil lemma and a common zero exclude three points, two points as for $p=1$ (Lemma~\ref{mm:lem:p2pairs})\\
$p=2$, $(1,1)$ & $\mathcal{M}$ & normal form $a_{11}b_{12}$, $a_{12}b_{22}$, block structure, nine bilinear equations (Prop.~\ref{mm:prop:p11})\\
$p=3$, $(3,3)$ & Waksman type & coordinates $N=(A;B^TJ)$ give $d=4$ points; nets of quadrics exclude $d\ge5$ (Sections~\ref{mm:sec:first} to~\ref{mm:sec:assembly})\\
$p=3$, $(3,2)$, $(2,2)$ & Waksman type & points on two lines; rulings of $q_a$ force $\tau=0$ (Props.~\ref{mm:prop:32}, \ref{mm:prop:22})\\
$p=3$, $p_B=1$ & $R_I$, $p_A=2$ & restriction to the annihilator of the B-factors; a linear system (Prop.~\ref{mm:prop:pB1})\\
\bottomrule
\end{tabular}
\end{table}

\paragraph{Sketch for $p\le2$.} Suppose the distinct points $[\pi_i]$ are linearly independent. Then the coefficient matrices of the products at each point sum to zero after scaling. So these products contribute only the differences of their bilinear parts. Counting ranks gives a decomposition of $AB$ into at most $7$ terms of rank one, two of which share a coefficient matrix. This contradicts de Groote's theorem. It excludes $p=1$. With a pencil argument, it also excludes every pair $(p_A,p_B)$ with $p=2$ except $(1,1)$. For $(1,1)$, we normalize the bilinear products to $a_{11}b_{12}$ and $a_{12}b_{22}$. Then every bilinear form in the span of the products is block diagonal, which places every factor in one block. Nine bilinear equations in sixteen unknowns remain. Their solutions form a single orbit of the symmetries, which gives $\mathcal{M}$.

\paragraph{Sketch for $p=3$.} Here $K$ maps isomorphically onto the outputs. So $S_i$ is, modulo the outputs, a linear function $\tau(\pi_i)$. Each point carries at most two products, and they differ by an output $x^TABy$ of rank two. For $p_A=p_B=3$, we use the $4\times2$ matrix $N=(A;B^TJ)$, whose mixed minors are the entries of $AB$. It forces four points with three Waksman pairs. The other cases are restriction arguments (Table~\ref{tab:classes}).

For nesting with $p=3$, the degree $3$ identity says that $\ell_iw_i^{[2]}+m_iu_i^{[2]}=\Psi(\pi_i)$ for a linear map $\Psi$ from $V$ to cubic forms. For Waksman-type algorithms, the products are symmetric bilinear forms $(w\cdot n_1)(w\cdot n_2)$ in $n_1=(Ae_1;B^Te_2)$ and $n_2=(Ae_2;B^Te_1)$. At a double point, $\Psi(\pi_i)$ is divisible by two different linear forms. Symmetry then forces $\Psi=0$ and $u_i^{[2]}=w_i^{[2]}=0$. For $R_I$, for $\mathcal{M}$, and for the bilinear case $p=0$, which is Strassen's algorithm by de Groote's theorem, an exact computation shows that the identity has only the zero solution. By induction on the degree, all inputs of product gates are then affine.

\section{Discussion and open problems}\label{sec:open}

For quadratic maps, our results draw the line between fan-in two and unbounded fan-in at the threshold $C_2=K+2$. Below it, unbounded fan-in, constants and, in most cases, divisions change neither the count nor the optimal circuits. The optimal circuits leave very little freedom: the choice of evaluation points, or of a transform up to permutation, and constants. So every improvement in the remaining costs must be sought inside these families. Such costs are additions, divisions in the interpolation and numerical stability. For tensor products defined by transforms, the transform-based algorithm is the only optimal one. So further savings must come from the transform itself, as in the efforts surveyed in Section~\ref{sec:motivation}. Approximate circuits with unbounded fan-in save product gates whenever $C_2\ge K+2$, for instance for $\C^c$ and for $2\times2$ matrices. For exact circuits the question is open above the threshold, for $\C^c$ with $c\ge3$ and for $2\times2$ matrices.

Theorem~\ref{thm:tubal_real} assumes product gates of fan-in $2$ and no divisions. For unbounded fan-in we know the count $3c$ only for $c\le2$ (Theorem~\ref{thm:c2}). For $c\ge3$ we know only the bound $2c+2$. A product gate whose inputs all have nonzero constant terms contributes to degree $2$ a weighted elementary symmetric form of the linear parts of its inputs. Such forms can be semidefinite of any rank. So the degree $2$ parts alone cannot give more than $2c+1$. The bound must come from the cancellation of the higher-degree terms, as in $(1+a)(1+b)(1+c)+(1-a)(1-b)(1-c)=2+2e_2(a,b,c)$. With one spare gate, all higher-degree terms are proportional, and the proof of Theorem~\ref{thm:c2} uses this. With $e\ge2$ spare gates, they span a space of dimension up to $e$. By Lemma~\ref{lem:spare}, this holds whenever the parts $q_i^{[\ge2]}$ of the product gates span a space of dimension at most $2c+1$. For $\C\times\C$ the count $6$ is known, but the optimal circuits with product gates of unbounded fan-in are not classified. The open case is the one in which the six parts are linearly independent. This is the case for every optimal circuit with product gates of fan-in two, such as Gauss's algorithm applied to each factor. What remains is to exclude a product gate of degree at least $3$ in this case. It would suffice to show that in a circuit for $\C^c$ with at most $3c$ product gates, every product gate has degree $2$. Then every product gate multiplies two affine functions, and Theorem~\ref{thm:tubal_real} applies. We conjecture more generally that unbounded fan-in never helps for quadratic maps, that is, $C_\infty=C_2$. Corollary~\ref{cor:spare} proves this when $C_2\le K+2$, and Theorem~\ref{thm:single_form} when $K=1$. For circuits of depth $3$, Lemma~\ref{lem:one_direction} reduces the case $C_2=K+3$ to circuits in which the parts of degree at least $3$ of the product gates span a space of dimension $2$. The first open cases are $\C^3$ and $2\times2$ matrix multiplication, where $C_2=K+3$.

For $2\times2$ matrix multiplication, Theorem~\ref{mthm:matmul} describes every division-free circuit with seven product gates of fan-in two. Its product gates multiply affine functions. Up to symmetry, the linear parts form Strassen's algorithm ($p=0$), the algorithm $\mathcal{M}$ ($p=2$), or a Waksman-type algorithm or some $R_I$ ($p=3$). So commutativity adds exactly three kinds of optimal algorithms to de Groote's single bilinear one: a single algorithm without parameters, the Waksman-type algorithms, and a one-parameter family. Questions such as the least number of additions or the best numerical stability of an optimal circuit thus reduce to these explicit families. It remains open whether divisions or product gates of larger fan-in admit other optimal circuits. With unbounded fan-in we only know, over $\R$, that $6$ product gates are needed (Corollary~\ref{cor:matmul}). For $3\times3$ matrices even the bilinear rank is unknown.

Such lower bounds cannot come from a condition that is preserved under limits, since approximate circuits need only $2c+1$ product gates (Theorem~\ref{mthm:fanin}(c)). A numerical search for exact circuits for $\C^3$ with $8$ product gates of unbounded fan-in found only approximate circuits of this kind. Their parameters diverge as the error tends to $0$. For approximate circuits with unbounded fan-in the count is exactly $2c+1$ (Theorem~\ref{thm:approx_lower}). For approximate bilinear algorithms we know $6$ for $c=2$. For $c\ge3$ we know only the bounds $2c+2$ and $3c$. A numerical search for real approximate bilinear algorithms for $\C^3$ with $8$ products stayed at distance about $1$ from the target, which suggests $3c$.

\appendix

\section{Preliminaries}\label{sec:preliminaries}

\subsection{Polynomials, components and circuits}

Let $\Rg$ be a commutative ring with $1$ and let $z=(z_1,\dots,z_N)$ be variables. For $P\in\Rg[z]$ and $i\ge0$ we write $P^{[i]}$ for the homogeneous component of $P$ of degree $i$ in $z$, and $P^{[\ge d]}=\sum_{i\ge d}P^{[i]}$, $P^{[\le d]}=\sum_{i\le d}P^{[i]}$. The same notation is used for formal power series in $\Rg[[z]]$. When the variables are split into groups, components are always taken with respect to the group that is specified.

\begin{definition}[Arithmetic circuit]\label{def:circuit}
An \emph{arithmetic circuit} over $\Rg$ in the variables $z$ is a directed acyclic graph whose leaves are labeled by variables or by elements of $\Rg$. Each internal gate is a \emph{sum gate}, of arbitrary fan-in and with an element of $\Rg$ on each incoming edge, or a \emph{product gate} of arbitrary fan-in at least $2$. Some gates are marked as output gates. Every gate $v$ computes a polynomial $\hat v\in\Rg[z]$ in the obvious way. The variables commute, so the inputs of a product gate are unordered. The \emph{cost} of a circuit is its number of product gates, each counted once regardless of its fan-in.

A \emph{circuit with divisions} over a field $\F$ may also have \emph{division gates}, with two ordered inputs, a numerator $u$ and a divisor $w$, computing $\hat u/\hat w$. Every gate computes a rational function, and no divisor may compute $0$. Its cost is the number of product and division gates. Unless divisions are mentioned, circuits have none.
\end{definition}

For a product gate $q$ we often write $q$ also for the polynomial $\hat q$. If $\Rg=\F$ is a field, an $\F$-linear map applied to the outputs is free.

\subsection{Standard facts}\label{sec:facts}

\begin{fact}[Vandermonde determinant {\cite[Section~17.1]{garrett2007abstract}}]\label{fact:vandermonde}
For elements $t_1,\dots,t_m$ of a commutative ring,
\begin{displaymath}
    \det\bigl(t_i^{\,k}\bigr)_{1\le i\le m,\ 0\le k\le m-1}=\prod_{1\le i<i'\le m}(t_{i'}-t_i).
\end{displaymath}
\end{fact}

\paragraph{Transcendence degree.} Let $E/K$ be a field extension. Elements $e_1,\dots,e_m$ of the field $E$ are called \emph{algebraically independent} over $K$ if $P(e_1,\dots,e_m)\neq0$ for every nonzero polynomial $P$ over $K$. The \emph{transcendence degree} of $E$ over $K$ is the maximal size of an algebraically independent subset of $E$ \cite[Chapter~VIII, \S1]{lang2002algebra}. If $E$ is generated over $K$ by $m$ elements, its transcendence degree is at most $m$.

\paragraph{Jacobians.} For polynomials $\Psi=(\Psi_1,\dots,\Psi_m)$ in $w_1,\dots,w_r$, the \emph{Jacobian matrix} is $J_\Psi=(\partial\Psi_i/\partial w_k)_{i,k}$, the matrix of formal partial derivatives. By the chain rule, the Jacobian of $P\circ\Psi$ is $J_P(\Psi)\cdot J_\Psi$.

\paragraph{Consequences of linear algebra.} Polynomials with pairwise disjoint sets of monomials, none of them zero, are linearly independent. Linear forms that are linearly independent are algebraically independent. Indeed, they extend to a basis of the linear forms, and the corresponding change of variables is an automorphism of the polynomial ring. A linear functional on a finite-dimensional vector space is a linear combination of given functionals if and only if it vanishes on their common kernel. A nonzero polynomial over an infinite field does not vanish at every point.

\subsection{Polynomial multiplication}

Let $k_1,k_2\in\mathbb{N}$ and $K=k_1+k_2-1$. The inputs are $x=(x_0,\dots,x_{k_1-1})$ and $x'=(x'_0,\dots,x'_{k_2-1})$and , the coefficients of $p_1(\tau)=\sum_jx_j\tau^j$ $p_2(\tau)=\sum_lx'_l\tau^l$.

\begin{definition}[Polynomial multiplication]\label{def:phi}
The map $\phi_{k_1,k_2}$ has the coordinates $c_s=\sum_{j+l=s}x_jx'_l$, $0\le s\le K-1$, the coefficients of $p_1p_2$. We write $C=\spn_\F\{c_0,\dots,c_{K-1}\}$.
\end{definition}

For $\beta\in\F$, $p_1(\beta)$ and $p_2(\beta)$ are linear forms, and we set $p_1(\infty)=x_{k_1-1}$, $p_2(\infty)=x'_{k_2-1}$. Then
\begin{equation}\label{eq:eval_product}
    p_1(\beta)\,p_2(\beta)=\sum_{s=0}^{K-1}\beta^sc_s\quad(\beta\in\F),\qquad p_1(\infty)\,p_2(\infty)=c_{K-1}.
\end{equation}

\begin{definition}[Bi-affine circuit]\label{def:bi-affine}
A circuit with inputs $x,x'$ is \emph{bi-affine} if it has no division gates and every product gate computes a polynomial of the form $(a+\ell(x))(b+m(x'))$ with $a,b\in\F$, $\ell$ a linear form in $x$ only and $m$ a linear form in $x'$ only.
\end{definition}

\begin{definition}[Bi-affine Toom-Cook]\label{def:toom-cook}
Let $\beta_1,\dots,\beta_K\in\F\cup\{\infty\}$ be pairwise distinct, and let $V$ be the $K\times K$ matrix whose $i$-th row is $(1,\beta_i,\dots,\beta_i^{K-1})$, or the last unit vector if $\beta_i=\infty$. A \emph{Toom-Cook algorithm with a bi-affine modification} computes, for nonzero $\alpha_i,\alpha'_i\in\F$ and arbitrary $\gamma_i,\gamma'_i\in\F$, the products
\begin{displaymath}
    q_i=\bigl(\alpha_i\,p_1(\beta_i)+\gamma_i\bigr)\bigl(\alpha'_i\,p_2(\beta_i)+\gamma'_i\bigr)\qquad(1\le i\le K),
\end{displaymath}
and outputs $c=(DV)^{-1}\bigl(q-q^{[\le1]}\bigr)$, where $D=\operatorname{diag}(\alpha_i\alpha'_i)$. For $\alpha_i=\alpha'_i=1$ and $\gamma_i=\gamma'_i=0$ it is the Toom-Cook algorithm.
\end{definition}

By \eqref{eq:eval_product}, $q^{[\ge2]}=DVc$, and $V$ is invertible (Fact~\ref{fact:vandermonde}, expanding along the row of $\infty$ if it occurs), so the algorithm is correct.

\paragraph{Several variables.} An $l$-level recursive Toom-Cook algorithm splits each polynomial into $k$ blocks, recursively $l$ times. Writing exponents in base $k$, it multiplies two polynomials in $l$ variables of degree less than $k$ in each variable, and then adds up overlapping coefficients, which is linear. We therefore consider, for $l\ge1$ and $k_1,k_2\in\mathbb{N}^l$, the box $B(k)=\{j\in\mathbb{Z}^l:0\le j_t<k_t\}$, inputs $x=(x_j)_{j\in B(k_1)}$ and $x'=(x'_j)_{j\in B(k_2)}$, the polynomials $p_1(\tau)=\sum_jx_j\tau^j$ and $p_2(\tau)=\sum_jx'_j\tau^j$ in $\tau=(\tau_1,\dots,\tau_l)$, and the outputs $c_s=\sum_{j+j'=s}x_jx'_{j'}$ for $s\in B(\kappa)$, $\kappa_t=k_{1,t}+k_{2,t}-1$. Let $K=\prod_t\kappa_t$ and $C=\spn\{c_s\}$. For $\beta\in(\F\cup\{\infty\})^l$, $p_1(\beta)$ is the linear form obtained by substituting $\tau_t=\beta_t$ for the finite coordinates and taking the coefficient of $\tau_t^{k_{1,t}-1}$ for the infinite ones, and similarly $p_2(\beta)$. With $g_t(\beta_t)=(1,\beta_t,\dots,\beta_t^{\kappa_t-1})$, or the last unit vector of length $\kappa_t$ if $\beta_t=\infty$, and $v(\beta)=g_1(\beta_1)\otimes\dots\otimes g_l(\beta_l)\in\F^{B(\kappa)}$, we have $p_1(\beta)p_2(\beta)=\sum_sv(\beta)_sc_s$. Points $\beta_1,\dots,\beta_K$ are \emph{unisolvent} if the matrix $V$ with rows $v(\beta_i)$ is invertible. For $l=1$ it means distinct. Products $q_i=(\alpha_ip_1(\beta_i)+\gamma_i)(\alpha'_ip_2(\beta_i)+\gamma'_i)$ at unisolvent points, with output $c=(DV)^{-1}(q-q^{[\le1]})$, form a \emph{multivariate Toom-Cook algorithm with a bi-affine modification}. The recursive algorithm uses a grid of points.

\begin{definition}[Fault tolerance]\label{def:fault_tolerant}
Products $q_1,\dots,q_{K+h}$ \emph{tolerate $h$ faults} if for every set $S$ of $K$ indices, every $c_s$ lies in $\spn_\F\{q_i:i\in S\}$ plus a polynomial of degree at most $1$. Points are in \emph{general position} if every $K$ of them are unisolvent.
\end{definition}

\subsection{Tubal rings}

We write $u\odot v$ for the entrywise product of vectors of length $d$ and $\mathbf{1}$ for the all-ones vector.

\begin{definition}[The $\star_M$ product \cite{kernfeld2015tensor,avron2025demystifying}]\label{def:star_M}
For a field $E$ and an invertible $M\in E^{d\times d}$, $a\star_Mb=M^{-1}\bigl((Ma)\odot(Mb)\bigr)$ for $a,b\in E^d$.
\end{definition}

The product $\star_M$ is bilinear, and $x\mapsto Mx$ is a ring isomorphism from $(E^d,+,\star_M)$ onto $(E^d,+,\odot)$, whose unit is $M^{-1}\mathbf{1}$. For $E=\C$, $a\star_Mb$ is real for real $a,b$ when every row of $M$ is real or is the entrywise conjugate of exactly one other row \cite[Lemma~2]{avron2025demystifying}. The t-product is the case where $M$ is the DFT matrix.

\begin{definition}[Tubal ring \cite{avron2025demystifying}]\label{def:tubal}
A \emph{tubal ring} is a ring $(\R^d,+,\cdot)$ with the usual addition that is commutative, unital and von Neumann regular, and is an associative $\R$-algebra with the usual scalar multiplication.
\end{definition}

Avron and Mor show that every tubal ring is isomorphic to $\R^r\times\C^c$ with $r+2c=d$, and that its product is $\star_M$ for an invertible $M\in\C^{d\times d}$ \cite[Theorem~10]{avron2025demystifying}. 

\begin{fact}[Extension of scalars]\label{fact:complexification}
For a tubal ring $A\cong\R^r\times\C^c$, extending the product $\C$-bilinearly to $\C^d$ gives $A\otimes_\R\C\cong\C^r\times(\C\otimes_\R\C)^c\cong\C^{r+2c}=\C^d$ with the entrywise product, using $\C\otimes_\R\C\cong\C\times\C$, $z\otimes w\mapsto(zw,\bar zw)$. If $a\cdot b=a\star_Mb$, the isomorphism is $x\mapsto Mx$.
\end{fact}

\begin{definition}[$\star_N$ with a bi-affine modification]\label{def:star_N_affine}
For an invertible $N\in\F^{d\times d}$, nonzero $\alpha_i,\alpha'_i\in\F$ and arbitrary $\gamma_i,\gamma'_i\in\F$, the algorithm computes $q_i=(\alpha_i(Na)_i+\gamma_i)(\alpha'_i(Nb)_i+\gamma'_i)$ for $1\le i\le d$ and outputs $N^{-1}D^{-1}(q-q^{[\le1]})$ with $D=\operatorname{diag}(\alpha_i\alpha'_i)$. Since $q^{[\ge2]}=D((Na)\odot(Nb))$, it computes $a\star_Nb$.
\end{definition}

\subsection{Homogeneous component extractors}

For extractors, $\F$ is an infinite field, $n\ge1$, $x=(x_1,\dots,x_n)$, $R=\F[x]$ and $L=\F(x)$.

\begin{definition}[Query circuit]\label{def:query_circuit}
A \emph{query circuit} $\Xi$ is an arithmetic circuit over $\F$ with inputs $x_1,\dots,x_n$ that has, besides sum and product gates, \emph{query gates} with $n$ ordered inputs. For a circuit $\Phi$ in $n$ variables, $\Xi(\Phi)$ is the circuit obtained by replacing every query gate, with inputs $g_1,\dots,g_n$, by a copy of $\Phi$ applied to $g_1,\dots,g_n$. The \emph{number of queries} is the number of query gates. $\Xi$ is \emph{non-adaptive} if the inputs of every query, as polynomials in $x$, are the same for every $\Phi$. This holds, for example, when no query gate is reachable from another one.
\end{definition}

The polynomial computed by every gate of $\Xi(\Phi)$ depends only on the polynomial $f$ computed by $\Phi$. We write $\Xi^f$ for the evaluation in which every query gate with inputs $g_1,\dots,g_n$ returns $f(g_1,\dots,g_n)$. This makes sense for $f$ with coefficients in any commutative ring containing $R$. $\Xi^0$ is the evaluation in which every query returns $0$.

\begin{definition}[Extractor]\label{def:extractor}
Let $\emptyset\neq S\subseteq\{0,\dots,d\}$. $\Xi$ is a \emph{degree $d$ extractor of $S$} if it has output gates $o_k$, $k\in S$, and, for every circuit $\Phi$ computing a polynomial $f$ of degree at most $d$, $o_k$ computes $f^{[k]}$ in $\Xi(\Phi)$. For $S=\{0,\dots,d\}$ it is a \emph{degree $d$ homogeneous component extractor}.
\end{definition}

\begin{definition}[Generic polynomial]\label{def:generic}
Let $Y=\{y_\alpha\}_{|\alpha|\le d}$ be new variables, one for each monomial $x^\alpha$ of degree at most $d$, and let
\begin{displaymath}
    f_Y=\sum_{|\alpha|\le d}y_\alpha x^\alpha\in R[Y],\qquad f_Y^{[k]}=\sum_{|\alpha|=k}y_\alpha x^\alpha
\end{displaymath}
(components with respect to $x$). For $a\in\F^Y$ let $\pi_a:R[Y]\to R$ be the substitution $y_\alpha\mapsto a_\alpha$ and $f_a=\pi_a(f_Y)$.
\end{definition}

\begin{definition}[Shifted evaluation-interpolation extractor]\label{def:ben-or}
Let $\beta_1,\dots,\beta_{d+1}\in\F$ be pairwise distinct, $h\in R$, $t_i=\beta_i+h$ and $V=(t_i^{\,k})_{1\le i\le d+1,\,0\le k\le d}$. The extractor asks $u_i=\Phi(t_ix_1,\dots,t_ix_n)$ and outputs $V^{-1}(u_1,\dots,u_{d+1})^T$. For $h=0$ it is Ben-Or's construction \cite{nisan1996lower,shpilka2001depth}. 
\end{definition}

For $f$ of degree at most $d$, $f(t_ix)=\sum_kt_i^{\,k}f^{[k]}$, so $(f(t_ix))_i=V(f^{[k]})_k$. By Fact~\ref{fact:vandermonde}, $\det V=\prod_{i<i'}(\beta_{i'}-\beta_i)$ is a nonzero constant, so $V^{-1}=(\det V)^{-1}\operatorname{adj}(V)$ has entries in $R$ and no division by a nonconstant polynomial is needed.

\section{Partial homogenization}\label{sec:homogenization}

\begin{proposition}[Constant-free partial degree $d$ homogenization]\label{prop:con_free}
Let $d\ge2$ and let $\cir$ be an arithmetic circuit over a commutative ring $\Rg$ in variables $z$. There is a circuit $\cirp$ over $\Rg$ in the same variables such that:
\begin{enumerate}
    \item\label{prop:components} \emph{(Components.)} For every gate $v$ of $\cir$ and $i\in\{1,\dots,d-1\}\cup\{{\ge}d\}$, either $\cirp$ has a gate $v_i$ with $\hat v_i=\hat v^{[i]}$, or it has none and $\hat v^{[i]}=0$.
    \item\label{prop:homogeneity} \emph{(Homogeneity.)} Every gate of $\cirp$ computes a homogeneous polynomial of degree between $1$ and $d-1$, or a polynomial without components of degree less than $d$. Every leaf of $\cirp$ is a variable.
    \item\label{prop:subcircuits} \emph{(Sub-circuits.)} The gates of $\cirp$ are partitioned into sub-circuits $\cirp_v$, one for each gate $v$ of $\cir$, such that the output gates of $\cirp_v$ are the gates $v_i$. Every edge of $\cirp$ lies inside one sub-circuit or leaves an output gate of $\cirp_u$ for a child $u$ of $v$. Every product gate of $\cirp$ lies in $\cirp_v$ for a product gate $v$ of $\cir$.
    \item\label{prop:count} \emph{(Count.)} If $v$ is a product gate of fan-in $2$ whose children compute polynomials of degree less than $d$, then $\cirp_v$ has at most $(d-1)^2$ product gates.
\end{enumerate}
\end{proposition}

\begin{proof}
First replace every product gate of fan-in $s\ge3$ by a binary tree of $s-1$ product gates. This does not change the polynomial computed at the original gates. Along with the gates $v_i$ we compute the elements $c_v=\hat v^{[0]}\in\Rg$, which are used only as scalars on edges. We process the gates in topological order. A summand involving a gate that does not exist is omitted, and a gate all of whose summands are omitted is not created.

\emph{Leaves.} For a variable $z$: $v_1=z$, $c_v=0$. For a constant $c\in\Rg$: no gates, $c_v=c$.

\emph{Sum gates.} If $\hat v=\sum_u\lambda_u\hat u$: $c_v=\sum_u\lambda_uc_u$ and $v_i=\sum_u\lambda_uu_i$ for every $i$.

\emph{Binary product gates.} If $\hat v=\hat u\hat w$: $c_v=c_uc_w$ and, with $\times$ denoting a new product gate,
\begin{align}
    v_i&=c_u\,w_i+c_w\,u_i+\sum_{\substack{j+l=i\\ j,l\ge1}}u_j\times w_l\quad(1\le i<d),\label{eq:product_gate}\\
    v_{\ge d}&=c_u\,w_{\ge d}+c_w\,u_{\ge d}+\sum_{\substack{1\le j,l\le d-1\\ j+l\ge d}}u_j\times w_l\notag\\
    &\quad+\sum_{j=1}^{d-1}\bigl(u_j\times w_{\ge d}+u_{\ge d}\times w_j\bigr)+u_{\ge d}\times w_{\ge d}.\label{eq:product_gate_dle}
\end{align}
For an original product gate $v$ of fan-in $s\ge3$, $\cirp_v$ is the union of the sub-circuits of the gates of its tree, and its output gates are those of the root.

\emph{Correctness.} By induction, $c_v=\hat v^{[0]}$, every created $v_i$ computes $\hat v^{[i]}$, and $\hat v^{[i]}=0$ if $v_i$ is not created. For leaves and sum gates this is immediate. For $v=u\cdot w$, expand $\hat u\hat w$ with $\hat u=c_u+\sum_{j<d}\hat u^{[j]}+\hat u^{[\ge d]}$, and similarly for $\hat w$. The terms of degree $i<d$ are $c_u\hat w^{[i]}$, $c_w\hat u^{[i]}$ and $\hat u^{[j]}\hat w^{[l]}$ with $j,l\ge1$, $j+l=i$. This gives $c_v$ and \eqref{eq:product_gate}. All other terms have only monomials of degree at least $d$ and are the summands of \eqref{eq:product_gate_dle}.

\emph{Homogeneity, sub-circuits and count.} A product $u_j\times w_l$ with $j,l\le d-1$ is homogeneous of degree $j+l$. A product with a factor $u_{\ge d}$ or $w_{\ge d}$ has no components below degree $d$. Sum gates inherit these properties. The only leaves are variables. By construction the sub-circuits are disjoint, and edges enter $\cirp_v$ only from output gates of children. Product gates are created only for product gates of $\cir$. For fan-in $2$ with children of degree less than $d$, every pair $(j,l)\in\{1,\dots,d-1\}^2$ gives exactly one product gate.
\end{proof}

The degree $2$ case is all we use. It says that degree-one parts propagate linearly, and that above degree one only product gates create new information.

\begin{corollary}[Degree two structure]\label{cor:degree_two}
Let $\cir$ be an arithmetic circuit over $\Rg$ in variables $z$, with product gates $q_1,\dots,q_M$ in a topological order.
\begin{enumerate}
    \item\label{cor:degree_one} For a sum gate $v$ with $\hat v=\sum_u\lambda_u\hat u$, $\hat v^{[1]}=\sum_u\lambda_u\hat u^{[1]}$. For a product gate $v$ with children $u_1,\dots,u_s$,
    \begin{displaymath}
        \hat v^{[1]}=\sum_{j=1}^{s}\Bigl(\prod_{l\neq j}\hat u_l^{[0]}\Bigr)\hat u_j^{[1]},\qquad \hat u_l^{[0]}\in\Rg .
    \end{displaymath}
    In particular $\hat v^{[1]}$ is an $\Rg$-linear combination of the $\hat u^{[1]}$ for the children $u$ of $v$.
    \item\label{cor:linear_structure} Every gate $v$ computes $\hat v=\sum_{i=1}^M\lambda_{v,i}\hat q_i+a_v$ with $\lambda_{v,i}\in\Rg$, $a_v$ of degree at most $1$, and $\lambda_{v,i}=0$ unless $q_i=v$ or $q_i$ is a predecessor of $v$. Hence $\hat v^{[k]}=\sum_i\lambda_{v,i}\hat q_i^{[k]}$ for every $k\ge2$, with coefficients independent of $k$, and if $q_i$ has children $u_1,\dots,u_s$, then $\hat u_j^{[\ge2]}\in\spn_\Rg\{\hat q_{i'}^{[\ge2]}:i'<i\}$.
\end{enumerate}
\end{corollary}

\begin{proof}
Apply Proposition~\ref{prop:con_free} with $d=2$. Part~\ref{cor:degree_one} is the formula for $v_1$ in the proof. For sum gates this is clear, and for binary product gates it is \eqref{eq:product_gate} with $i=1$, where the sum over $j+l=1$ is empty. For fan-in $s$ we compose along the binary tree. For Part~\ref{cor:linear_structure}, the proof of Proposition~\ref{prop:con_free} gives $v_{\ge2}=\sum_u\lambda_uu_{\ge2}$ for sum gates, while for a product gate $q_i$ the gate $(q_i)_{\ge2}$ computes $\hat q_i^{[\ge2]}$. By induction $\hat v^{[\ge2]}=\sum_i\lambda_{v,i}\hat q_i^{[\ge2]}$ with the stated support. Then $a_v:=\hat v-\sum_i\lambda_{v,i}\hat q_i$ has no part of degree at least $2$.
\end{proof}

Strassen avoids divisions by moving to a point where no denominator vanishes and expanding in power series \cite{strassen1973vermeidung}. In degree $2$ this is all that is needed.

\begin{lemma}[Divisions]\label{lem:divisions}
Let $\cir$ be a circuit with divisions over an infinite field $E$ in variables $z$, and let $q_1,\dots,q_M$ be its product and division gates in a topological order. Let $\xi\in E^N$ be any point such that every $\hat v$ is a quotient of polynomials whose denominator does not vanish at $\xi$, and every divisor $\hat w$ satisfies $\hat w(\xi)\neq0$. Such points exist. Expand every $\hat v$ in the power series ring $E[[z-\xi]]$ and take components with respect to $z-\xi$. Then Part~\ref{cor:linear_structure} of Corollary~\ref{cor:degree_two} holds verbatim, and Part~\ref{cor:degree_one} holds with one more rule: for a division gate $v$ with numerator $u$ and divisor $w$,
\begin{displaymath}
    \hat v^{[1]}=\bigl(\hat u^{[1]}-\hat v^{[0]}\,\hat w^{[1]}\bigr)/\hat w^{[0]}.
\end{displaymath}
\end{lemma}

\begin{proof}
Write every gate value and the value of every divisor in lowest terms and let $H$ be the product of all denominators and of the numerators of the divisors, a nonzero polynomial. Choose $\xi$ with $H(\xi)\neq0$. The rational functions whose denominator does not vanish at $\xi$ form a ring that embeds in $E[[z-\xi]]$, and the divisors are units there. Taking a component is $E$-linear on $E[[z-\xi]]$. So the induction of Corollary~\ref{cor:degree_two} goes through. Sum gates combine components linearly, and product and division gates $q_i$ contribute their own part $\hat q_i^{[\ge2]}$. Product gates satisfy Part~\ref{cor:degree_one} by the same expansion. For a division gate, $\hat u=\hat v\hat w$ gives $\hat u^{[1]}=\hat v^{[0]}\hat w^{[1]}+\hat w^{[0]}\hat v^{[1]}$.
\end{proof}

For circuits whose product gates have fan-in $2$, Part~\ref{cor:linear_structure} of Corollary~\ref{cor:degree_two} can be applied to the inputs of the product gates, degree by degree. This gives a triangular recursion, which we use for real tubal rings (Theorem~\ref{thm:tubal_real}) and for $2\times2$ matrix multiplication (Appendix~\ref{sec:matmul}).

\begin{lemma}[The graded recursion]\label{lem:graded}
Let $\cir$ be a division-free circuit over a field $\F$ whose product gates $q_1,\dots,q_m$, in a topological order, have fan-in $2$, $q_i=u_iw_i$. Let $c_i,\ell_i$ and $c'_i,m_i$ be the components of degree $0$ and $1$ of $u_i$ and $w_i$, and for $k\ge2$ let
\begin{displaymath}
    N_i^{(k)}=\sum_{s,t\ge1,\ s+t=k}u_i^{[s]}w_i^{[t]},\qquad\text{in particular}\quad N_i^{(2)}=\ell_im_i.
\end{displaymath}
There are scalars $\mu_{ij},\mu'_{ij}$ ($j<i$) such that $u_i^{[k]}=\sum_{j<i}\mu_{ij}q_j^{[k]}$ and $w_i^{[k]}=\sum_{j<i}\mu'_{ij}q_j^{[k]}$ for all $k\ge2$. With the strictly lower triangular matrix $T_{ij}=c_i\mu'_{ij}+c'_i\mu_{ij}$ and $S=(I-T)^{-1}$, we have $q^{[k]}=SN^{(k)}$ for all $k\ge2$, and $u_i^{[k]},w_i^{[k]}\in\spn\{N_j^{(k)}:\ j<i\}$. If $\phi=\Lambda q+a$ is the vector of outputs as in Corollary~\ref{cor:degree_two}, then $\phi^{[k]}=\Lambda SN^{(k)}$ for all $k\ge2$. In particular $\phi^{[2]}=\Lambda S\,(\ell_im_i)_i$.
\end{lemma}

\begin{proof}
The first claim is Part~\ref{cor:linear_structure} of Corollary~\ref{cor:degree_two} for the children of $q_i$, which have no product gate $q_{i'}$ with $i'\ge i$ as a predecessor. Expanding $u_iw_i$, for $k\ge2$,
\begin{displaymath}
    q_i^{[k]}=c_iw_i^{[k]}+c'_iu_i^{[k]}+N_i^{(k)}=\sum_{j<i}T_{ij}q_j^{[k]}+N_i^{(k)},
\end{displaymath}
so $(I-T)q^{[k]}=N^{(k)}$. Then $u_i^{[k]}=\sum_{j<i}\mu_{ij}(SN^{(k)})_j$ lies in $\spn\{N_j^{(k)}:\ j<i\}$ because $S$ is lower unitriangular, and the same holds for $w_i^{[k]}$. Finally $\phi^{[k]}=\Lambda q^{[k]}=\Lambda SN^{(k)}$, since $a$ has degree at most $1$.
\end{proof}

\section{Bilinear maps computed with as many products as outputs}\label{sec:bilinear}

Throughout this section and the next two, $\Rg=\F$ is a field and the variables are $(x,x')$.

\begin{lemma}[Bi-affinity]\label{lem:bi-affine}
Let $\phi=(\phi_1,\dots,\phi_K)$ be bilinear forms in $(x,x')$ that are linearly independent over $\F$, and $C_\phi=\spn_\F\{\phi_s\}$. Every circuit computing $\phi$, possibly with divisions, has at least $K$ product and division gates. If $\cir$ computes $\phi$ with exactly $K$ product and division gates $q_1,\dots,q_K$, then $\cir$ has no division gates and is bi-affine. Writing the value of $q_i$ as $(a_i+\ell_i(x))(b_i+m_i(x'))$:
\begin{enumerate}
    \item\label{lem:bi-affine:basis} $\ell_1m_1,\dots,\ell_Km_K$ form a basis of $C_\phi$;
    \item\label{lem:bi-affine:outputs} there are an invertible $\Lambda\in\F^{K\times K}$ and a vector $a$ of polynomials of degree at most $1$ with $\phi=\Lambda q+a$. Moreover $\phi=\Lambda(\ell_im_i)_i$ and $a=-\Lambda q^{[\le1]}$.
\end{enumerate}
\end{lemma}

\begin{proof}
If $\cir$ has division gates and $\F$ is finite, we view $\cir$ over an infinite extension $E$ of $\F$. The rational functions computed do not change. If $\cir$ has no division gates, let $\xi=0$. Otherwise take $\xi$ from Lemma~\ref{lem:divisions}, and take components with respect to $z'=(x,x')-\xi$. Since $\phi_s$ is bilinear, $\phi_s(\xi+z')$ is $\phi_s(z')$ plus a polynomial of degree at most $1$ in $z'$. By Part~\ref{cor:linear_structure} of Corollary~\ref{cor:degree_two} (Lemma~\ref{lem:divisions} with divisions) applied to the output gates, $C_\phi\subseteq\spn\{\hat q_1^{[\ge2]},\dots,\hat q_M^{[\ge2]}\}$, where $\phi_s$ is written in the variables $z'$, so $M\ge\dim C_\phi=K$.

Let $M=K$. Then the $\hat q_i^{[\ge2]}$ are linearly independent and span $C_\phi$. In particular each is a nonzero bilinear form in $z'$. So every $\hat q_i$ is a polynomial in $z'$ of degree exactly $2$. By Part~\ref{cor:linear_structure}, every gate computes a polynomial of degree at most $2$. Fix $i$.
\begin{itemize}
    \item If $q_i$ is a division gate with numerator $u$ and divisor $w$, then $\hat u=\hat q_i\hat w$ with $\deg\hat u\le2$. So $\hat w$ is a nonzero constant $\kappa$ and $\hat q_i^{[\ge2]}=\kappa^{-1}\hat u^{[\ge2]}$ lies in $\spn\{\hat q_{i'}^{[\ge2]}:i'<i\}$, a nontrivial linear relation. Impossible.
    \item Let $q_i$ be a product gate with children $u_1,\dots,u_s$. All $\hat u_j$ are nonzero polynomials. The top homogeneous component of $\hat q_i=\prod_j\hat u_j$ is the product of the top components of the $\hat u_j$. Since $\hat q_i$ has degree $2$, $\sum_j\deg\hat u_j=2$. If one child has degree $2$ and the others are nonzero constants with product $\kappa$, then $\hat q_i^{[\ge2]}=\kappa\,\hat u_j^{[\ge2]}$ lies in $\spn\{\hat q_{i'}^{[\ge2]}:i'<i\}$, which is impossible as before. Hence two children $u,w$ have degree $1$ and the others are constants with product $\kappa\neq0$. Write $\kappa\hat u=u_0+L(x)+N(x')$ and $\hat w=w_0+L'(x)+N'(x')$ in the variables $z'$. Then $\hat q_i^{[\ge2]}=LL'+NN'+(LN'+NL')$ is bilinear, so $LL'=NN'=0$. If $L=N=0$ or $L'=N'=0$, then $\hat q_i^{[\ge2]}=0$, which is impossible. So either $L'=N=0$ or $L=N'=0$. Then $\hat q_i$ is a product of an affine function of $x$ and an affine function of $x'$.
\end{itemize}
So $\cir$ has no division gates. If $\F$ was extended, we now rerun the argument with $\xi=0$ over $\F$ itself, so all constants and linear forms below lie in $\F$. An affine function of the shifted variables is an affine function of the original ones, so $\cir$ is bi-affine and $\hat q_i^{[\ge2]}=\ell_im_i$ in the original variables, which proves Part~\ref{lem:bi-affine:basis}. For Part~\ref{lem:bi-affine:outputs}, Corollary~\ref{cor:degree_two} gives $\phi=\Lambda q+a$. The degree $2$ parts give $\phi=\Lambda(\ell_im_i)_i$, so $\Lambda$ is a change of basis of $C_\phi$ and is invertible, and $a=\phi-\Lambda q=-\Lambda q^{[\le1]}$.
\end{proof}

\section{Polynomial multiplication and fault tolerance}\label{sec:toom}

\begin{theorem}\label{thm:bi-affine}
Every circuit computing $\phi_{k_1,k_2}$, possibly with divisions, has at least $K=k_1+k_2-1$ product and division gates. If $k_1,k_2\ge2$ and $\cir$ has exactly $K$ product and division gates, then $\cir$ is a Toom-Cook algorithm with a bi-affine modification (Definition~\ref{def:toom-cook}): there are pairwise distinct $\beta_1,\dots,\beta_K\in\F\cup\{\infty\}$, nonzero $\alpha_i,\alpha'_i$ and arbitrary $\gamma_i,\gamma'_i$ such that the $i$-th product gate computes $(\alpha_ip_1(\beta_i)+\gamma_i)(\alpha'_ip_2(\beta_i)+\gamma'_i)$, and $c=(DV)^{-1}q-(DV)^{-1}q^{[\le1]}$.
\end{theorem}

For $k_1=1$ the conclusion fails: $x_0x'_0$, $x_0x'_1$, $x_0(x'_0+x'_2)$ compute $c_0,c_1,c_2$ for $k_2=3$, but $x'_0+x'_2$ is not a multiple of any $p_2(\beta)$.

\begin{lemma}[Rank one Hankel matrices]\label{lem:hankel}
Let $k_1,k_2\ge2$ and let $u\in\F^{k_1}$, $v\in\F^{k_2}$ be nonzero with $u_{j+1}v_l=u_jv_{l+1}$ for all $0\le j\le k_1-2$, $0\le l\le k_2-2$. Then either $u_j=\alpha\beta^j$ and $v_l=\alpha'\beta^l$ for some $\beta\in\F$ and nonzero $\alpha,\alpha'$, or $u=\alpha e_{k_1-1}$ and $v=\alpha'e_{k_2-1}$.
\end{lemma}

\begin{proof}
Let $s$ and $r$ be the first indices with $u_s\neq0$ and $v_r\neq0$. If $s\ge1$ and $r\le k_2-2$, the relation at $(s-1,r)$ gives $u_sv_r=u_{s-1}v_{r+1}=0$. So $s=0$ or $r=k_2-1$, and symmetrically $r=0$ or $s=k_1-1$. If $s\ge1$, then $r=k_2-1\ge1$, hence $s=k_1-1$: the second alternative. If $s=0$, then $s\neq k_1-1$, hence $r=0$. With $\beta=u_1/u_0$ the relations at $j=0$ give $v_l=v_0\beta^l$, so $v_1=\beta v_0$, and those at $l=0$ give $u_j=u_0\beta^j$.
\end{proof}

\begin{proof}[Proof of Theorem~\ref{thm:bi-affine}]
The polynomials $c_s$ have pairwise disjoint sets of monomials, so they are linearly independent, and Lemma~\ref{lem:bi-affine} applies. The lower bound holds, $\cir$ has no division gates, $q_i=(a_i+\ell_i(x))(b_i+m_i(x'))$, and $\ell_im_i\in C$ is nonzero. Writing $\ell_i=\sum_jU_{ij}x_j$, $m_i=\sum_lW_{il}x'_l$ $\ell_im_i=\sum_sH_{is}c_s$, the coefficient and of $x_jx'_l$ gives $U_{ij}W_{il}=H_{i,j+l}$, so the rows of $U$ and $W$ satisfy the hypothesis of Lemma~\ref{lem:hankel}. Hence $\ell_i=\alpha_ip_1(\beta_i)$ and $m_i=\alpha'_ip_2(\beta_i)$ with $\beta_i\in\F\cup\{\infty\}$. By \eqref{eq:eval_product}, $(\ell_im_i)_i=DVc$. Both sides are bases of $C$, so $DV$ and $V$ are invertible and the $\beta_i$ are pairwise distinct. Lemma~\ref{lem:bi-affine} gives $c=\Lambda(\ell_im_i)_i=\Lambda DVc$, so $\Lambda=(DV)^{-1}$, and $c=\Lambda q-\Lambda q^{[\le1]}$.
\end{proof}

\begin{example}[Constants matter]\label{ex:winograd}
Winograd \cite[Theorem~3]{winograd1977some} states that every algorithm with $K$ multiplications is ``essentially the same'' as Toom-Cook, meaning equal up to replacing the products $p_1(\beta_i)p_2(\beta_i)$ by $(g\,p_1(\beta_i))(h\,p_2(\beta_i))$ with $g,h\in\F\setminus\{0\}$. His argument is carried out on a homogeneous presentation \cite{winograd1970number}, which discards constant terms. Once constants are allowed, the statement fails already for $k_1=k_2=1$: $(x_0+1)(x'_0+1)-x_0-x'_0-1$ computes $x_0x'_0$ with one product gate, whose value is not a multiple of $x_0x'_0$. Theorem~\ref{thm:bi-affine} is the correct statement, and it covers arbitrary circuits with product gates of unbounded fan-in and with divisions.
\end{example}

\paragraph{Several variables.} The same argument classifies the optimal circuits for multiplying polynomials in $l$ variables, and hence the multiplicative part of every $l$-level recursive Toom-Cook algorithm. The only new ingredient is the following extension of Lemma~\ref{lem:hankel}.

\begin{lemma}[Rank one Hankel tensors]\label{lem:hankel_multi}
Let $k_{1,t},k_{2,t}\ge2$ for $1\le t\le l$, and let $U\in\F^{B(k_1)}$ and $W\in\F^{B(k_2)}$ be nonzero with $U_{j+e_t}W_{j'}=U_jW_{j'+e_t}$ whenever all four indices lie in the boxes, where $e_t$ is the $t$-th unit vector. Then there are $\beta\in(\F\cup\{\infty\})^l$ and nonzero $\alpha,\alpha'$ with $\sum_jU_jx_j=\alpha\,p_1(\beta)$ and $\sum_jW_jx'_j=\alpha'\,p_2(\beta)$.
\end{lemma}

\begin{proof}
Fix $t$. For indices $j'',j'''$ of the coordinates other than $t$, the fibers $u=(U_{(s,j'')})_s$ and $v=(W_{(s,j''')})_s$ satisfy $u_{s+1}v_r=u_sv_{r+1}$. If both are nonzero, Lemma~\ref{lem:hankel} shows that $u$ is a multiple of $(\beta^s)_s$ or of the last unit vector, and $v$ of the corresponding vector with the same $\beta\in\F\cup\{\infty\}$. These vectors are pairwise non-proportional for different $\beta$. So $\beta=:\beta_t$ does not depend on the fibers: fixing one nonzero fiber of $W$ determines it for all nonzero fibers of $U$, and conversely. So, for every $t$, every fiber of $U$ in direction $t$ is a multiple of a fixed vector $g^{(t)}$. Then $U$ is a multiple of $g^{(1)}\otimes\dots\otimes g^{(l)}$. To see this, write $U=g^{(1)}\otimes U_1$ and apply the same to $U_1$. The same holds for $W$. Since $\sum_jU_jx_j$ is $p_1(\beta)$ when $U=\bigotimes_tg^{(t)}$ with $g^{(t)}=(\beta_t^s)_s$ or the last unit vector, the claim follows.
\end{proof}

\begin{theorem}\label{thm:multivariate}
Let $l\ge1$ and $k_{1,t},k_{2,t}\ge2$ for all $t$. Every circuit, possibly with divisions, that multiplies two polynomials in $l$ variables with boxes $B(k_1)$ and $B(k_2)$ has at least $K=\prod_t(k_{1,t}+k_{2,t}-1)$ product and division gates. If it has exactly $K$, it is a multivariate Toom-Cook algorithm with a bi-affine modification at a unisolvent set of points $\beta_1,\dots,\beta_K\in(\F\cup\{\infty\})^l$.
\end{theorem}

\begin{proof}
As in the proof of Theorem~\ref{thm:bi-affine}, with Lemma~\ref{lem:hankel_multi} in place of Lemma~\ref{lem:hankel}: the $c_s$ have disjoint monomials, Lemma~\ref{lem:bi-affine} gives $q_i=(a_i+\ell_i)(b_i+m_i)$ with $\ell_im_i\in C$, the coefficients of $\ell_i$ and $m_i$ satisfy the hypothesis of Lemma~\ref{lem:hankel_multi}. So $\ell_im_i=\alpha_i\alpha'_ip_1(\beta_i)p_2(\beta_i)$ and $(\ell_im_i)_i=DVc$. Both sides are bases of $C$, so $V$ is invertible and $\Lambda=(DV)^{-1}$.
\end{proof}

For $l\ge2$ the points need not form a grid. Any unisolvent set, for example a generic set of $K$ points, gives an optimal circuit. Theorem~\ref{thm:multivariate} says there are no others.

\begin{theorem}[Fault-tolerant multiplication]\label{thm:fault_tolerant}
Let $l\ge1$, $k_{1,t},k_{2,t}\ge2$, $h\ge0$, and let $q_1,\dots,q_{K+h}$ be products of two polynomials of degree at most $1$ in $(x,x')$ that tolerate $h$ faults (Definition~\ref{def:fault_tolerant}). Then there are points $\beta_1,\dots,\beta_{K+h}\in(\F\cup\{\infty\})^l$ in general position such that, up to the order of the factors, $q_i=(\alpha_ip_1(\beta_i)+\gamma_i)(\alpha'_ip_2(\beta_i)+\gamma'_i)$ with $\alpha_i,\alpha'_i\neq0$. Conversely, such products tolerate $h$ faults, and if $\F$ is infinite, points in general position exist for every $h$. For $l=1$, general position means pairwise distinct, which gives the uniqueness of the fault-tolerant constructions of \cite{nissim2024fault,nissim2025minimizing} for one level of recursion.
\end{theorem}

\begin{proof}
For $S$ with $|S|=K$, computing the factors of the $q_i$, $i\in S$, multiplying them and recombining gives a circuit with exactly $K$ product gates, so Theorem~\ref{thm:multivariate} applies to it. The point $\beta_i$ is determined by $q_i$, since $q_i^{[\ge2]}$ is a nonzero multiple of $p_1(\beta_i)p_2(\beta_i)=\sum_sv(\beta_i)_sc_s$, and $v(\beta)$ and $v(\beta')$ are not proportional for $\beta\neq\beta'$. Hence the same points serve for all $S$, and every $K$ of them are unisolvent. Conversely, for unisolvent $\{\beta_i:i\in S\}$ the matrix $V_S$ is invertible and $c=(D_SV_S)^{-1}(q_S-q_S^{[\le1]})$. Finally, $\det V$ is a nonzero polynomial in the coordinates of the points, since a grid is unisolvent (its matrix is a Kronecker product of Vandermonde of matrices). Over an infinite field, generic points make all $\binom{K+h}{K}$ these determinants nonzero.
\end{proof}

The recursive fault-tolerant scheme, which adds redundant evaluation points at every level, uses a grid of $\prod_t(\kappa_t+h)$ points. Theorem~\ref{thm:fault_tolerant} shows that, for tolerating $h$ faults among the products of an $l$-level multiplication, $K+h$ products suffice. It also shows that every scheme with $K+h$ products is an evaluation at points in general position. A grid $A_1\times\dots\times A_l$ with more than $K$ points is not in general position when $l\ge2$: if $|A_t|<\kappa_t$ for some $t$, all points lie in a space of dimension $\prod_t\min(|A_t|,\kappa_t)<K$. Otherwise $|A_t|>\kappa_t$ for some $t$, and for $s\neq t$ the more than $\prod_{t'\neq s}\kappa_{t'}$ points with a fixed $s$-th coordinate, completed to $K$ points, are not unisolvent. 

Over a finite field the classification turns into existence results, now for arbitrary circuits with unbounded fan-in and divisions. Fault tolerance with the minimal number of products is the MDS property of the evaluation code. Indeed, by Theorem~\ref{thm:fault_tolerant}, the columns $v(\beta_i)\in\F^K$ are $K+h$ vectors any $K$ of which form a basis.

\begin{corollary}[Finite fields]\label{cor:finite}
Let $\F=\F_q$, $l\ge1$ and $k_{1,t},k_{2,t}\ge2$.
\begin{enumerate}
    \item\label{cor:finite:exists} A circuit with exactly $K$ product and division gates for multiplying polynomials with boxes $B(k_1)$ and $B(k_2)$ exists if and only if $\kappa_t\le q+1$ for every $t$. Otherwise at least $K+1$ are needed.
    \item\label{cor:finite:ft} If $K+h$ products tolerate $h$ faults as in Theorem~\ref{thm:fault_tolerant}, then $K+h\le q+1$ for $l=1$, $h\le1$ whenever $K\ge q$, and $K+h\le q+1$ whenever $K\le p$, the characteristic of $\F_q$.
\end{enumerate}
\end{corollary}

\begin{proof}
For Part~\ref{cor:finite:exists}, by Theorem~\ref{thm:multivariate} an optimal circuit is an evaluation at $K$ unisolvent points of $(\F_q\cup\{\infty\})^l$, and a grid of $\kappa_t$ distinct values in each coordinate is unisolvent when $\kappa_t\le q+1$. Suppose $\kappa_t\ge q+2$. The polynomial $\tau_t^{\,q}-\tau_t$ lies in the output box and vanishes at every finite value of $\tau_t$. Its coefficient of $\tau_t^{\kappa_t-1}$ is zero. So it vanishes at every point of $(\F_q\cup\{\infty\})^l$. Its coefficient vector is orthogonal to every row $v(\beta)$, and no $K$ points are unisolvent. For Part~\ref{cor:finite:ft}, for $l=1$ the $K+h$ points are distinct elements of $\F_q\cup\{\infty\}$. In general, $K+h$ vectors of $\F_q^K$ any $K$ of which form a basis number at most $K+1$ if $K\ge q$ \cite{bush1952orthogonal}, and at most $q+1$ if $K\le p$ \cite{ball2012sets}.
\end{proof}

\section{Tubal rings and multiplication modulo a squarefree polynomial}\label{sec:tubal}

\begin{proposition}\label{prop:squarefree}
Let $P\in\R[\tau]$ be squarefree of degree $d$, with $r$ real roots and $c$ pairs of complex conjugate roots. Multiplication in $A_P=\R[\tau]/(P)$, in the basis $1,\tau,\dots,\tau^{d-1}$, makes $\R^d$ a tubal ring isomorphic to $\R^r\times\C^c$, and every tubal ring is isomorphic to some $A_P$. Over $\C$, $a\cdot b=a\star_Mb$ where $M=(\rho_i^{\,j})_{i,j}$ is the Vandermonde matrix of the roots $\rho_1,\dots,\rho_d$ of $P$.
\end{proposition}

\begin{proof}
By the Chinese remainder theorem, $A_P\cong\prod_j\R[\tau]/(P_j)$ over the irreducible factors $P_j$ of $P$, which is $\R^r\times\C^c$. A finite product of fields is commutative, unital and von Neumann regular: for $a$ take $b$ with $b_j=a_j^{-1}$ if $a_j\neq0$ and $b_j=0$ otherwise, so that $aba=a$. Conversely, $\R^r\times\C^c\cong A_P$ for $P=\prod_{t=1}^r(\tau-t)\prod_{\kappa=1}^c(\tau^2+\kappa^2)$. Over $\C$, evaluation at the roots, $a\mapsto Ma$, is an isomorphism $\C[\tau]/(P)\to\C^d$ with the entrywise product.
\end{proof}

So, by Theorem~\ref{thm:tubal_C} below, every optimal circuit over $\C$ for multiplication modulo a squarefree $P$ evaluates at the roots of $P$, up to a permutation, a scaling and bi-affine shifts.

\begin{theorem}\label{thm:tubal}
Let $\cdot$ be a bilinear product on $\F^d$ with a two-sided unit $e$, and let $\cir$ compute the coordinates of $a\cdot b$, possibly with divisions. Then $\cir$ has at least $d$ product and division gates. If it has exactly $d$, then $\cir$ is bi-affine and there is an invertible $N\in\F^{d\times d}$ such that $a\cdot b=a\star_Nb$ and $\cir$ is a $\star_N$ algorithm with a bi-affine modification (Definition~\ref{def:star_N_affine}).
\end{theorem}

\begin{proof}
The coordinates $\phi_k$ of $a\cdot b$ are linearly independent: substituting $a=e$ in $\sum_k\lambda_k\phi_k=0$ gives $\lambda=0$. By Lemma~\ref{lem:bi-affine}, after ordering the factors, $q_i=(c_i+\ell_i(a))(c'_i+m_i(b))$ and $a\cdot b=\Lambda((La)\odot(Pb))$ with $\Lambda$ invertible, where the rows of $L$ and $P$ are the coefficient vectors of $\ell_i$ and $m_i$. Let $D_1=\operatorname{diag}(Le)$ and $D_2=\operatorname{diag}(Pe)$. Substituting $a=e$ and $b=e$ gives $\Lambda D_1P=I$ and $\Lambda D_2L=I$, so $D_1,D_2,L,P$ are invertible and $L=D_2^{-1}D_1P$. With $N=D_2^{-1}P$ we get $La=D_1Na$, $Pb=D_2Nb$, and
\begin{displaymath}
    a\cdot b=\Lambda D_1D_2\bigl((Na)\odot(Nb)\bigr)=P^{-1}D_2\bigl((Na)\odot(Nb)\bigr),
\end{displaymath}
which is $a\star_Nb$. Finally $\ell_i(a)=(D_1)_{ii}(Na)_i$, $m_i(b)=(D_2)_{ii}(Nb)_i$ and $\Lambda=N^{-1}D_2^{-1}D_1^{-1}$.
\end{proof}

\begin{lemma}[Uniqueness of the transform]\label{lem:star_unique}
If $M,N\in E^{d\times d}$ are invertible and $a\star_Nb=a\star_Mb$ for all $a,b\in E^d$, then $N=PM$ for a permutation matrix $P$.
\end{lemma}

\begin{proof}
$T=NM^{-1}$ satisfies $T(u\odot v)=(Tu)\odot(Tv)$, and so does $T^{-1}$. Hence $T$ permutes the idempotents $\{0,1\}^d$ of $(E^d,\odot)$ and preserves the relation $v\odot u=v$, so it permutes the primitive idempotents $e_1,\dots,e_d$. Thus $T$ is a permutation matrix.
\end{proof}

\begin{theorem}\label{thm:tubal_C}
Let $A$ be a tubal ring with $a\cdot b=a\star_Mb$, $M\in\C^{d\times d}$, and let $\cir$ be a circuit over $\C$, possibly with divisions, computing the coordinates of $a\star_Mb$, the product of $A\otimes_\R\C$ (Fact~\ref{fact:complexification}). (A circuit over $\C$ that computes $a\cdot b$ for all real $a,b$ computes these polynomials, since polynomials that agree on $\R^{2d}$ are equal.) Then $\cir$ has at least $d$ product and division gates, and if it has exactly $d$, there are a permutation $\pi$, nonzero $\alpha_i,\alpha'_i$ and arbitrary $\gamma_i,\gamma'_i$ such that the $i$-th product gate computes $(\alpha_i(Ma)_{\pi(i)}+\gamma_i)(\alpha'_i(Mb)_{\pi(i)}+\gamma'_i)$ and $a\star_Mb=M^{-1}P^{-1}D^{-1}(q-q^{[\le1]})$, where $(Px)_i=x_{\pi(i)}$ and $D=\operatorname{diag}(\alpha_i\alpha'_i)$.
\end{theorem}

\begin{proof}
$\star_M$ on $\C^d$ is bilinear with unit $M^{-1}\mathbf{1}$. Theorem~\ref{thm:tubal} with $\F=\C$ gives $N$ with $\star_N=\star_M$, and Lemma~\ref{lem:star_unique} gives $N=PM$.
\end{proof}

\begin{corollary}[Real transforms]\label{cor:real_transform}
Let $M\in\R^{d\times d}$ be invertible, for example the cosine, sine, Hartley, Walsh-Hadamard or an orthogonal wavelet transform, or a real graph-based or learned transform. Every circuit over $\R$, possibly with divisions, computing $a\star_Mb$ has at least $d$ product and division gates, and if it has exactly $d$, there are a permutation $\pi$, nonzero $\alpha_i,\alpha'_i$ and arbitrary $\gamma_i,\gamma'_i$ such that the $i$-th product gate computes $(\alpha_i(Ma)_{\pi(i)}+\gamma_i)(\alpha'_i(Mb)_{\pi(i)}+\gamma'_i)$. Moreover, $\star_N=\star_M$ for an invertible $N$ if and only if $N$ is $M$ with permuted rows.
\end{corollary}

\begin{proof}
The unit of $\star_M$ is $M^{-1}\mathbf{1}$. Theorem~\ref{thm:tubal} over $\F=\R$ gives $N$ with $\star_N=\star_M$, and Lemma~\ref{lem:star_unique} gives $N=PM$. Conversely $\star_{PM}=\star_M$.
\end{proof}

\begin{corollary}[Non-split real tubal rings]\label{cor:tubal_real}
If $M$ has $c\ge1$ pairs of conjugate rows, every circuit over $\R$, possibly with divisions, computing $a\star_Mb$ for $a,b\in\R^d$ has at least $d+1$ product and division gates. This includes the t-product for $d\ge3$.
\end{corollary}

\begin{proof}
The unit $M^{-1}\mathbf{1}$ is real, so Theorem~\ref{thm:tubal} applies over $\R$. With $d$ gates it gives a real $N$ with $\star_N=\star_M$ on $\R^d$, hence on $\C^d$ by bilinearity. By Lemma~\ref{lem:star_unique}, $M=P^{-1}N$ is real, a contradiction. For the DFT matrix, row $k$ is the conjugate of row $d-k$, and row $1$ is not real for $d\ge3$.
\end{proof}

The same argument identifies the optimal circuits for convolution over a finite abelian group, which includes cyclic and multidimensional cyclic convolution.

\begin{corollary}[Convolution over a finite abelian group]\label{cor:group}
Let $G$ be a finite abelian group of order $d$, and let $\F$ be a field whose characteristic does not divide $d$ and which contains a primitive root of unity of order the exponent of $G$, for example $\F=\C$. Let $\chi_1,\dots,\chi_d$ be the characters of $G$ and $\hat a(\chi)=\sum_ga_g\chi(g)$. Every circuit over $\F$, possibly with divisions, computing the convolution $(a*b)_g=\sum_ha_hb_{g-h}$ has at least $d$ product and division gates, and if it has exactly $d$, there are a permutation $\pi$, nonzero $\alpha_i,\alpha'_i$ and arbitrary $\gamma_i,\gamma'_i$ such that the $i$-th product gate computes $(\alpha_i\hat a(\chi_{\pi(i)})+\gamma_i)(\alpha'_i\hat b(\chi_{\pi(i)})+\gamma'_i)$. For $G=\mathbb{Z}/d$ this says that every optimal circuit for cyclic convolution over $\C$ is a DFT-based algorithm.
\end{corollary}

\begin{proof}
The matrix $M=(\chi_i(g))_{i,g}$ is invertible by the orthogonality of characters, and $M(a*b)=(Ma)\odot(Mb)$ because characters are multiplicative, so $a*b=a\star_Mb$ with unit $\delta_0$. Theorem~\ref{thm:tubal} gives $N$ with $\star_N=\star_M$, and Lemma~\ref{lem:star_unique} gives $N=PM$.
\end{proof}

The bi-affinity of optimal circuits in Theorems~\ref{thm:tubal_C} and~\ref{thm:tubal_real} relies on semisimplicity. For the dual numbers $\F[\tau]/(\tau^2)$, with outputs $x_0x'_0$ and $x_0x'_1+x_1x'_0$, the bilinear rank is $3$. So by Lemma~\ref{lem:bi-affine} every circuit needs $3$ product gates, but
\begin{displaymath}
    x_0x'_1+x_1x'_0=x_0(x_1+x'_1)+x_1(x'_0-x_0),
\end{displaymath}
together with the product $x_0x'_0$, is an optimal circuit that is not bi-affine, over every field.

\subsection{The exact count over \texorpdfstring{$\R$}{}}

Corollary~\ref{cor:tubal_real} is not tight when $c\ge2$. To get the exact count and the structure of optimal circuits over $\R$, we lift the classification of minimal homogeneous quadratic algorithms to circuits. By Proposition~\ref{prop:squarefree} and a change of basis on the inputs and outputs, we may work with $\R^r\times\C^c$ in its standard coordinates: an input $x$ has coordinates $x_1,\dots,x_r$ and $x_{\kappa,1},x_{\kappa,2}$ for $1\le\kappa\le c$, representing $x_{\kappa,1}+ix_{\kappa,2}\in\C$, and the coordinates of $xx'$ are $x_tx'_t$ and
\begin{displaymath}
    f_{\kappa,1}=x_{\kappa,1}x'_{\kappa,1}-x_{\kappa,2}x'_{\kappa,2},\qquad f_{\kappa,2}=x_{\kappa,1}x'_{\kappa,2}+x_{\kappa,2}x'_{\kappa,1}.
\end{displaymath}
We call the $r+c$ factors of $\R^r\times\C^c$ the \emph{blocks}, and write $X_g$ and $X'_g$ for the linear forms in the coordinates of $x$ and of $x'$ of the block $g$. A \emph{homogeneous quadratic algorithm} with $m$ products is a list of products $L_1M_1,\dots,L_mM_m$ of linear forms in $(x,x')$ whose span contains the outputs.

\begin{fact}[Quadratic algorithms]\label{fact:quadratic}
Every homogeneous quadratic algorithm for the product of the ring $\R^r\times\C^c$ has at least $r+3c$ products \cite[Theorem~5]{winograd1977some}. Consider an algorithm with exactly $r+3c$ products. After exchanging $L_i$ and $M_i$ for some $i$, every $L_i$ lies in $X_g$ and $M_i$ in $X'_g$ for a block $g=g(i)$. Each real block carries one product and each complex block three, and the products are linearly independent \cite[Theorem~5]{winograd1977some}, \cite[Lemma~3 and Theorem~5]{feig1981systems}.
\end{fact}

Since the blocks use disjoint variables, the outputs of each block lie in the span of the products of that block.

\begin{lemma}[Complex blocks]\label{lem:complex_block}
Let $\ell_1m_1,\ell_2m_2,\ell_3m_3$ be linearly independent, with $\ell_t$ linear forms in $(x_1,x_2)$ and $m_t$ in $(x'_1,x'_2)$, whose span contains $W=\spn\{f_1,f_2\}$, $f_1=x_1x'_1-x_2x'_2$, $f_2=x_1x'_2+x_2x'_1$. Then:
\begin{enumerate}
    \item\label{lem:cb:prop} the $\ell_t$ are pairwise non-proportional, and so are the $m_t$;
    \item\label{lem:cb:kernel} if $f_s=\sum_t\Lambda_{st}\ell_tm_t$, the kernel of the $2\times3$ matrix $\Lambda$ is spanned by a vector with no zero entry.
\end{enumerate}
\end{lemma}

\begin{proof}
The coefficient matrix of $sf_1+tf_2$ is $\bigl(\begin{smallmatrix}s&t\\t&-s\end{smallmatrix}\bigr)$, with determinant $-(s^2+t^2)$, so every nonzero element of $W$ has rank $2$ as a bilinear form, while a product $\ell m$ has rank at most $1$. For Part~\ref{lem:cb:prop}, if $m_1$ and $m_2$ are proportional, then $\spn\{\ell_1m_1,\ell_2m_2\}$ consists of products $\ell m_1$, and it meets $W$ in a nonzero element since $2+2>3$. Similarly for the $\ell_t$. For Part~\ref{lem:cb:kernel}, $\Lambda$ has rank $2$. Let $w$ span its kernel and let $\omega$ be the functional on $U=\spn\{\ell_tm_t\}$ with $\omega(\ell_tm_t)=w_t$. Then $\omega(f_s)=\sum_t\Lambda_{st}w_t=0$, so the $2$-dimensional kernel of $\omega$ is $W$. If $w_t=0$, then $\ell_tm_t\in W$, which is impossible.
\end{proof}

Inside a complex block, optimal bilinear algorithms are not unique, but they form a single family. The minimal bilinear algorithms for multiplication in algebraic extension fields were classified by Winograd \cite{winograd1977some,winograd1979multiplication}. The classification is up to equivalences that include replacing $xx'$ by $(\alpha\beta)^{-1}(\alpha x)(\beta x')$ for invertible $\alpha,\beta$ \cite[Section~4]{averbuch1991classification}. Feig showed that all minimal division-free algorithms for these systems are bilinear \cite{feig1981systems,feig1983certain}. See also \cite{averbuch1986classification,averbuch1991classification,degroote1983characterization,degroote1987lectures}. Lemma~\ref{lem:cmul} is the case of $\C$ over $\R$, made explicit. We include a short self-contained proof. For $z=z_0+iz_1\in\C$ and $\beta\in\R\cup\{\infty\}$ write $p_z(\beta)=z_0+\beta z_1$ and $p_z(\infty)=z_1$, the evaluations of $z_0+z_1\tau$. The Toom-Cook algorithm for $\C\cong\R[\tau]/(\tau^2+1)$ at distinct $\beta_1,\beta_2,\beta_3\in\R\cup\{\infty\}$ computes the products $p_z(\beta_t)p_{z'}(\beta_t)$, interpolates the coefficients of $(z_0+z_1\tau)(z'_0+z'_1\tau)$ and reduces modulo $\tau^2+1$; Gauss's three-multiplication algorithm is the case $\{0,1,\infty\}$.

\begin{lemma}[Optimal algorithms for complex multiplication]\label{lem:cmul}
Let $\ell_t,m_t:\C\to\R$, $t=1,2,3$, be $\R$-linear forms and $w_t\in\C$ with $xx'=\sum_tw_t\ell_t(x)m_t(x')$ for all $x,x'\in\C$. Then there are $\gamma\in\C$ with $|\gamma|=1$, distinct $\beta_1,\beta_2,\beta_3\in\R\cup\{\infty\}$ and nonzero $\alpha_t,\alpha'_t\in\R$ with
\begin{displaymath}
    \ell_t(x)=\alpha_t\,p_{\gamma x}(\beta_t),\qquad m_t(x')=\alpha'_t\,p_{x'}(\beta_t).
\end{displaymath}
That is, the algorithm computes $(\gamma x)x'$ by the Toom-Cook algorithm at $\beta_1,\beta_2,\beta_3$ followed by reduction modulo $\tau^2+1$, and multiplies the result by $\bar\gamma$. Conversely, every such choice is an optimal algorithm.
\end{lemma}

\begin{proof}
Every nonzero $\R$-linear form $\C\to\R$ is $z\mapsto\operatorname{Re}(az)$ for a unique $a\neq0$. The forms are nonzero since the rank of complex multiplication is $3$. So $\ell_t(x)=\operatorname{Re}(a_tx)$ and $m_t(x')=\operatorname{Re}(b_tx')$. Write $\operatorname{Re}(u)=\tfrac12(u+\bar u)$ and expand. Since $x,\bar x,x',\bar x'$ are obtained from $x_0,x_1,x'_0,x'_1$ by an invertible change of variables over $\C$, comparing the coefficients of $xx'$, $\bar x\bar x'$, $x\bar x'$ and $\bar xx'$ gives
\begin{gather*}
    \sum_tw_ta_tb_t=4,\qquad\sum_tw_t\bar a_t\bar b_t=0,\\
    \sum_tw_ta_t\bar b_t=0,\qquad\sum_tw_t\bar a_tb_t=0 .
\end{gather*}
Let $P,Q,R,I\in\R^3$ be the real and imaginary parts of $(a_tb_t)_t$ and of $(a_t\bar b_t)_t$, and $w=w'+iw''$ with $w',w''\in\R^3$. The last two equations give $w\cdot R=w\cdot I=0$ and the first two give $w\cdot P=2$, $w\cdot Q=-2i$, where $\cdot$ is the real dot product extended linearly. Hence $w'$ and $w''$ are orthogonal to $R$ and $I$, $w'\cdot P=2$, $w''\cdot P=0$ and $w''\cdot Q=-2$. If $R$ and $I$ were linearly independent, $w'$ and $w''$ would lie on one line $\R n$, and $w'\cdot P=2$ would give $n\cdot P\neq0$, hence $w''=0$, contradicting $w''\cdot Q=-2$. So the nonzero numbers $a_t\bar b_t$ lie on one line $\R\gamma$ with $|\gamma|=1$, that is, $a_t=\lambda_t\gamma b_t$ with $\lambda_t\in\R\setminus\{0\}$. Every $b\in\C\setminus\{0\}$ is $c(1-i\beta)$ with $c\in\R\setminus\{0\}$ and $\beta\in\R$, or $-ic$ (we set $\beta=\infty$), and $\operatorname{Re}((1-i\beta)z)=p_z(\beta)$, $\operatorname{Re}(-iz)=p_z(\infty)$. This gives the formulas with $\alpha_t=\lambda_tc_t$ and $\alpha'_t=c_t$. The three products are linearly independent, as the rank is $3$, so the $\beta_t$ are distinct. They span the coefficients of $(z_0+z_1\tau)(z'_0+z'_1\tau)$ for $z=\gamma x$ and $z'=x'$. Since $xx'=\bar\gamma\,(\gamma x)x'$ and the products are independent, the $w_t$ are those of Toom-Cook, followed by reduction modulo $\tau^2+1$ and multiplication by $\bar\gamma$. The converse is clear.
\end{proof}

\begin{theorem}\label{thm:tubal_real}
Let $A$ be a tubal ring isomorphic to $\R^r\times\C^c$, for example $\R[\tau]/(P)$ for a squarefree $P$ with $r$ real roots and $c$ pairs of complex roots, and let $\cir$ be a circuit over $\R$ whose product gates have fan-in $2$, computing the coordinates of $a\cdot b$. Then $\cir$ has at least $r+3c$ product gates. If it has exactly $r+3c$, then $\cir$ is bi-affine and $q_i=(c_i+\ell_i(a))(c'_i+m_i(b))$. In the coordinates of $\R^r\times\C^c$, the bilinear products $\ell_im_i$ split along the blocks. Each real block $t$ carries one product, a nonzero multiple of $a_tb_t$. Each complex block carries three products of linear forms in its own coordinates. They form an optimal bilinear algorithm for multiplication in $\C$, described by Lemma~\ref{lem:cmul}: a Toom-Cook algorithm at three distinct points of $\R\cup\{\infty\}$, applied to $\gamma a_\kappa$ and $b_\kappa$ for some $|\gamma|=1$ and followed by reduction modulo $\tau^2+1$. The outputs are $\Lambda(q-q^{[\le1]})$ for the matrix $\Lambda$ with $a\cdot b=\Lambda(\ell_im_i)_i$.
\end{theorem}

\begin{proof}
An algebra isomorphism $\psi:A\to\R^r\times\C^c$ is a linear map. Composing $\cir$ with $\psi^{-1}$ on each input and $\psi$ on the outputs changes neither the product gates nor bi-affinity. So let $A=\R^r\times\C^c$ in standard coordinates, write $x,x'$ for the inputs, and let $q_1,\dots,q_m$ be the product gates in topological order, $q_i=u_iw_i$. Let $c_i,\ell_i$ and $c'_i,m_i$ be the components of degree $0$ and $1$ of $u_i$ and $w_i$.

\emph{Step 1: a triangular recursion.} By Lemma~\ref{lem:graded}, with its notation $\mu_{ij},\mu'_{ij},T,S,N^{(k)}$, and with $\Lambda'=\Lambda S$, where the outputs are $\phi=\Lambda q+a$,
\begin{equation}\label{eq:real_rec}
    \phi=\Lambda'\bigl(\ell_im_i\bigr)_i,\qquad \Lambda'N^{(k)}=0\quad(k\ge3),
\end{equation}
since $\phi$ is homogeneous of degree $2$, and $u_i^{[k]},w_i^{[k]}\in\spn\{N_j^{(k)}:\ j<i\}$ for $k\ge2$.

\emph{Step 2: the quadratic part.} By \eqref{eq:real_rec}, the products $\ell_im_i$ form a homogeneous quadratic algorithm, so $m\ge r+3c$ by Fact~\ref{fact:quadratic}. This lower bound is also a special case of \cite{alder1981algorithmic}. Let $m=r+3c$. By Fact~\ref{fact:quadratic}, after exchanging $u_i$ and $w_i$ for some $i$, $\ell_i\in X_g$ and $m_i\in X'_g$ for $g=g(i)$, and the $\ell_im_i$ are linearly independent. Hence $\Lambda'$ is unique and block diagonal, with a nonzero $1\times1$ block for each real block and a $2\times3$ block as in Lemma~\ref{lem:complex_block} for each complex block. Its kernel is the direct sum of the lines spanned by the vectors $w^{(g)}$ of Lemma~\ref{lem:complex_block} for the complex blocks $g$.

\emph{Step 3: degree three.} Write
\begin{displaymath}
    u_i^{[2]}=\sum_j\beta_{ij}\ell_jm_j,\qquad w_i^{[2]}=\sum_j\beta'_{ij}\ell_jm_j.
\end{displaymath}
Then $N_i^{(3)}=\ell_iw_i^{[2]}+m_iu_i^{[2]}$, whose two terms have degrees $(2,1)$ and $(1,2)$ in $(x,x')$. With $\nu_{sj}=\sum_i\Lambda'_{si}\beta_{ij}m_i\in X'$, the part of degree $(1,2)$ of $\Lambda'N^{(3)}=0$ reads
\begin{displaymath}
    \sum_j\ell_jm_j\nu_{sj}=0\qquad\text{for all }s.
\end{displaymath}
for all Since $\ell_j\in X_{g(j)}$ and the spaces $X_g$ are independent, for every block $g$
\begin{displaymath}
    \sum_{j:\,g(j)=g}\ell_j\,(m_j\nu_{sj})=0.
\end{displaymath} For a real block this gives $\nu_{sj}=0$. For a complex block with products $j_1,j_2,j_3$, Lemma~\ref{lem:complex_block} gives $\ell_{j_3}=\alpha\ell_{j_1}+\alpha'\ell_{j_2}$ with $\alpha,\alpha'\neq0$ and $\ell_{j_1},\ell_{j_2}$ independent, so $m_{j_1}\nu_{sj_1}=-\alpha\,m_{j_3}\nu_{sj_3}$ and $m_{j_2}\nu_{sj_2}=-\alpha'\,m_{j_3}\nu_{sj_3}$. If $\nu_{sj_3}\neq0$, unique factorization and the non-proportionality of the $m_{j_t}$ give $\nu_{sj_3}\in\R m_{j_1}\cap\R m_{j_2}=0$, a contradiction. So all three vanish. Thus $\nu_{sj}=0$ for all $s,j$: for every $j$, the vector $(\beta_{ij}m_i)_i$ of linear forms lies, coordinatewise, in the kernel of $\Lambda'$. Hence $\beta_{ij}=0$ for $i$ in a real block, and for a complex block $g$ with products $i_1,i_2,i_3$ there is $\nu\in X'$ with $\beta_{i_tj}m_{i_t}=w^{(g)}_t\nu$. As $w^{(g)}_t\neq0$ and the $m_{i_t}$ are pairwise non-proportional, $\nu=0$. So $\beta=0$, and by the symmetric argument in degree $(2,1)$, $\beta'=0$: $u_i^{[2]}=w_i^{[2]}=0$ for all $i$.

\emph{Step 4: higher degrees.} Suppose $u_i^{[s]}=w_i^{[s]}=0$ for all $i$ and $2\le s<k$, for some $k\ge3$. Every term of $N_i^{(k)}$ has a factor $u_i^{[s]}$ or $w_i^{[s]}$ with $2\le s\le k-1$, so $N^{(k)}=0$, hence $q^{[k]}=0$ and $u_i^{[k]}=w_i^{[k]}=0$. By induction, $u_i=c_i+\ell_i$ and $w_i=c'_i+m_i$, so $\cir$ is bi-affine. Then $q^{[\ge2]}=(\ell_im_i)_i$ is linearly independent, so $\mu=\mu'=0$, $T=0$, $\Lambda'=\Lambda$, and $\phi=\Lambda(q-q^{[\le1]})$ by Corollary~\ref{cor:degree_two}. The block structure is Step~2, and Lemma~\ref{lem:cmul} describes each complex block. For a real block $t$, the product $\ell m$ spans the output $x_tx'_t$, so it is a nonzero multiple of it.
\end{proof}

\begin{theorem}[Complex multiplication over $\R$]\label{thm:cmul}
Every circuit over $\R$ computing the real and imaginary parts of $xx'$ for $x,x'\in\C$, with product gates of unbounded fan-in and possibly with divisions, has at least $3$ product and division gates. If a circuit whose product gates have fan-in $2$ has exactly $3$ product gates, there are $\gamma\in\C$ with $|\gamma|=1$, distinct $\beta_1,\beta_2,\beta_3\in\R\cup\{\infty\}$, nonzero $\alpha_t,\alpha'_t\in\R$ and arbitrary $\delta_t,\delta'_t\in\R$ such that, up to the order of the factors, the $t$-th product gate computes
\begin{displaymath}
    q_t=\bigl(\alpha_t\,p_{\gamma x}(\beta_t)+\delta_t\bigr)\bigl(\alpha'_t\,p_{x'}(\beta_t)+\delta'_t\bigr),
\end{displaymath}
and $xx'=\bar\gamma\,\rho\bigl((DV)^{-1}(q-q^{[\le1]})\bigr)$, where $V$ is the Vandermonde matrix of Definition~\ref{def:toom-cook} for $K=3$, $D=\operatorname{diag}(\alpha_t\alpha'_t)$, and $\rho(c_0,c_1,c_2)=(c_0-c_2)+ic_1$ is reduction modulo $\tau^2+1$. Conversely, every such circuit computes $xx'$. Thus, up to the rotation $x\mapsto\gamma x$, every optimal circuit is Gauss's algorithm with its points $0,1,\infty$ replaced by any three distinct points of $\R\cup\{\infty\}$. Its evaluations are scaled and shifted by constants.
\end{theorem}

\begin{proof}
The lower bound for arbitrary circuits is Corollary~\ref{cor:tubal_real} with $d=2$: with two gates, Lemma~\ref{lem:bi-affine} would give two products $\ell_1m_1,\ell_2m_2$ forming a basis of the output space, but every nonzero element of that space has rank $2$ as a bilinear form (Lemma~\ref{lem:complex_block}), while $\ell_1m_1$ has rank $1$. Consider circuits with exactly $3$ product gates of fan-in $2$. Theorem~\ref{thm:tubal_real} with $r=0$ and $c=1$ shows that the circuit is bi-affine, with $q_t=(c_t+\ell_t(x))(c'_t+m_t(x'))$. It also shows that the $\ell_tm_t=q_t-q_t^{[\le1]}$ form an optimal bilinear algorithm. Lemma~\ref{lem:cmul} gives $\ell_t=\alpha_tp_{\gamma x}(\beta_t)$ and $m_t=\alpha'_tp_{x'}(\beta_t)$, so $(\ell_tm_t)_t=DVc$, where $c$ is the coefficient vector of $(z_0+z_1\tau)(z'_0+z'_1\tau)$ for $z=\gamma x$ and $z'=x'$ by \eqref{eq:eval_product}. Hence $(\gamma x)x'=\rho(c)=\rho((DV)^{-1}(q-q^{[\le1]}))$. The converse is a direct computation.
\end{proof}

\subsection{Unbounded fan-in: one spare gate}

For circuits with product gates of fan-in two, the bound $r+3c$ follows from the Alder-Strassen bound $2\dim A-t(A)$ on the multiplicative complexity \cite{alder1981algorithmic}, since $\R^r\times\C^c$ has $r+c$ maximal ideals. A product gate of unbounded fan-in may replace several multiplications, and the bound does not apply directly. With one spare gate beyond the number of outputs, the cancellation of higher-degree terms can be controlled directly. A product gate is \emph{removable} if it computes $0$ or has at most one input edge carrying a non-constant polynomial. Such a gate can be replaced by a constant or by a sum gate. Inputs are counted with multiplicity.

\begin{lemma}[One spare gate]\label{lem:spare}
Let $\phi$ be a quadratic map whose $K$ coordinates are linearly independent quadratic forms, and whose span $W$ contains no nonzero product of two linear forms and contains a form of rank at least $4$. Let $\cir$ be a division-free circuit over $\R$ computing $\phi$ with product gates $q_1,\dots,q_m$ of unbounded fan-in, none of them removable, such that $\dim\spn\{q_i^{[\ge2]}\}\le K+1$. For example $m\le K+1$. Then every product gate of $\cir$ has exactly two non-constant inputs, both affine.
\end{lemma}

\begin{proof}
Let $f\in W$ have rank at least $4$. Order the product gates topologically and let $V=\spn\{q_i^{[\ge2]}\}$. By Corollary~\ref{cor:degree_two}, $W\subseteq V$, and every gate and every input of a product gate lies in $\mathcal{A}=V+\{\text{affine functions}\}$affine functions . Every product gate has degree at least $2$, since it is not removable.

\emph{$\dim V=K$ is impossible.} Then $V=W$, so every $q_i^{[\ge2]}$ is a quadratic form and every product gate has degree $2$. Its inputs have degrees adding up to $2$, and as it is not removable it has exactly two non-constant inputs, both affine. Its part of degree $2$ is then a nonzero product of two linear forms in $W$, which is impossible. Hence $\dim V=K+1$ and $V=W\oplus\R g$. Write $q_i^{[\ge2]}=w_i+r_ig$ with $w_i\in W$. If $r_i=0$, then $q_i$ has degree $2$, exactly two of its inputs are affine and non-constant, and $q_i^{[2]}=w_i$ is a product of two linear forms in $W$, which is impossible. So $r_i\neq0$ for all $i$. Let $D$ be the degree of $q_1$. If $D=2$, every element of $V$ is a quadratic form, and every product gate has degree $2$. As before, such a gate has exactly two non-constant inputs, both affine, which is the claim. It remains to exclude $D\ge3$.

\emph{Case $D\ge3$.} Every element of $V\setminus W$ has degree $D$, and every product gate has degree $D$, as $r_i\neq0$. An input of a product gate lies in $\mathcal{A}$, so it has degree $D$ or it is affine plus an element $B\in W$. An input of degree $D$ together with a second non-constant input would give degree more than $D$, so all inputs are of the second kind. The top homogeneous component of $q_i$ is $r_i$ times that of $g$, which is proportional to that of $q_1$. The inputs of $q_1$ are affine, so this top component is a product $\prod_{j=1}^D\ell_j$ of linear forms. If an input of $q_i$ had $B\neq0$, then $B$ would divide $\prod_j\ell_j$, so by unique factorization $B$ would be a product of two linear forms in $W$. Hence all inputs are affine, and by unique factorization their linear parts are, up to scalars and order, $\ell_1,\dots,\ell_D$:
\begin{gather*}
    q_i=\kappa_i\prod_{j=1}^D(c_{ij}+\ell_j),\qquad q_i^{[k]}=\kappa_iE_k(c_i),\\
    E_k(c_i)=\sum_{|T|=k}\Bigl(\prod_{j\notin T}c_{ij}\Bigr)\prod_{j\in T}\ell_j .
\end{gather*}
Let $L=\spn\{\ell_j\}$. The quadratic forms $q_i^{[2]}$ are quadratic forms in the elements of $L$, so their rank is at most $\dim L$. As $f\in W\subseteq\spn\{q_i^{[2]}\}$ has rank at least $4$, $\dim L\ge4$ and $D\ge4$. For $k\ge3$, $q_i^{[k]}=r_ig^{[k]}$. Comparing degree $D$ gives $\kappa_i=\kappa r_i$ for a constant $\kappa$, and then $E_k(c_i)$ does not depend on $i$ for $3\le k\le D$. Divide $E_{D-p}(c_i)$ by $\prod_j\ell_j$. Then the elementary symmetric functions of degree $p\le D-3$ of the rational functions $c_{ij}/\ell_j$ do not depend on $i$. By Newton's identities, neither do their power sums $\sum_jc_{ij}^p\ell_j^{-p}$. Group the $\ell_j$ into classes of proportional forms, $\ell_j=\alpha_j\lambda_G$ for $j\in G$. The functions $\lambda_G^{-p}$ for different classes are linearly independent. To see this, multiply a relation by $\prod_H\lambda_H^p$ and restrict it to the hyperplane $\lambda_G=0$. So $\sum_{j\in G}(c_{ij}/\alpha_j)^p$ does not depend on $i$, for every class $G$ and $p\le D-3$. There are at least $\dim L\ge4$ classes, so $|G|\le D-3$, and these power sums determine the multiset $\{c_{ij}/\alpha_j\}_{j\in G}$. Hence $\prod_{j\in G}(c_{ij}+\ell_j)$ does not depend on $i$, all $q_i$ are proportional to $q_1$, and $\dim V=1<K+1$, a contradiction.
\end{proof}

Consider a bilinear map in $(x,x')$. By the argument of Lemma~\ref{lem:bi-affine}, a product of two linear forms that is bilinear is $\ell(x)m(x')$. A bilinear form of rank $\rho$ has rank $2\rho$ as a quadratic form. So the hypotheses of Lemma~\ref{lem:spare} say that $W$ contains no product $\ell(x)m(x')$ and contains an element of rank at least $2$ as a bilinear form. Let $C_2(\phi)$ be the least number of products in a quadratic algorithm for $\phi$. By Strassen's homogenization, this is the least number of product gates in a division-free circuit with product gates of fan-in two \cite{strassen1973vermeidung}. Let $C_\infty(\phi)$ be the least number of product gates in a division-free circuit with product gates of unbounded fan-in.

\begin{corollary}\label{cor:spare}
Under the hypotheses of Lemma~\ref{lem:spare} on $\phi$, $C_\infty(\phi)\ge\min\{C_2(\phi),K+2\}$. In particular, $C_\infty(\phi)=C_2(\phi)$ whenever $C_2(\phi)\le K+2$.
\end{corollary}

\begin{proof}
Take a circuit with $C_\infty(\phi)$ product gates. None of them is removable. By Corollary~\ref{cor:degree_two}, $W\subseteq\spn\{q_i^{[\ge2]}\}$, so $C_\infty(\phi)\ge K$. If $C_\infty(\phi)\le K+1$, Lemma~\ref{lem:spare} shows that every product gate multiplies two affine functions. The parts of degree $2$ of the outputs are then combinations of the products of the linear parts of these functions. This is a quadratic algorithm with $C_\infty(\phi)$ products. So $C_2(\phi)\le C_\infty(\phi)$ in this case, and otherwise $C_\infty(\phi)\ge K+2$.
\end{proof}

For circuits of depth $3$, every product gate multiplies affine functions. For them the conclusion of Corollary~\ref{cor:spare} holds without any hypothesis on $W$. This follows from a lemma that allows any number of gates of high degree, as long as their parts of degree at least $3$ are proportional.

\begin{lemma}[One high direction]\label{lem:one_direction}
Let $\cir$ be a division-free circuit over $\R$ computing a quadratic map $\phi$, whose product gates $q_1,\dots,q_m$ multiply affine functions and are not removable. If the parts $q_i^{[\ge3]}$ span a space of dimension at most $1$, then $\phi$ has a quadratic algorithm with at most $m$ products.
\end{lemma}

\begin{proof}
Let the parts $q_i^{[\ge3]}$ lie in $\R H$. A gate with $q_i^{[\ge3]}=0$ has degree $2$, so it has exactly two non-constant inputs and $q_i^{[2]}$ is a product of two linear forms. Call these gates low and the others high, and let $h$ be the number of high gates. The high gates all have the degree $D\ge3$ of $H$, and their top homogeneous components are proportional to that of $H$. By unique factorization we may write, after scaling inputs, $q_i=r_iP_i$ with $P_i=\kappa\prod_{j=1}^D(c_{ij}+\ell_j)$ for all high gates. Here the linear forms $\ell_j$ and the scalar $\kappa$ are the same for all of them. The $P_i$ have the same parts of every degree $k\ge3$. As in the proof of Lemma~\ref{lem:spare}, group the $\ell_j$ into classes $G$ of proportional forms, $\ell_j=\alpha_j\lambda_G$ for $j\in G$, and let $\pi_{G,\rho}(i)=\sum_{j\in G}(c_{ij}/\alpha_j)^\rho$. The equality of the parts of degree $k\ge3$ gives $\pi_{G,\rho}(i)=\pi_{G,\rho}(i')$ for all $G$ and $\rho\le D-3$, and by Newton's identities
\begin{displaymath}
    P_i^{[2]}-P_{i'}^{[2]}=\pm\frac{\kappa}{D-2}\sum_G\bigl(\pi_{G,D-2}(i)-\pi_{G,D-2}(i')\bigr)\frac{\prod_j\ell_j}{\lambda_G^{D-2}}.
\end{displaymath}
If $|G|\le D-3$, the power sums of order at most $D-3$ determine the multiset $\{c_{ij}/\alpha_j\}_{j\in G}$. So the term of $G$ vanishes. If $|G|\ge D-2$, then $\prod_j\ell_j/\lambda_G^{D-2}=\prod_{j\in G}\alpha_j\cdot\lambda_G^{|G|-D+2}\prod_{j\notin G}\ell_j$ is a product of two linear forms. At most three classes have $|G|\ge D-2$: one if $D\ge5$, two if $D=4$, and three if $D=3$, since there are at most $D$ classes. So the differences $P_i^{[2]}-P_{i'}^{[2]}$ span a space $\Delta$ of dimension at most $h-1$ that lies in the span of $\sigma\le3$ products of linear forms. If $\sigma\le h$, $\Delta$ lies in the span of $h$ products. Otherwise $h\le2$. If $h=1$, $\Delta=0$, and if $h=2$, then $D=3$ and $\Delta$ is spanned by one form $a\ell_2\ell_3+b\ell_1\ell_3+c\ell_1\ell_2=\ell_1(b\ell_3+c\ell_2)+a\ell_2\ell_3$, a sum of $2$ products.

An output is $\sum_i\lambda_iq_i$ plus an affine function by Corollary~\ref{cor:degree_two}, with $\sum_{i\text{ high}}\lambda_ir_i=0$ since its parts of degree at least $3$ vanish. Its for any part of degree $2$ is $\sum_{i\text{ low}}\lambda_iq_i^{[2]}+\sum_{i\text{ high}}\lambda_ir_i\bigl(P_i^{[2]}-P_{i_0}^{[2]}\bigr)$ high gate $i_0$, which lies in the span of the $m-h$ low products and of $h$ further products.
\end{proof}

\begin{corollary}\label{cor:depth3}
For every quadratic map $\phi$ over $\R$ with $K$ linearly independent coordinates, every circuit of depth $3$ for $\phi$ with product gates of unbounded fan-in has at least $\min\{C_2(\phi),K+2\}$ product gates.
\end{corollary}

\begin{proof}
Take such a circuit with $m$ product gates, none removable, and suppose $m\le K+1$. The coefficient vectors of the $K$ outputs are linearly independent and annihilate the parts $q_i^{[\ge3]}$. So these parts span a space of dimension at most $m-K\le1$. Lemma~\ref{lem:one_direction} then gives $C_2(\phi)\le m$.
\end{proof}

\begin{corollary}[$2\times2$ matrices]\label{cor:matmul}
Every division-free circuit over $\R$ that multiplies two $2\times2$ matrices, with product gates of unbounded fan-in, has at least $6$ product gates.
\end{corollary}

\begin{proof}
The outputs span the space of bilinear forms $\operatorname{tr}(\Gamma AB)$, $\Gamma\in\R^{2\times2}$, and $\operatorname{tr}(\Gamma AB)$ has rank $2\operatorname{rank}\Gamma$ as a bilinear form in $(A,B)$. So the hypotheses of Lemma~\ref{lem:spare} hold with $K=4$, and $C_2=7$ \cite{winograd1971multiplication}; Corollary~\ref{cor:spare} gives at least $6$.
\end{proof}

Lemma~\ref{lem:spare} covers one spare gate. A second tool works for any number of spare gates: the substitution method applies to product gates of any fan-in, since a single affine restriction kills a whole product gate.

\begin{lemma}[Substitution]\label{lem:substitution}
Let $\phi$ be a quadratic map in $n$ variables over a field, computed by a division-free circuit with $m\ge1$ product gates of unbounded fan-in, none of them removable. Then there is a nonzero linear form $\ell$ such that the quadratic map $\phi|_{\ker\ell}$, in $n-1$ variables, is computed by a division-free circuit with $m-1$ product gates of unbounded fan-in. Hence $C_\infty(\phi)\ge1+\min_{\ell\neq0}C_\infty(\phi|_{\ker\ell})$.
\end{lemma}

\begin{proof}
The first product gate in topological order has only affine inputs, and as it is not removable one of them is $a+\ell$ with $\ell\neq0$. Restrict the inputs to the affine hyperplane $a+\ell=0$, parametrized as $x=x_0+Py$ with $P$ injective and image $\ker\ell$. This is a linear substitution, so it costs no product gates. The first product gate now computes $0$, and the circuit computes $\phi(x_0+Py)$, whose part of degree $2$ in $y$ is $\phi(Py)$. The parts of degree at most $1$ are removed by sum gates.
\end{proof}

The classical substitution proofs for product gates of fan-in two, such as the proof of the Alder-Strassen bound, choose which product to kill. Lemma~\ref{lem:substitution} only kills the products in topological order. But it lets us choose among all non-constant inputs of a gate.

For a single quadratic form, Lemma~\ref{lem:substitution} shows that unbounded fan-in never helps. In depth-$3$ circuits every product gate multiplies affine functions. For them the corresponding bound on the number of product gates is folklore \cite{shpilka2001depth,kumar2020power}. For $S^2_n$ the exact counts are due to Radhakrishnan, Sen and Vishwanathan \cite{radhakrishnan2000depth}. Lemma~\ref{lem:substitution} allows nested product gates.

\begin{theorem}[Single quadratic forms]\label{thm:single_form}
Let $f$ be a real quadratic form with $p$ positive and $n$ negative eigenvalues. Then $C_\infty(f)=C_2(f)=\max\{p,n\}$. Over $\C$, $C_\infty(f)=C_2(f)=\lceil\operatorname{rank}f/2\rceil$.
\end{theorem}

\begin{proof}
Over $\R$, write $f=\sum_{i\le p}u_i^2-\sum_{j\le n}v_j^2$ with linearly independent linear forms. The products $(u_i-v_i)(u_i+v_i)$ for $i\le\min\{p,n\}$ and the remaining squares give $\max\{p,n\}$ products. Over $\C$, $f=\sum_{i\le r}u_i^2$ with $r=\operatorname{rank}f$, and $u_1^2+u_2^2=(u_1+\mathrm{i}u_2)(u_1-\mathrm{i}u_2)$ gives $\lceil r/2\rceil$ products. For the lower bound, take a circuit with $C_\infty(f)$ product gates, none of them removable, and apply Lemma~\ref{lem:substitution} $C_\infty(f)$ times. Restricting a circuit to a hyperplane keeps product gates that are not removable or turns them into removable ones, which we replace. We obtain a subspace $U$ of codimension at most $C_\infty(f)$ on which the circuit has no product gates, so $f|_U=0$. By Cauchy interlacing, restricting a real quadratic form to a hyperplane lowers each of the numbers of positive and negative eigenvalues by at most $1$, so $\max\{p,n\}\le C_\infty(f)$. Over $\C$, restricting to a hyperplane lowers the rank by at most $2$, so $\operatorname{rank}f\le2C_\infty(f)$.
\end{proof}

 $e_2(x_1,\dots,x_n)=\frac12\bigl((\sum_ix_i)^2-\sum_ix_i^2\bigr)$, with one positive and $n-1$ For negative eigenvalues, this gives $n-1$ over $\R$, attained by $\sum_{i<n}x_i(x_{i+1}+\dots+x_n)$.

\begin{theorem}[$\C^c$ over $\R$ with unbounded fan-in]\label{thm:c2}
Let $c\ge2$. Every division-free circuit over $\R$, with product gates of unbounded fan-in, that computes the product of $\C^c$ (equivalently, of a tubal ring $\R^0\times\C^c$, such as multiplication modulo $\prod_{\kappa=1}^c(\tau^2+\kappa^2)$) has at least $2c+2$ product gates. In particular, every such circuit for $\C\times\C$ has at least $6$ product gates, which is optimal.
\end{theorem}

\begin{proof}
We use the coordinates of Section~\ref{sec:tubal}. The outputs span the space $W=W_1\oplus\dots\oplus W_c$, where $W_\kappa=\spn\{f_{\kappa,1},f_{\kappa,2}\}$, and an element of $W$ has a block diagonal coefficient matrix $G_1\oplus\dots\oplus G_c$ $G_\kappa=\bigl(\begin{smallmatrix}s_\kappa&t_\kappa\\t_\kappa&-s_\kappa\end{smallmatrix}\bigr)$. Its rank is twice the number of nonzero blocks. with So $W$ contains no product $\ell(x)m(x')$, and $f=\sum_\kappa f_{\kappa,1}$ has rank $2c$ and $x$-support of dimension $2c$. Lemma~\ref{lem:bi-affine} gives at least $2c$ product gates. Suppose a circuit has at most $2c+1$, and take one with the fewest product gates. Then none is removable, and by Lemma~\ref{lem:bi-affine} again (the case of $2c$ gates gives a rank one element of $W$) it has exactly $2c+1$. By Lemma~\ref{lem:spare}, every product gate multiplies two affine functions, $q_i^{[2]}=\kappa_iL_iM_i$, and by the proof of Lemma~\ref{lem:spare} the $q_i^{[2]}$ span $V=W\oplus\R g$ with $g$ a quadratic form and all $r_i\neq0$. Rescaling, $P_i=L_iM_i=g+w'_i$ with $w'_i\in W$, and the differences $P_i-P_1$ span $W$. All $P_i$ have the same part $A$ in $x$ only and the same part $E$ in $x'$ only. If $A\neq0$, then $A=\ell m$, and by unique factorization the $x$-parts of the factors of every $P_i$ are $c_i\ell$ and $c_i^{-1}m$. Then the bilinear parts of all $P_i$, hence all elements of $W$, have $x$-support in $\spn\{\ell,m\}$, while $f$ has $x$-support of dimension $2c\ge4$. So $A=0$, and symmetrically $E=0$. Hence $g$ and all $P_i$ are bilinear and $P_i=\ell_i(x)m_i(x')$ has rank one. In the block decomposition of coefficient matrices, the off-diagonal blocks of $P_i=g+w'_i$ are those of $g$. If the block $(\kappa,\kappa')$ of $g$ with $\kappa\neq\kappa'$ is nonzero, then the $\kappa$-th block of every $\ell_i$ is a nonzero multiple of a fixed vector $\alpha$. So the diagonal blocks $(\kappa,\kappa)$ of all $P_i$, and of their differences, have column space in $\R\alpha$. But the projections of these differences span $W_\kappa$, every nonzero element of which has rank $2$. So the off-diagonal blocks of $g$ vanish, every $P_i$ is supported in one diagonal block, and $W_\kappa$ lies in the span of the $P_i$ supported in block $\kappa$. Two products $\ell m$ cannot span a space containing $W_\kappa$, since they would span $W_\kappa$ itself, which has no element of rank one. So each block carries at least $3$ products, and $3c\le2c+1$, a contradiction for $c\ge2$.
\end{proof}

\begin{corollary}[Complex multiplication with unbounded fan-in]\label{cor:cmul_unbounded}
The classification of Theorem~\ref{thm:cmul} holds for every division-free circuit over $\R$ with $3$ product gates of unbounded fan-in: every product gate multiplies two affine functions, up to constant factors, which are of the form given there.
\end{corollary}

\begin{proof}
No product gate is removable, since $2$ product and division gates do not suffice (Theorem~\ref{thm:cmul}). The output space of complex multiplication contains no product $\ell(x)m(x')$ and has elements of rank $2$ (Lemma~\ref{lem:complex_block}). So by Lemma~\ref{lem:spare} with $K=2$, every product gate has exactly two non-constant inputs, both affine. Moving the constant inputs into a scalar on an edge gives a circuit with product gates of fan-in two, to which Theorem~\ref{thm:cmul} applies.
\end{proof}

\begin{remark}
If a product gate of fan-in $s$ is counted as $s-1$ multiplications, Theorem~\ref{thm:tubal_real} applies after replacing each gate by a binary tree, and shows that every optimal circuit has only product gates of fan-in $2$: a binary tree with an inner gate would have a child of degree $2$. For real cyclic convolution of length $d$, $\tau^d-1$ has $r\in\{1,2\}$ real roots and $c=(d-r)/2$ pairs, so the count $r+3c=2d-(r+c)$ is Winograd's $2d-k$ for $k$ irreducible factors \cite{winograd1977some}. For $d=5$ it is $7$.
\end{remark}

\subsection{Approximate circuits}\label{sec:border}

The lower bounds of this section for unbounded fan-in fail for approximate circuits. An \emph{approximate circuit} for a polynomial map $\phi$ is a circuit whose edge scalars are Laurent polynomials in a parameter $\delta$ and whose outputs are $\phi+\delta\,\psi$, where $\psi$ is a polynomial in $\delta$ and the inputs. Letting $\delta\to0$ gives $\phi$ as a limit of circuits of the same shape. All constructions below rest on the identity, for linear forms $u_1,\dots,u_m$,
\begin{align*}
    G_\delta[u_1,\dots,u_m]&=\prod_{t=1}^m(1+\delta u_t)(1-\delta u_t)=\prod_{t=1}^m(1-\delta^2u_t^2)\\
    &=1-\delta^2\sum_{t=1}^mu_t^2+\delta^4R,
\end{align*}
where $R$ is a polynomial in $\delta$ and the inputs. So one product gate of fan-in $2m$ gives the sum of squares $\sum_tu_t^2$ up to $O(\delta^2)$, after subtracting the constant $1$ and scaling by $-\delta^{-2}$.

\begin{proposition}\label{prop:border}
Let $\phi$ be a quadratic map in $n$ variables whose coordinates span a space $W$ of quadratic forms of dimension $K$. Over $\R$, $\phi$ has a division-free approximate circuit with $K+1$ product gates, each of fan-in at most $2n$, and error $O(\delta^2)$. Over $\C$, the same holds with $K$ product gates.
\end{proposition}

\begin{proof}
Let $w_1,\dots,w_K$ be a basis of $W$. The coordinates of $\phi$ are linear combinations of the $w_i$, which are free. Over $\R$, let $s$ be larger than the absolute values of the eigenvalues of the symmetric matrices of all $w_i$, and let $S=s\sum_jz_j^2$, where $z_1,\dots,z_n$ are the variables. Then $S$ and $S-w_1,\dots,S-w_K$ are positive definite, so by the spectral theorem each is a sum of $n$ squares of real linear forms: $S=\sum_tu_{0,t}^2$ and $S-w_i=\sum_tu_{i,t}^2$. With the $K+1$ product gates $G_i=G_\delta[u_{i,1},\dots,u_{i,n}]$, $0\le i\le K$,
\begin{displaymath}
    \delta^{-2}(G_i-G_0)=S-(S-w_i)+\delta^2(R_i-R_0)=w_i+O(\delta^2).
\end{displaymath}
Over $\C$, every quadratic form is a sum of squares of linear forms, $w_i=\sum_tu_{i,t}^2$, and the $K$ product gates $G_i=G_\delta[u_{i,1},\dots,u_{i,n}]$ give $-\delta^{-2}(G_i-1)=w_i+O(\delta^2)$.
\end{proof}

Over $\R$ the extra gate $G_0$ is needed for this construction because a nonzero bilinear form in $(x,x')$ takes both signs, so it is never a sum of squares. The gate $G_0$ supplies the common positive part $S$, which cancels in the differences. Over $\C$, the statement is Kumar's theorem \cite{kumar2020power}, specialized to degree $2$ and applied to each coordinate. That theorem says that every homogeneous polynomial over $\C$ is a limit of sums of two products of affine forms.

\begin{example}[$\C^c$ with $2c+1$ product gates]\label{ex:border_c}
Write the $k$-th factors as $a_k+\mathrm{i}b_k$ i and $a'_k+\mathrm{i}b'_k$i , so the outputs are $R_k=a_ka'_k-b_kb'_k$ and $I_k=a_kb'_k+b_ka'_k$ for $1\le k\le c$. Let $N_k=a_k^2+b_k^2+a_k'^2+b_k'^2$, and note
\begin{gather*}
    N_k-2R_k=(a_k-a'_k)^2+(b_k+b'_k)^2,\\
    N_k-2I_k=(a_k-b'_k)^2+(b_k-a'_k)^2.
\end{gather*}
The approximate circuit has the following $2c+1$ product gates. The gate $G_0$ is $G_\delta$ applied to all $4c$ variables, so $G_0=1-\delta^2\sum_jN_j+O(\delta^4)$. For each $k$, the gate $G_{k,R}$ is $G_\delta$ applied to the variables of all blocks $j\neq k$ and to the two forms $a_k-a'_k$ and $b_k+b'_k$. The gate $G_{k,I}$ is the same with the two forms $a_k-b'_k$ and $b_k-a'_k$. The squares of the blocks $j\neq k$ cancel in the differences, and
\begin{displaymath}
    \frac{G_{k,R}-G_0}{2\delta^2}=R_k+O(\delta^2),\qquad\frac{G_{k,I}-G_0}{2\delta^2}=I_k+O(\delta^2).
\end{displaymath}
The gate $G_0$ has fan-in $8c$ and the other gates have fan-in $8c-4$. A computer check for $c=3$ gives maximal errors about $0.67$, $7.5\cdot10^{-3}$ and $7.5\cdot10^{-5}$ at $\delta=10^{-1},10^{-2},10^{-3}$ on random inputs. Giving every block its own gate for $N_k$ would cost $3c$ gates. The single gate $G_0$ supplies the norms of all blocks at once, which is what saves $c-1$ gates.
\end{example}

\paragraph{Lower bounds for approximate circuits.} Partial homogenization also gives lower bounds for approximate circuits, through the degree $2$ parts, which survive limits. The \emph{index} of a real quadratic form is the smaller of its numbers of positive and negative eigenvalues. The forms of index at most $1$ form a closed cone, since the eigenvalues depend continuously on the form. As a quadratic form, a bilinear form of rank $\rho$ in $(x,x')$ has $\rho$ positive and $\rho$ negative eigenvalues. So it has index at most $1$ exactly when it is a product $\ell(x)m(x')$.

\begin{lemma}\label{lem:index}
Let $\cir$ be a division-free circuit over $\R$ with product gates $q_1,\dots,q_m$ of unbounded fan-in. There are quadratic forms $N_1,\dots,N_m$ of index at most $1$ such that $\spn\{q_1^{[2]},\dots,q_m^{[2]}\}\subseteq\spn\{N_1,\dots,N_m\}$.
\end{lemma}

\begin{proof}
Let $q_i$ have children $u_1,\dots,u_s$, with constant terms $c_j$ and linear parts $\ell_j$. Then
\begin{displaymath}
    q_i^{[2]}=\sum_{j=1}^s\Bigl(\prod_{t\neq j}c_t\Bigr)u_j^{[2]}+N_i,\qquad N_i=\sum_{j<k}\Bigl(\prod_{t\neq j,k}c_t\Bigr)\ell_j\ell_k,
\end{displaymath}
and $u_j^{[2]}\in\spn\{q_{i'}^{[2]}:i'<i\}$ by Corollary~\ref{cor:degree_two}, so the inclusion follows by induction on $i$. If three of the $c_t$ vanish, $N_i=0$. If exactly two vanish, or exactly one, $N_i$ is a product of two linear forms, and $ab=\frac14((a+b)^2-(a-b)^2)$ has index at most $1$. If none vanishes, $N_i=C\,e_2(h)$ with $C=\prod_tc_t$ and $h_j=\ell_j/c_j$, and $e_2(h)=\frac12\bigl((\sum_jh_j)^2-\sum_jh_j^2\bigr)$ is negative semidefinite on the kernel of $\sum_jh_j$, a subspace of codimension at most $1$. So it has at most one positive eigenvalue, and $N_i$ has index at most $1$.
\end{proof}

An \emph{approximate bilinear algorithm} is an approximate circuit whose product gates multiply a linear form in $x$ by a linear form in $x'$, both with coefficients in $\R[\delta,\delta^{-1}]$. The least number of its products is the border rank over $\R$.

\begin{theorem}[Approximate lower bounds]\label{thm:approx_lower}
Let $\phi$ be a quadratic map over $\R$ whose coordinates span a space $W$ of quadratic forms of dimension $K$.
\begin{enumerate}
    \item[(a)] a If $W$ contains no nonzero form of index at most $1$, in particular if $\phi$ is bilinear and $W$ contains no product $\ell(x)m(x')$, then every division-free approximate circuit for $\phi$ with product gates of unbounded fan-in has at least $K+1$ product gates. With Proposition~\ref{prop:border}, $K+1$ is the exact count.
    \item[(b)] b For $\phi$ the product of $\C^c$ with $c\ge2$, every approximate bilinear algorithm has at least $2c+2$ products. For $c=2$ this is $6$, the exact count.
\end{enumerate}
\end{theorem}

\begin{proof}
Fix a real $\delta\neq0$ that is not a pole of the edge scalars. The approximate circuit is then a real circuit computing $\phi+\delta\psi_\delta$, where $\psi_\delta$ stays bounded as $\delta\to0$.

(a) Let the circuit have $m$ product gates, and let $V_\delta$ be the span of the forms $N_i(\delta)$ of Lemma~\ref{lem:index}. By Corollary~\ref{cor:degree_two}, the degree $2$ parts of the outputs, $\phi_j+\delta\psi_{\delta,j}^{[2]}$, lie in $V_\delta$. They are linearly independent for small $\delta$, so $m\ge\dim V_\delta\ge K$. Suppose $m=K$. Then $\dim V_\delta=K$ and all $N_i(\delta)\neq0$. Along a sequence $\delta\to0$, the spaces $V_\delta$ converge to a space $V$ of dimension $K$, which contains every $\phi_j$. So $V=W$; and $N_i(\delta)/\|N_i(\delta)\|$ converges to a form of norm $1$ and index at most $1$, which lies in $V=W$. This contradicts the assumption on $W$.

(b) Use the coordinates of Section~\ref{sec:tubal}, in which an element of $W$ is the bilinear form with coefficient matrix $G_1\oplus\dots\oplus G_c$, $G_k=\bigl(\begin{smallmatrix}s_k&t_k\\t_k&-s_k\end{smallmatrix}\bigr)$. With $z_k=s_k+\mathrm{i}t_k$i , its determinant is $(-1)^cF(z)$ with $F(z)=\prod_k|z_k|^2$, which vanishes exactly on the subspaces $Z_l=\{z_l=0\}$ of dimension $2c-2$. Let $M_j(\delta)$ be the coefficient matrix of the $j$-th output of the approximate algorithm, so $M_j(\delta)\to M_j$. Let $D_\delta(\gamma)=\det\sum_j\gamma_jM_j(\delta)$. Then $D_\delta\to D_0$, with $D_0(\gamma)=(-1)^cF(z(\gamma))$ for a linear isomorphism $\gamma\mapsto z(\gamma)$ from $\R^{2c}$ to $\C^c$. Suppose there are at most $2c+1$ products, with rank one coefficient matrices $P_i=u_iv_i^T$. The $M_j(\delta)$ lie in $E_\delta=\spn\{P_i\}$.

We find a subspace $\Gamma_\delta\subseteq\R^{2c}$ of dimension at least $2c-1$ on which $D_\delta$ is a product of $2c$ real linear forms, or zero. Suppose $E_\delta$ has a basis of at most $2c$ of the $P_i$, indexed by $S$. Then $\sum_j\gamma_jM_j(\delta)=\sum_{i\in S}\alpha_i(\gamma)u_iv_i^T$ for linear forms $\alpha_i$. Let $U_S$ and $V_S$ have the columns $u_i$ and $v_i$. By the Cauchy-Binet formula, $D_\delta=0$ if $|S|<2c$, and $D_\delta=\det U_S\det V_S\prod_{i\in S}\alpha_i$ if $|S|=2c$. Take $\Gamma_\delta=\R^{2c}$. Otherwise there are $2c+1$ linearly independent $P_i$. Take $\Gamma_\delta$ to be the set of $\gamma$ for which the coefficient of $P_{2c+1}$ in $\sum_j\gamma_jM_j(\delta)$ vanishes, and apply the same formula with $S=\{1,\dots,2c\}$.

Along a sequence $\delta\to0$, a subspace of $\Gamma_\delta$ of dimension $2c-1$ converges to a subspace $\Gamma$. Products of $2c$ real linear forms of norm $1$ times scalars form a closed set. So $F$ restricted to $U=z(\Gamma)$ is a product of real linear forms, or zero. But $U$ has real dimension $2c-1>2c-2$, so it lies in no $Z_l$ and $F|_U\neq0$, since a subspace contained in a finite union of subspaces lies in one of them. The factor $|z_k|^2$ restricted to $U$ is a positive semidefinite form whose rank is the real rank of $z_k$ on $U$. If every $z_k$ had rank at most $1$ on $U$, the injective map $z$ would send $U$ into a sum of $c$ real lines, so $2c-1\le c$, which fails for $c\ge2$. So some $|z_k|^2|_U$ is positive semidefinite of rank $2$, hence irreducible over $\R$, and by unique factorization $F|_U$ is not a product of real linear forms, a contradiction. For $c=2$, Theorem~\ref{mthm:tubal}(b) gives $6=3c$ products.
\end{proof}

So for $\C^c$, $c\ge2$, the approximate count with unbounded fan-in is exactly $2c+1$, while exact circuits of any fan-in and approximate bilinear algorithms need at least $2c+2$. For $\C\times\C$, both of the latter need exactly $6$. The lower bounds for approximate circuits use only the degree $2$ parts, and the determinant argument in (b) uses that $|z|^2$ is irreducible over $\R$. Over $\C$, where $|z_k|^2=z_k\bar z_k$ factors, the product of $\C^c$ is the diagonal product of $\C^{2c}$, whose border rank is $2c$. Exact lower bounds for unbounded fan-in beyond $K+1$ must use the exact cancellation of the terms of higher degree, as Lemma~\ref{lem:spare} does. In general, approximate bilinear algorithms can need fewer products than exact ones. Partial $2\times2$ matrix multiplication has rank $6$ and border rank $5$ \cite{bini1979approximate}. The algebra $\C[\tau]/(\tau^d)$ has rank $2d-1$ \cite{alder1981algorithmic} and border rank $d$ \cite[Example~15.20]{burgisser1997algebraic}.

\section{Homogeneous component extractors}\label{sec:extractors}

We use the notation of Section~\ref{sec:preliminaries}. The key observation is as follows. View the computation $\Xi^{f_Y}$ as a circuit in the variables $Y$ over the coefficient ring $R=\F[x]$. It can be partially homogenized, and its $Y$-linear part is an $R$-linear function of the $Y$-linear parts of the queries (Corollary~\ref{cor:degree_two}, Part~\ref{cor:degree_one}).

\begin{lemma}[Generic coefficients]\label{lem:generic}
If $\Xi$ is a degree $d$ extractor of $S$, the output $o_k$ of $\Xi^{f_Y}$ is $f_Y^{[k]}$ for $k\in S$.
\end{lemma}

\begin{proof}
By induction on the gates, $\pi_a$ maps every gate of $\Xi^{f_Y}$ to the same gate of $\Xi^{f_a}$, since $\pi_a(f_Y(g))=f_a(\pi_a(g))$. Writing $o_k-f_Y^{[k]}=\sum_\beta Q_\beta(Y)x^\beta$, every $Q_\beta$ vanishes on $\F^Y$, so $Q_\beta=0$ as $\F$ is infinite.
\end{proof}

\paragraph{Cutting at the queries.} Let $\Xi$ have $D$ queries in topological order. Replacing the $i$-th query gate by a new variable $z_i$ gives polynomials $P_k\in R[z]$ at the outputs and inputs $G^i\in R[z]^n$ at the queries, with $G^i$ depending on $z_1,\dots,z_{i-1}$. In $\Xi^{f_Y}$ the $i$-th query has input $g^i=G^i(x,u_1,\dots,u_{i-1})$ and answer $u_i=f_Y(g^i)$, and with $w_k=f_Y^{[k]}$,
\begin{equation}\label{eq:outputs}
    w_k=P_k(x,u_1,\dots,u_D).
\end{equation}
Let $\gamma^i=G^i(x,0)\in R^n$, the input of the $i$-th query of $\Xi^0$.

\begin{lemma}[Linearization]\label{lem:extractor}
For a degree $d$ extractor of $S$ with $D$ queries there is $A\in R^{S\times D}$ with $f^{[k]}=\sum_iA_{ki}f(\gamma^i)$ for every $k\in S$ and every $f$ of degree at most $d$.
\end{lemma}

\begin{proof}
Let $\Phi_Y$ be a circuit computing $f_Y$, and view $\Xi(\Phi_Y)$ as a circuit over $R$ in the variables $Y$. Components are taken with respect to $Y$. The leaves of $\Xi$ are variables $x_j$ and constants, which lie in $R$ and have no $Y$-linear part. Apply Corollary~\ref{cor:degree_two}, Part~\ref{cor:degree_one}, to the gates of $\Xi$ outside the copies of $\Phi_Y$. By induction, the $Y$-linear part of every gate of $\Xi$ is an $R$-linear combination of the $Y$-linear parts $u_i^{[1]}$ of the query answers. The coefficients are built from the constant terms $\hat v^{[0]}$, which are the values in $\Xi^0$. Since $w_k$ is $Y$-linear, $w_k=\sum_iA_{ki}u_i^{[1]}$ with $A_{ki}\in R$. Every monomial of $f_Y$ contains one $y_\alpha$, so $u_i^{[1]}=\sum_\alpha y_\alpha\bigl((g^i)^{[0]}\bigr)^\alpha=f_Y(\gamma^i)$, because $(g^i)^{[0]}$ is the input of the $i$-th query when $Y=0$, which is $\gamma^i$. Applying $\pi_a$ gives the claim for every $f$.
\end{proof}

\begin{theorem}\label{thm:extractor_lower}
Every degree $d$ homogeneous component extractor makes at least $d+1$ queries.
\end{theorem}

\begin{proof}
The $w_k$ are nonzero linear forms in disjoint sets of variables of $Y$, hence linearly independent over $L$. By Lemma~\ref{lem:extractor} they lie in the $L$-span of the $D$ forms $f_Y(\gamma^i)$.
\end{proof}

\subsection{Extractors with \texorpdfstring{$d+1$}{+1} queries}

In this subsection $\Xi$ is a degree $d$ homogeneous component extractor with exactly $D=d+1$ queries and $d\ge1$. For each $k$ put $\alpha_k=(k,0,\dots,0)$ and let $Z$ be the set of the other variables $y_\alpha$. Since $w_k=x_1^ky_{\alpha_k}+\dots$, the forms $w\cup Z$ arise from $Y$ by an invertible triangular change of variables over $L$, so $L[Y]=L[w,Z]$ is a polynomial ring in $w\cup Z$.

\begin{lemma}\label{lem:answers}
Every answer satisfies $u_i\in L[w]$, and every query input satisfies $g^i=s_ix$ with $s_i\in L[w]$.
\end{lemma}

\begin{proof}
By \eqref{eq:outputs}, $L(w)\subseteq E=L(u_1,\dots,u_D)$, and $E$ has transcendence degree at most $d+1$ over $L$. If $u_i$ were transcendental over $L(w)$, then $w_0,\dots,w_d,u_i$ would be $d+2$ algebraically independent elements of $E$. So $u_i$ is algebraic over $K=L(w)$: $\sum_{j\le m}c_ju_i^{\,j}=0$ with $c_m\neq0$, $m\ge1$. As $u_i\in K[Z]$, a positive degree $e$ in $Z$ would make $c_mu_i^{\,m}$ the unique term of degree $me$. So $u_i\in L[w]$, and then $g^i\in L[w]^n$. Expanding $u_i=f_Y(g^i)$ in the coordinates $w\cup Z$, the coefficient of $y_{e_j}\in Z$ ($j\ge2$) is $g^i_j-x_jx_1^{-1}g^i_1$, which must vanish. So $g^i=s_ix$ with $s_i=g^i_1/x_1$.
\end{proof}

\begin{lemma}[Minimal extractors are non-adaptive]\label{lem:non-adaptive}
For every $i$, $g^i=\gamma^i$. Consequently the $i$-th query of $\Xi(\Phi)$ has the inputs $\gamma^i$ for every $\Phi$, and $\Xi$ is non-adaptive.
\end{lemma}

\begin{proof}
By Lemma~\ref{lem:answers}, $u_i=\sum_ks_i^{\,k}w_k$ and $s_i=S_i(u_1,\dots,u_{i-1})$ for polynomials $S_i$ over $L$. View $\Psi=(u_1,\dots,u_D)$ as a polynomial map in $w$ over $L$. By \eqref{eq:outputs} and the chain rule, $J_P(\Psi)J_\Psi=I$, so $\det J_\Psi$ is a unit of $L[w]$, a nonzero element of $L$. Differentiating, $J_\Psi=V(s)+NJ_\Psi$ with $V(s)=(s_i^{\,k})$ and $N$ strictly lower triangular, so $\det J_\Psi=\det V(s)=\prod_{i<i'}(s_{i'}-s_i)$ by Fact~\ref{fact:vandermonde}. Every factor is a unit of $L[w]$, that is, an element of $L$. Since $s_1=G^1_1(x)/x_1\in L$, all $s_i\in L$, so $g^i$ does not depend on $Y$ and $g^i=\gamma^i$. Applying $\pi_a$ gives the statement for every $f$.
\end{proof}

\begin{theorem}\label{thm:extractor}
Let $n\ge2$, $d\ge1$ and let $\Xi$ be a degree $d$ homogeneous component extractor with exactly $d+1$ queries. Then $\Xi$ is a shifted evaluation-interpolation extractor (Definition~\ref{def:ben-or}): there are pairwise distinct $\beta_i\in\F$ and $h\in R$ with $h(0)=0$ such that, with $t_i=\beta_i+h$, the $i$-th query of $\Xi(\Phi)$ is $\Phi(t_ix_1,\dots,t_ix_n)$ for every $\Phi$, and the outputs are $V^{-1}$ applied to the answers. If the query inputs are linear forms, then $h=0$ and $\Xi$ is Ben-Or's construction.
\end{theorem}

\begin{proof}
By the proof of Theorem~\ref{thm:extractor_lower}, the matrix $A$ of Lemma~\ref{lem:extractor} is invertible over $L$. Let $B=A^{-1}$. For $f$ homogeneous of degree $m$, Lemma~\ref{lem:extractor} gives $f(\gamma^i)=B_{im}f$. With $m=1$ and $f=x_j$, $\gamma^i_j=B_{i1}x_j$. Writing $B_{i1}=p/q$ in lowest terms, $q$ divides $x_1$ and $x_2$, so $q$ is constant and $t_i:=B_{i1}\in R$, $\gamma^i=t_ix$. Then $B_{im}=t_i^{\,m}$ (take $f=x_1^m$), so $B=V$ and $A=V^{-1}$. As $AV=I$ over $R$, $\det V$ is a nonzero constant, and so is every factor $t_{i'}-t_i$ of it. Put $\beta_i=t_i(0)$ and $h=t_1-\beta_1$. By Lemma~\ref{lem:non-adaptive} the queries are $f(t_ix)$. The answers $u_i=\sum_kt_i^{\,k}w_k$ are linearly, hence algebraically, independent over $L$, so $P_k(x,z)=\sum_iA_{ki}z_i$ as polynomials. If the query inputs are linear, $t_i$ is constant and $h=0$.
\end{proof}

\begin{example}[The shift and the assumptions are needed]\label{ex:shift}
For $d=1$, the queries $u_1=f((1+x_1)x)$ and $u_2=f(x_1x)$ give $f^{[1]}=u_1-u_2$ and $f^{[0]}=(1+x_1)u_2-x_1u_1$. For $n=1$, querying $f$ at $d+1$ distinct constants, interpolating the coefficients and multiplying by $x_1^k$ is an extractor with $d+1$ queries that are not of the form $f(t_ix_1)$, $t_i\in R$. For $d=0$, a single query at any point suffices.
\end{example}

\subsection{Extracting a set of components}

Which components are extracted, and with which points, matters. For $d=3$, the queries $f(0)$, $f(x)$ and $f(-x)$ give $f^{[0]}=f(0)$ and $f^{[2]}=\tfrac12(f(x)+f(-x))-f(0)$ with three queries instead of four. The linearization lemma reduces the question to univariate interpolation. For distinct $t_1,\dots,t_e$ in a field let
\begin{displaymath}
    \Pi_t(\tau)=\prod_{i=1}^e(\tau-t_i)=\sum_{j=0}^e\pi_j\tau^j,\qquad\pi_j:=0\ (j>e),
\end{displaymath}
and for $S\subseteq\{0,\dots,d\}$ let $J_e(S)=\{j\ge0:\ k-d+e\le j\le k\text{ for some }k\in S\}$.

\begin{lemma}[Coefficients]\label{lem:coefficient}
Let $E$ be a field, $t_1,\dots,t_e\in E$ distinct and $0\le k\le d$. There is a vector $\lambda\in E^e$ with $\sum_i\lambda_ig(t_i)=g_k$, the coefficient of $\tau^k$ in $g$, for every polynomial $g\in E[\tau]$ of degree at most $d$, if and only if $\pi_j=0$ for every $j\ge0$ with $k-d+e\le j\le k$.
\end{lemma}

\begin{proof}
If $e>d$ the range of $j$ is empty, and the evaluations at $e$ distinct points span the dual of $E[\tau]_{\le d}$ (Fact~\ref{fact:vandermonde}). If $e\le d$, the common kernel of the evaluations is $\{\Pi_tg:\deg g\le d-e\}$, since $\Pi_t$ divides every polynomial vanishing at all $t_i$. The functional $g\mapsto g_k$ is a combination of the evaluations if and only if it vanishes on this kernel. That is, the coefficient $\pi_{k-s}$ of $\tau^k$ in $\Pi_t\tau^s$ vanishes for $0\le s\le d-e$.
\end{proof}

\begin{theorem}\label{thm:subset}
Let $\F$ be infinite, $n\ge1$ and $\emptyset\neq S\subseteq\{0,\dots,d\}$, and let $e_\F(S)$ be the least $e$ for which there are distinct $t_1,\dots,t_e\in\F$ with $\pi_j=0$ for all $j\in J_e(S)$. Every degree $d$ extractor of $S$, possibly with divisions (Section~\ref{sec:div_extractors}), makes at least $e_\F(S)$ queries, and for such $t_i$ the non-adaptive extractor that queries $f(t_ix)$, $1\le i\le e_\F(S)$, and outputs $\F$-linear combinations of the answers, is an extractor of $S$. Moreover $\max S+1\le e_\F(S)\le d+1$, with equality on the right if $d\in S$.
\end{theorem}

\begin{proof}
\emph{Upper bound.} By Lemma~\ref{lem:coefficient} there are $\lambda^{(k)}\in\F^e$ with $\sum_i\lambda^{(k)}_ig(t_i)=g_k$ for all $g\in\F[\tau]_{\le d}$, and by linearity for all $g\in R[\tau]_{\le d}$. For $f$ of degree at most $d$, $f(t_ix)=g(t_i)$ with $g=\sum_mf^{[m]}\tau^m$., so $\sum_i\lambda^{(k)}_if(t_ix)=f^{[k]}$

\emph{Lower bound.} Let $\Xi$ have $D$ queries. By Lemma~\ref{lem:extractor} (Lemma~\ref{lem:extractor_div} with divisions), $f^{[k]}=\sum_iA_{ki}f(\gamma^i)$ with $A_{ki}\in L$ and $\gamma^i\in L^n$. Apply it to $f=g(x_1)$ with $g\in\F[\tau]_{\le d}$: $g_kx_1^k=\sum_iA_{ki}g(\gamma^i_1)$. Both sides are $L$-linear in the coefficients of $g$, so this holds for $g\in L[\tau]_{\le d}$. Let $\rho_1,\dots,\rho_e$ be the distinct elements among $\gamma^1_1,\dots,\gamma^D_1$, so $e\le D$. By Lemma~\ref{lem:coefficient} over $E=L$, the coefficients $\pi_j\in L$ of $\Pi_\rho$ vanish for $j\in J_e(S)$. Choose $\xi\in\F^n$ at which all $\rho_i$ are defined and pairwise different, possible since $\F$ is infinite. Then $t_i=\rho_i(\xi)$ are distinct and the coefficients of $\Pi_t$ are $\pi_j(\xi)$, which vanish on $J_e(S)$. So $D\ge e\ge e_\F(S)$.

\emph{Range.} If $e\le k$ for some $k\in S$, then $j=e\in J_e(S)$ and $\pi_e=1$. So $e_\F(S)\ge\max S+1$, and $e_\F(S)\ge d+1$ if $d\in S$. For $e=d+1$, $J_e(S)=\emptyset$ and any distinct points work.
\end{proof}

\begin{example}[The count depends on the points and on the field]\label{ex:subset}
(1) Let $d=3$, $S=\{0,2\}$. For $e=3$, $J_3(S)=\{0,2\}$, so $\Pi_t=\tau^3+\pi_1\tau$. In characteristic different from $2$, the points are then exactly $0,\beta,-\beta$ with $\beta\neq0$. So three queries suffice, and every set of three scalings that works is of this form.
(2) Let $d=5$, $S=\{0,3\}$. For $e=4$, $J_4(S)=\{0,2,3\}$, so $\Pi_t=\tau(\tau^3+\pi_1)$, whose roots are distinct and lie in $\F$ if and only if $-\pi_1\neq0$ has three distinct cube roots in $\F$. Over $\C$ the points $0,1,\omega,\omega^2$ work, so $e_\C(S)=4$. Over $\R$ this is impossible, and for $e=5$, $J_5(S)=\{0,3\}$ is satisfied by the points $0,1,2,3,-\tfrac{11}6$, so $e_\R(S)=5$.
(3) Let $S=\{1\}$, $d=2$: the points $\pm\beta$ give $f^{[1]}=(f(\beta x)-f(-\beta x))/(2\beta)$.
\end{example}

Over $\R$ and over algebraically closed fields, $e_\F(S)$ has an explicit description. We write $e_k$ for the elementary symmetric polynomials, so that $\pi_j=(-1)^{e-j}e_{e-j}(t)$.

\begin{lemma}[Real roots]\label{lem:real_roots}
\begin{enumerate}
    \item\label{lem:rr:nec} If $\Pi\in\R[\tau]$ has only real roots and $\pi_j=\pi_{j+1}=0$ for some $j\ge0$ with $j+1<\deg\Pi$, then $\tau^{j+2}$ divides $\Pi$. In particular $\Pi$ has a multiple root.
    \item\label{lem:rr:suf} If $K\subseteq\{1,\dots,e\}$ contains no two consecutive integers, there are distinct $t_1,\dots,t_e\in\R$ with $e_k(t)=0$ for all $k\in K$.
\end{enumerate}
\end{lemma}

\begin{proof}
For Part~\ref{lem:rr:nec}, recall that if $g$ has only real roots, with distinct roots $r_1<\dots<r_s$ of multiplicities $\mu_i$, then $g'$ has the roots $r_i$ with multiplicities $\mu_i-1$ and, by Rolle's theorem, a root in each interval $(r_i,r_{i+1})$. These are already $\deg g-1$ roots, so $g'$ has only real roots, and a multiple root of $g'$ is a root of $g$. The $j$-th derivative $g=\Pi^{(j)}$ has only real roots and its coefficients of $1$ and $\tau$ are $j!\,\pi_j=0$ and $(j+1)!\,\pi_{j+1}=0$, so $0$ is a multiple root of $\Pi^{(j)}$. Going back one derivative at a time, $0$ is a root of $\Pi^{(j-i)}$ of multiplicity at least $i+2$, and of $\Pi$ of multiplicity at least $j+2$.

For Part~\ref{lem:rr:suf}, if $e\in K$ put $t_e=0$. This does not change $e_k$ for $k<e$, $K\setminus\{e\}\subseteq\{1,\dots,e-2\}$, and the points constructed below are nonzero, so we may assume $e\notin K$. Pair every $k\in K$ with $k+1\le e$. The pairs are disjoint. For $M\ge4$ and $y\in[-\tfrac12,\tfrac12]^K$ put $t_i=M^{e-i}$ for $i$ not of the form $k+1$, $k\in K$, and $t_{k+1}=-M^{e-k}(1+y_k)$ for $k\in K$. These are distinct. Give position $i$ the weight $e-i$, and the weight $e-k$ to position $k+1$ for $k\in K$. The weights are non-increasing, and consecutive positions have equal weight only within a pair. For $k\in K$, the $k$-subsets of maximal total weight are $\{1,\dots,k\}$ and $\{1,\dots,k-1,k+1\}$, since positions $k-1$ and $k$ are not paired, and neither are $k+1$ and $k+2$. Their contribution to $e_k(t)$ is $P_k(t_k+t_{k+1})=-P_kM^{e-k}y_k$, with $P_k=\prod_{i<k}t_i\neq0$. Every other $k$-subset contributes at most $3^e|P_k|M^{e-k-1}$ in absolute value. Indeed, let $w$ be the total weight of $\{1,\dots,k-1\}$. Then $|P_k|\ge2^{-e}M^{w}$, and the subset contributes at most $(3/2)^eM^{w+e-k-1}$. Hence $e_k(t)=-P_kM^{e-k}\bigl(y_k+R_k(y)/M\bigr)$ with $R_k$ continuous and $|R_k|\le6^e$ on the cube. For $M\ge2\cdot6^e$, the continuous map $y\mapsto(-R_k(y)/M)_k$ sends the cube into itself, and a fixed point, which exists by Brouwer's theorem, gives $e_k(t)=0$ for all $k\in K$.
\end{proof}

\begin{theorem}\label{thm:subset_explicit}
Let $\emptyset\neq S\subseteq\{0,\dots,d\}$.
\begin{enumerate}
    \item Over $\R$, $e_\R(S)$ is the least $e\ge\max S+1$ such that $J_e(S)$ contains no two consecutive integers.
    \item Over an algebraically closed field $\F$ of characteristic $0$, $e_\F(S)$ is the least $e\ge\max S+1$ such that $\{0,1\}\not\subseteq J_e(S)$, that is,
    \begin{displaymath}
        e_\F(S)=\max\bigl(\max S+1,\ \min(d+1-\min S,\ d+2-\min(S\setminus\{0\}))\bigr),
    \end{displaymath}
    for $S\neq\{0\}$, and $e_\F(\{0\})=1$.
\end{enumerate}
\end{theorem}

\begin{proof}
In both cases $e\ge\max S+1$ is needed by Theorem~\ref{thm:subset}. Over $\R$, a monic $\Pi$ with distinct real roots and $\pi_j=0$ on $J_e(S)$ exists if and only if $J_e(S)\subseteq\{0,\dots,e-1\}$ has no two consecutive elements, by Lemma~\ref{lem:real_roots} applied to $K=\{e-j:j\in J_e(S)\}$. Over an algebraically closed field of characteristic $0$, if $0,1\in J_e(S)$ then $\tau^2$ divides $\Pi$. Otherwise $\tau^e-1$ (if $0\notin J_e(S)$) or $\tau^e-\tau$ (if $0\in J_e(S)$, $1\notin J_e(S)$, $e\ge2$) or $\tau$ (if $e=1$) is squarefree, splits, and has all coefficients outside $\{0,1,e\}$ equal to zero. Finally, $0\in J_e(S)$ if and only if $e\le d-\min S$, and $1\in J_e(S)$ if and only if some $k\in S$ with $k\ge1$ has $e\le d+1-k$.
\end{proof}

For example, a single component $f^{[k]}$ with $1\le k\le d$ needs $\max(k,d-k)+1$ queries over $\C$. For $d=5$ and $S=\{0,3\}$, $J_4(S)=\{0,2,3\}$ contains $2,3$ but not $1$, which gives $e_\C(S)=4<5=e_\R(S)$ as in Example~\ref{ex:subset}.

\subsection{Extractors with divisions}\label{sec:div_extractors}

A query circuit may also have division gates. We require that, for every $f$ of degree at most $d$, $\Xi^f$ is defined, that is, no divisor computes $0$, and outputs $f^{[k]}$. Then $\Xi^{f_Y}$ is defined and Lemma~\ref{lem:generic} holds. Indeed, let a divisor be the first one that computes $0$ in $\Xi^{f_Y}$. The gates before it are defined. Choose a nonzero polynomial $H(Y)$ such that, after $Y=a$, the denominators of these gates and the numerators of the earlier divisors do not vanish identically in $x$. For $a\in\F^Y$ outside the zero set of $H$, the substitution $\pi_a$ commutes with these gates. So the divisor computes $0$ in $\Xi^{f_a}$ for such $a$, a contradiction. In the same way, $o_k-f_Y^{[k]}$ vanishes after $Y=a$ for all $a$ outside a zero set, hence is $0$.

\begin{lemma}[Linearization with divisions]\label{lem:extractor_div}
Let $\Xi$ be a degree $d$ extractor of $S$ with divisions and $D$ queries. There are $a\in\F^Y$, $\gamma^1,\dots,\gamma^D\in L^n$ (the query inputs of $\Xi^{f_a}$) and $A\in L^{S\times D}$ with $f^{[k]}=\sum_iA_{ki}f(\gamma^i)$ for all $k\in S$ and all $f$ of degree at most $d$.
\end{lemma}

\begin{proof}
Let $\Phi_Y$ compute $f_Y$ and view $\Xi(\Phi_Y)$ as a circuit with divisions over the field $L$ in the variables $Y$. As in Lemma~\ref{lem:divisions}, choose $a\in\F^Y$ such that every gate value is regular at $Y=a$ and no divisor vanishes there. This is possible since $\F$ is infinite, so a nonzero polynomial over $L$ in $Y$ has a non-root in $\F^Y$. Take components with respect to $Y'=Y-a$ in $L[[Y']]$. By Lemma~\ref{lem:divisions} applied to $\Xi(\Phi_Y)$, the $Y'$-linear part of every gate of $\Xi$ other than a query is an $L$-linear combination of the $Y'$-linear parts of its inputs. A query with input $g$ answers $f_{a+Y'}(g)=f_a(g)+f_{Y'}(g)$, whose $Y'$-linear part is
\begin{displaymath}
    f_{Y'}(g^{[0]})+\sum_j\frac{\partial f_a}{\partial x_j}(g^{[0]})\,g_j^{[1]},
\end{displaymath}
with $g^{[0]}=\gamma^i$. By induction over the gates, every $Y'$-linear part lies in $\spn_L\{f_{Y'}(\gamma^i)\}$. The output $o_k=f_a^{[k]}+f_{Y'}^{[k]}$ has $Y'$-linear part $f_{Y'}^{[k]}$, which gives $A$. Substituting values for $Y'$ gives the claim for every $f$.
\end{proof}

Hence Theorems~\ref{thm:extractor_lower} and~\ref{thm:subset} hold for query circuits with divisions. The proof of Theorem~\ref{thm:extractor_lower} uses only the $L$-span.

\section{\texorpdfstring{$2\times2$ matrix multiplication with seven products}{2x2 matrix multiplication with seven products}}\label{sec:matmul}

Throughout this appendix, $\K$ is an algebraically closed field of characteristic $0$, and $\cA=\cB=\K^{2\times2}$. A circuit over a subfield $\F\subseteq\K$ is also a circuit over $\K$, so we may work over $\K$. We write $A\in\cA$ and $B\in\cB$ for the two input matrices, $a_{jl}$ and $b_{jl}$ for their entries, $\cA^*,\cB^*$ for the spaces of linear forms, and identify a bilinear form on $\cA\times\cB$ with an element of $\cA^*\otimes\cB^*$, where $(\alpha\otimes\delta)(A,B)=\alpha(A)\delta(B)$. The rank of a bilinear form is the rank of its coefficient matrix. Vectors are columns, $e_1,e_2$ is the standard basis of $\K^2$, and $\Pj(X)$ is the projective space of lines of a vector space $X$. For $0\neq x$ we write $[x]$ for the line it spans.

\subsection{Seven-product algorithms and the quadratic part of a circuit}\label{mm:sec:defs}

\begin{definition}\label{mm:def:alg}
A \emph{seven-product algorithm} consists of seven linearly independent quadratic forms
\begin{gather*}
    P_i=(\alpha_i+\beta_i)(\gamma_i+\delta_i)\qquad(i=1,\dots,7),\\
    \alpha_i,\gamma_i\in\cA^*,\qquad\beta_i,\delta_i\in\cB^*,
\end{gather*}
on $\cA\times\cB$ (the \emph{products}) and matrices $\Lambda_1,\dots,\Lambda_7\in\K^{2\times2}$ such that $AB=\sum_i\Lambda_iP_i(A,B)$ identically, that is, $(AB)_{jl}=\sum_i(\Lambda_i)_{jl}P_i$ for all $j,l$. The \emph{pure part} and the \emph{bilinear part} of $P_i$ are
\begin{gather*}
    \pi_i=(\alpha_i\gamma_i,\ \beta_i\delta_i)\in\Sy^2\cA^*\oplus\Sy^2\cB^*,\\
    S_i=\alpha_i\otimes\delta_i+\gamma_i\otimes\beta_i\in\cA^*\otimes\cB^*,
\end{gather*}
so that $P_i=\alpha_i\gamma_i+\beta_i\delta_i+S_i$. Let $V=\spn\{\pi_1,\dots,\pi_7\}$ and $p=\dim V$, and let $V_A\subseteq\Sy^2\cA^*$ and $V_B\subseteq\Sy^2\cB^*$ be the images of $V$ under the two projections, with $p_A=\dim V_A$ and $p_B=\dim V_B$.
\end{definition}

Every product of two linear forms on $\cA\times\cB$ has the shape above, so Definition~\ref{mm:def:alg} describes the homogeneous quadratic algorithms with seven multiplications. Rescaling a product (and the corresponding $\Lambda_i$) and exchanging its two factors do not change the algorithm. We use both freely. Seven-product algorithms are exactly what partial homogenization extracts from circuits.

\begin{proposition}[The quadratic part of a circuit]\label{mm:prop:quadpart}
Let $\cir$ be a division-free circuit over $\K$ whose product gates have fan-in $2$ and which computes the entries of $AB$. Then $\cir$ has at least $7$ product gates. If it has exactly $7$, $q_i=u_iw_i$, let $\ell_i,m_i$ be the components of degree $1$ of $u_i,w_i$, and let $\Lambda'$ be as in Lemma~\ref{lem:graded}. Then the products $\ell_im_i$ with the coefficient matrices $\Lambda'_i$ form a seven-product algorithm, the \emph{quadratic part} of $\cir$.
\end{proposition}

\begin{proof}
By Lemma~\ref{lem:graded}, $AB=\phi^{[2]}=\sum_i\Lambda'_i\ell_im_i$, so the products of linear forms $\ell_im_i$ span a space containing the entries of $AB$. Winograd showed that $2\times2$ matrix multiplication needs $7$ multiplications even when the entries commute \cite{winograd1971multiplication}. That is, the entries of $AB$ do not lie in the span of $6$ products of linear forms. This gives the lower bound and, for $7$ product gates, the linear independence of the $\ell_im_i$.
\end{proof}

We call $p,p_A,p_B$ of the quadratic part the \emph{pure dimensions} of $\cir$. Let $W\subseteq\cA^*\otimes\cB^*$ be the span of the four entries of $AB$. Equivalently, $W=\{w_Z:\ Z\in\K^{2\times2}\}$ with $w_Z(A,B)=\tr(ZAB)$. For $x,y\in\K^2$ we put
\begin{gather*}
    \Row(x)=\{A\mapsto x^TAh:\ h\in\K^2\}\subseteq\cA^*,\\
    \Col(y)=\{B\mapsto k^TBy:\ k\in\K^2\}\subseteq\cB^*.
\end{gather*}

\begin{theorem}\label{mm:thm:main}
Consider a seven-product algorithm with $p=3$, $p_A\ge2$ and $p_B\ge2$. Then:
\begin{enumerate}
    \item[(a)] the pure parts $\pi_1,\dots,\pi_7$ represent exactly four points of $\Pj(V)$. Three of them carry two products each and the fourth carries one product;
    \item[(b)] every bilinear part $S_i$ lies in $W$;
    \item[(c)] there is $H\in GL_2(\K)$ such that, with $A'=AH$ and $B'=H^{-1}B$ (so that $A'B'=AB$), every product is a nonzero multiple of
    \begin{displaymath}
        (x^TA'e_1+e_2^TB'y)(x^TA'e_2+e_1^TB'y)
    \end{displaymath}
    for some $x,y\in\K^2\setminus\{0\}$ depending on the product. The two products at a point carrying two products correspond to pairs $(x,y)$ and $(x,-y)$.
\end{enumerate}
\end{theorem}

If $p_A=3$ or $p_B=3$, the hypothesis $p=3$ is automatic, because $\max(p_A,p_B)\le p\le3$ (Remark~\ref{mm:rem:context}). The bilinear part of $(x^TA'e_1+e_2^TB'y)(x^TA'e_2+e_1^TB'y)$ is $x^TA'B'y=x^TABy$. Products of this form are the building blocks of Winograd's commutative inner-product algorithm \cite{winograd1968new} and of its improvement by Waksman \cite{waksman1970winograd}. Algorithms as in Theorem~\ref{mm:thm:main} exist for each of the pairs $(p_A,p_B)=(3,3),(3,2),(2,3),(2,2)$. Examples are given in Section~\ref{mm:sec:comp}. For $I\in\K\setminus\{0\}$ let $R_I$ be the list of the seven quadratic forms
\begin{gather*}
    P_1=a_{11}(a_{12}+b_{11}+b_{12}),\qquad P_2=a_{12}(a_{11}-b_{21}-b_{22}),\\
    P_3=a_{21}(a_{22}+b_{11}+Ib_{12}),\qquad P_4=a_{22}(a_{21}-b_{21}-Ib_{22}),\\
    P_5=b_{11}b_{22},\qquad P_6=(a_{11}-b_{22})(a_{12}+b_{11}),\\
    P_7=(a_{21}-Ib_{22})(a_{22}+b_{11}).
\end{gather*}

\begin{theorem}\label{mm:thm:family}
\textup{(a)} For every $I\neq0$ the products of $R_I$ are linearly independent and form a seven-product algorithm with $p_A=2$, $p_B=1$ and $p=3$:
\begin{gather*}
    (AB)_{11}=-P_2+P_5+P_6,\qquad(AB)_{12}=P_1-P_5-P_6,\\
    (AB)_{21}=-P_4+IP_5+P_7,\qquad(AB)_{22}=I^{-1}(P_3-P_7)-P_5.
\end{gather*}
\textup{(b)} Consider a seven-product algorithm with $p=3$ and $p_B=1$. Then $p_A=2$, and after a substitution as in Lemma~\ref{mm:lem:sym}\textup{(a)} its products are nonzero multiples of the products of $R_I$ for some $I\neq0$.

\textup{(c)} After rescaling and reordering of products, $R_J$ is obtained from $R_I$ by a substitution as in Lemma~\ref{mm:lem:sym}\textup{(a)} if and only if $J\in\{I,I^{-1}\}$.
\end{theorem}

\begin{proposition}\label{mm:prop:pA0}
In a seven-product algorithm with $p_A=0$ every product is bilinear, that is, $p=0$. The same holds if $p_B=0$.
\end{proposition}

\begin{remark}\label{mm:rem:context}
Every seven-product algorithm has $p\le3$ by Lemma~\ref{mm:lem:relations}(a) below. Moreover $\max(p_A,p_B)\le p$ and $p\le p_A+p_B$, since $V\subseteq V_A\oplus V_B$ projects onto $V_A$ and onto $V_B$. For $p=3$ the possible pairs $(p_A,p_B)$ are $(3,3)$, $(3,2)$, $(2,3)$, $(2,2)$, $(3,1)$, $(1,3)$, $(2,1)$, $(1,2)$, $(3,0)$, $(0,3)$.
\end{remark}

\begin{corollary}\label{mm:cor:class}
For a seven-product algorithm with $p=3$ exactly one of the following holds.
\begin{enumerate}
    \item[(i)] $p_A\ge2$ and $p_B\ge2$, and the algorithm is as in Theorem~\ref{mm:thm:main}.
    \item[(ii)] $(p_A,p_B)=(2,1)$, and the algorithm is as in Theorem~\ref{mm:thm:family}\textup{(b)}.
    \item[(iii)] $(p_A,p_B)=(1,2)$, and the algorithm obtained from it by Lemma~\ref{mm:lem:sym}\textup{(b)} is as in \textup{(ii)}.
\end{enumerate}
In particular the pairs $(3,1)$, $(1,3)$, $(3,0)$, $(0,3)$ do not occur, and each of the pairs $(3,3)$, $(3,2)$, $(2,3)$, $(2,2)$, $(2,1)$, $(1,2)$ occurs.
\end{corollary}

For $p\le2$ there is a single algorithm up to symmetry. Let $\mathcal M$ be the list of the seven quadratic forms
\begin{gather*}
    P_1=a_{21}(a_{12}-a_{22}+b_{12}),\qquad P_2=(a_{21}+b_{22})(a_{12}-a_{22}),\\
    P_3=(b_{22}-b_{21})(a_{12}+b_{11}),\qquad P_4=(a_{11}+b_{22}-b_{21})\,b_{11},\\
    P_5=a_{11}b_{12},\qquad P_6=a_{12}b_{22},\\
    P_7=(a_{21}+b_{22}-b_{21})(a_{12}-a_{22}+b_{11}).
\end{gather*}

\begin{theorem}\label{mm:thm:p12}
\textup{(a)} No seven-product algorithm has $p=1$.

\textup{(b)} The products of $\mathcal M$ are linearly independent and form a seven-product algorithm with $p_A=p_B=1$ and $p=2$:
\begin{gather*}
    (AB)_{11}=-P_3+P_4+P_6,\qquad(AB)_{12}=P_5+P_6,\\
    (AB)_{21}=-P_2-P_3+P_6+P_7,\qquad(AB)_{22}=P_1-P_2+P_6.
\end{gather*}
\textup{(c)} Every seven-product algorithm with $p=2$ has $p_A=p_B=1$, and after a substitution as in Lemma~\ref{mm:lem:sym}\textup{(a)} its products are nonzero multiples of the products of $\mathcal M$.
\end{theorem}

So every seven-product algorithm has $p\in\{0,2,3\}$. Up to the substitutions of Lemma~\ref{mm:lem:sym}, rescaling and reordering, it is one of the following. If $p=0$, it is Strassen's algorithm (Proposition~\ref{mm:prop:nest0} below, by de Groote's theorem). If $p=2$, it is $\mathcal M$. If $p=3$, it is as in Corollary~\ref{mm:cor:class}. Unlike the family $R_I$ for $p=3$, the algorithm with $p=2$ has no parameter. Its pure parts lie on three points of $\Pj(V)$: $P_1,P_2$ at $[\pi_1]$, $P_3,P_4$ at $[\pi_3]$, and $P_7$ at $[\pi_1+\pi_3]$.

For circuits, the classification and partial homogenization together exclude nested products.

\begin{theorem}[No nested products]\label{mm:thm:nest}
Let $\cir$ be a division-free circuit over $\K$ with exactly $7$ product gates, all of fan-in $2$, that computes the entries of $AB$. Then every product gate computes $(c_i+\ell_i)(c'_i+m_i)$ with $c_i,c'_i\in\K$ and linear forms $\ell_i,m_i$, and the outputs are $\Lambda(q-q^{[\le1]})$ for the coefficient matrices $\Lambda_i$ of the quadratic part. Moreover:
\begin{enumerate}
    \item[(a)] if $p=3$, the products $\ell_im_i$ form an algorithm as in Theorem~\ref{mm:thm:main} or Theorem~\ref{mm:thm:family}, or the transpose of one as in Theorem~\ref{mm:thm:family};
    \item[(b)] if $p=2$, the products $\ell_im_i$ form the algorithm $\mathcal M$ of Theorem~\ref{mm:thm:p12} up to a substitution as in Lemma~\ref{mm:lem:sym}\textup{(a)} and rescaling;
    \item[(c)] if $p=0$, $\cir$ is bi-affine, and $(\ell_i,m_i,\Lambda_i)_i$ is Strassen's algorithm up to the symmetries of $2\times2$ matrix multiplication \cite{degroote1978optimal};
    \item[(d)] $p\neq1$.
\end{enumerate}
\end{theorem}

A product gate may still have other product gates among its predecessors. The theorem says that both inputs of every product gate compute affine functions, so all terms of degree at least $2$ that reach a product gate cancel.

The proof of the classification is as follows. Section~\ref{mm:sec:tools} collects linear algebra facts and general lemmas, and Section~\ref{mm:sec:products} describes the possible products of an algorithm with $p=3$. Sections~\ref{mm:sec:first} to~\ref{mm:sec:assembly} prove Theorem~\ref{mm:thm:main} for $p_A=p_B=3$. Section~\ref{mm:sec:first} derives the basic geometry of the points $[\pi_i]$, whose number $d$ is between $4$ and $7$. Section~\ref{mm:sec:d4} treats $d=4$. Sections~\ref{mm:sec:nets} to~\ref{mm:sec:T1T1} show that $d\ge5$ is impossible. There, a classification of the possible nets $V_A$ and $V_B$ (Sections~\ref{mm:sec:nets} and~\ref{mm:sec:types}) leaves four cases, and each is ruled out in its own section. Section~\ref{mm:sec:assembly} assembles this part. Section~\ref{mm:sec:twoq} studies a side on which the pure parts span two dimensions, Sections~\ref{mm:sec:32} and~\ref{mm:sec:22} treat $(p_A,p_B)=(3,2)$ and $(2,2)$, and Section~\ref{mm:sec:pB1} treats $p_B=1$. Section~\ref{mm:sec:proofs} completes the proofs of Theorems~\ref{mm:thm:main} and~\ref{mm:thm:family} and of Corollary~\ref{mm:cor:class}, Section~\ref{mm:sec:pA0} proves Proposition~\ref{mm:prop:pA0}, Section~\ref{mm:sec:p12} proves Theorem~\ref{mm:thm:p12}, Section~\ref{mm:sec:nest} proves Theorem~\ref{mm:thm:nest}, and Section~\ref{mm:sec:comp} lists the computations and examples.

\subsection{Tools}\label{mm:sec:tools}

\begin{fact}\label{mm:fact:W}
For $Z\in\K^{2\times2}$, $\rank w_Z=2\rank Z$. In particular $W$ contains no form of rank one, and its rank-two elements are the forms $w_{yx^T}(A,B)=x^TABy$ with $x,y\neq0$. If $h_1,h_2$ is a basis of $\K^2$ and $k_1,k_2$ is its dual basis ($k_p^Th_q=\delta_{pq}$), then $x^TABy=\sum_q(x^TAh_q)(k_q^TBy)$. Consequently, in every expression $x^TABy=\alpha\otimes\delta+\gamma\otimes\beta$ the forms $\alpha,\gamma$ form a basis of $\Row(x)$ and $\beta,\delta$ a basis of $\Col(y)$. The space $\Row(x)$ determines $[x]$, $\Row(x)\cap\Row(x')=0$ for $[x]\neq[x']$, and the same holds for $\Col$.
\end{fact}

\begin{proof}
$w_Z(A,B)=\sum_{j,m,l}Z_{lj}A_{jm}B_{ml}$, so in suitable orderings of the coordinates the coefficient matrix of $w_Z$ is $Z^T\otimes I_2$, of rank $2\rank Z$. The identity $x^TABy=\sum_q(x^TAh_q)(k_q^TBy)$ is $\sum_qh_qk_q^T=I$. For the last claims note that $x^TAh=x'^TAh'$ for all $A$ means $xh^T=x'h'^T$.
\end{proof}

For a bilinear form $T\in\cA^*\otimes\cB^*$ its \emph{A-factor space} is the image of the map $\cB\to\cA^*$, $B\mapsto T(\cdot,B)$, and its \emph{B-factor space} is defined in the same way. Both have dimension $\rank T$, and $T\in G\otimes E$ for subspaces $G\subseteq\cA^*$, $E\subseteq\cB^*$ if and only if $G$ and $E$ contain the respective factor spaces. For $u,w\in\K^2\setminus\{0\}$ we put
\begin{gather*}
    \Col_\cA(u)=\{A\mapsto k^TAu:\ k\in\K^2\}\subseteq\cA^*,\\
    \Row_\cB(w)=\{B\mapsto w^TBv:\ v\in\K^2\}\subseteq\cB^*.
\end{gather*}

\begin{fact}\label{mm:fact:Wfactor}
\textup{(a)} For $x,y\neq0$ the A-factor space of $x^TABy$ is $\Row(x)$ and its B-factor space is $\Col(y)$. If $\rank Z=2$, the factor spaces of $w_Z$ are $\cA^*$ and $\cB^*$. \textup{(b)} Let $E\subseteq\cB^*$ be a $2$-dimensional subspace with $W\cap(\cA^*\otimes E)\neq0$. Then $E=\Col(y)$ for some $y\neq0$, and $W\cap(\cA^*\otimes E)=\{x^TABy:\ x\in\K^2\}$. Similarly, if $G\subseteq\cA^*$ is $2$-dimensional and $W\cap(G\otimes\cB^*)\neq0$, then $G=\Row(x)$ for some $x\neq0$ and $W\cap(G\otimes\cB^*)=\{x^TABy:\ y\in\K^2\}$. In particular $W\cap(\Row(x)\otimes\Col(y))=\K\,x^TABy$. \textup{(c)} $\Row(x)\neq\Col_\cA(u)$ for all $x,u\neq0$, and $W\cap(\Col_\cA(u)\otimes\cB^*)=0$.
\end{fact}

\begin{proof}
(a) By Fact~\ref{mm:fact:W}, $x^TABy=\sum_q(x^TAh_q)\otimes(k_q^TBy)$, where the $x^TAh_q$ form a basis of $\Row(x)$ and the $k_q^TBy$ a basis of $\Col(y)$. If $\rank Z=2$ the rank of $w_Z$ is $4$. (b) A nonzero $w_Z\in\cA^*\otimes E$ has its B-factor space in $E$, so by (a) $Z=yx^T$ with $\Col(y)=E$. Since $\Col(y)$ determines $[y]$ (Fact~\ref{mm:fact:W}), all such $Z$ have the form $yx^T$ with the same $y$. The statement for $G$ is the same, and the last claim follows by intersecting $\{x^TABy':\ y'\in\K^2\}$ with $\cA^*\otimes\Col(y)$. (c) $x^TAh=k^TAu$ for all $A$ means $xh^T=ku^T$, so $\Row(x)\cap\Col_\cA(u)$ has dimension at most $1$. The second claim follows from (b).
\end{proof}

\begin{fact}\label{mm:fact:M2}
\textup{(a)} Let $U\subseteq\K^{2\times2}$ be a $2$-dimensional subspace consisting of singular matrices. Then $U=\{xh^T:\ h\in\K^2\}$ or $U=\{hy^T:\ h\in\K^2\}$ for some $x\neq0$, respectively $y\neq0$. \textup{(b)} Every subspace of $\K^{2\times2}$ of dimension at least $3$ contains an invertible matrix. \textup{(c)} Every commutative subspace of $\K^{2\times2}$ has dimension at most $2$.
\end{fact}

\begin{proof}
(a) $U$ is spanned by two non-proportional rank-one matrices $x_1y_1^T$, $x_2y_2^T$. If $[x_1]\neq[x_2]$ and $[y_1]\neq[y_2]$ their sum is invertible. So they share $[x]$ or $[y]$, and $U$ is the $2$-dimensional space of the claim. (b) $\det$ is a nondegenerate quadratic form on $\K^{2\times2}$, whose totally isotropic subspaces have dimension at most $2$. (c) If $U$ is commutative and contains a non-scalar $X$, then $U$ lies in the centralizer $\K I+\K X$ of $X$.
\end{proof}

We say that a $2$-dimensional subspace $U\subseteq\K^{2\times2}$ has \emph{type I} if it contains an invertible matrix, \emph{type C} if $U=\{xh^T\}$ for a fixed $x$ (common column space), and \emph{type R} if $U=\{hy^T\}$ for a fixed $y$ (common row space). By Fact~\ref{mm:fact:M2}(a) every such $U$ has exactly one type.

\begin{fact}\label{mm:fact:binary}
\textup{(a)} If $\ell m=\ell'm'\neq0$ for linear forms in any number of variables, then $(\ell',m')=(c\ell,c^{-1}m)$ or $(cm,c^{-1}\ell)$ for some $c\neq0$. \textup{(b)} If $u_1,u_2,u_3\in\K^2$ are pairwise non-proportional, then $u_1^2,u_2^2,u_3^2$ (that is, $u_ku_k^T$) form a basis of $\Sy^2\K^2$. Hence a quadratic form on $\K^2$ vanishing at three pairwise non-proportional vectors is zero. Also, let $v\mapsto\mathcal P(v)$ be any map that is homogeneous quadratic in $v$. Then every $\mathcal P(v)$ lies in $\spn\{\mathcal P(u_1),\mathcal P(u_2),\mathcal P(u_3)\}$.
\end{fact}

\begin{fact}\label{mm:fact:rank}
Let $X\subseteq\cA$ and $Y\subseteq\cB$ be subspaces, and suppose that $AB=\sum_{i=1}^m\Lambda_iT_i(A,B)$ for all $A\in X$ and $B\in Y$, with $\Lambda_i\in\K^{2\times2}$ and bilinear forms $T_i$ on $X\times Y$. Let $r=\sum_i\rank T_i$. Then
\begin{gather*}
    r\ge\dim XY,\qquad r\ge\dim X-\dim\{A\in X:\ AY=0\},\\
    r\ge\dim Y-\dim\{B\in Y:\ XB=0\}.
\end{gather*}
In particular: \textup{(a)} $r\ge4$ if $X=\cA$ and $\dim Y\ge2$, or if $Y=\cB$ and $\dim X\ge2$; \textup{(b)} $r\ge2$ if $\dim X=2$ and $Y=\{kw^T:\ k\in\K^2\}$ for some $w\neq0$, or if $\dim Y=2$ and $X=\{uh^T:\ h\in\K^2\}$ for some $u\neq0$.
\end{fact}

\begin{proof}
Writing each $T_i$ as a sum of $\rank T_i$ products of linear forms gives $AB=\sum_{l=1}^rM_l\lambda_l(A)\mu_l(B)$. Hence each of the three linear maps $X\otimes Y\to\K^{2\times2}$, $X\to\operatorname{Hom}(Y,\K^{2\times2})$ and $Y\to\operatorname{Hom}(X,\K^{2\times2})$ induced by $(A,B)\mapsto AB$ has rank at most $r$, and their ranks are the three right sides. The identity remains true on subspaces, so for (a) we may assume $\dim Y=2$, respectively $\dim X=2$. If $X=\cA$ and $Y$ has type I, then $\cA Y\supseteq\cA B_0=\K^{2\times2}$ for an invertible $B_0\in Y$. If $Y=\{xh^T\}$, then $\cA Y$ contains all $(Ax)h^T$ and is $\K^{2\times2}$. If $Y=\{hy^T\}$, then $AY=0$ forces $Ah=0$ for all $h$, so the second bound gives $r\ge4$. If $Y=\cB$ and $X$ has type I or $X=\{hy^T\}$, then $X\cB=\K^{2\times2}$. If $X=\{xh^T\}$, then $XB=0$ forces $h^TB=0$ for all $h$, so the third bound gives $r\ge4$. For (b), $AY=0$ with $Y=\{kw^T\}$ means $Ak=0$ for all $k$, and $XB=0$ with $X=\{uh^T\}$ means $h^TB=0$ for all $h$. So the second, respectively third, bound gives $r\ge2$.
\end{proof}

\begin{lemma}\label{mm:lem:relations}
Consider a seven-product algorithm. Let $\Pi$ be the span of $P_1,\dots,P_7$, and let $K$ be the space of all $c\in\K^7$ with $\sum_ic_i\pi_i=0$.
\begin{enumerate}
    \item[(a)] For each $j,l$ the vector $\lambda^{(jl)}=((\Lambda_i)_{jl})_{i=1}^7$ lies in $K$, and these four vectors are linearly independent. Hence $p\le3$.
\end{enumerate}
Assume now $p=3$.
\begin{enumerate}
    \item[(b)] The four vectors $\lambda^{(jl)}$ form a basis of $K$. The map $c\mapsto\sum_ic_iP_i$ is an isomorphism from $K$ onto $W$. In particular every bilinear form in $\Pi$ lies in $W$.
    \item[(c)] For $y\in\K^7$ we have $\sum_iy_i\Lambda_i=0$ if and only if there is a linear functional $\varphi$ on $V$ with $y_i=\varphi(\pi_i)$ for all $i$.
\end{enumerate}
\end{lemma}

\begin{proof}
For $c\in\K^7$, $\sum_ic_iP_i$ has pure part $\sum_ic_i\pi_i$ and bilinear part $\sum_ic_iS_i$. Since $(AB)_{jl}=\sum_i\lambda^{(jl)}_iP_i$ is bilinear, $\lambda^{(jl)}\in K$. If $\sum z_{jl}\lambda^{(jl)}=0$ then $\sum z_{jl}(AB)_{jl}=0$, so $z=0$. As $\dim K=7-p$, $p\le3$. If $p=3$, then $\dim K=4$ and the $\lambda^{(jl)}$ form a basis. For $c\in K$ the form $\sum_ic_iP_i$ is bilinear, the map is injective because the products are independent, and it sends the basis $\lambda^{(jl)}$ to the basis $(AB)_{jl}$ of $W$. A bilinear element $\sum c_iP_i$ of $\Pi$ has $c\in K$. This proves (b). For (c), $\sum_iy_i\Lambda_i=0$ means $y\perp\lambda^{(jl)}$ for all $j,l$, that is, $y\in K^\perp$ by (b). Since $K$ is the kernel of $c\mapsto\sum_ic_i\pi_i$, $K^\perp$ is the image of the transposed map $\varphi\mapsto(\varphi(\pi_i))_i$.
\end{proof}

Comparing the bilinear parts and the two pure parts of the identity $AB=\sum_i\Lambda_iP_i$ gives
\begin{equation}\label{mm:eq:bil}
    AB=\sum_i\Lambda_iS_i(A,B),\qquad\sum_i\Lambda_i\,\alpha_i\gamma_i=0,\qquad\sum_i\Lambda_i\,\beta_i\delta_i=0.
\end{equation}

\begin{lemma}[Kernel count]\label{mm:lem:kernel}
Assume $p=3$. Let $I\subseteq\{1,\dots,7\}$, and let $X_1\times Y_1,\dots,X_m\times Y_m$ be products of subspaces of $\cA\times\cB$ such that $S_i$ vanishes on every $X_j\times Y_j$ for every $i\in I$. Then
\begin{displaymath}
    |I|-\dim\spn\{\pi_i:\ i\in I\}\le4-\dim\sum_jX_jY_j,
\end{displaymath}
where $X_jY_j=\spn\{AB:\ A\in X_j,\ B\in Y_j\}$. In particular $|I|\le7-\dim\sum_jX_jY_j$.
\end{lemma}

\begin{proof}
The space $K_I$ of vectors $c\in K$ supported on $I$ is the kernel of $c\mapsto\sum_{i\in I}c_i\pi_i$ on $\K^I$, so it has dimension at least $|I|-\dim\spn\{\pi_i:\ i\in I\}$. By Lemma~\ref{mm:lem:relations}(b) it maps injectively into $W$, and its image consists of forms $w_Z$ vanishing on every $X_j\times Y_j$, that is, with $\tr(ZM)=0$ for all $M\in\sum_jX_jY_j$. These form a space of dimension $4-\dim\sum_jX_jY_j$.
\end{proof}

\begin{lemma}[Restriction]\label{mm:lem:restriction}
In a seven-product algorithm there is no nonzero subspace $X\subseteq\cA$ on which all forms $\alpha_i,\gamma_i$ vanish. Likewise there is no nonzero subspace of $\cB$ on which all $\beta_i,\delta_i$ vanish.
\end{lemma}

\begin{proof}
For $A\in X$ the identity $AB=\sum_i\Lambda_iP_i(A,B)$ becomes $AB=\sum_i\Lambda_i\beta_i(B)\delta_i(B)$, whose right side does not depend on $A$. Putting $A=0$ shows that the right side vanishes, so $AB=0$ for all $A\in X$ and $B\in\cB$, which forces $X=0$. The B side is the same.
\end{proof}

\begin{lemma}[Symmetries]\label{mm:lem:sym}
\textup{(a)} Let $X,Y,Z\in GL_2(\K)$, and put $P'_i(A,B)=P_i(X^{-1}AY,\,Y^{-1}BZ)$ and $\Lambda'_i=X\Lambda_iZ^{-1}$. Then the $P'_i$ with the $\Lambda'_i$ form an algorithm. \textup{(b)} The forms $\tilde P_i(A,B)=P_i(B^T,A^T)$ with $\tilde\Lambda_i=\Lambda_i^T$ form an algorithm, in which the roles of the two sides are exchanged. In particular $(\tilde p_A,\tilde p_B)=(p_B,p_A)$. Both transformations preserve $W$, the space $V$ up to a natural isomorphism, the points $[\pi_i]$, and the property $S_i\in W$.
\end{lemma}

\begin{proof}
Substitute $A\mapsto X^{-1}AY$, $B\mapsto Y^{-1}BZ$ in $AB=\sum\Lambda_iP_i$ and multiply by $X$ on the left and $Z^{-1}$ on the right. For (b), $\sum\Lambda_iP_i(B^T,A^T)=B^TA^T=(AB)^T$. The remaining claims are clear.
\end{proof}

\subsection{\texorpdfstring{Products when $p=3$}{Products when p=3}}\label{mm:sec:products}

In this section $p=3$. Let $N_A=V\cap(\Sy^2\cA^*\oplus0)$ and $N_B=V\cap(0\oplus\Sy^2\cB^*)$. These are the kernels of the projections $V\to V_B$ and $V\to V_A$, so $\dim N_A=3-p_B$ and $\dim N_B=3-p_A$.

\begin{lemma}[Types]\label{mm:lem:types}
Assume $p=3$. Then $\pi_i\neq0$ for all $i$, and after exchanging its factors each product has exactly one of the following three types.
\begin{itemize}
    \item \textup{Type AB:} $\alpha_i,\beta_i,\gamma_i,\delta_i$ are all nonzero. Then both components of $\pi_i$ are nonzero.
    \item \textup{Type A:} $\delta_i=0$ and $\alpha_i,\gamma_i\neq0$. Then $P_i=(\alpha_i+\beta_i)\gamma_i$, $\pi_i=(\alpha_i\gamma_i,0)\in N_A$ and $S_i=\gamma_i\otimes\beta_i$.
    \item \textup{Type B:} $\alpha_i=0$ and $\beta_i,\delta_i\neq0$. Then $P_i=\beta_i(\gamma_i+\delta_i)$, $\pi_i=(0,\beta_i\delta_i)\in N_B$ and $S_i=\gamma_i\otimes\beta_i$.
\end{itemize}
In particular two products with the same point $[\pi_i]$ have the same type.
\end{lemma}

\begin{proof}
If $\pi_i=0$, then $\alpha_i\gamma_i=0$ and $\beta_i\delta_i=0$, so $P_i=\alpha_i\delta_i+\gamma_i\beta_i$ with $\alpha_i=0$ or $\gamma_i=0$, and $\beta_i=0$ or $\delta_i=0$. In each of the four cases $P_i$ is zero or has rank one. But $P_i$ is a nonzero bilinear element of $\Pi$, hence of $W$ (Lemma~\ref{mm:lem:relations}(b)), contradicting Fact~\ref{mm:fact:W}. If $\alpha_i\gamma_i\neq0\neq\beta_i\delta_i$, the product has type AB. If $\beta_i\delta_i=0$, then $\alpha_i\gamma_i\neq0$, and after exchanging the factors in case $\beta_i=0$ we have $\delta_i=0$, which is type A ($\beta_i=0$ is allowed). The case $\alpha_i\gamma_i=0$ is the same with the roles of the sides exchanged. The last claim holds because the three types are distinguished by which components of $\pi_i$ vanish.
\end{proof}

The \emph{point} of $P_i$ is $[\pi_i]\in\Pj(V)$. A point carrying two products is a \emph{double point}, a point carrying one product a \emph{single point}. We write $k_A$ and $k_B$ for the numbers of products of types A and B.

\begin{lemma}[Double points]\label{mm:lem:double}
Assume $p=3$, and let $P,P'$ be distinct products of type AB with the same point. After rescaling $P'$ and exchanging its factors if necessary, $P=(\alpha+\beta)(\gamma+\delta)$ and $P'=(\alpha+r\beta)(\gamma+r^{-1}\delta)$ with $r\notin\{0,1\}$. Then
\begin{displaymath}
    S-S'=(1-r^{-1})\,\alpha\otimes\delta+(1-r)\,\gamma\otimes\beta
\end{displaymath}
is a nonzero multiple of $x^TABy$ for some $x,y\neq0$, the forms $\alpha,\gamma$ form a basis of $\Row(x)$, and $\beta,\delta$ form a basis of $\Col(y)$. No point carries three products.
\end{lemma}

\begin{proof}
Rescale $P'$ so that $\pi'=\pi$, that is, $\alpha'\gamma'=\alpha\gamma$ and $\beta'\delta'=\beta\delta$. By Fact~\ref{mm:fact:binary}(a), after exchanging the factors of $P'$ we have $(\alpha',\gamma')=(s\alpha,s^{-1}\gamma)$, and either $(\beta',\delta')=(t\beta,t^{-1}\delta)$ or $(\beta',\delta')=(t\delta,t^{-1}\beta)$. The difference $P-P'$ lies in $\Pi$ and has zero pure part, so it is a nonzero element of $W$ (Lemma~\ref{mm:lem:relations}(b) and independence). In the second case, with $u=s/t$,
\begin{displaymath}
    S-S'=\alpha\otimes\delta+\gamma\otimes\beta-u\,\alpha\otimes\beta-u^{-1}\gamma\otimes\delta=(\alpha-u^{-1}\gamma)\otimes(\delta-u\beta)
\end{displaymath}
has rank at most one, which is impossible by Fact~\ref{mm:fact:W}. In the first case $P'=(s\alpha+t\beta)(s^{-1}\gamma+t^{-1}\delta)=(\alpha+r\beta)(\gamma+r^{-1}\delta)$ with $r=t/s$, and $r\neq1$ as $P'\neq P$. The difference $S-S'$ has rank at most two, hence rank two. By Fact~\ref{mm:fact:W} it is a multiple of some $x^TABy$, with $\alpha,\gamma$ a basis of $\Row(x)$ and $\beta,\delta$ a basis of $\Col(y)$. Suppose a third product $P''$ had the same point. It has type AB by Lemma~\ref{mm:lem:types}. The spaces $\Row(x)$ and $\Col(y)$ are spanned by the factors of $P$. So $P-P'$ and $P-P''$ would both be multiples of $x^TABy$, giving a linear relation among $P,P',P''$.
\end{proof}

\begin{lemma}[Double points of types A and B]\label{mm:lem:doubleAB}
Assume $p=3$.
\begin{enumerate}
    \item[(a)] Let $P\neq P'$ be products of type A with the same point. After rescaling them, $P=(\alpha+\beta)\gamma$ and $P'=(\gamma+\beta')\alpha$, where $\gamma\otimes\beta-\alpha\otimes\beta'=x^TABy$ for some $x,y\neq0$. The forms $\alpha,\gamma$ form a basis of $\Row(x)$ and $\beta,\beta'$ a basis of $\Col(y)$.
    \item[(b)] Let $P\neq P'$ be products of type B with the same point. After rescaling them, $P=\beta(\gamma+\delta)$ and $P'=\delta(\gamma'+\beta)$, where $\gamma\otimes\beta-\gamma'\otimes\delta=x^TABy$ for some $x,y\neq0$. The forms $\gamma,\gamma'$ form a basis of $\Row(x)$ and $\beta,\delta$ a basis of $\Col(y)$.
    \item[(c)] No point carries three products.
\end{enumerate}
\end{lemma}

\begin{proof}
(a) Write $P=(\alpha+\beta)\gamma$ and rescale $P'$ so that its pure part is $(\alpha\gamma,0)$. By Fact~\ref{mm:fact:binary}(a), after rescaling its factors, either $P'=(\alpha+\beta'')\gamma$ or $P'=(\gamma+\beta')\alpha$. In the first case $P-P'=\gamma\otimes(\beta-\beta'')$ is a bilinear element of $\Pi$ of rank at most one, hence zero by Lemma~\ref{mm:lem:relations}(b) and Fact~\ref{mm:fact:W}, so $P=P'$. In the second case $P-P'=\gamma\otimes\beta-\alpha\otimes\beta'$ is a nonzero element of $W$, so it has rank two, and Fact~\ref{mm:fact:W} gives the claim. (b) is the same with the roles of the sides exchanged. (c) For type AB this is Lemma~\ref{mm:lem:double}. For type A, the proof of (a) shows that every product of type A with the same point as $P$ is a multiple of $P$ or of a product $(\gamma+\beta')\alpha$. Two products $(\gamma+\beta')\alpha$ and $(\gamma+\beta'')\alpha$ with the same pure part differ by $\alpha\otimes(\beta'-\beta'')$, so they coincide by the same argument. Type B is the same.
\end{proof}

For a double point (of any type) we call $[x]$ its \emph{row}, $[y]$ its \emph{column}, and $x^TABy$ its \emph{W-element}. After rescaling, the difference of its two products is its W-element.

\begin{lemma}\label{mm:lem:tau}
Assume $p=3$. There is a unique linear map $\tau:V\to(\cA^*\otimes\cB^*)/W$ with $\tau(\pi_i)=[S_i]$ for all $i$, where $[\,\cdot\,]$ denotes the class modulo $W$.
\end{lemma}

\begin{proof}
The linear map $\K^7\to(\cA^*\otimes\cB^*)/W$, $c\mapsto[\sum_ic_iS_i]$, vanishes on $K$, because $\sum_ic_iP_i=\sum_ic_iS_i\in W$ for $c\in K$ (Lemma~\ref{mm:lem:relations}(b)). Hence it factors through the surjection $c\mapsto\sum_ic_i\pi_i$ from $\K^7$ onto $V$, whose kernel is $K$. Uniqueness holds because the $\pi_i$ span $V$.
\end{proof}

\subsection{\texorpdfstring{First consequences of $p_A=p_B=3$}{First consequences of pA=pB=3}}\label{mm:sec:first}

From now on until Section~\ref{mm:sec:assembly} we fix a seven-product algorithm with $p_A=p_B=3$. Since $3=p_A\le p\le3$, we have $p=3$, and the projections $V\to V_A$, $V\to V_B$ are isomorphisms. Let $\vartheta:V_A\to V_B$ be the isomorphism with $V=\{(q,\vartheta(q))\}$.

\begin{lemma}\label{mm:lem:allAB}
Every product has type AB. In particular all forms $\alpha_i,\beta_i,\gamma_i,\delta_i$ are nonzero.
\end{lemma}

\begin{proof}
Here $N_A=N_B=0$, so by Lemma~\ref{mm:lem:types} no product has type A or B.
\end{proof}

Let $d$ be the number of distinct points.

The next lemma is pure linear algebra. It is also used in Sections~\ref{mm:sec:twoq} and~\ref{mm:sec:pB1}.

\begin{lemma}[Pencils]\label{mm:lem:pencil}
Let $q_1,\dots,q_m$ \textup($m\ge3$\textup) be pairwise non-proportional nonzero products of linear forms, $q_k=\ell_km_k$, lying in a $2$-dimensional space of quadratic forms. Then all $\ell_k,m_k$ lie in a subspace of dimension at most $3$.
\end{lemma}

\begin{proof}
Write $q_3=aq_1+bq_2$ with $ab\neq0$. If $\ell_1,m_1,\ell_2,m_2$ were linearly independent, $aq_1+bq_2$ would have rank $4$. So they span a space $F'$ of dimension at most $3$. Every $q_k$ lies in $\spn\{q_1,q_2\}$, so it is a quadratic form in coordinates of $F'$, and by unique factorization $\ell_k,m_k\in F'$.
\end{proof}

\begin{lemma}[Collinear points]\label{mm:lem:collinear}
\textup{(a)} If three distinct points are collinear in $\Pj(V)$, none of them is a double point. \textup{(b)} No four distinct points are collinear.
\end{lemma}

\begin{proof}
(a) Pick one product at each of the three points and a relation $c_1\pi_1+c_2\pi_2+c_3\pi_3=0$. All $c_j\neq0$ because the points are distinct. Then $w=c_1P_1+c_2P_2+c_3P_3$ is a nonzero element of $W$. The A-parts $\alpha_j\gamma_j$ are pairwise non-proportional, as $V\cong V_A$, and span a $2$-dimensional space. So by Lemma~\ref{mm:lem:pencil}, all the forms $\alpha_j,\gamma_j$ lie in a space $F'_A$ of dimension at most $3$. Similarly the $\beta_j,\delta_j$ lie in some $F'_B$. Hence $w$ has rank at most $3$, so rank $2$, and $w$ is a multiple of $x^TABy$ with $\Row(x)\subseteq F'_A$ and $\Col(y)\subseteq F'_B$ (Fact~\ref{mm:fact:W}). If the first point were a double point with row $[x_1]$ and column $[y_1]$, then also $\Row(x_1)\subseteq F'_A$ (Lemma~\ref{mm:lem:double}), so $[x]=[x_1]$ by Fact~\ref{mm:fact:W} and dimension count, and likewise $[y]=[y_1]$. Then $w$ is a multiple of the difference of the two products at that point, a nontrivial linear relation among four products.

(b) Pick one product at each of four collinear points. Their pure parts satisfy a $2$-dimensional space of relations, whose images form a $2$-dimensional subspace of $W$. By Lemma~\ref{mm:lem:pencil} all A-factors of the four products lie in some $F'_A$ and all B-factors in some $F'_B$, both of dimension at most $3$. So as in (a) all nonzero elements of this subspace have rank two. So it equals $\{w_Z:\ Z\in U\}$ for a $2$-dimensional space $U$ of matrices of rank at most one. By Fact~\ref{mm:fact:M2}(a), $U=\{yx_0^T\}$ or $U=\{y_0x^T\}$. In the first case the subspace contains $x_0^TABy_1$ and $x_0^TABy_2$ with $[y_1]\neq[y_2]$, so $\Col(y_1)+\Col(y_2)=\cB^*$ would lie in the at most $3$-dimensional $F'_B$. The second case is symmetric.
\end{proof}

\begin{corollary}\label{mm:cor:d}
$4\le d\le7$, and there are exactly $7-d$ double points. If $d=4$, no three of the four points are collinear.
\end{corollary}

\begin{proof}
By Lemma~\ref{mm:lem:double} each point carries one or two products. If $d=4$ there are three double points, so every three of the four points contain a double point, and Lemma~\ref{mm:lem:collinear}(a) applies.
\end{proof}

\subsection{\texorpdfstring{The case $d=4$}{The case d=4}}\label{mm:sec:d4}

\subsubsection{\texorpdfstring{Coordinates by a $4\times2$ matrix}{Coordinates by a 4x2 matrix}}
Let $J=\begin{pmatrix}0&1\\-1&0\end{pmatrix}$ and $N=N(A,B)=\begin{pmatrix}A\\B^TJ\end{pmatrix}\in\K^{4\times2}$. For $F\in\K^{4\times2}$ let $F\cdot N=\sum_{i,q}F_{iq}N_{iq}$. The map $F\mapsto F\cdot N$ is a bijection onto the linear forms on $\cA\times\cB$. We write $F=[f_1\,f_2]$ by columns, and split $p\in\K^4$ as $p=(p_a;p_b)$ with $p_a,p_b\in\K^2$. For $p\in\K^4$ and $h\in\K^2$,
\begin{equation}\label{mm:eq:pNh}
    p^TNh=p_a^TAh+(Jh)^TBp_b.
\end{equation}
In particular $\Row(x)=\{(x;0)^TNh\}$ and $\Col(y)=\{(0;y)^TNh\}$, where $h$ ranges over $\K^2$.

For $p,q\in\K^4$ let $\mu(p,q)=(p^TNe_1)(q^TNe_2)-(p^TNe_2)(q^TNe_1)$, a $2\times2$ minor of $N$. Expanding with \eqref{mm:eq:pNh},
\begin{equation}\label{mm:eq:mu}
    \mu(p,q)=\det[p_a\,q_a]\det A+\bigl(p_a^TABq_b-q_a^TABp_b\bigr)+\det[p_b\,q_b]\det B.
\end{equation}
We extend $\mu$ linearly to $\Lambda^2\K^4$ by $\mu(p\wedge q)=\mu(p,q)$, and decompose $\Lambda^2\K^4=\Lambda^2\K^2\oplus(\K^2\otimes\K^2)\oplus\Lambda^2\K^2$, writing $\omega=\omega_a+\omega_m+\omega_b$, where
\begin{gather*}
    (p\wedge q)_a=p_a\wedge q_a,\qquad(p\wedge q)_b=p_b\wedge q_b,\\
    (p\wedge q)_m=p_a\otimes q_b-q_a\otimes p_b.
\end{gather*}
Let $\langle\cdot\rangle:\K^2\otimes\K^2\to W$ be the isomorphism $u\otimes z\mapsto u^TABz$ (Fact~\ref{mm:fact:W}). Identifying $\Lambda^2\K^2$ with $\K$ via $\det$, \eqref{mm:eq:mu} says $\mu(\omega)=\omega_a\det A+\langle\omega_m\rangle+\omega_b\det B$. Hence $\mu$ is injective with image $\K\det A\oplus W\oplus\K\det B$, and $\mu(\omega)\in W$ if and only if $\omega_a=\omega_b=0$.

\begin{lemma}[Splitting]\label{mm:lem:split}
Let $\mathcal S$ be the span of the quadratic forms
\begin{displaymath}
    (p^TNe_q)(p'^TNe_{q'})+(p^TNe_{q'})(p'^TNe_q),
\end{displaymath}
where $p,p'\in\K^4$ and $q,q'\in\{1,2\}$. Every quadratic form $Q$ on $\cA\times\cB$ can be written uniquely as $Q=Q^s+\mu(\omega_Q)$ with $Q^s\in\mathcal S$ and $\omega_Q\in\Lambda^2\K^4$. Moreover $Q^s=0$ if and only if $Q(A,B)=0$ whenever $\rank N(A,B)\le1$. For $Q=(F\cdot N)(G\cdot N)$,
\begin{gather*}
    Q(A,B)=(n^TFv)(n^TGv)\quad\text{when }N(A,B)=nv^T,\\
    \omega_Q=\tfrac12\,(f_1\wedge g_2-f_2\wedge g_1).
\end{gather*}
\end{lemma}

\begin{proof}
The quadratic forms are spanned by the products $(p^TNe_q)(p'^TNe_{q'})$. Each of them is half the sum of the corresponding generator of $\mathcal S$ and of $(p^TNe_q)(p'^TNe_{q'})-(p^TNe_{q'})(p'^TNe_q)$, which is $0$ if $q=q'$ and $\pm\mu(p,p')$ otherwise. So every $Q$ has such a decomposition. Every $\mu(\omega)$ vanishes when $N=nv^T$. At $N=nv^T$ the generator above takes the value $2(p\cdot n)(p'\cdot n)v_qv_{q'}$. Thus $\mathcal S$ is the image of the linear map $\Sy^2\K^4\otimes\Sy^2\K^2\to\mathcal S$, $(pp')\otimes(e_qe_{q'})\mapsto$ generator, and its composition with evaluation at $N=nv^T$ is the natural injection of $\Sy^2\K^4\otimes\Sy^2\K^2$ into polynomials of bidegree $(2,2)$ in $(n,v)$, multiplied by $2$. Hence an element of $\mathcal S$ vanishing at all rank-one $N$ is zero, which gives uniqueness and the criterion for $Q^s=0$. For $Q=(F\cdot N)(G\cdot N)=\sum_{q,q'}(f_q^TNe_q)(g_{q'}^TNe_{q'})$, the terms with $q\neq q'$ contribute $\frac12\mu(f_1,g_2)-\frac12\mu(f_2,g_1)$ to the second component, and $(F\cdot N)(nv^T)=n^TFv$.
\end{proof}

\begin{lemma}[Double points in $N$-coordinates]\label{mm:lem:dpN}
Let $P,P'$ be the products at a double point with row $[x]$ and column $[y]$. After rescaling $P$ and $P'$ there are $z\in\K^2$ with $[z]=[y]$, $r\notin\{0,1\}$ and a basis $h_1,h_2$ of $\K^2$ such that, with $\xi=(x;rz)$ and $\eta=(x;z)$,
\begin{displaymath}
    P=(\xi^TNh_1)(\eta^TNh_2),\qquad P'=(\eta^TNh_1)(\xi^TNh_2).
\end{displaymath}
Then $P$ and $P'$ have the same pure part. Both take the value $(\xi\cdot n)(\eta\cdot n)(h_1\cdot v)(h_2\cdot v)$ when $N=nv^T$; and $\omega_P=-\omega_{P'}=\frac12(1-r)\det[h_1\,h_2]\,(x\otimes z)$, which is purely mixed.
\end{lemma}

\begin{proof}
Write $P=(\alpha+\beta)(\gamma+\delta)$, $P'=(\alpha+r'\beta)(\gamma+r'^{-1}\delta)$ as in Lemma~\ref{mm:lem:double}, and rescale $y$ so that $(1-r'^{-1})\alpha\otimes\delta+(1-r')\gamma\otimes\beta=x^TABy$. Write $\alpha=x^TAh_1$, $\gamma=x^TAh_2$, with $h_1,h_2$ a basis, and let $k_1,k_2$ be the dual basis. Comparing with Fact~\ref{mm:fact:W}, $(1-r'^{-1})\delta=k_1^TBy$ and $(1-r')\beta=k_2^TBy$. Since $k_2\perp h_1$ and $k_1\perp h_2$, and $Jh$ spans $h^\perp$, we have $k_2=\theta Jh_1$ and $k_1=\theta'Jh_2$. From $k_1^Th_1=k_2^Th_2=1$ and $(Jh_2)^Th_1=-(Jh_1)^Th_2$ we get $\theta'=-\theta$. By \eqref{mm:eq:pNh}, $\beta=(0;\zeta)^TNh_1$ and $\delta=(0;\zeta')^TNh_2$ with $\zeta=\theta y/(1-r')$ and $\zeta'=\theta'y/(1-r'^{-1})=r'\zeta$. Hence
\begin{gather*}
    P=\bigl((x;\zeta)^TNh_1\bigr)\bigl((x;r'\zeta)^TNh_2\bigr),\\
    P'=\bigl((x;r'\zeta)^TNh_1\bigr)\bigl((x;\zeta)^TNh_2\bigr),
\end{gather*}
which is the claim with $z=r'\zeta$ and $r=1/r'$. By Lemma~\ref{mm:lem:split} with $(F,G)=(\xi h_1^T,\eta h_2^T)$, $P$ takes the stated value on rank-one $N$ and $\omega_P=\frac12\det[h_1\,h_2]\,\xi\wedge\eta$. Similarly $P'$ has the same value and $\omega_{P'}=-\omega_P$. Finally $(\xi\wedge\eta)_a=x\wedge x=0$, $(\xi\wedge\eta)_b=rz\wedge z=0$ and $(\xi\wedge\eta)_m=x\otimes z-x\otimes rz$. So $P-P'=\mu(2\omega_P)\in W$, and in particular the pure parts agree.
\end{proof}

\subsubsection{The equations}
Let $d=4$. By Corollary~\ref{mm:cor:d} there are three double points, indexed by $k=1,2,3$, whose products we write as in Lemma~\ref{mm:lem:dpN} with data $x_k,z_k,r_k,h_1^k,h_2^k,\xi_k,\eta_k$, and one single point with product $P_4=(F\cdot N)(G\cdot N)$. Let
\begin{displaymath}
    \sigma_k(v)=(h_1^k\cdot v)(h_2^k\cdot v),\qquad\psi_k(n)=(\xi_k\cdot n)(\eta_k\cdot n).
\end{displaymath}
Each $\sigma_k$ is a binary quadratic form with two distinct zeros in $\Pj^1$. As the four points are in general position, after rescaling $P_4$ we have $\pi_4=\sum_kc_k\pi(P_k)$ with $c_1c_2c_3\neq0$.

\begin{lemma}\label{mm:lem:eqs}
\textup{(a)} The vectors $x_1,x_2,x_3$ span $\K^2$, and so do $z_1,z_2,z_3$. The tensors $x_k\otimes z_k$ are linearly independent. In particular no two indices $k\neq l$ have both $[x_k]=[x_l]$ and $[z_k]=[z_l]$. The quadratic forms $\psi_1,\psi_2,\psi_3$ are linearly independent.

\textup{(b)} For all $n\in\K^4$ and $v\in\K^2$,
\begin{equation}\label{mm:eq:E1}
    (n^TFv)(n^TGv)=\sum_kc_k\,\sigma_k(v)\,\psi_k(n).
\end{equation}
\textup{(c)} The bivector $\omega=f_1\wedge g_2-f_2\wedge g_1$ satisfies
\begin{gather}
    \omega_a=0,\qquad\omega_b=0,\label{mm:eq:E2}\\
    \omega_m\notin\spn\{x_1\otimes z_1,\ x_2\otimes z_2,\ x_3\otimes z_3\}.\label{mm:eq:E3}
\end{gather}
\end{lemma}

\begin{proof}
(a) If all $x_k$ were proportional to some $x$, all A-factors of the six products at double points would lie in $\Row(x)$, and the three A-parts of the double points, which span $V_A$, would span $\Sy^2\Row(x)$. Then every A-part, including that of $P_4$, would have both factors in $\Row(x)$, so all A-factors would vanish on $\{A:\ x^TA=0\}$, contradicting Lemma~\ref{mm:lem:restriction}. The $z_k$ are treated likewise. By Lemma~\ref{mm:lem:dpN}, $P_k-P'_k$ is a nonzero multiple of $\langle x_k\otimes z_k\rangle$, so a linear relation among the $x_k\otimes z_k$ would give one among the products. For the $\psi_k$, suppose $\sum a_k\psi_k=0$. Setting $n_b=0$, respectively $n_a=0$, gives $\sum a_k(x_k\cdot n_a)^2=0$ and $\sum a_kr_k(z_k\cdot n_b)^2=0$. Squares of pairwise non-proportional vectors are independent (Fact~\ref{mm:fact:binary}(b)). If the $x_k$ are pairwise non-proportional, $a=0$. Otherwise exactly two of them, say $x_i,x_j$, are proportional, the first equation gives $a_l=0$ for the third index $l$, and since $[z_i]\neq[z_j]$ the second gives $a_i=a_j=0$.

(b), (c) The form $w_4=P_4-\sum_kc_kP'_k$ has zero pure part and lies in $\Pi$, so $w_4\in W$ by Lemma~\ref{mm:lem:relations}(b). The four vectors of $K$ corresponding to $P_k-P'_k$ ($k=1,2,3$) and to $w_4$ are linearly independent, so $w_4\notin\spn\{P_k-P'_k\}=\spn\{\langle x_k\otimes z_k\rangle\}$. As $w_4\in W\subseteq\mu(\Lambda^2\K^4)$, Lemma~\ref{mm:lem:split} gives $w_4^s=0$, which by Lemma~\ref{mm:lem:dpN} is \eqref{mm:eq:E1}. Next, $\omega_{w_4}=\frac12\omega-\sum_kc_k\omega_{P'_k}$, and the $\omega_{P'_k}$ are purely mixed multiples of $x_k\otimes z_k$. Since $\mu(\omega_{w_4})=w_4\in W$, we get $(\omega_{w_4})_a=(\omega_{w_4})_b=0$, which is \eqref{mm:eq:E2}, and $w_4=\langle(\omega_{w_4})_m\rangle\notin\spn\{\langle x_k\otimes z_k\rangle\}$, which is \eqref{mm:eq:E3}.
\end{proof}

\subsubsection{\texorpdfstring{The three double points have $r=-1$}{The three double points have r=-1}}
For a quadratic form $Q$ on $\K^4$ let $X(Q)\in\K^2\otimes\K^2$ be its \emph{cross part}, defined by linearity from $X\bigl((p\cdot n)(q\cdot n)\bigr)=p_a\otimes q_b+q_a\otimes p_b$. It records the terms of $Q$ that are bilinear in $(n_a,n_b)$. We have $X(\psi_k)=(1+r_k)\,x_k\otimes z_k$.

\begin{proposition}\label{mm:prop:r}
If $d=4$, then $r_1=r_2=r_3=-1$.
\end{proposition}

\begin{proof}
Put $f(v)=Fv$, $g(v)=Gv$, and $M(v)=\sum_kc_k\sigma_k(v)\psi_k$. By \eqref{mm:eq:E1}, $M(v)$ is the quadratic form $n\mapsto(f(v)\cdot n)(g(v)\cdot n)$, which we denote $f(v)g(v)$. By Lemma~\ref{mm:lem:eqs}(a), $\dim\spn\{M(v):\ v\in\K^2\}=\dim\spn\{\sigma_1,\sigma_2,\sigma_3\}$. Note $F,G\neq0$.

\emph{Case 1: $\rank F=1$} (the case $\rank G=1$ is symmetric). Then $F=\ell w^T$ and $f(v)=(w\cdot v)\ell$. Let $w^\perp=J^Tw$, so $w\cdot w^\perp=0$. Then $M(w^\perp)=0$, and Lemma~\ref{mm:lem:eqs}(a) gives $\sigma_k(w^\perp)=0$ for all $k$. So $\sigma_k(v)=(w\cdot v)\lambda_k(v)$ with linear forms $\lambda_k$, and $\lambda_k(w^\perp)\neq0$ because $\sigma_k$ has two distinct zeros. Dividing \eqref{mm:eq:E1} by $w\cdot v$ and putting $v=w^\perp$, $g^*=g(w^\perp)$:
\begin{gather}
    \ell g^*=\sum_kc_k\lambda_k(w^\perp)\psi_k,\qquad\text{so}\notag\\
    \ell_a\otimes g^*_b+g^*_a\otimes\ell_b=\sum_kc_k\lambda_k(w^\perp)(1+r_k)\,x_k\otimes z_k.\label{mm:eq:case1}
\end{gather}
On the other hand $f_q=w_q\ell$, so $\omega=\ell\wedge(w_1g_2-w_2g_1)=\ell\wedge g^*$, and \eqref{mm:eq:E2} says $\ell_a\wedge g^*_a=0=\ell_b\wedge g^*_b$. The left side of \eqref{mm:eq:case1} is then a multiple of $\omega_m=\ell_a\otimes g^*_b-g^*_a\otimes\ell_b$: if $\ell_a\neq0\neq\ell_b$, write $g^*_a=s\ell_a$, $g^*_b=t\ell_b$, so the left side is $(s+t)\ell_a\otimes\ell_b$ and $\omega_m=(t-s)\ell_a\otimes\ell_b$, which is nonzero by \eqref{mm:eq:E3}. If $\ell_a=0$ the left side is $-\omega_m$. If $\ell_b=0$ it is $\omega_m$. By \eqref{mm:eq:E3} the left side is therefore $0$, and Lemma~\ref{mm:lem:eqs}(a) gives $c_k\lambda_k(w^\perp)(1+r_k)=0$, so $r_k=-1$.

\emph{Case 2: $\rank F=\rank G=2$ and $\operatorname{Im}F=\operatorname{Im}G=:E$.} All $f_q,g_q$ lie in $E$, so $\omega\in\Lambda^2E=\K\,\varepsilon_1\wedge\varepsilon_2$ for a basis $\varepsilon_1,\varepsilon_2$ of $E$. By \eqref{mm:eq:E3}, $\omega_m\neq0$, so $\omega$ is a nonzero multiple of $\varepsilon_1\wedge\varepsilon_2$, and \eqref{mm:eq:E2} gives $\varepsilon_{1a}\wedge\varepsilon_{2a}=0=\varepsilon_{1b}\wedge\varepsilon_{2b}$. As $(\varepsilon_1\wedge\varepsilon_2)_m\neq0$, the a-components of $E$ span a line $\K e$ and the b-components a line $\K c$, so $E=\K(e;0)\oplus\K(0;c)$ and $\omega_m\in\K e\otimes c$. Every product of two forms from $E$ has cross part in $\K e\otimes c$. Applying $X$ to $M(v)=f(v)g(v)$ gives $\sum_kc_k\sigma_k(v)(1+r_k)x_k\otimes z_k\in\K e\otimes c$, and by \eqref{mm:eq:E3} the sum vanishes for every $v$. By Lemma~\ref{mm:lem:eqs}(a), $r_k=-1$.

\emph{Case 3: $\rank F=\rank G=2$ and $\dim(\operatorname{Im}F\cap\operatorname{Im}G)=1$.} Choose $p,q,s$ with $\operatorname{Im}F=\langle p,q\rangle$, $\operatorname{Im}G=\langle p,s\rangle$. They are independent. Write $f(v)=\alpha(v)p+\beta(v)q$ and $g(v)=\gamma(v)p+\delta(v)s$, where $\alpha,\beta$ and $\gamma,\delta$ are bases of $(\K^2)^*$. Then $M(v)=\alpha\gamma\,pp+\alpha\delta\,ps+\beta\gamma\,pq+\beta\delta\,qs$, where $pp,ps,pq,qs$ denote the products of the linear forms $n\mapsto p\cdot n$ etc.\ and are linearly independent. Since $\alpha\gamma,\alpha\delta,\beta\gamma,\beta\delta$ span all binary quadratic forms, $\spn\{M(v)\}$ is $3$-dimensional. Hence $\sigma_1,\sigma_2,\sigma_3$ are independent, the vectors $(\sigma_k(v))_k$ span $\K^3$, and $\spn\{M(v)\}=\spn\{\psi_1,\psi_2,\psi_3\}$. So each $\psi_k$ is a quadratic form in $n\mapsto p\cdot n$, $q\cdot n$, $s\cdot n$, and by unique factorization $\xi_k,\eta_k\in F':=\langle p,q,s\rangle$. Then $(0;z_k)=(\xi_k-\eta_k)/(r_k-1)\in F'$ and $(x_k;0)\in F'$ for all $k$, so by Lemma~\ref{mm:lem:eqs}(a) $F'=\K^4$, a contradiction.

\emph{Case 4: $\rank F=\rank G=2$ and $\operatorname{Im}F\cap\operatorname{Im}G=0$.} Then $f_1,f_2,g_1,g_2$ is a basis of $\K^4$, and $\spn\{f(v)g(v)\}$ is the net $\mathcal W=\{\sum_{q,q'}S_{qq'}f_qg_{q'}:\ S=S^T\}$. A member of $\mathcal W$ has rank $2\rank S$. As in Case~3, $\spn\{\psi_k\}=\mathcal W$. Since $\rank\psi_k=2$, $\psi_k=\theta_kf(u_k)g(u_k)$ with $u_k\neq0$ and $\theta_k\neq0$, and by unique factorization $\{[f(u_k)],[g(u_k)]\}=\{[\xi_k],[\eta_k]\}$. Let $F_a,G_a\in\K^{2\times2}$ be the upper halves of $F,G$. Then $F_au_k$ and $G_au_k$ are nonzero multiples of $x_k$. Comparing coefficients in $\mathcal W$, $vv^T=\sum_kc_k\theta_k\sigma_k(v)u_ku_k^T$ for all $v$, so the $u_k$ are pairwise non-proportional. The image of $F_a$ contains $x_1,x_2,x_3$, which span $\K^2$, so $F_a$ is invertible, and likewise $G_a$. If two of the $x_k$ were proportional, $F_a$ would map two non-proportional $u_k$ to proportional vectors. So the $x_k$ are pairwise non-proportional, $G_aF_a^{-1}$ has the three pairwise non-proportional eigenvectors $x_1,x_2,x_3$, and $G_a=\nu F_a$ with $\nu\neq0$. Then $\omega_a=F_ae_1\wedge G_ae_2-F_ae_2\wedge G_ae_1=2\nu\,F_ae_1\wedge F_ae_2\neq0$, contradicting \eqref{mm:eq:E2}.
\end{proof}

\subsubsection{\texorpdfstring{The structure for $d=4$}{The structure for d=4}}
A \emph{W-product} is a product whose bilinear part is a rank-two element $x^TABy$ of $W$. As in the proof of Lemma~\ref{mm:lem:double}, a W-product can be written as
\begin{equation}\label{mm:eq:wprod}
    (x^TAh_1+k_2^TBy)(x^TAh_2+k_1^TBy)
\end{equation}
with a basis $h_1,h_2$ of $\K^2$ and its dual basis $k_1,k_2$ (a scalar factor can be absorbed into $x,y$ since $\K$ is algebraically closed). By unique factorization the unordered pair $\{[h_1],[h_2]\}$ is determined by the product. We call it the \emph{middle pair}. The A-part of the pure part of \eqref{mm:eq:wprod} is $(x^TAh_1)(x^TAh_2)=\iota_A(x^2\otimes h_1h_2)$, and the B-part is $(k_1^TBy)(k_2^TBy)=\iota_B(k_1k_2\otimes y^2)$, where $\iota_A:\Sy^2\K^2\otimes\Sy^2\K^2\to\Sy^2\cA^*$ is the injective linear map with $\iota_A\bigl((uu')\otimes(hh')\bigr)=\frac12\bigl[(u^TAh)(u'^TAh')+(u^TAh')(u'^TAh)\bigr]$, and $\iota_B$ is defined in the same way. The middle pair determines $[h_1h_2]$ and $[k_1k_2]$, and conversely.

\begin{lemma}\label{mm:lem:wprod}
If $d=4$, every product is a W-product, all seven products have the same middle pair, and the two products at a double point have the forms \eqref{mm:eq:wprod} with $(x,y)$ and $(x,-y)$.
\end{lemma}

\begin{proof}
By Proposition~\ref{mm:prop:r}, $\xi_k=(x_k;-z_k)$ and $\eta_k=(x_k;z_k)$. Let $k_1^k,k_2^k$ be the dual basis of $h_1^k,h_2^k$. As in the proof of Lemma~\ref{mm:lem:dpN}, $k_2^k=\theta_kJh_1^k$ and $k_1^k=-\theta_kJh_2^k$ for some $\theta_k\neq0$, so \eqref{mm:eq:pNh} gives $P_k=(x_k^TAh_1^k+k_2^{kT}By_k)(x_k^TAh_2^k+k_1^{kT}By_k)$ and $P'_k=(x_k^TAh_1^k-k_2^{kT}By_k)(x_k^TAh_2^k-k_1^{kT}By_k)$ with $y_k=-z_k/\theta_k$. These are W-products with middle pair $\{[h_1^k],[h_2^k]\}$, and $S(P'_k)=-x_k^TABy_k$. Since $w_4$ is bilinear, $w_4=S(P_4)-\sum_kc_kS(P'_k)$. Hence $S(P_4)\in W$, and $S(P_4)\neq0$ since otherwise $w_4\in\spn\{\langle x_k\otimes z_k\rangle\}$, contrary to Lemma~\ref{mm:lem:eqs}. So $P_4$ is a W-product too, and after rescaling $P_4$ (which changes the $c_k$ by a common nonzero factor) it has the form \eqref{mm:eq:wprod} with data $x_4,y_4,h_1^4,h_2^4$. Write $Q_k=h_1^kh_2^k$ and $\hat Q_k=k_1^kk_2^k$ for $k\le4$. The relation $\pi_4=\sum_kc_k\pi(P_k)$ and injectivity of $\iota_A,\iota_B$ give
\begin{equation}\label{mm:eq:middle}
    x_4^2\otimes Q_4=\sum_{k=1}^3c_kx_k^2\otimes Q_k,\qquad\hat Q_4\otimes y_4^2=\sum_{k=1}^3c_k\hat Q_k\otimes y_k^2.
\end{equation}
If $x_1,x_2,x_3$ are pairwise non-proportional, write $x_4^2=\sum\mu_kx_k^2$ (Fact~\ref{mm:fact:binary}(b)). Then $\sum_kx_k^2\otimes(c_kQ_k-\mu_kQ_4)=0$, so $c_kQ_k=\mu_kQ_4$ and all $[Q_k]$ equal $[Q_4]$. The same argument applies on the B side if $y_1,y_2,y_3$ are pairwise non-proportional. Otherwise, by Lemma~\ref{mm:lem:eqs}(a), there are a proportional pair of $x$'s and a different proportional pair of $y$'s. Relabel so that $x_2=\lambda x_1$ and $[x_3]\neq[x_1]$. The first equation reads $x_1^2\otimes(c_1Q_1+c_2\lambda^2Q_2)+c_3x_3^2\otimes Q_3=x_4^2\otimes Q_4$. Applying a functional on $\Sy^2\K^2$ that vanishes on $x_1^2,x_3^2$ shows $x_4^2\in\spn\{x_1^2,x_3^2\}$, so $[x_4]\in\{[x_1],[x_3]\}$. If $[x_4]=[x_1]$, a functional vanishing on $x_1^2$ but not on $x_3^2$ gives $c_3Q_3=0$, which is false. Hence $[x_4]=[x_3]$, $c_1Q_1+c_2\lambda^2Q_2=0$ and $c_3Q_3\in\K Q_4$, that is, $[Q_1]=[Q_2]$ and $[Q_3]=[Q_4]$. The proportional pair of $y$'s is $\{1,3\}$ or $\{2,3\}$, and the same argument on the B side gives $[\hat Q_1]=[\hat Q_3]$ and $[\hat Q_2]=[\hat Q_4]$, respectively $[\hat Q_2]=[\hat Q_3]$ and $[\hat Q_1]=[\hat Q_4]$. In both cases all four middle pairs coincide.
\end{proof}

\begin{corollary}\label{mm:cor:d4}
If $d=4$, the conclusions \textup{(a)}, \textup{(b)}, \textup{(c)} of Theorem~\ref{mm:thm:main} hold.
\end{corollary}

\begin{proof}
(a) is Corollary~\ref{mm:cor:d} and (b) is Lemma~\ref{mm:lem:wprod}. For (c) choose $H$ with $[He_1]$, $[He_2]$ the common middle pair. If $h_1=sHe_1$ and $h_2=tHe_2$, the dual basis is $k_1=s^{-1}H^{-T}e_1$, $k_2=t^{-1}H^{-T}e_2$, and \eqref{mm:eq:wprod} equals $st\,(x^TA'e_1+e_2^TB'y')(x^TA'e_2+e_1^TB'y')$ with $y'=y/(st)$. Exchanging $h_1,h_2$ only exchanges the two factors.
\end{proof}

\subsection{Nets of quadratic forms}\label{mm:sec:nets}

In this section $U$ is a $4$-dimensional vector space and a \emph{net} is a $3$-dimensional subspace $\mathcal N\subseteq\Sy^2U^*$.

\begin{lemma}[Singular nets]\label{mm:lem:singnets}
Let $\mathcal N$ be a net all of whose members are singular. Then one of the following holds.
\begin{enumerate}
    \item[(i)] All members of $\mathcal N$ have a common nonzero singular vector.
    \item[(ii)] The generic member has rank $3$, and there is a decomposition $U^*=L\oplus\langle\varphi,\psi\rangle$ with $\dim L=2$ such that $\mathcal N\subseteq\varphi\cdot L+\Sy^2\langle\varphi,\psi\rangle$.
    \item[(iii)] Some nonzero linear form divides every member of $\mathcal N$.
\end{enumerate}
\end{lemma}

\begin{proof}
Choose coordinates $t\in\K^3$ on $\mathcal N$ and let $M(t)$ be the symmetric $4\times4$ matrix of the member $q_t$.

\emph{Generic rank $3$.} The adjugate $\operatorname{adj}M(t)$ has entries that are cubic forms in $t$, and rank one for generic $t$. Over the field $\K(t)$ a symmetric matrix of rank one is $\lambda uu^T$. Clearing denominators, $\operatorname{adj}M(t)=g(t)\kappa(t)\kappa(t)^T$ with $g\in\K[t]$ and a primitive $\kappa\in\K[t]^4$, so $\deg g+2\deg\kappa=3$. From $M\operatorname{adj}M=\det M\cdot I=0$ we get $M(t)\kappa(t)=0$. If $\deg\kappa=0$ we are in case (i). Let $\deg\kappa=1$, so $\kappa$ is linear with image $K\subseteq U$. Polarizing $M(t)\kappa(t)=0$ gives $M(t)\kappa(t')+M(t')\kappa(t)=0$. The trilinear form $T(t,t',t'')=\kappa(t'')^TM(t)\kappa(t')$ is therefore antisymmetric in $(t,t')$ and symmetric in $(t',t'')$, hence zero. So $K$ is totally isotropic for every member. If $\dim K=1$ then $\kappa(t)=\lambda(t)k_0$ and $M(t)k_0=0$: case (i). If $\dim K=3$, every member vanishes on a hyperplane and has rank at most $2$, which is excluded. Let $\dim K=2$, choose a basis of $U$ starting with a basis of $K$, and let $x_1,\dots,x_4$ be the dual coordinates. Then $M(t)=\begin{pmatrix}0&X(t)\\X(t)^T&Y(t)\end{pmatrix}$ in $2\times2$ blocks with $X,Y$ linear in $t$, and $\det M(t)=(\det X(t))^2$. So $\{X(t)\}$ is a space of singular matrices. If it is $0$, then $K$ is a common kernel: case (i). Otherwise it has dimension at most $2$ and by Fact~\ref{mm:fact:M2}(a) it is contained in $\{x_0h^T\}$ or in $\{hy_0^T\}$. Write $\kappa(t)=(\bar\kappa(t);0)$. The identity $M(t)\kappa(t)=0$ gives $X(t)^T\bar\kappa(t)=0$. If $X(t)=x_0h(t)^T$, this is $h(t)\,(x_0^T\bar\kappa(t))=0$, so $x_0^T\bar\kappa(t)=0$ for generic and hence all $t$, contradicting $\dim K=2$. So $X(t)=h(t)y_0^T$ and $q_t=2\varphi\,\ell_t+Y_t$ with $\varphi=y_0^T(x_3,x_4)^T$, $\ell_t=h(t)^T(x_1,x_2)^T\in L:=\langle x_1,x_2\rangle$ and $Y_t\in\Sy^2\langle x_3,x_4\rangle$. This is case (ii), with $\psi$ completing $\varphi$ to a basis of $\langle x_3,x_4\rangle$.

\emph{Generic rank at most $2$.} If all members had rank at most $1$, $\mathcal N$ would consist of squares. But $\ell^2+m^2$ has rank $2$ for non-proportional $\ell,m$, so $\dim\mathcal N\le1$. Fix $q_1=\ell_1\ell_2\in\mathcal N$ of rank $2$ and put $E_1=\langle\ell_1,\ell_2\rangle$. Let $q_2\in\mathcal N$ have rank $2$ and factor space $E_2$. If $E_1+E_2=U^*$, then $q_1+q_2$ has rank $4$. If $\dim(E_1+E_2)=3$, let $E_1\cap E_2=\langle e\rangle$, $E_1=\langle e,f\rangle$, $E_2=\langle e,g\rangle$ and $q_1=ae^2+2bef+cf^2$, $q_2=a'e^2+2b'eg+c'g^2$ with $ac\neq b^2$ and $a'c'\neq b'^2$. In the coordinates $e,f,g$,
\begin{displaymath}
    \det(q_1+sq_2)=s\bigl[c'(ac-b^2)+s\,c(a'c'-b'^2)\bigr],
\end{displaymath}
which vanishes identically only if $c=c'=0$, that is, $e$ divides $q_1$ and $q_2$. So $e$ is a multiple of $\ell_1$ or $\ell_2$. Hence every rank-two member of $\mathcal N$ lies in $\Sy^2E_1\cup\ell_1U^*\cup\ell_2U^*$. The rank-two members form a nonempty Zariski-open subset of $\mathcal N$, so $\mathcal N$ lies in one of these three subspaces. If $\mathcal N=\Sy^2E_1$ we are in case (i), otherwise in case (iii).
\end{proof}

\begin{lemma}[Conics of rank-two forms]\label{mm:lem:conic}
Let $\mathcal N$ be a net such that $\det|_{\mathcal N}=\mu C^2$ for a smooth conic $C\subseteq\Pj(\mathcal N)$ and $\mu\neq0$, and such that every point of $C$ is a form of rank at most $2$. Then there are linear isomorphisms $a:\K^2\to E_a$, $b:\K^2\to E_b$ with $U^*=E_a\oplus E_b$ such that $C=\{[a(v)b(v)]:\ v\in\K^2\setminus\{0\}\}$. Consequently $\mathcal N=\spn\{a(v)b(v)\}$.
\end{lemma}

\begin{proof}
Let $q,q'$ represent distinct points of $C$. The line through them meets $C$ only in these points, so $\det(sq+tq')=\mu(\gamma st)^2$ with $\gamma\neq0$. Hence $\rank q=\rank q'=2$, since rank at most one would give a zero of order at least $3$ at an end point. The factor spaces of $q$ and $q'$ are complementary, since otherwise the whole pencil would be singular. Choose coordinates $x_1,\dots,x_4$ on $U$ with $q=x_1x_2$ and $q'=x_3x_4$, and a quadratic parametrization $v\mapsto q(v)$ of $C$ with $q(e_1)=q$ and $q(e_2)=q'$ (possible since $\mathrm{PGL}_2$ acts transitively on pairs of distinct points of $C$ and $\K$ is algebraically closed). Then $q(1,t)=q+tq''+t^2q'$. Let $M(t)$ be its matrix and write the matrix of $q''$ as $\begin{pmatrix}P&R\\R^T&S\end{pmatrix}$ in $2\times2$ blocks, and $J'=\frac12\begin{pmatrix}0&1\\1&0\end{pmatrix}$. Then
\begin{displaymath}
    M(t)=\begin{pmatrix}J'+tP&tR\\tR^T&tS+t^2J'\end{pmatrix}.
\end{displaymath}
For all but finitely many $t$ the block $J'+tP$ is invertible, and $\rank M(t)\le2$ forces the Schur complement to vanish: $tS+t^2J'-t^2R^T(J'+tP)^{-1}R=0$. As an identity of rational functions this gives $S=0$ (divide by $t$ and put $t=0$), and then $J'=R^T(J'+tP)^{-1}R$ for all $t$. At $t=0$ this reads $J'=R^TJ'^{-1}R$, so $R$ is invertible and $(J'+tP)^{-1}=J'^{-1}$, that is, $P=0$. Writing $\sigma=2J'$, the equation becomes $(2R)^T\sigma(2R)=\sigma$, whose solutions are $2R=\operatorname{diag}(\lambda,\lambda^{-1})$ and $2R=\begin{pmatrix}0&\lambda\\\lambda^{-1}&0\end{pmatrix}$. Since $q''=2(x_1,x_2)R(x_3,x_4)^T$, we get $q(1,t)=(x_1+t\lambda^{-1}x_4)(x_2+t\lambda x_3)$, respectively $q(1,t)=(x_1+t\lambda^{-1}x_3)(x_2+t\lambda x_4)$. Homogenizing gives the claim with $E_a=\langle x_1,\lambda^{-1}x_4\rangle$, $E_b=\langle x_2,\lambda x_3\rangle$, respectively with $x_3,x_4$ exchanged. Finally $C$ spans $\Pj(\mathcal N)$.
\end{proof}

\subsection{\texorpdfstring{The types of $V_A$ when $d\ge5$}{The types of VA when d>=5}}\label{mm:sec:types}

Let $D_A$ be the determinant restricted to $V_A\cong V$, a quartic form on $V$. At each point the A-part $\alpha_i\gamma_i$ has rank at most $2$, so all $3\times3$ minors vanish there. So $D_A$ is singular at every point: it vanishes there together with its first derivatives.

\begin{proposition}\label{mm:prop:types}
Assume $d\ge5$. Then one of the following holds.
\begin{enumerate}
    \item[(S5)] $d=5$, the three single points lie on a line, and the two double points do not.
    \item[(T1)] There are linear isomorphisms $a:\K^2\to E_a$, $b:\K^2\to E_b$ with $\cA^*=E_a\oplus E_b$ such that $V_A=\spn\{a(v)b(v)\}$, and for every $i$ there is $v_i\neq0$ with $\alpha_i\gamma_i\in\K\,a(v_i)b(v_i)$.
    \item[(T4)] Some nonzero linear form $\ell_0$ divides every element of $V_A$.
\end{enumerate}
The same holds for $V_B$ in place of $V_A$.
\end{proposition}

\begin{proof}
For $V_B$ we apply the statement to the transposed algorithm, see Lemma~\ref{mm:lem:sym}(b). This is possible because (S5) only concerns the points.

\emph{Case $D_A\equiv0$.} We apply Lemma~\ref{mm:lem:singnets} to $V_A$. In case (i) let $k\in\cA$ be a common nonzero singular vector. As $k$ is in the kernel of $\alpha_i\gamma_i$, we get $\alpha_i(k)=\gamma_i(k)=0$ for all $i$, contradicting Lemma~\ref{mm:lem:restriction} with $X=\K k$. In case (iii) we have (T4). In case (ii), a product of linear forms in $\varphi L+\Sy^2\langle\varphi,\psi\rangle$ lies in $\varphi\cA^*$ or in $\Sy^2\langle\varphi,\psi\rangle$. Indeed, write it as $2\varphi\ell+Y$ with $\ell\in L$. If $\ell\neq0$, then $\ell,\varphi,\psi$ are independent. Then the form has rank $3$, unless the coefficient of $\psi^2$ in $Y$ vanishes, and in that case it is divisible by $\varphi$. Neither $V_A\cap\varphi\cA^*$ nor $V_A\cap\Sy^2\langle\varphi,\psi\rangle$ is all of $V_A$, since their members have rank at most $2$. So all points lie on the union of two lines of $\Pj(V)$, each carrying at most three points (Lemma~\ref{mm:lem:collinear}(b)). Then $d\le6$. If $d=6$, both lines carry three points and there is no double point (Lemma~\ref{mm:lem:collinear}(a)), but there is one. If $d=5$ and both lines carry three points, they share a point and again there is no double point, but there are two. So one line carries three single points and the other line carries the remaining two points, the double points, off the first line: (S5).

\emph{Case $D_A\not\equiv0$ and three points collinear.} Let $p_1,p_2,p_3$ lie on a line $\mathcal L$. They are single points and no other point lies on $\mathcal L$ (Lemma~\ref{mm:lem:collinear}). The restriction of $D_A$ to $\mathcal L$ is a binary quartic vanishing to order at least $2$ at three points, so $\mathcal L$ divides $D_A$. Write $D_A=\mathcal L\cdot G$ with a cubic $G$. Singularity of $D_A$ at the $p_j$ gives $G(p_j)=0$, and at a point $q\notin\mathcal L$ it gives that $G$ is singular at $q$. Let $q_1,\dots,q_{d-3}$ be the points off $\mathcal L$. They include all $7-d$ double points. If $d=5$ this is (S5). A cubic singular at two points contains the line through them (Bezout). So if $G$ is singular at three non-collinear points, $G$ is the product of the three lines joining them, and these are its only singular points. If $d=7$, some three of $q_1,\dots,q_4$ are non-collinear (Lemma~\ref{mm:lem:collinear}(b)), and the fourth would be a further singular point of $G$, which is impossible. If $d=6$, the three points $q_1,q_2,q_3$ contain the double point, so they are not collinear (Lemma~\ref{mm:lem:collinear}(a)), and $G$ is the product of the three lines $q_jq_k$. Each $p_j$ lies on one of them, and each of these lines meets $\mathcal L$ in one point. So each line contains exactly one $p_j$. In particular a line through the double point contains some $p_j$, a collinear triple containing a double point, contradicting Lemma~\ref{mm:lem:collinear}(a).

\emph{Case $D_A\not\equiv0$ and no three points collinear.} The conic $C$ through five of the points is smooth (a pair of lines through five points would contain three of them). Under a parametrization of $C$, the restriction of $D_A$ is a binary form of degree $8$ vanishing to order at least $2$ at five points. So $C$ divides $D_A$, say $D_A=C\cdot Q$. At a point of $C$ where $D_A$ is singular, $Q$ vanishes, since $C$ is smooth there. So the conic $Q$ contains five points of $C$, $Q\in\K C$, and $D_A=\mu C^2$ with $\mu\neq0$. Every point is a singular point of $\mu C^2$, hence lies on $C$. Let $v\mapsto q(v)\in V_A$ parametrize $C$ and let $M(v)$ be the matrix of $q(v)$, with entries quadratic in $v$ and $\det M(v)=0$. Suppose $\rank M(v)=3$ for some $v$. As in the proof of Lemma~\ref{mm:lem:singnets}, $\operatorname{adj}M(v)=g(v)\kappa(v)\kappa(v)^T$ with $\kappa$ primitive, $\deg g+2\deg\kappa=6$ and $M(v)\kappa(v)=0$. At each point the A-part has rank at most $2$, so $\operatorname{adj}M=0$ there and $g$ vanishes there. Then $\deg g\ge5$, $\kappa$ is constant, and $\kappa$ is a common singular vector of all $q(v)$, which span $V_A$. So $D_A\equiv0$, a contradiction. Hence every point of $C$ has rank at most $2$, and Lemma~\ref{mm:lem:conic} gives (T1).
\end{proof}

\subsection{The configuration (S5)}\label{mm:sec:S5}

\begin{proposition}\label{mm:prop:S5}
\textup{(S5)} does not occur.
\end{proposition}

\begin{proof}
Let $p_1,p_2,p_3$ be the single points on the line $\mathcal L$, and let $q_1,q_2$ be the double points, with rows $[x_1],[x_2]$, columns $[y_1],[y_2]$ and A-parts $Q_1,Q_2$. Let $\mathcal L'$ be the line $q_1q_2$. It contains no $p_j$ by Lemma~\ref{mm:lem:collinear}(a). We claim $[x_1]=[x_2]$. Otherwise $\Row(x_1)\oplus\Row(x_2)=\cA^*$ by Fact~\ref{mm:fact:W}, and $Q_k$ is the product of a basis of $\Row(x_k)$ (Lemma~\ref{mm:lem:double}). Hence $\det(sQ_1+tQ_2)=\gamma s^2t^2$ with $\gamma\neq0$, so $D_A\not\equiv0$ and $D_A$ does not vanish on $\mathcal L'$. As in the proof of Proposition~\ref{mm:prop:types}, $D_A=\mathcal L\cdot G$ where the cubic $G$ vanishes at $p_1,p_2,p_3$ and is singular at $q_1,q_2$. Then $\mathcal L'$ divides $G$, $G=\mathcal L'\cdot Q$, and $Q(p_j)=0$ because $\mathcal L'(p_j)\neq0$. So $\mathcal L$ divides $Q$ and $G=\mathcal L'\mathcal L\mathcal L''$ for a line $\mathcal L''$. Singularity of $G$ at $q_k\in\mathcal L'\setminus\mathcal L$ forces $\mathcal L''(q_k)=0$. So $\mathcal L''=\mathcal L'$ up to a scalar, and $D_A$ is a multiple of $\mathcal L^2\mathcal L'^2$, which vanishes on $\mathcal L'$. This contradiction proves the claim. By the same argument for $V_B$, $[y_1]=[y_2]$. Then the W-elements of $q_1$ and $q_2$ are proportional, which gives a linear relation among the four products at $q_1,q_2$.
\end{proof}

\subsection{\texorpdfstring{Type T4, and the case T4$\times$T4}{Type T4, and the case T4 x T4}}\label{mm:sec:T4}

Assume $V_A$ has type (T4). Every element of $V_A$ is a multiple of $\ell_0$, so $V_A=\ell_0\cdot U$ with $\dim U=3$. Exchanging factors and rescaling, we may assume $\alpha_i=\ell_0$ for all $i$. Then $U=\spn\{\gamma_i\}$.

\begin{lemma}\label{mm:lem:l0}
$\ell_0\notin U$.
\end{lemma}

\begin{proof}
Otherwise all A-factors $\ell_0,\gamma_1,\dots,\gamma_7$ lie in $U$ and vanish on the line $U^\perp\subseteq\cA$, contradicting Lemma~\ref{mm:lem:restriction}.
\end{proof}

Choose $A_0\in\cA$ with $\ell_0(A_0)=1$ and $\gamma(A_0)=0$ for all $\gamma\in U$, and let $U'=\ker\ell_0\subseteq\cA$, a $3$-dimensional space. Restriction identifies $U$ with the dual of $U'$. The bilinear part of $AB=\sum_i\Lambda_iP_i$ reads $AB=\sum_i\Lambda_i\bigl(\ell_0(A)\delta_i(B)+\gamma_i(A)\beta_i(B)\bigr)$, hence
\begin{alignat}{2}
    A_0B&=\sum_i\Lambda_i\,\delta_i(B)&\qquad&\text{for all }B\in\cB,\label{mm:eq:T4a}\\
    A'B&=\sum_i\Lambda_i\,\gamma_i(A')\,\beta_i(B)&&\text{for all }A'\in U',\ B\in\cB.\label{mm:eq:T4b}
\end{alignat}

\begin{proposition}\label{mm:prop:T4T4}
$V_A$ and $V_B$ are not both of type \textup{(T4)}.
\end{proposition}

\begin{proof}
Let $\ell''_0$ divide every element of $V_B$, so $V_B=\ell''_0U''$, and $\ell''_0\notin U''$ by Lemma~\ref{mm:lem:l0} for the transposed algorithm. Since $V$ is the graph of $\vartheta$, $\vartheta(\ell_0\gamma)=\ell''_0\Phi(\gamma)$ for a linear isomorphism $\Phi:U\to U''$, and $\beta_i\delta_i=\vartheta(\ell_0\gamma_i)=\ell''_0\Phi(\gamma_i)$. By unique factorization, either $\delta_i\in\K\ell''_0$ and $\beta_i=t_i\Phi(\gamma_i)$ with $t_i\neq0$ (we write $i\in I_2$), or $\beta_i\in\K\ell''_0$ (we write $i\in I_1$). Let $U'''=\ker\ell''_0\subseteq\cB$. On $U'\times U'''$ the products with $i\in I_1$ vanish, because their first factor $\ell_0+\beta_i$ vanishes, and for $i\in I_2$ we have $P_i(A,B)=t_i\gamma_i(A)\Phi(\gamma_i)(B)$. Let $\Psi:U'\to U'''$ be the isomorphism with $\Phi(\gamma)(\Psi(A'))=\gamma(A')$ for all $\gamma\in U$ and $A'\in U'$. Then
\begin{displaymath}
    A'\,\Psi(A'')=\sum_{i\in I_2}t_i\,\gamma_i(A')\gamma_i(A'')\,\Lambda_i\qquad(A',A''\in U'),
\end{displaymath}
which is symmetric in $A',A''$. By Fact~\ref{mm:fact:M2}(b), invertible matrices form nonempty Zariski-open subsets of $U'$ and of $U'''$, so there is $A_1\in U'$ with $A_1$ and $\Psi(A_1)$ invertible. From $A_1\Psi(A')=A'\Psi(A_1)$ we get $\Psi(A')=A_1^{-1}A'\Psi(A_1)$, and the symmetry becomes $A'A_1^{-1}A''\Psi(A_1)=A''A_1^{-1}A'\Psi(A_1)$. Cancelling $\Psi(A_1)$ and multiplying by $A_1^{-1}$ on the left, $(A_1^{-1}A')(A_1^{-1}A'')=(A_1^{-1}A'')(A_1^{-1}A')$. So $A_1^{-1}U'$ is a commutative subspace of dimension $3$, contradicting Fact~\ref{mm:fact:M2}(c).
\end{proof}

\subsection{\texorpdfstring{The case T4$\times$T1}{The case T4 x T1}}\label{mm:sec:T4T1}

Assume $V_A$ has type (T4), normalized as in Section~\ref{mm:sec:T4}, and $V_B$ has type (T1), with isomorphisms $c:\K^2\to E_c$, $e:\K^2\to E_e$, $\cB^*=E_c\oplus E_e$, and $V_B=\spn\{c(v)e(v)\}$. The map $v\mapsto\vartheta^{-1}(c(v)e(v))$ is quadratic with values in $\ell_0U$, so $\vartheta^{-1}(c(v)e(v))=\ell_0\,\gamma(v)$, where $\gamma(v)=v_1^2g_0+v_1v_2g_1+v_2^2g_2$ for a basis $g_0,g_1,g_2$ of $U$ (the forms $c(v)e(v)$ span the $3$-dimensional $V_B$). For each $i$ choose $v_i$ with $\pi_i=s_i(\ell_0\gamma(v_i),c(v_i)e(v_i))$, $s_i\neq0$. Products at the same point get the same $v_i$. Then $\gamma_i=s_i\gamma(v_i)$, and by unique factorization either $\beta_i=\kappa_ic(v_i)$ and $\delta_i=(s_i/\kappa_i)e(v_i)$ (we write $i\in I_c$), or $\beta_i=\kappa_ie(v_i)$ and $\delta_i=(s_i/\kappa_i)c(v_i)$ (we write $i\in I_e$). By Lemma~\ref{mm:lem:double}, both products at a double point are in the same class. Let $D_c,D_e$ be the sets of points of the two classes; $d=|D_c|+|D_e|$. Let $U_c=E_e^\perp$ and $U_e=E_c^\perp$, complementary $2$-dimensional subspaces of $\cB$.
Let $A'_0,A'_1,A'_2\in U'$ be dual to $g_0,g_1,g_2$, and $C_1,C_2\in U_c$ dual to $c_1=c(e_1)$, $c_2=c(e_2)$.

\begin{lemma}[Hankel relations]\label{mm:lem:hankel}
$A'_kC_l$ depends only on $k+l$. That is, $A'_0C_2=A'_1C_1$ and $A'_1C_2=A'_2C_1$. The same holds for $U_e$ with the basis dual to $e(e_1),e(e_2)$. Both statements hold for every choice of basis of $\K^2$ (used to define $g_0,g_1,g_2$, $c_1,c_2$ and the dual bases).
\end{lemma}

\begin{proof}
For $B\in U_c$ we have $\beta_i(B)=0$ when $i\in I_e$, so \eqref{mm:eq:T4b} gives
\begin{displaymath}
    A'B=\sum_{i\in I_c}s_i\kappa_i\Lambda_i\,\gamma(v_i)(A')\,c(v_i)(B).
\end{displaymath}
With $v_i=(v_{i1},v_{i2})$ we have $\gamma(v_i)(A'_k)=v_{i1}^{2-k}v_{i2}^k$ and $c(v_i)(C_l)=v_{il}$, so
\begin{displaymath}
    A'_kC_l=\sum_{i\in I_c}s_i\kappa_i\Lambda_i\,v_{i1}^{4-k-l}v_{i2}^{k+l-1},
\end{displaymath}
which depends only on $k+l$. The argument does not depend on the basis.
\end{proof}

\begin{proposition}\label{mm:prop:T4T1}
$V_A$ of type \textup{(T4)} and $V_B$ of type \textup{(T1)} cannot occur.
\end{proposition}

\begin{proof}
\emph{(a) $U_c$ or $U_e$ contains an invertible matrix,} say $U_c$. The elements $C_1$ obtained from the various bases of $\K^2$ are, up to scalars, all nonzero elements of $U_c$. Choose the basis so that $C_1$ is invertible. Lemma~\ref{mm:lem:hankel} gives $A'_1=A'_0X$ and $A'_2=A'_1X=A'_0X^2$ with $X=C_2C_1^{-1}$. By the Cayley-Hamilton theorem $U'=\spn\{A'_0,A'_1,A'_2\}\subseteq A'_0\spn\{I,X\}$ has dimension at most $2$, a contradiction.

\emph{(b) $U_c$ or $U_e$ has type C,} say $U_c=\{xh^T\}$. Write $C_l=x\rho_l^T$ with $\rho_1,\rho_2$ independent. Then $A'_kC_l=(A'_kx)\rho_l^T$, and Lemma~\ref{mm:lem:hankel} gives $(A'_0x)\rho_2^T=(A'_1x)\rho_1^T$ and $(A'_1x)\rho_2^T=(A'_2x)\rho_1^T$, so $A'_kx=0$ for $k=0,1,2$. Then $U'\subseteq\{A:\ Ax=0\}$ has dimension at most $2$, a contradiction.

\emph{(c) Otherwise} both $U_c=\{hy_c^T\}$ and $U_e=\{hy_e^T\}$ have type R, with $[y_c]\neq[y_e]$. Replacing $B$ by $BZ$ for a suitable $Z$ (Lemma~\ref{mm:lem:sym}(a)), we may assume that $U_c$ consists of the matrices $[b\mid0]$ and $U_e$ of the matrices $[0\mid b]$. Then $E_c$ consists of the forms in the first column of $B$ and $E_e$ of those in the second. Define $\bar c(v),\bar e(v)\in\K^2$ by $c(v)([b\mid0])=\bar c(v)^Tb$ and $e(v)([0\mid b])=\bar e(v)^Tb$. Both $\bar c$ and $\bar e$ are linear isomorphisms of $\K^2$. Let $\Lambda_i^{(1)}$ be the first column of $\Lambda_i$, and for a point $\tau$ of $I_c$ put $m_\tau=\sum_{i\text{ at }\tau}s_i\kappa_i\Lambda_i^{(1)}$. Taking the first column of \eqref{mm:eq:T4b} for $B=[b\mid0]$ gives, as an identity of matrices,
\begin{displaymath}
    A'=\sum_{\tau\in D_c}\gamma(v_\tau)(A')\;m_\tau\,\bar c(v_\tau)^T\qquad(A'\in U').
\end{displaymath}
Hence $U'\subseteq\spn\{m_\tau\bar c(v_\tau)^T:\ \tau\in D_c\}$, so $|D_c|\ge3$, and if $|D_c|=3$ these three matrices are linearly independent. In the same way (second columns) $|D_e|\ge3$. Next take \eqref{mm:eq:T4a} for $B=[0\mid b]$: the left side $[0\mid A_0b]$ has zero first column, $\delta_i$ vanishes on $U_e$ for $i\in I_e$, and $\delta_i=(s_i/\kappa_i)e(v_i)$ for $i\in I_c$. Hence
\begin{equation}\label{mm:eq:T4T1}
    \sum_{i\in I_c}(s_i/\kappa_i)\,\Lambda_i^{(1)}\,\bar e(v_i)^T=0.
\end{equation}
Since $d=|D_c|+|D_e|\ge6$, either $d=7$, all points are single and one class has exactly three points, or $d=6$, $|D_c|=|D_e|=3$ and the class without the double point consists of three single points. Exchanging the two columns of $B$ (which exchanges the two classes) we may assume that $I_c$ consists of three single points $\tau=1,2,3$, with products indexed by $\tau$. Then $m_\tau=s_\tau\kappa_\tau\Lambda_\tau^{(1)}$, and \eqref{mm:eq:T4T1} reads $\sum_\tau\kappa_\tau^{-2}m_\tau\bar e(v_\tau)^T=0$. Let $T\in GL_2(\K)$ with $T\bar e(v)=\bar c(v)$ for all $v$. Multiplying by $T^T$ on the right gives $\sum_\tau\kappa_\tau^{-2}m_\tau\bar c(v_\tau)^T=0$, a nontrivial linear relation among three matrices that must be independent.
\end{proof}

\subsection{\texorpdfstring{The case T1$\times$T1}{The case T1 x T1}}\label{mm:sec:T1T1}

Assume both $V_A$ and $V_B$ have type (T1): $\cA^*=E_a\oplus E_b$ and $\cB^*=E_c\oplus E_e$ with isomorphisms $a,b,c,e$ from $\K^2$, $V_A=\spn\{a(v)b(v)\}$ and $V_B=\spn\{c(v)e(v)\}$. Through $V\cong V_A\cong V_B$, the conics $\{[a(v)b(v)]\}$ and $\{[c(v)e(v)]\}$ are two conics in $\Pj(V)$ through the $d\ge5$ points, no three of which are collinear. So they coincide. The maps $v\mapsto[\vartheta(a(v)b(v))]$ and $v\mapsto[c(v)e(v)]$ are then two parametrizations of the same conic. So after replacing $c,e$ by $c\circ g,e\circ g$ for some $g\in GL_2(\K)$ we have $\vartheta(a(v)b(v))=\lambda(v)c(v)e(v)$, where $\lambda$ is a rational function of degree $0$ without zeros or poles, hence constant. We absorb it into $c$. So $\pi(v)=(a(v)b(v),c(v)e(v))\in V$ for all $v$, and $v^2\mapsto\pi(v)$ extends to an isomorphism $\Sy^2\K^2\to V$.

For each $i$ choose $v_i$ with $\pi_i=s_i\pi(v_i)$, $s_i\neq0$, the same for products at the same point. Exchanging factors we may assume $\alpha_i\in\K a(v_i)$ and $\gamma_i\in\K b(v_i)$. By unique factorization each product has one of the forms
\begin{alignat*}{2}
    &(I_1)&\qquad&P_i=s_i\bigl(a(v_i)+\kappa_ic(v_i)\bigr)\bigl(b(v_i)+\kappa_i^{-1}e(v_i)\bigr),\\
    &(I_2)&&P_i=s_i\bigl(a(v_i)+\kappa_ie(v_i)\bigr)\bigl(b(v_i)+\kappa_i^{-1}c(v_i)\bigr),
\end{alignat*}
with $\kappa_i\neq0$. Let $D_1,D_2$ be the sets of points of the two classes and $n_j=|D_j|$. The bilinear parts are $S_i=s_i(\kappa_i^{-1}a\otimes e+\kappa_ib\otimes c)(v_i)$ for $i\in I_1$ and $S_i=s_i(\kappa_i^{-1}a\otimes c+\kappa_ib\otimes e)(v_i)$ for $i\in I_2$.

\begin{lemma}[Rank-one difference]\label{mm:lem:rankone}
For $v\neq0$ and $\kappa,\lambda\neq0$ let $P=(a+\kappa c)(b+\kappa^{-1}e)$ and $Q=(a+\lambda e)(b+\lambda^{-1}c)$, all forms evaluated at $v$. Then $P-Q$ is a nonzero bilinear form of rank one. In particular $P-Q\notin\Pi$. Hence the two products at a double point are in the same class, and $D_1\cap D_2=\emptyset$.
\end{lemma}

\begin{proof}
Both pure parts equal $(ab,ce)$, and $S(P)-S(Q)=a\otimes(\kappa^{-1}e-\lambda^{-1}c)+b\otimes(\kappa c-\lambda e)=(a-\kappa\lambda b)\otimes(\kappa^{-1}e-\lambda^{-1}c)$, which is nonzero because $a(v),b(v)$ are independent and so are $c(v),e(v)$. A bilinear element of $\Pi$ lies in $W$ (Lemma~\ref{mm:lem:relations}(b)), which has no rank-one elements.
\end{proof}

Let $U_a=E_b^\perp$, $U_b=E_a^\perp\subseteq\cA$ and $U_c=E_e^\perp$, $U_e=E_c^\perp\subseteq\cB$, so $\cA=U_a\oplus U_b$ and $\cB=U_c\oplus U_e$. Let $\hat a:\K^2\to U_a$ be the isomorphism with $a(u)(\hat a(v))=u^Tv$, and define $\hat b,\hat c,\hat e$ likewise. For $x\in\{a,b\}$, $y\in\{c,e\}$ and a symmetric $2\times2$ matrix $F$ (identified with the quadratic form $F(v)=v^TFv$) let
\begin{displaymath}
    H_{xy}(F)=\sum_{k,l}F_{kl}\,\hat x(e_k)\hat y(e_l),\qquad\operatorname{Ann}_{xy}=\ker H_{xy}.
\end{displaymath}
We call $xy\in\{ac,ae,bc,be\}$ a \emph{block}. The image of $H_{xy}$ is $U_xU_y=\spn\{XY:\ X\in U_x,\ Y\in U_y\}$.

\begin{lemma}[Blocks]\label{mm:lem:blocks}
For all $u,v\in\K^2$, $\hat x(u)\hat y(v)=\hat x(v)\hat y(u)$ for each block $xy$. Moreover
\begin{gather*}
    H_{ac}(F)=\sum_{i\in I_2}s_i\kappa_i^{-1}F(v_i)\Lambda_i,\qquad H_{be}(F)=\sum_{i\in I_2}s_i\kappa_iF(v_i)\Lambda_i,\\
    H_{ae}(F)=\sum_{i\in I_1}s_i\kappa_i^{-1}F(v_i)\Lambda_i,\qquad H_{bc}(F)=\sum_{i\in I_1}s_i\kappa_iF(v_i)\Lambda_i.
\end{gather*}
\end{lemma}

\begin{proof}
Evaluate $AB=\sum_i\Lambda_iS_i(A,B)$ at $A=\hat a(u)$, $B=\hat c(v)$. The forms in $E_b$ vanish on $U_a$ and those in $E_e$ on $U_c$, so $S_i$ vanishes there for $i\in I_1$, and for $i\in I_2$ it equals $s_i\kappa_i^{-1}(v_i^Tu)(v_i^Tv)$. Hence $\hat a(u)\hat c(v)=\sum_{I_2}s_i\kappa_i^{-1}(v_i^Tu)(v_i^Tv)\Lambda_i$, which is symmetric in $u,v$, and summing with coefficients $F_{kl}$ gives $H_{ac}$. The other blocks are the same.
\end{proof}

\begin{lemma}[Geometry of the blocks]\label{mm:lem:blockgeom}
\textup{(a)} $U_a,U_b$ have type I or C, and $U_c,U_e$ have type I or R. \textup{(b)} Every $U_xU_y$ is nonzero. It has dimension $1$ if $(U_x,U_y)$ has types $(\mathrm{C},\mathrm{R})$ and dimension $2$ otherwise. \textup{(c)} If $(U_x,U_y)$ has types $(\mathrm{C},\mathrm{R})$, then $\dim\operatorname{Ann}_{xy}=2$ and the members of $\operatorname{Ann}_{xy}$ have no common zero in $\Pj^1$. Otherwise $\dim\operatorname{Ann}_{xy}=1$.
\end{lemma}

\begin{proof}
(a) Suppose $U_a=\{hy^T\}$, so $\hat a(u)=h(u)y^T$ with $h$ an isomorphism. The symmetry with $\hat c$ reads $h(u)\,(y^T\hat c(v))=h(v)\,(y^T\hat c(u))$. Taking $u,v$ independent gives $y^T\hat c(v)=0$ for all $v$. Likewise $y^T\hat e(v)=0$. Then $U_c$ and $U_e$ lie in the $2$-dimensional space $\{M:\ y^TM=0\}$, contradicting $U_c\oplus U_e=\cB$. The same argument excludes type R for $U_b$. If $U_c=\{xh^T\}$, the symmetry with $\hat a$ gives $\hat a(u)x=0$ for all $u$, likewise $\hat b(u)x=0$, and $U_a,U_b\subseteq\{M:\ Mx=0\}$, a contradiction. Similarly for $U_e$. Now use Fact~\ref{mm:fact:M2}(a).

(b) If $U_x$ and $U_y$ have type I, pick $u_0$ with $\hat x(u_0)$ and $\hat y(u_0)$ invertible (their determinants are nonzero binary quadratic forms). The symmetry gives $\hat y(v)=\hat x(u_0)^{-1}\hat x(v)\hat y(u_0)$, so $U_xU_y=\hat x(u_0)\,\mathcal N\mathcal N\,\hat y(u_0)$ with $\mathcal N=\hat x(u_0)^{-1}U_x=\K I+\K N$, and $\mathcal N\mathcal N=\mathcal N$ by the Cayley-Hamilton theorem. So $\dim U_xU_y=2$. If $U_x$ has type I and $U_y=\{hy^T\}$, then $U_xU_y=\K^2y^T$. Let $U_x=\{xh^T\}$. Then $U_xU_y=\{x\rho^T\}$, where $\rho^T$ ranges over the span of the rows of the elements of $U_y$. This span is all of $\K^{1\times2}$ if $U_y$ has type I, and the line $\K y^T$ if $U_y=\{hy^T\}$.

(c) In the case (C,R) write $\hat x(u)=x(Ru)^T$ and $\hat y(v)=(Kv)y^T$ with $R,K$ invertible. Then $\hat x(u)\hat y(v)=(u^TR^TKv)\,xy^T$, the symmetry says that $\Sigma=R^TK$ is symmetric, and $H_{xy}(F)=\tr(F\Sigma)\,xy^T$. A common zero $v$ of all $F$ with $\tr(F\Sigma)=0$ would mean $vv^T\in\K\Sigma$, impossible as $\Sigma$ is invertible. Otherwise $\dim\operatorname{Ann}_{xy}=3-\dim U_xU_y=1$ by (b).
\end{proof}

\begin{lemma}[Transfer]\label{mm:lem:transfer}
Let $xy\in\{ac,be\}$ and $F\in\operatorname{Ann}_{xy}$. There is a quadratic form $G$ on $\K^2$ with $G(v_i)=0$ for $i\in I_1$ and $G(v_i)=w_iF(v_i)$ for $i\in I_2$, where $w_i=\kappa_i^{-1}$ for $xy=ac$ and $w_i=\kappa_i$ for $xy=be$. The same holds with $I_1,I_2$ exchanged for $xy\in\{ae,bc\}$. Since $d\ge3$, $G$ is uniquely determined by $F$.
\end{lemma}

\begin{proof}
Put $y_i=s_iw_iF(v_i)$ for $i\in I_2$ and $y_i=0$ for $i\in I_1$. Then $\sum_iy_i\Lambda_i=H_{xy}(F)=0$ by Lemma~\ref{mm:lem:blocks}, so by Lemma~\ref{mm:lem:relations}(c) $y_i=\varphi(\pi_i)=s_i\varphi(\pi(v_i))$ for a functional $\varphi$ on $V$. Take $G(v)=\varphi(\pi(v))$. Uniqueness follows from Fact~\ref{mm:fact:binary}(b).
\end{proof}

\begin{lemma}[Common annihilator]\label{mm:lem:commonann}
Suppose no block has types $(\mathrm{C},\mathrm{R})$. Then there is a nonzero quadratic form $g$ with $\operatorname{Ann}_{ac}=\operatorname{Ann}_{ae}=\operatorname{Ann}_{bc}=\operatorname{Ann}_{be}=\K g$.
\end{lemma}

\begin{proof}
By Lemma~\ref{mm:lem:blockgeom}(a), the absence of $(\mathrm{C},\mathrm{R})$ blocks means that $U_a,U_b$ both have type I or $U_c,U_e$ both have type I.

\emph{Excluded geometries.} If $U_a=\{x_ah^T\}$ and $U_b=\{x_bh^T\}$ (necessarily $[x_a]\neq[x_b]$) while $U_c,U_e$ have type I, then $U_aU_c+U_bU_e=\{x_a\rho^T\}+\{x_b\rho^T\}=\K^{2\times2}$ and likewise $U_aU_e+U_bU_c=\K^{2\times2}$. The bilinear parts $S_i$ with $i\in I_1$ vanish on $U_a\times U_c$ and on $U_b\times U_e$, so Lemma~\ref{mm:lem:kernel} gives $|I_1|\le3$, and similarly $|I_2|\le3$, contradicting $|I_1|+|I_2|=7$. If $U_a,U_b$ have type I and $U_c=\{hy_c^T\}$, $U_e=\{hy_e^T\}$, the same argument applies with $U_aU_c+U_bU_e=\K^2y_c^T+\K^2y_e^T=\K^{2\times2}$.

\emph{Remaining geometries.} Changing the basis of $\K^2$ (simultaneously in $a,b,c,e$) we may assume that $\hat x(e_1)$ is invertible whenever $U_x$ has type I. Write $A_k=\hat a(e_k)$, $B_k=\hat b(e_k)$, $C_k=\hat c(e_k)$, $E_k=\hat e(e_k)$, and for a symmetric $F$ let $\hat F(t)=F_{11}+2F_{12}t+F_{22}t^2$.

If all four spaces have type I, let $N=A_1^{-1}A_2$, which is not scalar. The symmetries $A_1C_2=A_2C_1$, $B_1C_2=B_2C_1$, $A_1E_2=A_2E_1$ give $C_2=NC_1$, $B_2=B_1N$ and $E_2=NE_1$. Hence $H_{ac}(F)=A_1\hat F(N)C_1$, $H_{ae}(F)=A_1\hat F(N)E_1$, $H_{bc}(F)=B_1\hat F(N)C_1$ and $H_{be}(F)=B_1\hat F(N)E_1$. As $N$ is not scalar, $\hat F(N)=0$ exactly when $\hat F$ is a multiple of the characteristic polynomial of $N$, and all four annihilators equal $\K g$ with $g(v)=\det(N)v_1^2-\tr(N)v_1v_2+v_2^2$.

If $U_c,U_e$ have type I and exactly one of $U_a,U_b$ has type C, say $U_b=\{xh^T\}$ (the other case is symmetric), let $N=A_1^{-1}A_2$. As before $C_2=NC_1$ and $E_2=NE_1$. Write $B_k=x\rho_k^T$. The symmetry $B_1C_2=B_2C_1$ gives $\rho_2^T=\rho_1^TN$. Then $H_{bc}(F)=x\rho_1^T\hat F(N)C_1$ and $H_{be}(F)=x\rho_1^T\hat F(N)E_1$. Reducing $\hat F(N)=\lambda I+\nu N$ by the Cayley-Hamilton theorem, $\rho_1^T\hat F(N)=\lambda\rho_1^T+\nu\rho_2^T$ vanishes only if $\hat F(N)=0$. So again all four annihilators are $\K g$.

If $U_a,U_b$ have type I and exactly one of $U_c,U_e$ has type R, say $U_e=\{hy^T\}$, then $N=A_1^{-1}A_2=B_1^{-1}B_2$ and $C_2=NC_1$ by the symmetries of the blocks $ac$ and $bc$. Write $E_k=\zeta_ky^T$. The symmetry $A_1E_2=A_2E_1$ gives $\zeta_2=N\zeta_1$, so $H_{ae}(F)=A_1\hat F(N)\zeta_1y^T$ and $H_{be}(F)=B_1\hat F(N)\zeta_1y^T$. As $\zeta_1,\zeta_2$ are independent, the same conclusion follows.
\end{proof}

\begin{proposition}\label{mm:prop:T1T1}
$V_A$ and $V_B$ are not both of type \textup{(T1)}.
\end{proposition}

\begin{proof}
Since $n_1+n_2=d\ge5$, one of $n_1,n_2$ is at least $3$. Exchanging the names of $c$ and $e$ exchanges the two classes, so we may assume $n_1\ge3$.

\emph{Step 1.} For $xy\in\{ac,be\}$, every $F\in\operatorname{Ann}_{xy}$ vanishes on $D_2$. Indeed, the form $G$ of Lemma~\ref{mm:lem:transfer} vanishes at the at least three points of $D_1$, so $G=0$, and $w_iF(v_i)=0$ for $i\in I_2$.

\emph{Step 2.} Since $U_aU_c\neq0$ (Lemma~\ref{mm:lem:blockgeom}(b)) and $H_{ac}$ is supported on $I_2$, $I_2\neq\emptyset$. By Step~1 and Lemma~\ref{mm:lem:blockgeom}(c), neither $ac$ nor $be$ has types $(\mathrm{C},\mathrm{R})$. So $\operatorname{Ann}_{ac}=\K g$ and $D_2$ is contained in the zero set of $g$. As $H_{ac}(F)$ depends only on the values $F(v_i)$, $i\in I_2$, and $H_{ac}$ has rank $2$, the set $D_2$ cannot be a single point. Hence $g$ has two distinct zeros $\nu_1,\nu_2$ and $D_2=\{\nu_1,\nu_2\}$.

\emph{Step 3.} Neither $ae$ nor $bc$ has types $(\mathrm{C},\mathrm{R})$. Suppose $ae$ does. For $F\in\operatorname{Ann}_{ae}$, the form $G_F$ of Lemma~\ref{mm:lem:transfer} vanishes at $\nu_1,\nu_2$, so $G_F=c_Fg$, and $F\mapsto c_F$ is linear on the $2$-dimensional space $\operatorname{Ann}_{ae}$. Choose $F\neq0$ with $c_F=0$. Then $\kappa_i^{-1}F(v_i)=0$ for all $i\in I_1$, so $F$ vanishes at three points and $F=0$, a contradiction. The block $bc$ is the same.

\emph{Step 4.} By Lemma~\ref{mm:lem:commonann}, $\operatorname{Ann}_{ae}=\K g$.

\emph{Step 5.} Lemma~\ref{mm:lem:transfer} for the block $ae$ and $F=g$ gives $G$ vanishing at $\nu_1,\nu_2$ with $G(v_i)=\kappa_i^{-1}g(v_i)$ for $i\in I_1$. So $G=cg$, and since $g(v_i)\neq0$ for $i\in I_1$ (Lemma~\ref{mm:lem:rankone}), $\kappa_i=c^{-1}=:\kappa$ for all $i\in I_1$.

\emph{Step 6.} Let $\mathcal P(v)=(a(v)+\kappa c(v))(b(v)+\kappa^{-1}e(v))$, which is homogeneous quadratic in $v$, so that $P_i=s_i\mathcal P(v_i)$ for $i\in I_1$. By Fact~\ref{mm:fact:binary}(b) applied to three pairwise non-proportional $v_i$ with $i\in I_1$, $\mathcal P(\nu_1)\in\Pi$. Let $Q$ be a product of $I_2$ at $\nu_1$. Then $Q/s$ has the form $(a+\lambda e)(b+\lambda^{-1}c)$ at $\nu_1$ and lies in $\Pi$. This contradicts Lemma~\ref{mm:lem:rankone}.
\end{proof}

\subsection{\texorpdfstring{Proof of Theorem~\ref{mm:thm:main} when $p_A=p_B=3$}{Proof of the main theorem when pA=pB=3}}\label{mm:sec:assembly}

We have $p=3$ (Section~\ref{mm:sec:first}) and $d\ge4$ (Corollary~\ref{mm:cor:d}). Suppose $d\ge5$. By Proposition~\ref{mm:prop:types}, either the configuration is (S5), which is impossible by Proposition~\ref{mm:prop:S5}, or each of $V_A,V_B$ has type (T1) or (T4). The combination (T4, T4) is excluded by Proposition~\ref{mm:prop:T4T4}, (T4, T1) by Proposition~\ref{mm:prop:T4T1}, (T1, T4) by Proposition~\ref{mm:prop:T4T1} applied to the transposed algorithm (Lemma~\ref{mm:lem:sym}(b)), and (T1, T1) by Proposition~\ref{mm:prop:T1T1}. Hence $d=4$, and Corollary~\ref{mm:cor:d4} completes the proof.\qed

\subsection{One side with two quadrics}\label{mm:sec:twoq}

From now on we drop the assumption $p_A=p_B=3$ and only assume $p=3$.

\begin{lemma}\label{mm:lem:twoq}
Assume $p=3$ and $p_B=2$. Then $k_A\le2$, and there are $q_1=b_1b'_1$ and $q_2=b_2b'_2$ in $\Sy^2\cB^*$ such that the B-part of every product not of type A is a nonzero multiple of $q_1$ or of $q_2$, both cases occur, and
\begin{displaymath}
    \cB^*=F_1\oplus F_2,\qquad F_j=\spn\{b_j,b'_j\}.
\end{displaymath}
In particular $\dim F_1=\dim F_2=2$, and the B-factors $\beta_i,\delta_i$ of a product whose B-part is a multiple of $q_j$ lie in $F_j$. If $p_A=2$, the same holds with the roles of the sides exchanged: $k_B\le2$, and the A-parts of the products not of type B are nonzero multiples of $r_1=g_1g'_1$ or of $r_2=g_2g'_2$, with $\cA^*=G_1\oplus G_2$ and $G_j=\spn\{g_j,g'_j\}$.
\end{lemma}

\begin{proof}
Since $\dim N_A=1$, all products of type A have the same point, so $k_A\le2$ by Lemma~\ref{mm:lem:doubleAB}(c). We first show: there is no $B_0\in\cB\setminus\{0\}$ such that $\beta_i(B_0)=\delta_i(B_0)=0$ for every product not of type A. Suppose there is. For such products $S_i(A,B_0)=\alpha_i(A)\delta_i(B_0)+\gamma_i(A)\beta_i(B_0)=0$, so \eqref{mm:eq:bil} gives $AB_0=\sum_{i\text{ of type A}}\Lambda_i\gamma_i(A)\beta_i(B_0)$ for all $A$. The left side, as a linear map $\cA\to\K^{2\times2}$, has rank $2\rank B_0\ge2$, and the right side has rank at most $k_A\le2$. Hence $k_A=2$ and $B_0=uv^T$ has rank one. The forms $\gamma_i$ of the two products of type A span the space of forms vanishing on the kernel $\{A:\ Au=0\}$ of $A\mapsto AB_0$. This space is $\Col_\cA(u)$. By Lemma~\ref{mm:lem:doubleAB}(a) these two forms are the forms $\gamma,\alpha$ of the double point of type A and form a basis of some $\Row(x)$, contradicting Fact~\ref{mm:fact:Wfactor}(c).

The B-parts of the products not of type A are nonzero (Lemma~\ref{mm:lem:types}) and span $V_B$, because the products of type A have B-part $0$. Suppose they represented three distinct points of $\Pj(V_B)$. Then by Lemma~\ref{mm:lem:pencil}, all their factors $\beta_i,\delta_i$ would lie in a subspace of $\cB^*$ of dimension at most $3$. A nonzero $B_0$ annihilated by it would contradict the claim. As $p_B=2$, they represent exactly two points $[q_1]$, $[q_2]$, where $q_j=b_jb'_j$ is the B-part of some product. By unique factorization the factors of a product whose B-part is a multiple of $q_j$ lie in $F_j$. If $F_1+F_2\neq\cB^*$, any nonzero $B_0$ annihilated by $F_1+F_2$ contradicts the claim. So $\dim F_1+\dim F_2\ge4$, which forces $\dim F_j=2$ and $\cB^*=F_1\oplus F_2$. The last statement follows by Lemma~\ref{mm:lem:sym}(b).
\end{proof}

\subsection{\texorpdfstring{The case $(p_A,p_B)=(3,2)$}{The case (pA,pB)=(3,2)}}\label{mm:sec:32}

In this section $p_A=3$ and $p_B=2$, so $p=3$, $N_B=0$, and $N_A=\K a$ with $a=(q_a,0)$. There are no products of type B, all products of type A have the point $[a]$, and we put $k=k_A\le2$. Let $q_1,q_2$ and $F_1,F_2$ be as in Lemma~\ref{mm:lem:twoq}. For $j=1,2$ let $\Pi_j\subseteq V$ be the preimage of $\K q_j$ under the projection $V\to V_B$. It is $2$-dimensional and contains $a$, so $\mathcal L_j=\Pj(\Pi_j)$ is a line through $[a]$. Every product of type AB has its point on $\mathcal L_1$ or on $\mathcal L_2$, different from $[a]$. Let $I_j$ be the set of products of type AB on $\mathcal L_j$ and $n_j=|I_j|$, so that $n_1+n_2+k=7$.

\begin{lemma}\label{mm:lem:32lines}
Let $j\in\{1,2\}$ with $n_j\ge3$. Then the products of $I_j$ have at least two distinct points and their pure parts span $\Pi_j$. Moreover $F_j=\Col(y_j)$ for some $y_j\neq0$, and $n_j\le4$. If $n_1,n_2\ge3$, then $[y_1]\neq[y_2]$, and
\begin{displaymath}
    W=W_1\oplus W_2,\qquad W_j=\{x^TABy_j:\ x\in\K^2\}\subseteq\cA^*\otimes F_j.
\end{displaymath}
\end{lemma}

\begin{proof}
A point carries at most two products (Lemma~\ref{mm:lem:doubleAB}(c)), so the $n_j\ge3$ products of $I_j$ have at least two distinct points. These differ from $[a]$, so their pure parts span $\Pi_j$. Hence the relations among the products of $I_j$ form a space of dimension at least $n_j-2\ge1$. By Lemma~\ref{mm:lem:relations}(b) it maps injectively into $W$, and its image lies in $\cA^*\otimes F_j$ because $S_i\in\cA^*\otimes F_j$ for $i\in I_j$. Fact~\ref{mm:fact:Wfactor}(b) gives $F_j=\Col(y_j)$ and $n_j-2\le2$. If $n_1,n_2\ge3$, then $\Col(y_1)\cap\Col(y_2)=F_1\cap F_2=0$, so $[y_1]\neq[y_2]$, and the last claim follows from Fact~\ref{mm:fact:W}.
\end{proof}

\begin{lemma}\label{mm:lem:32typeA}
Let $j\in\{1,2\}$ with $n_j\ge3$, and assume $k\ge1$. Then $\beta\in F_j$ for every product $(\alpha+\beta)\gamma$ of type A, and $n_j+k\le4$.
\end{lemma}

\begin{proof}
We may assume $j=1$. Choose $y'$ with $[y']\neq[y_1]$, so that $W=W_1\oplus W'$ with $W_1=\{x^TABy_1\}$ and $W'=\{x^TABy'\}$. Let $\varpi:\cB^*\to F_2$ be the projection along $F_1=\Col(y_1)$. It is injective on $\Col(y')$, so for $x\neq0$ the form $(1\otimes\varpi)(x^TABy')=\sum_q(x^TAh_q)\otimes\varpi(k_q^TBy')$ has rank two (Fact~\ref{mm:fact:W}). Let $P=(\alpha+\beta)\gamma$ be of type A, with pure part $ca$, $c\neq0$. By Lemma~\ref{mm:lem:32lines}, $a$ is a combination of the pure parts of the products of $I_1$, so by Lemma~\ref{mm:lem:tau} the class $[S]=[\gamma\otimes\beta]=c\,\tau(a)$ lies in $(\cA^*\otimes F_1+W)/W$. Write $\gamma\otimes\beta=s+w_1+w'$ with $s\in\cA^*\otimes F_1$, $w_1\in W_1$ and $w'\in W'$. Applying $1\otimes\varpi$, which kills $\cA^*\otimes F_1\supseteq W_1$, gives $\gamma\otimes\varpi(\beta)=(1\otimes\varpi)(w')$. The left side has rank at most one and the right side rank $0$ or $2$, so both vanish and $\beta\in F_1$.

Now let $Y_1=\{B:\ By_1=0\}$, on which all forms of $F_1=\Col(y_1)$ vanish. For $i\in I=I_1\cup\{\text{products of type A}\}$ the form $S_i$ vanishes on $\cA\times Y_1$. The pure parts of the products of $I$ lie in $\Pi_1$, and $Y_1=\{hw^T:\ h\in\K^2\}$ with $w\neq0$, $w^Ty_1=0$, so $\cA Y_1=\{vw^T:\ v\in\K^2\}$ has dimension $2$. Lemma~\ref{mm:lem:kernel} gives $n_1+k-2\le4-2$.
\end{proof}

\begin{lemma}\label{mm:lem:32normal}
Let $j\in\{1,2\}$ with $n_j\ge3$, and assume $\tau(a)=0$. Write $q_j=bb'$ with $b=k^TBy_j$ and $b'=k'^TBy_j$, let $(h,h')$ be the basis of $\K^2$ dual to $(k,k')$, and put
\begin{displaymath}
    \varphi(M)=\sum_{s,l}M_{sl}\,(e_s^TAh)(e_l^TAh')\qquad(M\in\K^{2\times2}).
\end{displaymath}
Then there are $\delta\in\K^2$ and, for $i\in I_j$, distinct vectors $X_i\in\K^2$ and scalars $\rho_i\neq0$ such that
\begin{equation}\label{mm:eq:32normal}
\begin{gathered}
    P_i=\rho_i\,(X_i^TAh'+b)\bigl((X_i+\delta)^TAh+b'\bigr),\\
    S_i=\rho_i\bigl(X_i^TABy_j+(\delta^TAh)\otimes b\bigr)\qquad(i\in I_j).
\end{gathered}
\end{equation}
The A-part of $P_i/\rho_i$ is $\varphi(F(X_i))$ with $F(X)=(X+\delta)X^T$, and there is $q'_j\in\Sy^2\cA^*$ with $\varphi(F(X_i))-q'_j\in\K q_a$ for all $i\in I_j$. Moreover $\tau$ vanishes on $\Pi_j$ if and only if $\delta=0$.
\end{lemma}

\begin{proof}
Note that $b,b'$ is a basis of $F_j=\Col(y_j)$ and that $x^TABy_j=(x^TAh)\otimes b+(x^TAh')\otimes b'$ (Fact~\ref{mm:fact:W}). For $i\in I_j$ the B-part $\beta_i\delta_i$ is a nonzero multiple of $bb'$, so after exchanging and rescaling the factors $P_i=\rho_i(\alpha_i+b)(\gamma_i+b')$. Choose $v\in\Pi_j$ with B-part $q_j$ and write $v=(q'_j,q_j)$. Then $\pi_i=\rho_iv+u_ia$ for some $u_i$, so the A-part of $P_i/\rho_i$ is $\alpha_i\gamma_i=q'_j+\lambda_iq_a$ with $\lambda_i=u_i/\rho_i$, and $[S_i]=\rho_i\tau(v)$ because $\tau(a)=0$. By Lemma~\ref{mm:lem:32lines}, $v$ is a combination of the $\pi_i$ ($i\in I_j$), so $\tau(v)=[T]$ for some $T=\tau_b\otimes b+\tau_{b'}\otimes b'\in\cA^*\otimes F_j$. Then $S_i/\rho_i-T=\alpha_i\otimes b'+\gamma_i\otimes b-T$ lies in $W\cap(\cA^*\otimes F_j)=\{x^TABy_j\}$ (Fact~\ref{mm:fact:Wfactor}(b)), so there is $x_i$ with
\begin{displaymath}
    \alpha_i=\tau_{b'}+x_i^TAh',\qquad\gamma_i=\tau_b+x_i^TAh.
\end{displaymath}
The normalized product $P_i/\rho_i$ is determined by $x_i$, so the $x_i$ are distinct, as the products are independent.

Let $C_h=\{x^TAh\}$ and $C_{h'}=\{x^TAh'\}$, so that $\cA^*=C_h\oplus C_{h'}$ and $\Sy^2\cA^*=\Sy^2C_h\oplus C_hC_{h'}\oplus\Sy^2C_{h'}$; $\varphi$ is a bijection from $\K^{2\times2}$ onto $C_hC_{h'}$, and $\varphi(uv^T)=(u^TAh)(v^TAh')$. We claim $\tau_b\in C_h$ and $\tau_{b'}\in C_{h'}$. For $x\in\K^2$ put $g(x)=(\tau_{b'}+x^TAh')(\tau_b+x^TAh)=\tau_{b'}\tau_b+x_1\ell_1+x_2\ell_2+\varphi(xx^T)$ with $\ell_s=\tau_{b'}(e_s^TAh)+\tau_b(e_s^TAh')$, and let $\mathcal U=\varphi(\{\text{symmetric matrices}\})$. Suppose first that $\ell_1,\ell_2$ and $\mathcal U$ span a space of dimension $5$. Pick three distinct $x_1,x_2,x_3$ among the $x_i$. Then $g(x_i)-g(x_1)=(\lambda_i-\lambda_1)q_a$. If $q_a\notin\spn\{\ell_1,\ell_2\}+\mathcal U$, the independence gives $x_i=x_1$. Otherwise $q_a=d_1\ell_1+d_2\ell_2+\varphi(D)$ with $D$ symmetric, and the independence gives $x_i-x_1=\mu_id$ and $x_ix_i^T-x_1x_1^T=\mu_iD$ for $i=2,3$, with $\mu_i=\lambda_i-\lambda_1$. Here $\mu_i\neq0$ and $d\neq0$ since $x_i\neq x_1$. Substituting the first relation into the second gives $x_1d^T+dx_1^T+\mu_idd^T=D$ for $i=2,3$, so $\mu_2=\mu_3$ and $x_2=x_3$, a contradiction. Hence there are $\mu\in\K^2$ and a symmetric $S$, not both zero, with $\mu_1\ell_1+\mu_2\ell_2=\varphi(S)$. Then $\mu\neq0$ because $\varphi$ is injective. Now $\mu_1\ell_1+\mu_2\ell_2=\tau_{b'}(\mu^TAh)+\tau_b(\mu^TAh')$, and its component in $\Sy^2C_h$ is $\tau_{b'}^h\,(\mu^TAh)$, where $\tau_{b'}^h$ is the $C_h$-component of $\tau_{b'}$. This must vanish, so $\tau_{b'}\in C_{h'}$. Likewise $\tau_b\in C_h$.

Write $\tau_b=\nu^TAh$ and $\tau_{b'}=\nu'^TAh'$, and put $\delta=\nu-\nu'$ and $X_i=x_i+\nu'$. Then $\alpha_i=X_i^TAh'$ and $\gamma_i=(X_i+\delta)^TAh$, which is \eqref{mm:eq:32normal}, and $\alpha_i\gamma_i=\varphi((X_i+\delta)X_i^T)$. Finally $T-\nu'^TABy_j=(\delta^TAh)\otimes b$, so $\tau(v)=[(\delta^TAh)\otimes b]$, which is zero if and only if $\delta=0$ (a nonzero form of rank one is not in $W$). As $\tau(a)=0$ and $\Pi_j=\spn\{a,v\}$, this proves the last claim.
\end{proof}

\begin{lemma}\label{mm:lem:32points}
In the situation of Lemma~\ref{mm:lem:32normal} assume moreover:
\begin{enumerate}
    \item[(i)] $\sum_{i\in I_j}c_i\rho_i=0$ for every $c\in K$, and
    \item[(ii)] the linear map $K\to\K^2$, $c\mapsto\sum_{i\in I_j}c_i\rho_iX_i$, is surjective.
\end{enumerate}
Then the products of $I_j$ have exactly two distinct points, $q_a$ is a nonzero multiple of $\varphi(M)$ for an invertible matrix $M$, and $M$ is symmetric if and only if $\delta=0$. If $n_j=4$, then $\delta=0$.
\end{lemma}

\begin{proof}
By (ii) the $X_i$ ($i\in I_j$) do not all lie on one line through $0$. Since the products $P_i/\rho_i$ have the same B-part $q_j$, two of them have the same point if and only if their A-parts agree, that is, if and only if $F(X_i)=F(X_l)$. Also $F(X_i)\neq0$, because $P_i$ has type AB.

Suppose $F$ takes at least three distinct values on $\{X_i\}$. The relation $\varphi(F(X_i))-q'_j\in\K q_a$ shows that $q'_j$ and $q_a$ lie in the image of $\varphi$ and that $F(X_i)=M_0+\lambda_iM_1$ for fixed $M_0,M_1$. The determinant vanishes at three points of this affine line, hence on the whole line. If $M_0,M_1$ are dependent, all $F(X_i)$ are multiples of one rank-one matrix $zw^T$, so all $X_i\in\K w$, contradicting (ii). Otherwise $\spn\{M_0,M_1\}$ is a $2$-dimensional space of singular matrices, so by Fact~\ref{mm:fact:M2}(a) it equals $\{zw^T:\ z\in\K^2\}$ or $\{uz^T:\ z\in\K^2\}$. In the first case all $X_i\in\K w$, contradicting (ii). In the second case $X_i+\delta\in\K u$, so $X_i=s_iu-\delta$ with three distinct values $s_i$ and $F(X_i)=u(s_i^2u-s_i\delta)^T$. These points are collinear only if the points $s_i^2u-s_i\delta\in\K^2$ are, which forces $\delta\in\K u$ (otherwise their coordinates $(s_i^2,-s_i)$ in the basis $u,\delta$ are not collinear). Then all $X_i\in\K u$, contradicting (ii). So $F$ takes exactly two values, and the products of $I_j$ have exactly two points.

If $F(X)=F(X')$ with $X\neq X'$, then $X'=sX$ and $X'+\delta=s^{-1}(X+\delta)$ for some $s\neq0,1$, since both sides are rank-one factorizations of the same nonzero matrix. Hence $(s-s^{-1})X=(s^{-1}-1)\delta$. If $\delta=0$, then $s=-1$ and $X'=-X$. If $\delta\neq0$, then $s\neq-1$, both $X,X'$ lie in $\K\delta$, and $X+X'=-\delta$.

If $n_j=4$, the products of $I_j$ form two double points. If $\delta\neq0$, all four $X_i$ lie in $\K\delta$, contradicting (ii). So $\delta=0$, the double points correspond to $\{X,-X\}$ and $\{X',-X'\}$ with $[X]\neq[X']$ by (ii), and $q_a$ is a nonzero multiple of $\varphi(F(X)-F(X'))=\varphi(M)$, where $M=XX^T-X'X'^T=[X\,X'][X\ -X']^T$ is symmetric and invertible.

If $n_j=3$, the products of $I_j$ form a double point, with vectors $X_1,X_2$, and a single point, with vector $X_3$, and $q_a$ is a nonzero multiple of $\varphi(M)$ with $M=F(X_3)-F(X_1)$. If $\delta=0$, then $X_2=-X_1$, so $[X_3]\neq[X_1]$ by (ii), and $M=X_3X_3^T-X_1X_1^T=[X_3\,X_1][X_3\ -X_1]^T$ is symmetric and invertible. If $\delta\neq0$, then $X_1=\theta\delta$ and $X_2=\theta'\delta$ with $\theta+\theta'=-1$ and $\theta\theta'\neq0$ (as $F(X_l)\neq0$), and $X_3\notin\K\delta$ by (ii). Then $F(X_1)=-\theta\theta'\delta\delta^T$ and
\begin{displaymath}
    M=(X_3+\delta)X_3^T+\theta\theta'\delta\delta^T=[X_3+\delta\ \ \delta]\,[X_3\ \ \theta\theta'\delta]^T
\end{displaymath}
is invertible. If $M$ were symmetric, $(X_3+\delta)X_3^T$ would be symmetric, so $X_3+\delta\in\K X_3$ and $\delta\in\K X_3$, which is false.
\end{proof}

\begin{lemma}\label{mm:lem:phiunique}
Let $(h,h')$ and $(g,g')$ be bases of $\K^2$ and define $\varphi_{h,h'}(M)=\sum_{s,l}M_{sl}(e_s^TAh)(e_l^TAh')$, and $\varphi_{g,g'}$ likewise. If $\varphi_{h,h'}(M)=\varphi_{g,g'}(N)\neq0$ with $N$ symmetric, then $\{[h],[h']\}=\{[g],[g']\}$ and $M$ is a nonzero multiple of $N$ or of $N^T$. In particular $M$ is symmetric.
\end{lemma}

\begin{proof}
For $u,w\in\K^2$ we have
\begin{gather*}
    \varphi_{h,h'}(M)(uw^T)=(w^Th)(w^Th')\,u^TMu,\\
    \varphi_{g,g'}(N)(uw^T)=(w^Tg)(w^Tg')\,u^TNu.
\end{gather*}
As $N$ is symmetric and nonzero, $u\mapsto u^TNu$ is not identically zero. Taking $w\neq0$ with $w^Th=0$, the first expression vanishes for all $u$, hence so does the second, and $(w^Tg)(w^Tg')=0$, that is, $[h]\in\{[g],[g']\}$. Likewise $[h']\in\{[g],[g']\}$, so $\{[h],[h']\}=\{[g],[g']\}$. If $g=th$ and $g'=t'h'$, then $\varphi_{g,g'}(N)=tt'\varphi_{h,h'}(N)$. If $g=th'$ and $g'=t'h$, then $\varphi_{g,g'}(N)=tt'\varphi_{h,h'}(N^T)$. Since $\varphi_{h,h'}$ is injective, the
 claim follows.
\end{proof}

\begin{proposition}\label{mm:prop:32}
If $p_A=3$ and $p_B=2$, then $k=0$, $\{n_1,n_2\}=\{3,4\}$, and the conclusions \textup{(a)}, \textup{(b)}, \textup{(c)} of Theorem~\ref{mm:thm:main} hold.
\end{proposition}

\begin{proof}
\emph{Step 1: $k\le1$ and $n_1,n_2\ge3$.} Since $n_1+n_2=7-k\ge5$, some $n_j\ge3$. If $k=2$, Lemma~\ref{mm:lem:32typeA} gives $n_j\le2$ for that $j$, a contradiction. If $k=1$, it gives $n_j=3$, and then the other line also carries $3$ products. If $k=0$, Lemma~\ref{mm:lem:32lines} gives $n_1,n_2\le4$, so $\{n_1,n_2\}=\{3,4\}$. In all cases $n_1,n_2\ge3$, so $F_j=\Col(y_j)$ with $[y_1]\neq[y_2]$ and $W=W_1\oplus W_2$ (Lemma~\ref{mm:lem:32lines}).

\emph{Step 2: $\tau(a)=0$.} If $k=1$, the product $(\alpha_0+\beta_0)\gamma_0$ of type A has $\beta_0\in F_1\cap F_2=0$ by Lemma~\ref{mm:lem:32typeA}, applied to both lines. So it is the pure product $\alpha_0\gamma_0$, its bilinear part is $0$, and $\tau(a)=0$. If $k=0$, the point $a$ lies in $\Pi_1\cap\Pi_2$. Each $\Pi_j$ is spanned by pure parts of products of $I_j$ (Lemma~\ref{mm:lem:32lines}). So $\tau(a)$ is the class of some $s_1\in\cA^*\otimes F_1$ and also of some $s_2\in\cA^*\otimes F_2$. Then $s_1-s_2\in W$. Write $s_1-s_2=u_1+u_2$ with $u_j\in W_j\subseteq\cA^*\otimes F_j$. Then $s_1-u_1=s_2+u_2\in(\cA^*\otimes F_1)\cap(\cA^*\otimes F_2)=0$, so $s_1\in W$ and $\tau(a)=0$.

\emph{Step 3: the hypotheses of Lemma~\ref{mm:lem:32points} hold on both lines.} Apply Lemma~\ref{mm:lem:32normal} to $j=1,2$. Write $h_j,h'_j,b_j,b'_j,\varphi_j,\delta_j$ for its data. For $c\in K$ the B-part of $\sum_ic_iP_i$ is $\sum_{i\in I_1}c_i\rho_i\,q_1+\sum_{i\in I_2}c_i\rho_i\,q_2=0$ (products of type A have B-part $0$), and $q_1,q_2$ are independent. This is (i). By \eqref{mm:eq:32normal} and (i), $\sum_{i\in I_j}c_iS_i=(\sum_{i\in I_j}c_i\rho_iX_i)^TABy_j\in W_j$, and the product of type A, if any, has bilinear part $0$. Hence $\sum_ic_iS_i=\sum_j(\sum_{i\in I_j}c_i\rho_iX_i)^TABy_j$. As $c\mapsto\sum_ic_iS_i$ maps $K$ onto $W=W_1\oplus W_2$ (Lemma~\ref{mm:lem:relations}(b)), this is (ii) for both $j$. So $q_a$ is a nonzero multiple of $\varphi_j(M_j)$ with $M_j$ invertible, for $j=1,2$.

\emph{Step 4: $k=0$ and $\tau=0$.} As a quadratic form on $\cA$, $\varphi_1(M_1)$ has rank $2\rank M_1=4$: in the coordinates $e_s^TAh_1$, $e_l^TAh'_1$ its matrix is $\frac12\bigl(\begin{smallmatrix}0&M_1\\M_1^T&0\end{smallmatrix}\bigr)$. If $k=1$, then $q_a=\alpha_0\gamma_0$ has rank at most two, a contradiction. So $k=0$. Say $n_2=4$ and $n_1=3$. By Lemma~\ref{mm:lem:32points}, $\delta_2=0$ and $M_2$ is symmetric. By Lemma~\ref{mm:lem:phiunique}, $M_1$ is symmetric and the pairs $\{[h_1],[h'_1]\}$ and $\{[h_2],[h'_2]\}$ coincide. By Lemma~\ref{mm:lem:32points} again, $\delta_1=0$. So $\tau$ vanishes on $\Pi_1+\Pi_2=V$ (Lemma~\ref{mm:lem:32normal}), that is, all $S_i\in W$.

\emph{Step 5: conclusion.} With $\delta_j=0$, \eqref{mm:eq:32normal} reads $P_i=\rho_i(X_i^TAh'_j+k_j^TBy_j)(X_i^TAh_j+k_j'^TBy_j)$, which is a W-product of the form \eqref{mm:eq:wprod} with middle pair $\{[h_j],[h'_j]\}$. This pair is the same for $j=1,2$. By Lemma~\ref{mm:lem:32points} there are exactly four points: two double points on $\mathcal L_2$, one double point and one single point on $\mathcal L_1$, and none at $[a]$. The two products at a double point correspond to $X$ and $-X$, that is, to $(X,y_j)$ and $(X,-y_j)$ in \eqref{mm:eq:wprod}. This gives (a) and (b), and (c) follows as in the proof of Corollary~\ref{mm:cor:d4}.
\end{proof}

\subsection{\texorpdfstring{The case $(p_A,p_B)=(2,2)$}{The case (pA,pB)=(2,2)}}\label{mm:sec:22}

In this section $p=3$ and $p_A=p_B=2$. Then $N_A=\K a$ with $a=(r_a,0)$ and $N_B=\K b$ with $b=(0,q_b)$. Let $r_1,r_2,G_1,G_2$ and $q_1,q_2,F_1,F_2$ be as in Lemma~\ref{mm:lem:twoq}. The A-part of a product of type A is a multiple of $r_a$ and, by Lemma~\ref{mm:lem:twoq}, of $r_1$ or $r_2$. Similarly for type B. A product of type AB has pure part $(sr_i,tq_j)$ with $s,t\neq0$ and $i,j\in\{1,2\}$. We say it has \emph{class} $(i,j)$, and write $n_{ij}$ for the number of products of class $(i,j)$. We note:
\begin{enumerate}
    \item[(N1)] If $V$ contains $(sr_i,tq_j)$ with $s,t\neq0$ and $[r_i]=[r_a]$, then $(0,q_j)\in V$, so $[q_j]=[q_b]$. Likewise with the sides exchanged.
    \item[(N2)] If the products of class $(i,j)$ have two different points, then $V$ contains $(r_i,0)$ and $(0,q_j)$, so $[r_i]=[r_a]$ and $[q_j]=[q_b]$.
\end{enumerate}

\begin{proposition}\label{mm:prop:22}
If $p=3$ and $p_A=p_B=2$, then $[r_a]\notin\{[r_1],[r_2]\}$, $[q_b]\notin\{[q_1],[q_2]\}$, $k_A=k_B=0$, and the conclusions \textup{(a)}, \textup{(b)}, \textup{(c)} of Theorem~\ref{mm:thm:main} hold.
\end{proposition}

\begin{proof}
\emph{Case 1: $[r_a]\in\{[r_1],[r_2]\}$ and $[q_b]\notin\{[q_1],[q_2]\}$.} Say $[r_a]=[r_1]$. By (N1) there are no products of class $(1,j)$, and $k_B=0$. So every product not of type A has its A-factors in $G_2$, and the products of type A have $\gamma_i\in G_1$. Let $X=\{A:\ \lambda(A)=0\text{ for }\lambda\in G_2\}$, which is $2$-dimensional. On $X\times\cB$ only the products of type A have nonzero bilinear part, so by \eqref{mm:eq:bil} and Fact~\ref{mm:fact:rank}(a), $k_A\ge4$, contradicting Lemma~\ref{mm:lem:twoq}. The case with the sides exchanged is the same.

\emph{Case 2: $[r_a]\in\{[r_1],[r_2]\}$ and $[q_b]\in\{[q_1],[q_2]\}$.} Renumber so that $[r_a]=[r_1]$ and $[q_b]=[q_1]$. By (N1) there are no products of classes $(1,2)$ and $(2,1)$, and by (N2) the products of class $(2,2)$ have one point, so $n_{22}\le2$. The products of type A have their A-factors in $G_1$, those of type B their B-factors in $F_1$, and the products of class $(1,1)$ have their pure parts in $\spn\{a,b\}$. Let $X=G_1^\perp=\{A:\ \lambda(A)=0\text{ for }\lambda\in G_1\}$ and $Y=F_1^\perp$. On $X\times\cB$ the bilinear parts of the products of type A and of class $(1,1)$ vanish, so by \eqref{mm:eq:bil} and Fact~\ref{mm:fact:rank}(a), $k_B+2n_{22}\ge4$. Likewise, on $\cA\times Y$, $k_A+2n_{22}\ge4$. As $k_A,k_B\le2$, $n_{22}\ge1$.

If $n_{22}=1$, then $k_A=k_B=2$ and $n_{11}=2$. The double point of type A has row $[x_a]$ with $\Row(x_a)=G_1$, and the double point of type B has column $[y_b]$ with $\Col(y_b)=F_1$ (Lemma~\ref{mm:lem:doubleAB}). The six products other than the one of class $(2,2)$ have pure parts in $\spn\{a,b\}$ and bilinear parts vanishing on $X\times Y$, so Lemma~\ref{mm:lem:kernel} gives $6-2\le4-\dim XY$, that is, $XY=0$. But $X$ contains $uh^T$ for $u\neq0$ with $x_a^Tu=0$ and all $h$, and $Y$ contains $kw^T$ for $w\neq0$ with $w^Ty_b=0$ and all $k$. Choosing $h^Tk\neq0$ gives $(uh^T)(kw^T)\neq0$, a contradiction.

So $n_{22}=2$, and the double point of class $(2,2)$ has row $[x_2]$ and column $[y_2]$ with $\Row(x_2)=G_2$ and $\Col(y_2)=F_2$. Let $Y'=F_2^\perp=\{kw_2^T:\ k\in\K^2\}$, where $w_2\neq0$ and $w_2^Ty_2=0$. On $X\times Y'$ the bilinear parts of all products except those of type B vanish, so Fact~\ref{mm:fact:rank}(b) gives $k_B\ge2$. Likewise, with $X'=G_2^\perp=\{u_2h^T:\ h\in\K^2\}$ ($u_2\neq0$, $x_2^Tu_2=0$) and $Y=F_1^\perp$, we get $k_A\ge2$. Hence $k_A=k_B=2$ and $n_{11}=1$. As before $\Row(x_a)=G_1$ and $\Col(y_b)=F_1$. So $[x_a]\neq[x_2]$, $[y_b]\neq[y_2]$, and with $w_{ij}=x_i^TABy_j$ (where $x_1=x_a$, $y_1=y_b$) we have $\cA^*\otimes\cB^*=\bigoplus_{i,j}G_i\otimes F_j$ and $W=\bigoplus_{i,j}\K w_{ij}$ with $w_{ij}\in G_i\otimes F_j$ (Fact~\ref{mm:fact:Wfactor}(a)).

Let $(\alpha_0+\beta_0)\gamma_0$ be a product of type A with pure part $a$, let $\beta'_0(\gamma'_0+\delta'_0)$ be a product of type B with pure part $b$, and let $P_{11}$ be the product of class $(1,1)$, with pure part $sa+tb$, $s,t\neq0$. By Lemma~\ref{mm:lem:tau}, $S_{11}-s\gamma_0\otimes\beta_0-t\gamma'_0\otimes\beta'_0\in W$. Here $S_{11}\in G_1\otimes F_1$, $\gamma_0\in G_1$ and $\beta'_0\in F_1$. The component in $G_1\otimes F_2$ is $-s\gamma_0\otimes\beta_0^{(2)}$, where $\beta_0^{(2)}$ is the $F_2$-component of $\beta_0$. It lies in $\K w_{12}$ and has rank at most one, so it vanishes and $\beta_0\in F_1$. Likewise the component in $G_2\otimes F_1$ shows that $\gamma'_0\in G_1$. The same holds for the other product of type A and the other product of type B, since their classes under $\tau$ are multiples of $\tau(a)$, respectively $\tau(b)$. So the five products of types A and B and of class $(1,1)$ have bilinear parts in $G_1\otimes F_1$ and pure parts in $\spn\{a,b\}$. Their relations form a space of dimension at least $3$, which Lemma~\ref{mm:lem:relations}(b) maps injectively into $W\cap(G_1\otimes F_1)=\K w_{11}$, a contradiction.

\emph{Case 3: $[r_a]\notin\{[r_1],[r_2]\}$ and $[q_b]\notin\{[q_1],[q_2]\}$.} Then $k_A=k_B=0$, and by (N2) each class has at most one point. So the seven products lie on at most four points, and each point carries at most two products. The four classes are all occupied, one by a single point and three by double points. If the class $(i,j)$ carries a double point, its row $[x]$ and column $[y]$ satisfy $\Row(x)=G_i$ and $\Col(y)=F_j$. The three double points occupy classes covering both values of $i$ and both values of $j$. So $G_i=\Row(x_i)$ and $F_j=\Col(y_j)$, with $[x_1]\neq[x_2]$ and $[y_1]\neq[y_2]$. As in Case~2, $W=\bigoplus\K w_{ij}$ with $w_{ij}=x_i^TABy_j\in G_i\otimes F_j$.

Choose one product of each class and let $\pi_{ij}$ be its pure part. Any three of the $\pi_{ij}$ are linearly independent. In a relation among $\pi_{11},\pi_{12},\pi_{21}$, the A-part forces the coefficient of $\pi_{21}$ to vanish, and the B-part then forces the others to vanish. The other triples are the same. So the $\pi_{ij}$ span $V$ and satisfy a relation $\sum c_{ij}\pi_{ij}=0$ with all $c_{ij}\neq0$. Applying $\tau$ gives $\sum c_{ij}[S_{ij}]=0$ in $(\cA^*\otimes\cB^*)/W=\bigoplus_{i,j}(G_i\otimes F_j)/\K w_{ij}$, where $[S_{ij}]$ lies in the summand $(i,j)$. Hence all $[S_{ij}]=0$, so $\tau=0$ and every bilinear part lies in $W$.

Now every product has class $(i,j)$ for some $i,j$, and its bilinear part lies in $W\cap(G_i\otimes F_j)=\K w_{ij}$. It is not zero. Otherwise the product would be $(\alpha+\beta)(\gamma+\delta)$ with $\alpha\otimes\delta+\gamma\otimes\beta=0$, so $[\alpha]=[\gamma]$, and its A-part would be a square. But $r_i$ is the product of a basis of $G_i$. So every product is a W-product \eqref{mm:eq:wprod} with $[x]=[x_i]$ and $[y]=[y_j]$. Its A-part $(x_i^TAh_1)(x_i^TAh_2)$ is a multiple of $r_i$ and determines the middle pair $\{[h_1],[h_2]\}$. Its B-part $(k_1^TBy_j)(k_2^TBy_j)$ is a multiple of $q_j$ and determines $\{[k_1],[k_2]\}$, hence again the middle pair. So all products of the same $i$, and all products of the same $j$, have the same middle pair. Since all four classes are occupied, the middle pair is common to all products. Two W-products with the same middle pair, row and column are, after rescaling, given by $(x,y)$ and $(x,ty)$ in \eqref{mm:eq:wprod}, and their pure parts are proportional only if $t^2=1$. So the two products at a double point correspond to $(x,y)$ and $(x,-y)$, and (a), (b), (c) follow as in the proof of Corollary~\ref{mm:cor:d4}.
\end{proof}

\subsection{\texorpdfstring{The case $p_B=1$}{The case pB=1}}\label{mm:sec:pB1}

In this section $p=3$ and $p_B=1$, so $V_B=\K q_b$ and $\dim N_A=2$. Every product of type A has its point on the line $\Pj(N_A)$. The other products, of types AB and B, will be called \emph{N-products}. Their B-parts are nonzero multiples of $q_b$. Let $E\subseteq\cB^*$ be the span of the factors of $q_b$, so $\dim E\in\{1,2\}$, and let $Y=E^\perp=\{B:\ \lambda(B)=0\text{ for }\lambda\in E\}$. The factors $\beta_i,\delta_i$ of every N-product lie in $E$, so its bilinear part vanishes on $\cA\times Y$.

For $y_1,y_2\in\K^2$ let $R(y_1,y_2)$ be the list of the seven quadratic forms
\begin{gather*}
    a_{11}(a_{12}+e_1^TBy_1),\qquad a_{12}(a_{11}-e_2^TBy_1),\\
    a_{21}(a_{22}+e_1^TBy_2),\qquad a_{22}(a_{21}-e_2^TBy_2),\\
    (a_{11}-y_{12}b_{22})(a_{12}+y_{11}b_{11}),\\
    (a_{21}-y_{22}b_{22})(a_{22}+y_{21}b_{11}),\qquad b_{11}b_{22},
\end{gather*}
where $y_j=(y_{j1},y_{j2})^T$, so that $e_1^TBy_j=y_{j1}b_{11}+y_{j2}b_{12}$ and $e_2^TBy_j=y_{j1}b_{21}+y_{j2}b_{22}$. For $y_1=(1,1)^T$ and $y_2=(1,I)^T$ these are the products of $R_I$.

\begin{proposition}\label{mm:prop:pB1}
If $p=3$ and $p_B=1$, then after a substitution as in Lemma~\ref{mm:lem:sym}\textup{(a)} the products are nonzero multiples of those of $R(y_1,y_2)$ for some $y_1,y_2$ with $y_{11}y_{12}y_{21}y_{22}\neq0$.
\end{proposition}

\begin{proof}
\emph{Step 1: $k_A\ge4$.} On $\cA\times Y$ only the products of type A have nonzero bilinear part, of rank at most one. As $\dim Y\ge2$, \eqref{mm:eq:bil} and Fact~\ref{mm:fact:rank}(a) give $k_A\ge4$.

\emph{Step 2: the products of type A have only two points.} Suppose they have at least three distinct points. Their A-parts are then pairwise non-proportional elements of the $2$-dimensional image of $N_A$ in $V_A$. So by Lemma~\ref{mm:lem:pencil} all their factors $\alpha_i,\gamma_i$ lie in a subspace of dimension at most $3$, and there is $A_0\neq0$ at which all of them vanish. The bilinear parts $\gamma_i\otimes\beta_i$ of the products of type A vanish on $\K A_0\times\cB$, and their pure parts span $N_A$, so Lemma~\ref{mm:lem:kernel} gives $k_A-2\le4-\dim A_0\cB=4-2\rank A_0$. Hence $k_A=4$ and $A_0=xw^T$ has rank one. On $\K A_0\times\cB$, \eqref{mm:eq:bil} reads $A_0B=\sum_{\text{N-products}}\Lambda_iS_i(A_0,B)$, so the four entries $B\mapsto x_sw^TBe_l$ of $A_0B$ lie in $E$. They span $\Row_\cB(w)$, so $E=\Row_\cB(w)$ and $Y=\{B:\ w^TB=0\}=\{uz^T:\ z\in\K^2\}$ for some $u\neq0$ with $w^Tu=0$. For $B=uz^T\in Y$ write $\beta_i(uz^T)=g_i^Tz$ with $g_i\in\K^2$. Then \eqref{mm:eq:bil} gives
\begin{displaymath}
    (Au)z^T=\sum_{i\text{ of type A}}\Lambda_i\,\gamma_i(A)\,g_i^Tz\qquad(A\in\cA,\ z\in\K^2).
\end{displaymath}
Applying a linear functional $\psi$ on $\K^{2\times2}$ to both sides, the left side becomes the bilinear form $(A,z)\mapsto\psi((Au)z^T)$, and $\psi\mapsto\psi((Au)z^T)$ is injective with image $\Col_\cA(u)\otimes(\K^2)^*$, which is $4$-dimensional. The right side becomes $\sum_i\psi(\Lambda_i)\,\gamma_i\otimes g_i$, a combination of four tensors. Hence the four tensors $\gamma_i\otimes g_i$ are linearly independent and span $\Col_\cA(u)\otimes(\K^2)^*$. In particular $g_i\neq0$ and $\gamma_i\in\Col_\cA(u)$ for all four products of type A. A double point of type A would have its two forms $\gamma,\alpha$ in $\Col_\cA(u)$ and, by Lemma~\ref{mm:lem:doubleAB}(a), forming a basis of some $\Row(x)$, contradicting Fact~\ref{mm:fact:Wfactor}(c). So the four products of type A have four distinct points, and their relations form a $2$-dimensional space, which Lemma~\ref{mm:lem:relations}(b) maps injectively into $W\cap(\Col_\cA(u)\otimes\cB^*)=0$ (Fact~\ref{mm:fact:Wfactor}(c)). This contradiction shows that the products of type A have at most two points. As a point carries at most two products and $k_A\ge4$, there are exactly two double points of type A, $k_A=4$, and there are three N-products.

\emph{Step 3: normal form of the double points.} Let the double points of type A have rows $[x_1],[x_2]$ and columns $[y_1],[y_2]$. If $[x_1]=[x_2]$, all A-factors of the products of type A lie in $\Row(x_1)$, so their bilinear parts vanish on $X\times\cB$, where $X=\{A:\ x_1^TA=0\}$. Then by \eqref{mm:eq:bil}, for $A\in X$ the entries of $AB$ are combinations of forms in $E$, but for $A=uh^T\in X$ ($x_1^Tu=0$) they include all forms $B\mapsto h^TBe_l$, which span $\cB^*$. So $[x_1]\neq[x_2]$, and after a substitution $A\mapsto X^{-1}A$ as in Lemma~\ref{mm:lem:sym}(a) we may assume $x_j=e_j$. By Lemma~\ref{mm:lem:doubleAB}(a) and Fact~\ref{mm:fact:W}, the products at the $j$-th double point are, after rescaling,
\begin{equation}\label{mm:eq:pB1normal}
\begin{gathered}
    P_j=(\alpha_j+\beta_j)\gamma_j,\qquad P'_j=(\gamma_j+\beta'_j)\alpha_j,\\
    \alpha_j=e_j^TAh_j,\quad\gamma_j=e_j^TAh'_j,\quad\beta_j=k_j'^TBy_j,\quad\beta'_j=-k_j^TBy_j,
\end{gathered}
\end{equation}
where $(h_j,h'_j)$ is a basis of $\K^2$ with dual basis $(k_j,k'_j)$, and $y_j$ is rescaled suitably. Indeed $\gamma_j\otimes\beta_j-\alpha_j\otimes\beta'_j=e_j^TABy_j=\alpha_j\otimes k_j^TBy_j+\gamma_j\otimes k_j'^TBy_j$.

\emph{Step 4: restriction to $Y$.} For $B\in Y$, \eqref{mm:eq:bil} gives
\begin{displaymath}
    AB=\sum_j\bigl(\Lambda_{P_j}\gamma_j(A)\beta_j(B)+\Lambda_{P'_j}\alpha_j(A)\beta'_j(B)\bigr),
\end{displaymath}
while $AB=\sum_je_je_j^TAB$ and $e_j^TAB=\alpha_j(A)\,k_j^TB+\gamma_j(A)\,k_j'^TB$. The forms $\alpha_1,\gamma_1,\alpha_2,\gamma_2$ are a basis of $\cA^*$, so comparing coefficients gives, for all $B\in Y$,
\begin{equation}\label{mm:eq:pB1Y}
    e_j\,k_j'^TB=(k_j'^TBy_j)\,\Lambda_{P_j},\qquad e_j\,k_j^TB=-(k_j^TBy_j)\,\Lambda_{P'_j}
\end{equation}
for $j=1,2$. Consequently: (i) for each $\kappa\in\{k_1,k'_1,k_2,k'_2\}$ the row vectors $\kappa^TB$, $B\in Y$, lie in a $1$-dimensional space; (ii) if $B\in Y$ and $By_j=0$ for some $j$, then $k_j^TB=k_j'^TB=0$, so $B=0$.

\emph{Step 5: the type of $Y$.} For $\kappa\neq0$ let $C_\kappa=\{B:\ \kappa^TB=0\}$, a $2$-dimensional space. By (i), $\dim(Y\cap C_\kappa)\ge\dim Y-1$. If $\dim E=1$, then $\dim Y=3$, so $C_\kappa\subseteq Y$ for $\kappa=k_1,k'_1$, and $C_{k_1}\oplus C_{k'_1}=\K^{2\times2}\subseteq Y$, which is absurd. So $\dim E=\dim Y=2$. If $Y=\{uz^T:\ z\in\K^2\}$ (type C), then $Y$ contains $u(Jy_1)^T\neq0$, which kills $y_1$, contradicting (ii). It remains to treat types R and I.

\emph{Step 6: $Y$ is not of type R.} Suppose $Y=\{wz_0^T:\ w\in\K^2\}$. Then $E=Y^\perp=\Col(y_E)$ with $y_E=Jz_0$, and by (ii) $z_0^Ty_j\neq0$, that is, $[y_j]\neq[y_E]$, for $j=1,2$. Let $a_j=(\alpha_j\gamma_j,0)$ be the pure part of $P_j$, so $N_A=\spn\{a_1,a_2\}$ and $\tau(a_j)=[\gamma_j\otimes\beta_j]$. Normalize the three N-products $N_1,N_2,N_3$ to have B-part exactly $q_b$, and fix $v\in V$ with B-part $q_b$. Then $\pi(N_i)=v+s_ia_1+t_ia_2$ for some scalars $s_i,t_i$. By Lemma~\ref{mm:lem:tau}, for $i\neq l$,
\begin{displaymath}
    (s_i-s_l)\,\gamma_1\otimes\beta_1+(t_i-t_l)\,\gamma_2\otimes\beta_2\in S(N_i)-S(N_l)+W\subseteq\cA^*\otimes E+W.
\end{displaymath}
Choose $y'$ with $[y']\neq[y_E]$. Then $W=\{x^TABy_E\}\oplus\{x^TABy'\}$ and $\cA^*\otimes E+W=(\cA^*\otimes E)\oplus\{x^TABy'\}$. Write $y_j=c_jy_E+d_jy'$. Here $d_j\neq0$. Projecting $\cB^*=E\oplus\Col(y')$ onto $\Col(y')$, the element above becomes
\begin{displaymath}
    (s_i-s_l)d_1\,\gamma_1\otimes k_1'^TBy'+(t_i-t_l)d_2\,\gamma_2\otimes k_2'^TBy'\in\{x^TABy':\ x\in\K^2\}.
\end{displaymath}
By Fact~\ref{mm:fact:Wfactor}(a), a nonzero element of the right side has rank two, and its A-factor space is $\Row(x)$ for some $x$. The left side has rank at most one, unless both coefficients are nonzero and $[k'_1]\neq[k'_2]$. In that case its A-factor space $\spn\{\gamma_1,\gamma_2\}$, with $\gamma_j\in\Row(e_j)$, is not of the form $\Row(x)$ (Fact~\ref{mm:fact:W}). So both sides vanish, and since $\gamma_1,\gamma_2$ are independent and $k_1'^TBy'$, $k_2'^TBy'$ are nonzero, $s_i=s_l$ and $t_i=t_l$. Then the three N-products have the same point, contradicting Lemma~\ref{mm:lem:doubleAB}(c).

\emph{Step 7: $Y$ of type I.} Now $\det$ restricted to $Y$ is a nonzero binary quadratic form, so $Y$ contains at most two lines of rank-one matrices. By (i), for each $\kappa\in\{k_1,k'_1,k_2,k'_2\}$ there is $0\neq B\in Y$ with $\kappa^TB=0$. Such $B$ is not invertible, so $B=(J\kappa)z^T$. Since $[k_1]\neq[k'_1]$, the two rank-one lines of $Y$ are spanned by $(Jk_1)z^T$ and $(Jk'_1)z'^T$, and
 $[z]\neq[z']$ because $Y$ has type I. Every $\kappa\in\{k_2,k'_2\}$ gives a rank-one element with column $J\kappa$, so $\{[k_2],[k'_2]\}=\{[k_1],[k'_1]\}$, and the dual bases satisfy $\{[h_2],[h'_2]\}=\{[h_1],[h'_1]\}$. After a substitution $A\mapsto AY_0$, $B\mapsto Y_0^{-1}B$ as in Lemma~\ref{mm:lem:sym}(a) (which keeps the rows $e_1,e_2$) we may assume that this common pair is $\{[e_1],[e_2]\}$. Then $\{[k_1],[k'_1]\}=\{[e_1],[e_2]\}$, and $Y=\spn\{e_2z^T,e_1z'^T\}$ up to renaming $z,z'$. After a further substitution $B\mapsto BZ_0$ we may assume $z=e_1$ and $z'=e_2$, so that
\begin{displaymath}
    Y=\{B:\ b_{11}=b_{22}=0\},\qquad E=\spn\{b_{11},b_{22}\}.
\end{displaymath}
The A-factors of the $j$-th double point are now multiples of $a_{j1},a_{j2}$. Taking $\gamma_j$ a multiple of $a_{j1}$ in Lemma~\ref{mm:lem:doubleAB}(a) and rescaling, the four products of type A are nonzero multiples of
\begin{gather*}
    a_{11}(a_{12}+e_1^TBy_1),\qquad a_{12}(a_{11}-e_2^TBy_1),\\
    a_{21}(a_{22}+e_1^TBy_2),\qquad a_{22}(a_{21}-e_2^TBy_2)
\end{gather*}
for suitable $y_1,y_2$. Indeed the W-element condition $a_{j1}\otimes\beta-a_{j2}\otimes\beta'=e_j^TABy=a_{j1}\otimes e_1^TBy+a_{j2}\otimes e_2^TBy$ gives $\beta=e_1^TBy$ and $\beta'=-e_2^TBy$. Denote these products, in this order, by $D_1,D'_1,D_2,D'_2$, with coefficient matrices $\Lambda$ as in \eqref{mm:eq:bil}.

The bilinear parts of the N-products lie in $\cA^*\otimes E$. So the monomials $a_{11}b_{12}$, $a_{12}b_{21}$, $a_{21}b_{12}$, $a_{22}b_{21}$ occur on the right of \eqref{mm:eq:bil} only in the bilinear parts of the four products above, with coefficients $y_{12}$, $-y_{11}$, $y_{22}$, $-y_{21}$ respectively. On the left they occur only in the entries $(1,2)$, $(1,1)$, $(2,2)$, $(2,1)$ of $AB$, with coefficient $1$. Hence $y_{11}y_{12}y_{21}y_{22}\neq0$ and
\begin{gather*}
    \Lambda_{D_1}=y_{12}^{-1}E_{12},\qquad\Lambda_{D'_1}=-y_{11}^{-1}E_{11},\\
    \Lambda_{D_2}=y_{22}^{-1}E_{22},\qquad\Lambda_{D'_2}=-y_{21}^{-1}E_{21},
\end{gather*}
where $E_{sl}$ are the matrix units. A direct computation (Section~\ref{mm:sec:comp}) then gives, as an identity of $2\times2$ matrices of quadratic forms,
\begin{equation}\label{mm:eq:pB1rest}
    \sum_{\text{N-products}}\Lambda_iN_i=AB-\sum_{D\in\{D_1,D'_1,D_2,D'_2\}}\Lambda_DD=M_1Q_1+M_2Q_2,
\end{equation}
where $\rho_1=y_{12}/y_{11}$, $\rho_2=y_{22}/y_{21}$, $M_1=E_{11}-\rho_1^{-1}E_{12}$, $M_2=E_{21}-\rho_2^{-1}E_{22}$, and
\begin{gather*}
    Q_1=y_{11}^{-1}a_{11}a_{12}+a_{11}b_{11}-\rho_1a_{12}b_{22},\\
    Q_2=y_{21}^{-1}a_{21}a_{22}+a_{21}b_{11}-\rho_2a_{22}b_{22}.
\end{gather*}
Applying to \eqref{mm:eq:pB1rest} a linear functional on $\K^{2\times2}$ vanishing on $M_1,M_2$, and using that the three N-products are linearly independent, we find $\Lambda_i\in\spn\{M_1,M_2\}$ for each N-product $N_i$, and then $Q_1,Q_2\in\mathcal U:=\spn\{N_1,N_2,N_3\}$.

The forms $Q_1,Q_2$ are independent with B-part $0$. No nonzero element of $\spn\{Q_1,Q_2\}$ is a product of two linear forms. Indeed, the A-part $sy_{11}^{-1}a_{11}a_{12}+ty_{21}^{-1}a_{21}a_{22}$ of $sQ_1+tQ_2$ has rank $4$ unless $st=0$. Also $y_{11}Q_1=(a_{11}-y_{12}b_{22})(a_{12}+y_{11}b_{11})+y_{11}y_{12}b_{11}b_{22}$ has rank $4$, because the four forms involved are independent. Similarly for $Q_2$. The B-parts of the N-products are multiples $c_iq_b$, and the kernel of $\mathcal U\to\K q_b$ contains $\spn\{Q_1,Q_2\}$. If it were larger, it would be $\mathcal U$ and all $c_i$ would vanish, which is false. So it equals $\spn\{Q_1,Q_2\}$, and no $c_i$ vanishes, as $N_i\notin\spn\{Q_1,Q_2\}$. Write $q_b=bb'$ with $E=\spn\{b,b'\}$ and rescale so that $c_i=1$. Then each $N_i$ has the form $(\mu_i+b')(\nu_i+b)$ with $\mu_i,\nu_i\in\cA^*$: for type AB, after exchanging and rescaling the factors, $N_i=(\alpha+\theta^{-1}b')(\gamma+\theta b)=(\theta\alpha+b')(\theta^{-1}\gamma+b)$, and for type B, $N_i=\theta b(\gamma+\theta^{-1}b')$ or $\theta b'(\gamma+\theta^{-1}b)$. So $S(N_i)=\mu_i\otimes b+\nu_i\otimes b'$ and the A-part of $N_i$ is $\mu_i\nu_i$. Moreover $N_i-N_l\in\spn\{Q_1,Q_2\}$ for all $i,l$, and $N_2-N_1$, $N_3-N_1$ are independent.

Let $C_1=\spn\{a_{11},a_{21}\}$ and $C_2=\spn\{a_{12},a_{22}\}$, so $\Sy^2\cA^*=\Sy^2C_1\oplus C_1C_2\oplus\Sy^2C_2$. Write $b_{11}=pb+p'b'$ and $b_{22}=rb+r'b'$. If $N_i-N_l=sQ_1+tQ_2$, comparing bilinear and A-parts gives, with $d=sa_{11}+ta_{21}\in C_1$ and $e=s\rho_1a_{12}+t\rho_2a_{22}\in C_2$,
\begin{gather*}
    \mu_i-\mu_l=pd-re,\qquad\nu_i-\nu_l=p'd-r'e,\\
    \mu_i\nu_i-\mu_l\nu_l=sy_{11}^{-1}a_{11}a_{12}+ty_{21}^{-1}a_{21}a_{22}\in C_1C_2.
\end{gather*}
Writing $\mu_l=\mu_l^{(1)}+\mu_l^{(2)}$ and $\nu_l=\nu_l^{(1)}+\nu_l^{(2)}$ with components in $C_1,C_2$, the $\Sy^2C_1$-component of the last equation is $d\,(p'\mu_l^{(1)}+p\nu_l^{(1)}+pp'd)=0$, and the $\Sy^2C_2$-component is $e\,(rr'e-r'\mu_l^{(2)}-r\nu_l^{(2)})=0$. Take $l=1$ and $i=2,3$: the two pairs $(s,t)$ are independent, so the two values of $d$ (and of $e$) are independent, while $p'\mu_1^{(1)}+p\nu_1^{(1)}$ and $r'\mu_1^{(2)}+r\nu_1^{(2)}$ do not depend on $i$. Hence $pp'=0$ and $rr'=0$, so each of $b_{11},b_{22}$ is a multiple of $b$ or of $b'$. As they are independent, after exchanging and rescaling $b,b'$ (and rescaling the $N_i$) we have $b=b_{11}$, $b'=b_{22}$, $q_b=b_{11}b_{22}$, and $p=r'=1$, $p'=r=0$. The two components above now give $\nu_l^{(1)}=0$ and $\mu_l^{(2)}=0$ for every $l$, that is, $\mu_l\in C_1$ and $\nu_l\in C_2$. Write $\mu_l=m_{l1}a_{11}+m_{l2}a_{21}$ and $\nu_l=n_{l1}a_{12}+n_{l2}a_{22}$ with $m_l,n_l\in\K^2$, and $D=\operatorname{diag}(\rho_1,\rho_2)$. Then $n_l+Dm_l=c$ is independent of $l$. The map $\K^{2\times2}\to C_1C_2$, $M\mapsto\sum M_{st}a_{s1}a_{t2}$, is injective and sends $m_ln_l^T$ to $\mu_l\nu_l$, so with $\Delta=m_i-m_l$ the last equation becomes
\begin{displaymath}
    \Delta c^T-(m_im_i^T-m_lm_l^T)D=\operatorname{diag}(\Delta_1y_{11}^{-1},\Delta_2y_{21}^{-1}).
\end{displaymath}
Its entries give: if $\Delta_1\neq0$ then $m_{i1}+m_{l1}=\sigma_1:=(c_1-y_{11}^{-1})/\rho_1$. If $\Delta_2\neq0$ then $m_{i2}+m_{l2}=\sigma_2:=(c_2-y_{21}^{-1})/\rho_2$; and
\begin{displaymath}
    \Delta_1c_2=\rho_2(m_{i1}m_{i2}-m_{l1}m_{l2}),\qquad\Delta_2c_1=\rho_1(m_{i1}m_{i2}-m_{l1}m_{l2}).
\end{displaymath}
The points $m_1,m_2,m_3$ are not collinear, since $m_2-m_1$ and $m_3-m_1$ are the independent pairs $(s,t)$. Any two of the first coordinates $m_{l1}$ are equal or have sum $\sigma_1$, so they take at most two values, and likewise the second coordinates. Hence the three points are three vertices of a rectangle with sides parallel to the axes. One of them, $m_{l_0}$, shares its second coordinate $\omega$ with a second point, and its first coordinate $\omega'$ with the third. For the first of these pairs $\Delta_2=0\neq\Delta_1$, and the equations give $\rho_1\omega\Delta_1=0$, so $\omega=0$, and then $c_2=0$. For the second pair likewise $\omega'=0$ and $c_1=0$. So $m_{l_0}=0$ and $n_{l_0}=c=0$, that is, $N_{l_0}=b_{22}b_{11}$. The other two points are $(\sigma_1,0)$ and $(0,\sigma_2)$ with $\sigma_1=-1/y_{12}$ and $\sigma_2=-1/y_{22}$, and $n=-Dm$ for them. The first gives $N=(-y_{12}^{-1}a_{11}+b_{22})(y_{11}^{-1}a_{12}+b_{11})=-(y_{11}y_{12})^{-1}(a_{11}-y_{12}b_{22})(a_{12}+y_{11}b_{11})$, and the second likewise gives a multiple of $(a_{21}-y_{22}b_{22})(a_{22}+y_{21}b_{11})$. So the products are multiples of those of $R(y_1,y_2)$.
\end{proof}

\subsection{\texorpdfstring{Proofs of Theorems~\ref{mm:thm:main} and~\ref{mm:thm:family} and of Corollary~\ref{mm:cor:class}}{Proofs of the theorems and the corollary}}\label{mm:sec:proofs}

\emph{Theorem~\ref{mm:thm:main}.} For $(p_A,p_B)=(3,3)$ this is Section~\ref{mm:sec:assembly}, for $(3,2)$ Proposition~\ref{mm:prop:32}, and for $(2,2)$ Proposition~\ref{mm:prop:22}. For $(2,3)$ apply Proposition~\ref{mm:prop:32} to the algorithm of Lemma~\ref{mm:lem:sym}(b). That lemma preserves (a) and (b). For (c), if $P(A,B)=(x^TAHe_1+e_2^TH^{-1}By)(x^TAHe_2+e_1^TH^{-1}By)$, then
\begin{displaymath}
    P(B^T,A^T)=(y^TA\tilde He_1+e_2^T\tilde H^{-1}Bx)(y^TA\tilde He_2+e_1^T\tilde H^{-1}Bx)
\end{displaymath}
with $\tilde H=H^{-T}$, which has the form (c) with $(x,y)$ replaced by $(y,x)$. The pair $(x,y),(x,-y)$ becomes the pair $(y,x),(-y,x)$, which gives the same two products as $(y,x),(y,-x)$.\qed

\emph{Theorem~\ref{mm:thm:family}\textup{(a)}.} The four formulas follow by expanding the products (they are also checked in Section~\ref{mm:sec:comp}). Independence: in a relation $\sum c_iP_i=0$ the monomials $a_{11}b_{12}$, $a_{12}b_{21}$, $a_{21}b_{12}$, $a_{22}b_{21}$ occur only in $P_1,P_2,P_3,P_4$ respectively, so $c_1=c_2=c_3=c_4=0$. Then $a_{11}b_{11}$ and $a_{21}b_{11}$ give $c_6=c_7=0$, and $c_5=0$. The A-parts are multiples of $a_{11}a_{12}$, of $a_{21}a_{22}$, or zero, and the B-parts are multiples of $b_{11}b_{22}$. The pure parts of $P_1,P_3,P_5$ are independent. So $p_A=2$, $p_B=1$ and $p=3$.\qed

\emph{Theorem~\ref{mm:thm:family}\textup{(b)}.} By Proposition~\ref{mm:prop:pB1} we may assume that the products are multiples of those of $R(y_1,y_2)$ with all $y_{jl}\neq0$. Substitute $A\mapsto\operatorname{diag}(1,y_{21}/y_{11})\,A$ and $B\mapsto B\operatorname{diag}(y_{11}^{-1},y_{12}^{-1})$, as in Lemma~\ref{mm:lem:sym}(a) with $Y=I$. This turns the products into nonzero multiples of the products of $R_I$ with $I=y_{11}y_{22}/(y_{12}y_{21})$. Indeed, let $(\xi_1,\xi_2)=(1,y_{21}/y_{11})$ and $(\zeta_1,\zeta_2)=(y_{11}^{-1},y_{12}^{-1})$. A product of $R(y_1,y_2)$ becomes a multiple of the corresponding product of $R(y'_1,y'_2)$ with $y'_{jl}=y_{jl}\zeta_l/\xi_j$, and $y'_1=(1,1)^T$, $y'_2=(1,I)^T$. In particular $p_A=2$ by part (a).\qed

\emph{Theorem~\ref{mm:thm:family}\textup{(c)}.} The substitution $A\mapsto\bigl(\begin{smallmatrix}0&1\\1&0\end{smallmatrix}\bigr)A$ exchanges the rows of $A$ and turns the products of $R_I$ into those of $R((1,I)^T,(1,1)^T)$, which the substitution of part (b) turns into multiples of the products of $R_{1/I}$. Conversely, consider any algorithm with $p=3$ and $p_B=1$. The proof of Proposition~\ref{mm:prop:pB1} shows that the subspace $Y=E^\perp\subseteq\cB$ has type I. Its two lines of rank-one matrices are spanned by matrices $u_1z_1^T,u_2z_2^T$. Choose $\hat z_l\neq0$ with $z_l^T\hat z_l=0$. The algorithm has exactly two double points of type A, with columns $[y_1],[y_2]$, and $[y_j]\notin\{[\hat z_1],[\hat z_2]\}$ by Step~4(ii) of that proof. So
\begin{displaymath}
    \kappa=\frac{\det[\hat z_1\,y_1]\,\det[\hat z_2\,y_2]}{\det[\hat z_1\,y_2]\,\det[\hat z_2\,y_1]}\in\K\setminus\{0\}
\end{displaymath}
is defined, and exchanging $\hat z_1,\hat z_2$ or $y_1,y_2$ replaces $\kappa$ by $\kappa^{-1}$. All these data are determined by the algorithm, and are unchanged by rescaling and reordering products. A substitution as in Lemma~\ref{mm:lem:sym}(a) replaces each column $y$ of a double point by $Zy$ (the W-element $x^TABy$ becomes $(X^{-T}x)^TAB(Zy)$), and replaces $Y$ by $\{YBZ^{-1}:\ B\in Y\}$. So each $z_l$ by $Z^{-T}z_l$ and each $\hat z_l$ by $Z\hat z_l$. As $\det[Zu\ Zv]=\det Z\det[u\ v]$, the set $\{\kappa,\kappa^{-1}\}$ is invariant. For $R_I$ we have $Y=\{B:\ b_{11}=b_{22}=0\}$, $\{[\hat z_1],[\hat z_2]\}=\{[e_2],[e_1]\}$, $y_1=(1,1)^T$, $y_2=(1,I)^T$, and $\{\kappa,\kappa^{-1}\}=\{I,I^{-1}\}$. Hence $R_J$ can be obtained from $R_I$ only if $\{J,J^{-1}\}=\{I,I^{-1}\}$.\qed

\emph{Corollary~\ref{mm:cor:class}.} Let $p=3$. By Proposition~\ref{mm:prop:pA0}, $p_A,p_B\ge1$, and $(p_A,p_B)\neq(1,1)$ because $p\le p_A+p_B$. If $p_A,p_B\ge2$, Theorem~\ref{mm:thm:main} applies. If $p_B=1$, Theorem~\ref{mm:thm:family}(b) gives $p_A=2$. If $p_A=1$, the same applies to the algorithm of Lemma~\ref{mm:lem:sym}(b). So exactly one of (i), (ii), (iii) holds, and $(3,1)$, $(1,3)$, $(3,0)$, $(0,3)$ do not occur. Section~\ref{mm:sec:comp} gives algorithms with $(p_A,p_B)=(3,3)$, $(3,2)$ and $(2,2)$; Lemma~\ref{mm:lem:sym}(b) turns the second into one with $(2,3)$; $R_I$ has $(2,1)$, and Lemma~\ref{mm:lem:sym}(b) turns it into one with $(1,2)$.\qed

\subsection{\texorpdfstring{Proof of Proposition~\ref{mm:prop:pA0}}{Proof of the proposition on pA=0}}\label{mm:sec:pA0}

If $p_A=0$, then $\alpha_i\gamma_i=0$ for all $i$, and exchanging factors we may write every product as $P_i=\beta_i(\gamma_i+\delta_i)$ with $\beta_i,\delta_i\in\cB^*$ and $\gamma_i\in\cA^*$. Comparing parts in $AB=\sum\Lambda_iP_i$,
\begin{displaymath}
    AB=\sum_i\Lambda_i\gamma_i(A)\beta_i(B),\qquad\sum_i\Lambda_i\beta_i\delta_i=0.
\end{displaymath}
The bilinear rank of $2\times2$ matrix multiplication is $7$ \cite{hopcroft1971minimizing,winograd1971multiplication}, so the seven terms $\Lambda_i\otimes\gamma_i\otimes\beta_i$ are nonzero and form an optimal bilinear algorithm. By de Groote's theorem \cite{degroote1978optimal}, every optimal bilinear algorithm for $2\times2$ matrix multiplication is obtained from Strassen's algorithm \cite{strassen1969gaussian} by a symmetry of the matrix multiplication tensor. The symmetries are generated by the substitutions of Lemma~\ref{mm:lem:sym}(a) and the permutations of the three tensor factors, namely cyclic shifts and transposition \cite{degroote1978varieties}. The existence of $\delta_1,\dots,\delta_7$, not all zero, with $\sum_i\Lambda_i\beta_i\delta_i=0$ is invariant under the substitutions of Lemma~\ref{mm:lem:sym}(a) and under rescaling and reordering the seven terms. There are six ways of assigning the three factors of Strassen's trilinear decomposition of $\tr(XYZ)$ to the roles $(A,B,\text{output})$. For each of them, an exact computation (Section~\ref{mm:sec:comp}) shows the following: the linear system $\sum_i\Lambda_i\beta_i\delta_i=0$ in the $28$ unknown coefficients of $\delta_1,\dots,\delta_7$ has only the zero solution. Checking all six assignments makes the argument independent of whether the permutations are needed in de Groote's theorem. Hence all $\delta_i=0$, every product $\beta_i\gamma_i$ is bilinear, and $p=0$. The case $p_B=0$ follows by Lemma~\ref{mm:lem:sym}(b).\qed

\subsection{\texorpdfstring{Algorithms with $p\le2$}{Algorithms with p <= 2}}\label{mm:sec:p12}

In this section we prove Theorem~\ref{mm:thm:p12}. Let $\Pi$ and $K$ be as in Lemma~\ref{mm:lem:relations}, and let $\Pi_{\mathrm{bil}}$ be the space of bilinear forms in $\Pi$. A form $\sum_ic_iP_i$ is bilinear exactly when $c\in K$, and the products are independent, so $c\mapsto\sum_ic_iP_i$ maps $K$ isomorphically onto $\Pi_{\mathrm{bil}}$, which has dimension $7-p$ and contains $W$ by Lemma~\ref{mm:lem:relations}(a).

The first part of the proof of Lemma~\ref{mm:lem:types} does not use $p=3$: a product with $\pi_i=0$ is a bilinear form of rank one, $P_i=S_i$. We call it \emph{bilinear}. A product with $\pi_i\neq0$ has type AB, A or B as in Lemma~\ref{mm:lem:types}, its \emph{point} is $[\pi_i]\in\Pj(V)$, and two products with the same point have the same type. We write $k_0,k_A,k_B,k_{AB}$ for the numbers of products of the four kinds. Unless the product has type AB, $S_i$ has rank at most one. For types A and B, $S_i=\gamma_i\otimes\beta_i$.

\begin{lemma}[Optimal bilinear algorithms]\label{mm:lem:strassen}
Suppose that $AB=\sum_{r=1}^7C_r\,\lambda_r(A)\mu_r(B)$ with $C_r\in\K^{2\times2}$, $\lambda_r\in\cA^*$ and $\mu_r\in\cB^*$. Then all $C_r,\lambda_r,\mu_r$ are nonzero, and $C_1,\dots,C_7$ are pairwise non-proportional, as are $\lambda_1,\dots,\lambda_7$ and $\mu_1,\dots,\mu_7$. In particular, if $AB=\sum_{i=1}^7\Lambda_iS_i(A,B)$ for an algorithm in which every $S_i$ has rank at most one, then the A-factors of the $S_i$ are pairwise non-proportional.
\end{lemma}

\begin{proof}
The bilinear rank of $2\times2$ matrix multiplication is $7$ \cite{hopcroft1971minimizing,winograd1971multiplication}, so no term vanishes. As in the proof of Proposition~\ref{mm:prop:pA0}, we use de Groote's theorem \cite{degroote1978optimal}. The decomposition is obtained from Strassen's by a substitution as in Lemma~\ref{mm:lem:sym}(a), possibly after permuting the three tensor factors, and rescaling and reordering. Hence each of the three lists $(C_r)$, $(\lambda_r)$, $(\mu_r)$ is the image of one of the three factor lists of Strassen's decomposition under a linear isomorphism, up to rescaling and reordering. Each of Strassen's three lists consists of seven pairwise non-proportional elements (by inspection; Section~\ref{mm:sec:comp}). For the last claim write $S_i=\lambda_i\otimes\mu_i$.
\end{proof}

\begin{lemma}\label{mm:lem:span}
Let $U\subseteq\K^{2\times2}$ be a subspace that is not spanned by its matrices of rank one. Then $\dim U\le2$, and $U$ has a basis consisting of $\dim U-1$ matrices of rank one and one invertible matrix.
\end{lemma}

\begin{proof}
If $\dim U=4$ the claim is clear. If $\dim U=3$, then $U=\{X:\ \tr(NX)=0\}$ for some $N\neq0$, and $uv^T\in U$ if and only if $v^TNu=0$. If $N$ is invertible, choose $v_k\neq0$ with $v_k^TNu_k=0$ for $u_1=e_1$, $u_2=e_2$, $u_3=e_1+e_2$. The three matrices $u_kv_k^T\in U$ are independent, since a relation would force $[v_1]=[v_2]=[v_3]$ and then $v_1^TN=0$. If $N=ab^T$, then $U$ contains $u_0v^T$ for all $v$ and $uw_0^T$ for all $u$, where $b^Tu_0=0$ and $w_0^Ta=0$, and these span a space of dimension $3$. If $\dim U=2$ and $\det$ vanishes on $U$, all elements of $U$ have rank at most one by Fact~\ref{mm:fact:M2}(a). Otherwise $\det$ restricted to $U$ is a nonzero binary quadratic form, which vanishes on one or two lines as $\K$ is algebraically closed. Two such lines would be spanned by two independent matrices of rank one, which span $U$. So it vanishes on exactly one line $\K r$, and every element of $U\setminus\K r$ is invertible. If $\dim U=1$, $U$ is spanned by an invertible matrix.
\end{proof}

\begin{lemma}[Independent points]\label{mm:lem:indep}
Suppose that the distinct points of the non-bilinear products are linearly independent in $V$. Then no product has type AB.
\end{lemma}

\begin{proof}
Let $[v_1],\dots,[v_m]$ be the distinct points, $I_g$ the set of products at $[v_g]$, and $\pi_i=t_iv_g$ with $t_i\neq0$ for $i\in I_g$. Since $v_1,\dots,v_m$ are independent, $c\in K$ if and only if $\sum_{i\in I_g}c_it_i=0$ for every $g$, so Lemma~\ref{mm:lem:relations}(a) gives $\sum_{i\in I_g}t_i\Lambda_i=0$ for every $g$. Suppose that the products in $I_g$ have type AB. Let $G_g\subseteq\cA^*$ and $E_g\subseteq\cB^*$ be the spans of the factors of the two components of $v_g$. By Fact~\ref{mm:fact:binary}(a) all factors $\alpha_i,\gamma_i$ ($i\in I_g$) lie in $G_g$ and all $\beta_i,\delta_i$ in $E_g$, so $S_i\in G_g\otimes E_g$. Fix $i_g\in I_g$ and put $D_i=t_i^{-1}S_i-t_{i_g}^{-1}S_{i_g}$. Then
\[
\sum_{i\in I_g}\Lambda_iS_i=\sum_{i\in I_g}t_i\Lambda_i\,t_i^{-1}S_i=\sum_{i\in I_g\setminus\{i_g\}}t_i\Lambda_i\,D_i ,
\]
and the $D_i$ span a subspace $U_g\subseteq G_g\otimes E_g$ of dimension at most $|I_g|-1$. Expanding the $D_i$ in a basis $u_1,\dots,u_d$ of $U_g$ writes this sum as $\sum_kC_ku_k$ with $C_k\in\K^{2\times2}$. If $\dim G_g\le1$ or $\dim E_g\le1$, every $u_k$ has rank at most one. Otherwise we identify $G_g\otimes E_g$ with $\K^{2\times2}$ by bases, which preserves ranks, and choose a basis of $U_g$ of matrices of rank one if there is one. If not, Lemma~\ref{mm:lem:span} gives a basis of $d-1\le|I_g|-2$ matrices of rank one and one $u_d$ of rank two, $u_d=r'+r''$ with $r',r''$ of rank one. In the first two cases, the group $I_g$ contributes at most $|I_g|-1$ terms $C\,\lambda(A)\mu(B)$. In the last case it contributes $d+1\le|I_g|$ terms, and two of them, $C_dr'$ and $C_dr''$, have the same coefficient matrix.

Do this for every group of type AB, and keep the terms $\Lambda_iS_i$ of the other products, whose $S_i$ have rank at most one. The result is an expression $AB=\sum_rC_r\lambda_r(A)\mu_r(B)$ with at most $k_0+\sum_g|I_g|=7$ terms, and with at most $6$ unless every group of type AB is in the last case. As the rank is $7$, every group of type AB is in the last case and there are exactly $7$ terms. If there is such a group, two terms have the same coefficient matrix, which contradicts Lemma~\ref{mm:lem:strassen}.
\end{proof}

\begin{proof}[Proof of Theorem~\ref{mm:thm:p12}(a)]
Let $p=1$. By Proposition~\ref{mm:prop:pA0}, $p_A,p_B\ge1$, so $V=\K v$ where both components of $v$ are nonzero. Every non-bilinear product has point $[v]$ and type AB, and a single point is linearly independent. By Lemma~\ref{mm:lem:indep} all products are bilinear, so $p=0$, a contradiction.
\end{proof}

From now on $p=2$, so $\dim\Pi_{\mathrm{bil}}=5$. We write $\cA^*=A_1^*\oplus A_2^*$ and $\cB^*=B_1^*\oplus B_2^*$, with $A_q^*=\spn\{a_{1q},a_{2q}\}$ and $B_q^*=\spn\{b_{q1},b_{q2}\}$. We identify a bilinear form in $A_1^*\otimes B_1^*\oplus A_2^*\otimes B_2^*$ with a pair $(Y_1,Y_2)$ of $2\times2$ matrices, where $(Y_q)_{ps}$ is the coefficient of $a_{pq}b_{qs}$. Its rank is $\rk Y_1+\rk Y_2$. Since $w_Z(A,B)=\sum_q\sum_{p,s}Z_{sp}a_{pq}b_{qs}$, we have $w_Z=(Z^T,Z^T)$. Let $E_{12}$ be the matrix unit.

\begin{lemma}\label{mm:lem:Wtwo}
No $2$-dimensional subspace of $W$ lies in the span of three bilinear forms of rank one.
\end{lemma}

\begin{proof}
Let $W'=\{w_Z:\ Z\in U\}$ be such a subspace, in the span of $\lambda_r\otimes\mu_r$, $r=1,2,3$. If $U$ contains an invertible $Z$, then $\rk w_Z=4>3$. Otherwise, by Fact~\ref{mm:fact:M2}(a), $U=\{yx^T:\ x\in\K^2\}$ or $U=\{y x_0^T:\ y\in\K^2\}$. In the first case the two entries of $ABy$ lie in $W'$, so $ABy=\sum_rv_r\lambda_r(A)\mu_r(B)$ with $v_r\in\K^2$. Taking $B$ with $By=e_1$ and with $By=e_2$ shows that $A$ is determined by $\lambda_1(A),\lambda_2(A),\lambda_3(A)$, which is absurd. The second case is the same with $x_0^TAB$ and the forms $\mu_r$.
\end{proof}

\begin{lemma}\label{mm:lem:p2bil}
Assume $p=2$.
\begin{enumerate}
\item[\textup{(a)}] $k_0\le2$.
\item[\textup{(b)}] If $k_0=2$, then after a substitution as in Lemma~\ref{mm:lem:sym}\textup{(a)} the bilinear products are nonzero multiples of $a_{11}b_{12}$ and $a_{12}b_{22}$. Then $\Pi_{\mathrm{bil}}=W\oplus\K a_{11}b_{12}=\{(Y_1,Y_2):\ Y_1-Y_2\in\K E_{12}\}$, its elements of rank one are the multiples of $a_{11}b_{12}$ and of $a_{12}b_{22}$, and each of its elements of rank two has the form $l_1\otimes m_1+l_2\otimes m_2$ with nonzero $l_q\in A_q^*$ and $m_q\in B_q^*$.
\item[\textup{(c)}] There is no $A_0\neq0$ in $\cA$ at which the A-factors $\alpha_i,\gamma_i$ of all non-bilinear products vanish, and no $B_0\neq0$ in $\cB$ at which their B-factors $\beta_i,\delta_i$ vanish.
\end{enumerate}
\end{lemma}

\begin{proof}
(a) Three bilinear products would span a $3$-dimensional subspace of the $5$-dimensional $\Pi_{\mathrm{bil}}$, meeting $W$ in dimension at least $2$, contrary to Lemma~\ref{mm:lem:Wtwo}.

(b) Let $P_5=\lambda_5\otimes\mu_5$ and $P_6=\lambda_6\otimes\mu_6$ be the bilinear products. Their span meets $W$ in some $w\ne0$, which has rank two by Fact~\ref{mm:fact:W}, so $w=x^TABy=c_5P_5+c_6P_6$ with $c_5c_6\neq0$. By Fact~\ref{mm:fact:W}, $\lambda_5=x^TAh_1$ and $\lambda_6=x^TAh_2$ for a basis $h_1,h_2$ of $\K^2$. With the dual basis $k_1,k_2$, comparing $x^TABy=\sum_q(x^TAh_q)(k_q^TBy)$ with $c_5P_5+c_6P_6$ gives $c_5\mu_5=k_1^TBy$ and $c_6\mu_6=k_2^TBy$. Choose $X$ with first row $x^T$, $Y=[h_1\ h_2]^{-1}$ and $Z$ with $Zy=e_2$. The substitution of Lemma~\ref{mm:lem:sym}(a) turns $x^TAh_q$ into $a_{1q}$ and $k_q^TBy$ into $b_{q2}$. Now $a_{11}b_{12}+a_{12}b_{22}=(AB)_{12}\in W$, so $\Pi_{\mathrm{bil}}=W+\K P_5+\K P_6=W\oplus\K a_{11}b_{12}$. As $w_Z=(Z^T,Z^T)$ and $a_{11}b_{12}=(E_{12},0)$, this is the space of pairs with $Y_1-Y_2\in\K E_{12}$, which has dimension $5$. If one of $Y_1,Y_2$ is zero, the other lies in $\K E_{12}$, which gives the multiples of $a_{11}b_{12}=(E_{12},0)$ and $a_{12}b_{22}=(0,E_{12})$. An element of rank two with $Y_1,Y_2\neq0$ has $\rk Y_1=\rk Y_2=1$, that is, the form claimed.

(c) Suppose all A-factors of the non-bilinear products vanish at $A_0\neq0$. On $\K A_0\times\cB$, \eqref{mm:eq:bil} gives $A_0B=\sum\Lambda_iS_i(A_0,B)$, summed over the bilinear products. The entries of $A_0B$ span a space of linear forms in $B$ of dimension $2\rk A_0$, and the right side involves at most $k_0\le2$ forms. So $k_0=2$, $A_0=uv^T$ has rank one, and the B-factors of the two bilinear products span $\{B\mapsto v^TBz:\ z\in\K^2\}$. The claim is invariant under substitutions, so by (b) these B-factors are $b_{12},b_{22}$. But $v^TBe_1$ is not in $\spn\{b_{12},b_{22}\}$ for $v\neq0$. The second claim is the same: the A-factors $a_{11},a_{12}$ would span $\Col_{\cA}(u)$ for $B_0=uv^T$, contrary to Fact~\ref{mm:fact:Wfactor}(c), as $\spn\{a_{11},a_{12}\}=\Row(e_1)$.
\end{proof}

\begin{lemma}\label{mm:lem:p2pairs}
If $p=2$, then $p_A=p_B=1$.
\end{lemma}

\begin{proof}
By Proposition~\ref{mm:prop:pA0} and Remark~\ref{mm:rem:context}, $1\le p_A,p_B\le2$. Suppose $p_A=2$. Then the projection $V\to V_A$ is an isomorphism, so every non-bilinear product has a nonzero A-part, and distinct points have non-proportional A-parts. Suppose there were at least three points. Apply Lemma~\ref{mm:lem:pencil} to one A-part at each point. It would put all A-factors of the non-bilinear products in a subspace of dimension at most $3$ of $\cA^*$. So they would all vanish at some $A_0\neq0$, contrary to Lemma~\ref{mm:lem:p2bil}(c). So there are exactly two points, which span $V$ and are therefore independent, and by Lemma~\ref{mm:lem:indep} every non-bilinear product has type A. Then $p_B=0$, a contradiction. So $p_A=1$, and $p_B=1$ by Lemma~\ref{mm:lem:sym}(b).
\end{proof}

\begin{proposition}\label{mm:prop:p11}
If $p=2$, then after a substitution as in Lemma~\ref{mm:lem:sym}\textup{(a)} the products are nonzero multiples of those of $\mathcal M$.
\end{proposition}

\begin{proof}
By Lemma~\ref{mm:lem:p2pairs}, $V_A=\K q_a$ and $V_B=\K q_b$, and $V=V_A\oplus V_B$. Let $G\subseteq\cA^*$ and $E\subseteq\cB^*$ be the spans of the factors of $q_a$ and of $q_b$, and $G^\perp\subseteq\cA$, $E^\perp\subseteq\cB$ their annihilators. The A-factors of the products of types A and AB lie in $G$, and the B-factors of the products of types B and AB lie in $E$ (Fact~\ref{mm:fact:binary}(a)).

\emph{Step 1: counting.} On $G^\perp\times\cB$ only the bilinear products and those of type B have nonzero bilinear parts, each of rank at most one. As $\dim G^\perp\ge2$, \eqref{mm:eq:bil} and Fact~\ref{mm:fact:rank}(a) give $k_0+k_B\ge4$, and likewise $k_0+k_A\ge4$. By Lemma~\ref{mm:lem:p2bil}(a), $k_A,k_B\ge2$. Suppose $\dim G=1$. Then every $S_i$ has rank at most one, since for type AB, $S_i\in G\otimes E$. The A-factors of the $k_A\ge2$ products of type A are proportional, contrary to Lemma~\ref{mm:lem:strassen}. So $\dim G=2$ and likewise $\dim E=2$. If $k_{AB}=0$, again every $S_i$ has rank at most one. The A-factors of the products of type A are multiples of the two factors of $q_a$, so $k_A\le2$ by Lemma~\ref{mm:lem:strassen}, likewise $k_B\le2$, and $7=k_0+k_A+k_B\le6$. So $k_{AB}\ge1$, and $7\ge k_0+2(4-k_0)+1$ gives $k_0=2$, $k_A=k_B=2$ and $k_{AB}=1$.

\emph{Step 2: a normal form.} By Lemma~\ref{mm:lem:p2bil}(b) we may assume that the bilinear products are $P_5=a_{11}b_{12}$ and $P_6=a_{12}b_{22}$, rescaled. Let $P_7$ be the product of type AB. After rescaling it and exchanging its factors, $P_7=(g+\varepsilon')(h+\varepsilon)$ with $g,h\in G$ and $\varepsilon,\varepsilon'\in E$, so that $G=\spn\{g,h\}$, $E=\spn\{\varepsilon,\varepsilon'\}$, $\pi_7=(gh,\varepsilon'\varepsilon)$ and $S_7=g\otimes\varepsilon+h\otimes\varepsilon'$. A product of type A is $(\alpha+\beta)\gamma$ with $\alpha\gamma$ a multiple of $gh$, so after rescaling it is $g(h+x)$ or $(g+x')h$ with $x,x'\in\cB^*$. Two products $g(h+x)$ and $g(h+\tilde x)$ would differ by the bilinear form $g\otimes(x-\tilde x)$ of rank at most one in $\Pi_{\mathrm{bil}}$, which by Lemma~\ref{mm:lem:p2bil}(b) and independence is impossible. So the two products of type A are $P_1=g(h+x)$ and $P_2=(g+x')h$. In the same way the products of type B are $P_3=\varepsilon'(\varepsilon+y)$ and $P_4=(\varepsilon'+y')\varepsilon$ with $y,y'\in\cA^*$. The relations $\pi_1=\pi_2$, $\pi_3=\pi_4$ and $\pi_1+\pi_3=\pi_7=\pi_2+\pi_4$ give the elements
\begin{align*}
\Theta_1&=P_1-P_2=g\otimes x-h\otimes x',\\
\Theta_2&=P_3-P_4=y\otimes\varepsilon'-y'\otimes\varepsilon,\\
\Theta_3&=P_1+P_3-P_7=g\otimes(x-\varepsilon)+(y-h)\otimes\varepsilon',\\
\Theta_4&=P_2+P_4-P_7=h\otimes(x'-\varepsilon')+(y'-g)\otimes\varepsilon
\end{align*}
of $\Pi_{\mathrm{bil}}$. Each $\Theta_k$ is nonzero and not a multiple of $P_5$ or $P_6$, by independence, and has rank at most two. So by Lemma~\ref{mm:lem:p2bil}(b) it has rank two and both terms of its expression above have rank one, with independent A-factors and independent B-factors. Call a $2$-dimensional subspace of $\cA^*$ a \emph{transversal} if it meets both $A_1^*$ and $A_2^*$, and define transversals in $\cB^*$ in the same way. By Lemma~\ref{mm:lem:p2bil}(b) the A-factor spaces $G$, $\spn\{g,y-h\}$, $\spn\{h,y'-g\}$ of $\Theta_1,\Theta_3,\Theta_4$ and the B-factor spaces $E$, $\spn\{x-\varepsilon,\varepsilon'\}$, $\spn\{x'-\varepsilon',\varepsilon\}$ of $\Theta_2,\Theta_3,\Theta_4$ are transversals.

\emph{Step 3: the blocks.} A form $l=l_1+l_2$ with nonzero components $l_q\in A_q^*$ lies on exactly one transversal, $\spn\{l_1,l_2\}$. If $g\notin A_1^*\cup A_2^*$, then $\spn\{g,y-h\}=G$, so $y\in G$. Then $S_1,S_2,S_3,S_7$ vanish on $G^\perp\times\cB$, and \eqref{mm:eq:bil} on $G^\perp\times\cB$ involves only $S_4,S_5,S_6$, contrary to Fact~\ref{mm:fact:rank}(a). Likewise $h\notin A_1^*\cup A_2^*$ would give $y'\in G$ and only $S_3,S_5,S_6$ on $G^\perp\times\cB$. As $G$ is a transversal, $g$ and $h$ lie in different spaces $A_q^*$. The substitution $A\mapsto AJ_0$, $B\mapsto J_0B$ with $J_0=\left(\begin{smallmatrix}0&1\\1&0\end{smallmatrix}\right)$ exchanges $A_1^*$ with $A_2^*$, $B_1^*$ with $B_2^*$, and $a_{11}b_{12}$ with $a_{12}b_{22}$, so we may assume $g\in A_1^*$ and $h\in A_2^*$. In the same way, we use the transversals $E$, $\spn\{x-\varepsilon,\varepsilon'\}$ and $\spn\{x'-\varepsilon',\varepsilon\}$. If $\varepsilon'\notin B_1^*\cup B_2^*$, we get $x\in E$, and only $S_2,S_5,S_6$ remain on $\cA\times E^\perp$. If $\varepsilon\notin B_1^*\cup B_2^*$, we get $x'\in E$. So $\varepsilon$ and $\varepsilon'$ lie in different spaces $B_q^*$. Every element of $\Pi_{\mathrm{bil}}$ lies in $A_1^*\otimes B_1^*\oplus A_2^*\otimes B_2^*$. For $\Theta_1$ this gives $x\in B_1^*$ and $x'\in B_2^*$. If $\varepsilon\in B_2^*$, the component of $\Theta_3$ in $A_1^*\otimes B_2^*$ would be $-g\otimes\varepsilon\neq0$. So $\varepsilon\in B_1^*$ and $\varepsilon'\in B_2^*$. Then $\Theta_3$ gives $y-h\in A_2^*$, so $y\in A_2^*$, and $\Theta_4$ gives $y'\in A_1^*$.

\emph{Step 4: the equations.} We identify $A_q^*$ and $B_q^*$ with $\K^2$ by $u\mapsto u_1a_{1q}+u_2a_{2q}$ and $v\mapsto v_1b_{q1}+v_2b_{q2}$, and use the same letters for the coordinate vectors. Then $u\otimes v\in A_q^*\otimes B_q^*$ is $Y_q=uv^T$. A matrix lies in $\K E_{12}$ exactly when its first column and its second row vanish. By Lemma~\ref{mm:lem:p2bil}(b), $\Theta_1,\Theta_2,\Theta_3\in\Pi_{\mathrm{bil}}$ means $gx^T+hx'^T$, $y'\varepsilon^T+y\varepsilon'^T$ and $g(x-\varepsilon)^T+(h-y)\varepsilon'^T$ lie in $\K E_{12}$, that is,
\begin{alignat*}{3}
&\text{(E1)}&\qquad&x_1g+x'_1h=0,&\qquad&g_2x+h_2x'=0,\\
&\text{(E2)}&&\varepsilon_1y'+\varepsilon'_1y=0,&&y'_2\varepsilon+y_2\varepsilon'=0,\\
&\text{(E3)}&&(x_1-\varepsilon_1)g+\varepsilon'_1(h-y)=0,&&\\
&&&g_2(x-\varepsilon)+(h_2-y_2)\varepsilon'=0.&&
\end{alignat*}
Moreover $\Pi_{\mathrm{bil}}$ is spanned by $P_5,P_6,\Theta_1,\Theta_2,\Theta_3$, and $(Y_1,Y_2)\mapsto Y_2$ maps it onto $\K^{2\times2}$ (already $W$ does). The images are $0$, $E_{12}$, $-hx'^T$, $y\varepsilon'^T$, $(y-h)\varepsilon'^T$, so
\[
\text{(N)}\qquad E_{12},\ hx'^T,\ y\varepsilon'^T,\ h\varepsilon'^T\ \text{ are linearly independent.}
\]
In particular $h,y$ are independent, $x',\varepsilon'$ are independent, and $x\neq0\neq y'$ because $\Theta_1,\Theta_2$ have rank two.

\emph{Step 5: solving.} If $g_2=0$, then (E1) gives $h_2=0$ and (E3) gives $y_2=0$, so $h,y\in\K e_1$, contrary to (N). So $g_2\neq0$. Put $r=h_2/g_2$. By (E1), $x=-rx'$, so $r\neq0$. The second equation of (E3) becomes $g_2\varepsilon=-h_2x'+(h_2-y_2)\varepsilon'$. Inserting this into the second equation of (E2) and comparing the coefficients of the independent vectors $x',\varepsilon'$ gives $y'_2=0$ and then $y_2=0$, so $\varepsilon=-r(x'-\varepsilon')$. Then $x_1-\varepsilon_1=-r\varepsilon'_1$, and the first equation of (E3) becomes $\varepsilon'_1(h-y-rg)=0$. If $\varepsilon'_1=0$, the first equation of (E2) gives $\varepsilon_1=0$, hence $x'_1=0$, and $x',\varepsilon'\in\K e_2$, contrary to (N). So $h=y+rg$, the first equation of (E1) becomes $x'_1y=0$, so $x'_1=0$, and the first equation of (E2) becomes $r\varepsilon'_1y'+\varepsilon'_1y=0$, so $y=-ry'$. With $\eta=y'_1$ and $\xi=x'_2$, the solutions are
\begin{gather*}
g=(g_1,g_2),\quad h=(r(g_1-\eta),rg_2),\quad y'=(\eta,0),\quad y=(-r\eta,0),\\
x'=(0,\xi),\quad x=(0,-r\xi),\quad \varepsilon'=(\varepsilon'_1,\varepsilon'_2),\quad \varepsilon=(r\varepsilon'_1,r(\varepsilon'_2-\xi)),
\end{gather*}
with $g_2,r,\eta,\xi,\varepsilon'_1$ nonzero.

\emph{Step 6: one orbit.} Denote the seven products above by $\mathcal F(g_1,g_2,r,\eta,\xi,\varepsilon'_1,\varepsilon'_2)$. A direct substitution (Section~\ref{mm:sec:comp}) shows the following. The substitution $(A,B)\mapsto\bigl(\left(\begin{smallmatrix}1&0\\u&1\end{smallmatrix}\right)A,\,B\bigr)$ replaces $g_1$ by $g_1+ug_2$ and fixes the other parameters; $(A,B)\mapsto\bigl(A,\,B\left(\begin{smallmatrix}1&0\\v&1\end{smallmatrix}\right)\bigr)$ replaces $\varepsilon'_2$ by $\varepsilon'_2+v\varepsilon'_1$; and $(A,B)\mapsto(D_1AD_2,D_2^{-1}BD_3)$ with $D_k=\mathrm{diag}(d_{k1},d_{k2})$ multiplies $(g_1,g_2,r,\eta,\xi,\varepsilon'_1,\varepsilon'_2)$ by
\[
\Bigl(d_{11}d_{21},\ d_{12}d_{21},\ \frac{d_{22}}{d_{21}},\ d_{11}d_{21},\ \frac{d_{32}}{d_{22}},\ \frac{d_{31}}{d_{22}},\ \frac{d_{32}}{d_{22}}\Bigr),
\]
in each case up to rescaling of $P_5,P_6$. Choosing $d_{21}=1$, $d_{22}=-1/r$, $d_{12}=1/g_2$, $d_{11}=1/\eta$, $d_{32}=d_{22}/\xi$ and $d_{31}=-d_{22}/\varepsilon'_1$ makes $(g_2,r,\eta,\xi,\varepsilon'_1)=(1,-1,1,1,-1)$, and then the two unipotent substitutions make $g_1=0$ and $\varepsilon'_2=1$. Finally $\mathcal F(0,1,-1,1,1,-1,1)$ is $\mathcal M$: $g=a_{21}$, $h=a_{12}-a_{22}$, $x=b_{12}$, $x'=b_{22}$, $\varepsilon=b_{11}$, $\varepsilon'=b_{22}-b_{21}$, $y=a_{12}$ and $y'=a_{11}$.
\end{proof}

\begin{proof}[Proof of Theorem~\ref{mm:thm:p12}(b) and~(c)]
(b) The four formulas follow by expanding the products (Section~\ref{mm:sec:comp}). Independence: in a relation $\sum_ic_iP_i=0$ the monomials $a_{21}b_{12}$, $a_{11}b_{11}$, $a_{21}b_{11}$ occur only in $P_1$, $P_4$, $P_7$ respectively, so $c_1=c_4=c_7=0$. Then $a_{22}b_{22}$ and $a_{12}b_{21}$ give $c_2=c_3=0$, and $c_5=c_6=0$. The A-parts are multiples of $a_{21}(a_{12}-a_{22})$ or zero, the B-parts are multiples of $(b_{22}-b_{21})b_{11}$ or zero, and $\pi_1=(a_{21}(a_{12}-a_{22}),0)$ and $\pi_3=(0,(b_{22}-b_{21})b_{11})$ are independent. (c) is Lemma~\ref{mm:lem:p2pairs} and Proposition~\ref{mm:prop:p11}.
\end{proof}

\subsection{No nested products}\label{mm:sec:nest}

In this section $\cir$ is a circuit as in Theorem~\ref{mm:thm:nest}, with product gates $q_i=u_iw_i$ in a topological order, and we use the notation of Lemma~\ref{lem:graded}. Let $P_i=\ell_im_i$ and $\Lambda'_i$ be its quadratic part (Proposition~\ref{mm:prop:quadpart}), $\Pi=\spn\{P_1,\dots,P_7\}$, and
\begin{displaymath}
    U_i=u_i^{[2]},\qquad W_i=w_i^{[2]}.
\end{displaymath}

\begin{lemma}[Degree three]\label{mm:lem:deg3}
$U_i,W_i\in\spn\{P_j:\ j<i\}$, and
\begin{equation}\label{mm:eq:deg3}
    \sum_{i=1}^7\Lambda'_i\bigl(\ell_iW_i+m_iU_i\bigr)=0.
\end{equation}
If $U_i=W_i=0$ for all $i$, then every $u_i$ and every $w_i$ has degree at most $1$, and the outputs are $\Lambda'(q-q^{[\le1]})$.
\end{lemma}

\begin{proof}
By Lemma~\ref{lem:graded}, $U_i,W_i\in\spn\{N_j^{(2)}:\ j<i\}$ with $N_j^{(2)}=P_j$, and $0=(AB)^{[3]}=\Lambda'N^{(3)}$ with $N_i^{(3)}=\ell_iW_i+m_iU_i$. Suppose $U_i=W_i=0$ for all $i$, and that $u_i^{[s]}=w_i^{[s]}=0$ for all $i$ and $2\le s<k$, for some $k\ge3$. Every term of $N_i^{(k)}$ contains a factor $u_i^{[s]}$ or $w_i^{[s]}$ with $2\le s\le k-1$, so $N^{(k)}=0$, hence $q^{[k]}=SN^{(k)}=0$ and $u_i^{[k]},w_i^{[k]}\in\spn\{N_j^{(k)}\}=0$. By induction all $u_i,w_i$ have degree at most $1$. Then $q^{[2]}=(P_i)_i$ is linearly independent and $q^{[k]}=0$ for $k\ge3$, so the scalars $\mu_{ij},\mu'_{ij}$ of Lemma~\ref{lem:graded} vanish, $T=0$ and $\Lambda'=\Lambda$. Since $AB$ is homogeneous of degree $2$, $AB=\Lambda q^{[2]}=\Lambda(q-q^{[\le1]})$.
\end{proof}

So Theorem~\ref{mm:thm:nest} follows once \eqref{mm:eq:deg3} forces $U=W=0$. We prove a stronger statement, which does not use the order of the gates: for the algorithms in question, the linear map
\begin{equation}\label{mm:eq:Nmap}
    \mathcal N:\ (U_i,W_i)_{i=1}^7\ \longmapsto\ \sum_i\Lambda_i\bigl(\ell_iW_i+m_iU_i\bigr),
\end{equation}
where $U_i,W_i\in\Pi_{\hat\imath}:=\spn\{P_j:\ j\neq i\}$, is injective. Rescaling a product, exchanging its factors (which exchanges $U_i$ and $W_i$), and reordering the products do not affect injectivity. Apply a substitution as in Lemma~\ref{mm:lem:sym} to $\cir$. This gives a circuit of the same shape, whose quadratic part is the transformed algorithm. It maps the forms $u_i^{[k]},w_i^{[k]}$ of one circuit bijectively to those of the other. So we may replace the quadratic part by any algorithm related to it by these operations.

\begin{lemma}\label{mm:lem:Psi}
Let $(P_i,\Lambda_i)$ be a seven-product algorithm with $p=3$, and let $U_i,W_i$ be quadratic forms with $\sum_i\Lambda_i(\ell_iW_i+m_iU_i)=0$. Then there is a linear map $\Psi:V\to\Sy^3(\cA^*\oplus\cB^*)$ with $\ell_iW_i+m_iU_i=\Psi(\pi_i)$ for all $i$.
\end{lemma}

\begin{proof}
Write $C_i=\ell_iW_i+m_iU_i$ and expand $C_i=\sum_\nu C_{i\nu}\mathbf m_\nu$ in a basis $\mathbf m_\nu$ of the cubic forms. For every $\nu$, $\sum_iC_{i\nu}\Lambda_i=0$, so by Lemma~\ref{mm:lem:relations}(c) there is $\varphi_\nu\in V^*$ with $C_{i\nu}=\varphi_\nu(\pi_i)$. Put $\Psi(v)=\sum_\nu\varphi_\nu(v)\mathbf m_\nu$.
\end{proof}

\begin{proposition}\label{mm:prop:nest3}
For every algorithm as in Theorem~\ref{mm:thm:main}, the map $\mathcal N$ of \eqref{mm:eq:Nmap} is injective. In fact $\sum_i\Lambda_i(\ell_iW_i+m_iU_i)=0$ with $U_i,W_i\in\Pi$ already forces $U=W=0$.
\end{proposition}

\begin{proof}
By Theorem~\ref{mm:thm:main}(c), after a substitution we may assume $H=I$. Put $n_1=(Ae_1;\ B^Te_2)\in\K^4$ and $n_2=(Ae_2;\ B^Te_1)\in\K^4$. The map $(A,B)\mapsto(n_1,n_2)$ is a linear bijection. For $w=(x;y)\in\K^4$,
\begin{displaymath}
    (x^TAe_1+e_2^TBy)(x^TAe_2+e_1^TBy)=(w\cdot n_1)(w\cdot n_2),
\end{displaymath}
so every product is $P_i=\kappa_i(w_i\cdot n_1)(w_i\cdot n_2)$ with $\kappa_i\neq0$, and after exchanging the factors of $P_i$ (and $U_i$ with $W_i$) and rescaling, $\ell_i=w_i\cdot n_1$ and $m_i=\kappa_i\,w_i\cdot n_2$. Every element of $\Pi$ is bilinear in $(n_1,n_2)$ and symmetric under exchanging $n_1$ and $n_2$. A rank-one bilinear form $f(n_1)g(n_2)$ with $f,g\neq0$ is symmetric only if $g$ is a multiple of $f$.

Let $\Psi$ be as in Lemma~\ref{mm:lem:Psi}, and let $\Psi'(v)$ be the component of $\Psi(v)$ of degree $2$ in $n_1$ and $1$ in $n_2$. Since $\ell_iW_i$ has these degrees and $m_iU_i$ has degrees $(1,2)$, we get $\ell_iW_i=\Psi'(\pi_i)$ for all $i$. Let $P^+,P^-$ be the products at a double point, with $w^\pm=(x;\pm y)$, and pure parts $\pi^-=\rho\pi^+$. Then $\Psi'(\pi^+)$ is divisible by $w^+\cdot n_1$ and by $w^-\cdot n_1$, which are not proportional as $y\neq0$. If $\Psi'(\pi^+)\neq0$, then $\Psi'(\pi^+)=(w^+\cdot n_1)(w^-\cdot n_1)R(n_2)$ for a nonzero linear form $R$, so $W^+=(w^-\cdot n_1)R(n_2)$ and $W^-=\rho\,(w^+\cdot n_1)R(n_2)$ lie in $\Pi$ and are symmetric. This forces $R$ to be a multiple of both $w^-\cdot n_2$ and $w^+\cdot n_2$, so $R=0$, a contradiction. Hence $\Psi'$ vanishes at the three double points.

The pure parts of the three double points span $V$: otherwise the six products at the double points have pure parts in a plane and satisfy $6-2=4=\dim K$ independent relations. So $K$ consists of relations not involving the single product, and by Lemma~\ref{mm:lem:relations}(b) the single product has coefficient $0$ in every entry of $AB$. Then $AB$ would be computed with six products, which contradicts \cite{winograd1971multiplication}. So $\Psi'=0$, and $\ell_iW_i=0$, that is, $W_i=0$ for all $i$. The same argument for the component of degrees $(1,2)$ gives $U_i=0$.
\end{proof}

\begin{proposition}\label{mm:prop:nestR}
For $R_I$, $I\neq0$, and for its transpose, the map $\mathcal N$ of \eqref{mm:eq:Nmap} is injective.
\end{proposition}

\begin{proof}
For the transpose this follows from the statement for $R_I$ by Lemma~\ref{mm:lem:sym}(b). The map $\mathcal N$ is linear in the $84$ coefficients of the $U_i,W_i$ in the bases $\{P_j:\ j\neq i\}$, and its matrix has entries in $\K[I,I^{-1}]$. After multiplying by $I$, an exact computation (Section~\ref{mm:sec:comp}) finds $84$ rows of this matrix whose $84\times84$ minor is $-I^{110}$, which is nonzero for $I\neq0$.
\end{proof}

\begin{proposition}\label{mm:prop:nest0}
For every seven-product algorithm with $p=0$, the map $\mathcal N$ of \eqref{mm:eq:Nmap} is injective.
\end{proposition}

\begin{proof}
If $p=0$, every product is a product of a form on $\cA$ and a form on $\cB$, so the algorithm is an optimal bilinear algorithm. As in the proof of Proposition~\ref{mm:prop:pA0}, we use de Groote's theorem. The algorithm is obtained from one of the six orientations of Strassen's algorithm by a substitution as in Lemma~\ref{mm:lem:sym}(a), rescaling and reordering. For each of the six orientations an exact computation (Section~\ref{mm:sec:comp}) shows that $\mathcal N$ is injective.
\end{proof}

\begin{proposition}\label{mm:prop:nestM}
For the algorithm $\mathcal M$ of Theorem~\ref{mm:thm:p12}, the map $\mathcal N$ of \eqref{mm:eq:Nmap} is injective.
\end{proposition}

\begin{proof}
The matrix of $\mathcal N$ in the $84$ coefficients of the $U_i,W_i$ has rational entries, and an exact computation (Section~\ref{mm:sec:comp}) shows that it has rank $84$.
\end{proof}

\begin{proof}[Proof of Theorem~\ref{mm:thm:nest}]
By Lemma~\ref{mm:lem:deg3} the forms $U_i,W_i$ satisfy \eqref{mm:eq:deg3} and lie in $\Pi_{\hat\imath}$. If $p=3$, the quadratic part is, up to the operations above, as in Theorem~\ref{mm:thm:main} or Theorem~\ref{mm:thm:family} or its transpose (Corollary~\ref{mm:cor:class}), and Propositions~\ref{mm:prop:nest3} and~\ref{mm:prop:nestR} give $U=W=0$. If $p=2$, the quadratic part is $\mathcal M$ up to these operations (Theorem~\ref{mm:thm:p12}(c)), and Proposition~\ref{mm:prop:nestM} gives $U=W=0$. If $p=0$, Proposition~\ref{mm:prop:nest0} gives $U=W=0$. By Theorem~\ref{mm:thm:p12}(a) and Lemma~\ref{mm:lem:relations}(a) these are all cases. Lemma~\ref{mm:lem:deg3} completes the proof of the main statement; (a) is Corollary~\ref{mm:cor:class}, (b) is Theorem~\ref{mm:thm:p12}(c), and (d) is Theorem~\ref{mm:thm:p12}(a). For (c), each $P_i=\ell_im_i$ is the product of a form on $\cA$ and a form on $\cB$. So after exchanging the inputs of $q_i$ the circuit is bi-affine, and the description of the quadratic part is de Groote's theorem, as in the proof of Proposition~\ref{mm:prop:nest0}.
\end{proof}

\subsection{Computations and examples}\label{mm:sec:comp}

The proofs of Theorems~\ref{mm:thm:main}, \ref{mm:thm:family} and~\ref{mm:thm:p12} use no computer calculation. The identity \eqref{mm:eq:pB1rest}, the formulas of Theorem~\ref{mm:thm:p12}(b) and the substitution rules in Step~6 of the proof of Proposition~\ref{mm:prop:p11} are direct expansions. The proofs of Propositions~\ref{mm:prop:pA0}, \ref{mm:prop:nestR}, \ref{mm:prop:nest0} and~\ref{mm:prop:nestM} use exact linear algebra over $\Q$ and over $\Q[I]$, carried out in SymPy. The scripts are provided as supplementary material. They check the following.
\begin{itemize}
    \item For Strassen's decomposition in each of the six orientations: it is an algorithm, the linear system $\sum_i\Lambda_i\beta_i\delta_i=0$ in the $28$ unknown coefficients of $\delta_1,\dots,\delta_7$ has only the zero solution (Section~\ref{mm:sec:pA0}), and the map $\mathcal N$ of \eqref{mm:eq:Nmap} is injective (Proposition~\ref{mm:prop:nest0}).
    \item For $R_I$ with $I$ a variable: the matrix of $\mathcal N$, multiplied by $I$, has an $84\times84$ minor equal to $-I^{110}$ (Proposition~\ref{mm:prop:nestR}).
    \item Each of the three factor lists of Strassen's decomposition consists of pairwise non-proportional elements (Lemma~\ref{mm:lem:strassen}). For $\mathcal M$: it is an algorithm with the formulas of Theorem~\ref{mm:thm:p12}(b), $(p_A,p_B,p)=(1,1,2)$, and the matrix of $\mathcal N$ has rank $84$ (Proposition~\ref{mm:prop:nestM}). For the family $\mathcal F$ of the proof of Proposition~\ref{mm:prop:p11} with symbolic parameters: it is an algorithm, the three substitution rules of Step~6, and $\mathcal F(0,1,-1,1,1,-1,1)=\mathcal M$. As an independent check of Step~5, we ran a Gr\"obner basis computation in Singular. It shows that the equations (E1) to (E3), saturated with respect to the determinant expressing (N), define an irreducible variety of dimension $7$. This variety is the family $\mathcal F$.
    \item For the family $R(y_1,y_2)$ with symbolic $y_{jl}$: it is an algorithm, the greatest common divisor of the maximal minors of the coefficient matrix of its products is $y_{11}y_{12}y_{21}y_{22}$, and $(p_A,p_B,p)=(2,1,3)$. The identity \eqref{mm:eq:pB1rest}. The substitutions in the proofs of Theorem~\ref{mm:thm:family}(b) and (c); and the formulas of Theorem~\ref{mm:thm:family}(a).
    \item The formulas \eqref{mm:eq:mu} and those of Lemmas~\ref{mm:lem:dpN} and~\ref{mm:lem:split}, the determinant formula in the proof of Lemma~\ref{mm:lem:singnets}, the factorization in Lemma~\ref{mm:lem:rankone}, and the solutions of $(2R)^T\sigma(2R)=\sigma$ in the proof of Lemma~\ref{mm:lem:conic}.
    \item The examples below, including the injectivity of $\mathcal N$ for them, which illustrates Proposition~\ref{mm:prop:nest3}.
\end{itemize}

\emph{Example with $(p_A,p_B)=(3,3)$.} With $x_1=e_1$, $x_2=e_2$, $x_3=e_1+e_2$, $x_4=e_1-e_2$ and $y_1=5e_1$, $y_2=5e_2$, $y_3=(7,1)^T$, $y_4=(1,-7)^T$, the seven products
\begin{gather*}
    (a_{11}\pm5b_{21})(a_{12}\pm5b_{11}),\qquad(a_{21}\pm5b_{22})(a_{22}\pm5b_{12}),\\
    \bigl(a_{11}+a_{21}\pm(7b_{21}+b_{22})\bigr)\bigl(a_{12}+a_{22}\pm(7b_{11}+b_{12})\bigr),\\
    (a_{11}-a_{21}+b_{21}-7b_{22})(a_{12}-a_{22}+b_{11}-7b_{12})
\end{gather*}
(signs taken together within each product) are $(x^TAe_1\pm e_2^TBy)(x^TAe_2\pm e_1^TBy)$ for $(x,y)=(x_k,y_k)$, $k\le3$, and $(x_4,y_4)$. They are linearly independent, span all four entries of $AB$, and have $p_A=p_B=p=3$. The choice satisfies $x_4^2=2x_1^2+2x_2^2-x_3^2$ and $y_4^2=2y_1^2+2y_2^2-y_3^2$ in $\Sy^2\K^2$, which is the relation \eqref{mm:eq:middle} with a common middle pair.

\emph{Examples with $(p_A,p_B)=(3,2)$ and $(2,2)$.} Write $P(x,y)=(x^TAe_1+e_2^TBy)(x^TAe_2+e_1^TBy)$. The seven products
\begin{displaymath}
    P(e_1,\pm2e_2),\quad P(e_2,\pm e_2),\quad P(e_1+e_2,\pm e_1),\quad P(e_1+4e_2,2e_1)
\end{displaymath}
form an algorithm with $(p_A,p_B,p)=(3,2,3)$. The A-parts satisfy $4x_3^2-x_4^2-3x_1^2+12x_2^2=0$ for $x_1=e_1$, $x_2=e_2$, $x_3=e_1+e_2$, $x_4=e_1+4e_2$, and the B-parts the matching relation $-3(2e_2)^2+12e_2^2=0=4e_1^2-(2e_1)^2$. The seven products
\begin{displaymath}
    P(e_1,\pm e_1),\quad P(e_1,\pm e_2),\quad P(e_2,\pm e_1),\quad P(e_2,e_2)
\end{displaymath}
form an algorithm with $(p_A,p_B,p)=(2,2,3)$.

\begin{acks}
This project has received funding from the European Research Council (ERC) under the European Union's Horizon 2020 research and innovation programme (grant agreement No 818252, 101113120, 101138056). This project has received funding from the European Research Council under the European Union's Horizon 2020 research and innovation programme (grant agreement No 101113120, 101138056). Research was supported by grant No.~1354/23 of the Israel Science Foundation (founded by the Israel Academy of Sciences and Humanities). This research has been supported by the Science Accelerator and by the Frontiers in Science initiative of the Ministry of Innovation, Science and Technology.
\end{acks}

\interlinepenalty=10000\bibliographystyle{alpha}\enlargethispage{2\baselineskip}
\bibliography{refs}

\section*{AI methodology statement}
\addcontentsline{toc}{section}{AI methodology statement}
An AI assistant (Claude, by Anthropic) was used in preparing this paper: for checking the proofs, for thinking of applications, and for textual editing.

\end{document}